\documentclass[a4paper,12pt,times,twoside,index,custommargin,custombib]{PhDThesisPSnPDF}
\input{preamble}

\title{On the Hydrodynamic Approximation of Quantum Integrable Models}

\subtitle{An Illustration via the repulsive Lieb-Liniger Model}

\author{Friedrich H\"ubner}

\dept{Department of Mathematics}

\university{King's College London}
\crest{\includegraphics[width=0.4\textwidth]{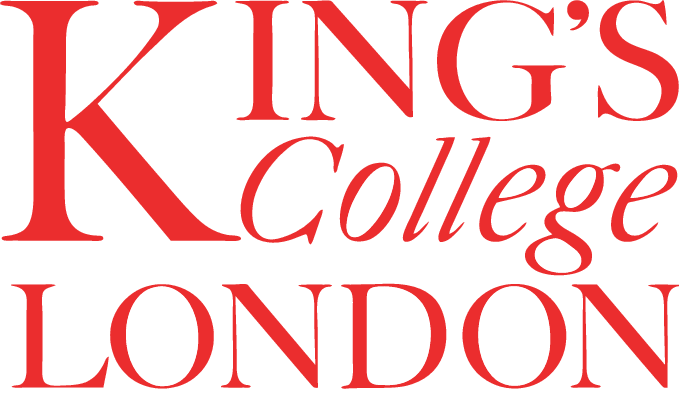}}
\supervisor{Prof.\ Benjamin Doyon\newline Prof.\ Yan Fyodorov}

\supervisorlinewidth{0.4\textwidth}

\degreetitle{Doctor of Philosophy}

\subject{LaTeX} \keywords{{LaTeX} {PhD Thesis} {Mathematics} {King's College London}}

\ifdefineAbstract
\fi

\ifdefineChapter
\fi

\begin{document}

\renewcommand{\hbar}{\mathchar'26\mkern-7mu h}

\mainmatter

\maketitle

%\include{dedication}
% ******************************* Thesis Declaration ***************************

\begin{declaration}

I hereby declare that except where specific reference is made to the work of 
others, the contents of this dissertation are original and have not been 
submitted in whole or in part for consideration for any other degree or 
qualification in this, or any other university. This dissertation is my own 
work and contains nothing which is the outcome of work done in collaboration 
with others, except as specified in the text and Acknowledgements. This 
dissertation contains fewer than 100,000 words including appendices, 
bibliography, footnotes, tables and equations and has fewer than 150 figures.

% Author and date will be inserted automatically from thesis.tex \author \degreedate

\end{declaration}

% ************************** Thesis Acknowledgements **************************

\begin{acknowledgements}

First and foremost, I would like to thank my supervisor Benjamin for the four years full of interesting scientific discussions and the exciting new developments that started from them. I will also always remember the crucial support I received from him during important steps towards becoming an independent researcher, such as publishing my first solo author paper, presenting my first talks, being invited to discussions/conferences and for all the help with designing/writing/preparing postdoc fellowship applications.

Next, let me thank my other collaborators Takato, Lorenzo, Eric, Leonardo, Jacopo, Juan and Sun Woo. Not only is research always more enjoyable in a group, but it also has been a real pleasure with all of you. Apart from the scientific side, thank you also for the useful advice you gave me and all the fun we had outside of science. More generally, it is a great experience to be part of the communities working on integrability, statistical physics and mathematical physics. All of them were exceptionally welcoming and open, especially towards PhD students. There are two special shoutouts: one to Olalla, who -- in a profession where mental health is often overlooked -- is always watching out for the well-being of others. And another one to Alvise, a brilliant scientist who gave me loads of useful advice preparing me for my postdoc years (and thanks to whom I applied to post doc fellowships). 

Speaking of the scientific community, I find it important to acknowledge the time and effort researchers spend into shaping the research environment. This includes being editors and reviewers for journals (special shoutout here to J.S. who alongside his research also runs the two-way open access journals Scipost), running in person or online seminars, and also as organizers of schools, workshops and conferences. Throughout my PhD such events allowed me to have many exciting journeys all over Europe. I especially remember the winter schools at GGI in Florence, the mathematical physics schools in L’Aquila, the soliton gas-generalized hydrodynamics conference 2023 in Marseille, the ICFT 2024 at City University London and the recent workshop at SwissMap in Les Diablerets. Events like these were important for me to meet other researchers, but also to establish myself as an independent researcher. In that regard I am particularly thankful that many of these events took special care to give space to young researchers; decisions from which I greatly profited. Based on these experiences I decided to join the organization committee of the Student Workshop on Integrability (a conference for PhD students organized by PhD students). I think many researchers of my generation on integrability owe a lot to Sascha who organized this event since 2022, first alone(!) and then in 2023 with Rebekka. In 2024 and 2025 Anastasiia and I joined him (also thanks to Balász for dealing with the local organization in Budapest). I would have never imagined how much fun it would be to organize with the two of you: we were such a good team, always enjoying ourselves no matter how many unexpected problems came along and eventually solving all of them.

Let me also thank the staff in the department office, especially Sandra, Gage and Shayon. While you are not involved in the science part, without your support, all the scientific advances and the publications of the department would not be possible. The same holds for all the other staff at KCL who make things work in the background.

Since a four-year PhD obviously cannot only evolve around science, I was super lucky to find so many good friends along the way, either in London or on the various scientific journeys. Cheers to our fun times, of which I hope will be many more in the future! In the context of the thesis, let me especially thank Noemi, Sophie, Sun Woo, Lee, Gabriele, Fabio, Adarsh, Harmeet, David, Oskar, Leonardo and Jane for giving comments on my thesis. I am grateful to Thekla, for all the love and for being there when I need it, and especially for the ongoing practical help and support. This PhD would have not been possible without you! Lastly, let me thank my family, for always providing me a home. In particular, shoutout to my grandfather Jochen, who got me interested in physics all those years ago. Look what you made me do!

I would like to finish these acknowledgments with some additional comments

\begin{itemize}
	\item I would like to thank my examiners Andreas Fring and Bruno Bertini whose thorough feedback meaningfully helped to improve the thesis.
	\item In addition to the PhD funding from the KCL Department of Mathematics, I am also grateful to have received the KCL Department of Mathematics Doris Chen scholarship 2021/2022. 
	\item Some special characters used in this thesis were designed by Sun Woo.
	\item In case you happen to be looking for a drink in Florence, I recommend you have a Campari Spritz at the Bar d’Angolo (meaning corner bar) next to Porta Romana. 
	\item I was shattered after hearing that Wigan Athletics sacked Shaun Maloney, one of their most talented managers. Please reinstate him immediately. I am vowing to boycott watching any more Wigan Athletics games until he is back.
\end{itemize}

\end{acknowledgements}

%!TEX root = thesis.tex

% ************************** Thesis Abstract *****************************
% Use `abstract' as an option in the document class to print only the titlepage and the abstract.
\begin{abstract}
	Generalized hydrodynamics is a framework to study the large scale dynamics of integrable models, special fine-tuned one-dimensional many-body systems that possess an infinite number of local conserved quantities. Unlike classical models, where the microscopic origins of generalized hydrodynamics are better understood, in quantum models it can only be derived using the hydrodynamic formalism. Using the paradigmatic and experimentally relevant repulsive Lieb-Liniger model as an example, this thesis introduces a new viewpoint on the dynamics of quantum integrable models by introducing so-called semi-classical Bethe models. These classical integrable models act as an intermediate description between the microscopic quantum realm and the macroscopic generalized hydrodynamics. After introducing these models and discussing their properties, we study the generalized hydrodynamics equation using new tools and show that solutions to the Euler generalized hydrodynamics equation of the Lieb-Liniger model exist, are unique and do not develop gradient catastrophes. Finally, we discuss new insights into the physics governing the diffusive correction, which, contrary to prior belief, is not described by a Navier-Stokes-like equation. Focusing on the main intuitive ideas, the thesis aims to provide a self-contained overview over these exciting new developments on generalized hydrodynamics.
\end{abstract}

% *********************** Adding TOC and List of Figures ***********************

\tableofcontents

%\listoffigures

%\listoftables

% \printnomenclature[space] space can be set as 2em between symbol and description
%\printnomenclature[3em]

%\printnomenclature

% ******************************** Main Matter *********************************

%!TEX root = thesis.tex

\chapter{Introduction}

Understanding physical systems and developing mathematical descriptions allowing to predict their evolution or properties is at the heart of theoretical physics. Unless one is working with extremely high energetic particles, the fundamental laws of physics are extremely well understood and tested. Unfortunately, theoretical physics is still far from being able to predict, say, the conductivity of a piece of metal or the friction exerted on a car by air the flowing around it, based only on those fundamental laws. The reason is that many-body systems simply have too many degrees of freedom to be able to capture them analytically or numerically. However, if one additionally makes very intuitive statistical assumptions (in a nutshell: average over all microscopic degrees of freedom which cannot be observed macroscopically), one is able to derive (universal) theories like thermodynamics~\cite{Ansermet_Brechet_2019}, fluid dynamics/hydrodynamics~\cite{landau2013fluid}, Boltzmann equations~\cite{kremer2010introduction}, stochastic thermodynamics~\cite{peliti2021stochastic}, etc. These theories (called emergent theories) are, all of sudden, able to give meaningful predictions for real world systems and are the basis for further refined descriptions used in applications.

This means that in our theoretical understanding of the world there is a gap: what happens when we make these statistical assumptions? Are they justified? What are their limitations? The universality of emergent theories suggests that the principles behind their emergence should not depend on the specific model. Hence, from the perspective of theoretical physics (in particular statistical physics), it makes sense to first try to study these problems in simplified toy models. Unfortunately, even this is extremely hard. As a starting point, over the past decades, there had been incredible progress in systems that are explicitly noisy, leading to descriptions like macroscopic fluctuation theory~\cite{RevModPhys.87.593} and KPZ (Kardar–Parisi–Zhang)-like equations~\cite{Spohn2014}. While this might be a good model for a system strongly coupled to a bath (e.g. for thin materials), it is unclear how well it describes the bulk of a material. In a body of water, for instance, the noise observed by a water molecule comes from the interaction with other water molecules moving (seemingly) at random. On a fundamental level, such closed many-body systems are fully deterministic and time-reversible. In particular, they are entropy conserving\footnote{Entropy is a nuanced concept, with many (inequivalent) definitions. For this discussion we will leave it abstract, focusing on its intuitive meaning.}. The second law of thermodynamics, i.e.\ that entropy can never decrease (and in fact will always increase unless the system is in thermal equilibrium), seems to be the fundamental gap. If it is assumed, it leads to thermalization and thus justifies emergent theories. If not, then entropy is fundamentally constant. Systems with noise satisfy the second law by construction: any random jump of the system increases it. One might think that therefore noise from a bath is crucial and that substances like water appear thermal because they are always somehow coupled to a bath, for instance through their boundary, through impurities or through radiation penetrating them. However, it is now understood that also closed (and fully clean) systems thermalize on their own, at least in the sense that local observables appear thermal for a (very) long time~\cite{Gogolin_2016, PhysRevA.89.053608,Calabrese_2016,Essler_2016,Cazalilla_2016,Caux_2016,Vidmar_2016,Eisert2015,DAlessio03052016,Mori_2018}. As information cannot be lost, it must still be hidden in complicated long range correlations throughout the system. Hence, also the noise that a water molecule feels will be extremely complicated and space-time correlated. If one introduces explicit noise into systems, it destroys these correlations and leads to unphysical uncorrelated white noise.

Hence, we would like to study non-noisy interacting systems. However, those are significantly more complex compared to noisy systems and thus still very little is known about them. The problems start already at a very basic level: we are not able to compute the theoretical thermal expectation values we would need to check thermalization. To our knowledge there are only 3 exceptions to this: a) classical 1D\footnote{By 1D we mean 1+1D, i.e.\ one space and one time dimension.} systems with local interactions (like anharmonic chains~\cite[Sec 2]{Spohn2014}), b) integrable models and c) more recently certain circuit systems~\cite{Klobas_2022,PhysRevE.102.062107,Klobas2019,Prosen_2017,kim2025circuitssimpleplatformemergence,sharipov2025ergodicbehaviorsreversible3state}. The systems of a) have been very influential and led, for instance, to the development of non-linear fluctuation hydrodynamics~\cite{Spohn2014} (establishing that 1D systems should be KPZ-superdiffusive rather than diffusive). However, we lack tools to understand their microscopic evolution.

This brings us to integrable systems~\cite{arutyunov2020elements,Korepin_Bogoliubov_Izergin_1993}. Those are a collection of (mostly) 1D systems which are exactly solvable\footnote{In mathematics and physics there are many systems/equations called integrable or exactly solvable. Here, we mean integrability in the many-body physics context. More precisely, we will consider models that have an infinite number of conserved quantities with local densities.}. Such exactly solvable models have been found in all major fields of physics and have always been highly influential as starting points to gain intuition, compute exact results and to develop approximations. For instance, they have been used to establish the thermalization-of-closed-systems picture above (see e.g.~\cite{PhysRevLett.98.050405,Calabrese_2016,Essler_2016,Cazalilla_2016,Caux_2016,Vidmar_2016,Ilievski_2016}). The meaning of ``exactly solvable'' varies across fields (and from model to model). Apart from few models, such as classical hard rods or the rule 54 cellular automata~\cite{Buca_2021,Klobas2019}, there is no explicit formula for their time-dynamics. In quantum models, one is typically able to diagonalize the Hamiltonian, i.e. to write down the exact eigenstates of the system and to compute thermal states, but not the evolution\footnote{Of course, once all eigenstates are known, one can formally write the evolution in terms of a sum over all eigenstates. But in practice, this requires to compute the overlaps of the eigenstates with the initial state, which are not known, except for special cases, e.g.~\cite{PhysRevA.89.033601,PhysRevLett.113.117203,PhysRevLett.113.117202}.}\,\footnote{There are some models where exact computations are possible: examples include quantum versions of cellular automata like rule 54~\cite{PhysRevLett.126.160602,10.21468/SciPostPhys.11.6.107} or the folded XXZ chain~\cite{10.21468/SciPostPhysCore.4.2.010,10.21468/SciPostPhys.10.5.099,PhysRevE.104.044106,fujimoto2024quantumtransportinteractingspin}.}. Generally, however, it is understood that integrability is connected to the existence of an infinite number of conservation laws (with local densities), and that this drastic constraint on the evolution is responsible for the ``exact solvability''. Since conservation laws play a major role in many-body physics, various formalisms have to be adapted. This does not mean, however, that the physics of integrable models is not relevant. In fact, in many 1D non-integrable systems are close to integrable systems (see the famous Fermi–Pasta–Ulam–Tsingou problem problem~\cite{10.1063/1.1855036}) and thus the additional conserved quantities are often almost conserved (in particular for short times and at low densities).

The central theme of this thesis is generalized hydrodynamics (GHD), a theory generalizing hydrodynamics to integrable models~\cite{BenGHD,PhysRevX.15.010501,ESSLER2023127572,Bouchoule_2022,Kerr2023,Bastianello_2022,RevModPhys.93.025003,Alba_2021,El_2021,doi:10.1142/13600}. In its present form it was first established in 2016~\cite{PhysRevX.6.041065,PhysRevLett.117.207201} on quantum systems, but it had already been established independently in hard rods in~\cite{10.1063/1.1692711,Boldrighini1983} and in integrable PDEs in~\cite{EL2003374}. Hydrodynamics aims to describe the large scale evolution of charges, which are believed to be the quantities that take longest to thermalize. As such, it is an extremely relevant theory in any sufficiently large many-body system out-of-equilibrium. Compared to the microscopic dynamics, hydrodynamic equations are an immense simplification that can still describe a rich plethora of complex behavior\footnote{See, for instance, the \href{https://gfm.aps.org/}{Gallery of Fluid Motion} by the American Physical Society.} (like shocks~\cite{10.1093/oso/9780198507000.001.0001,Huicheng_2022,doi:10.1142/S0219891623500261,10.1063/5.0244792} and turbulence~\cite{Frisch_1995,10.1093/acprof:oso/9780199689385.001.0001,ZHOU20101,ZHOU20211,Galtier_2022}). In particular, the hydrodynamic theories for usual systems (Euler and Navier-Stokes equation for fluids, Fick's laws for diffusion, Fourier law for heat transport, etc.) are successfully applied throughout physics and beyond (like chemistry and engineering, see e.g.~\cite{elger2020engineering,raju2011fluid,incropera2011fundamentals}). However, we still understand very little about their emergence and their limitations.

\begin{figure}[!h]
	\centering
	\includegraphics[scale=0.5]{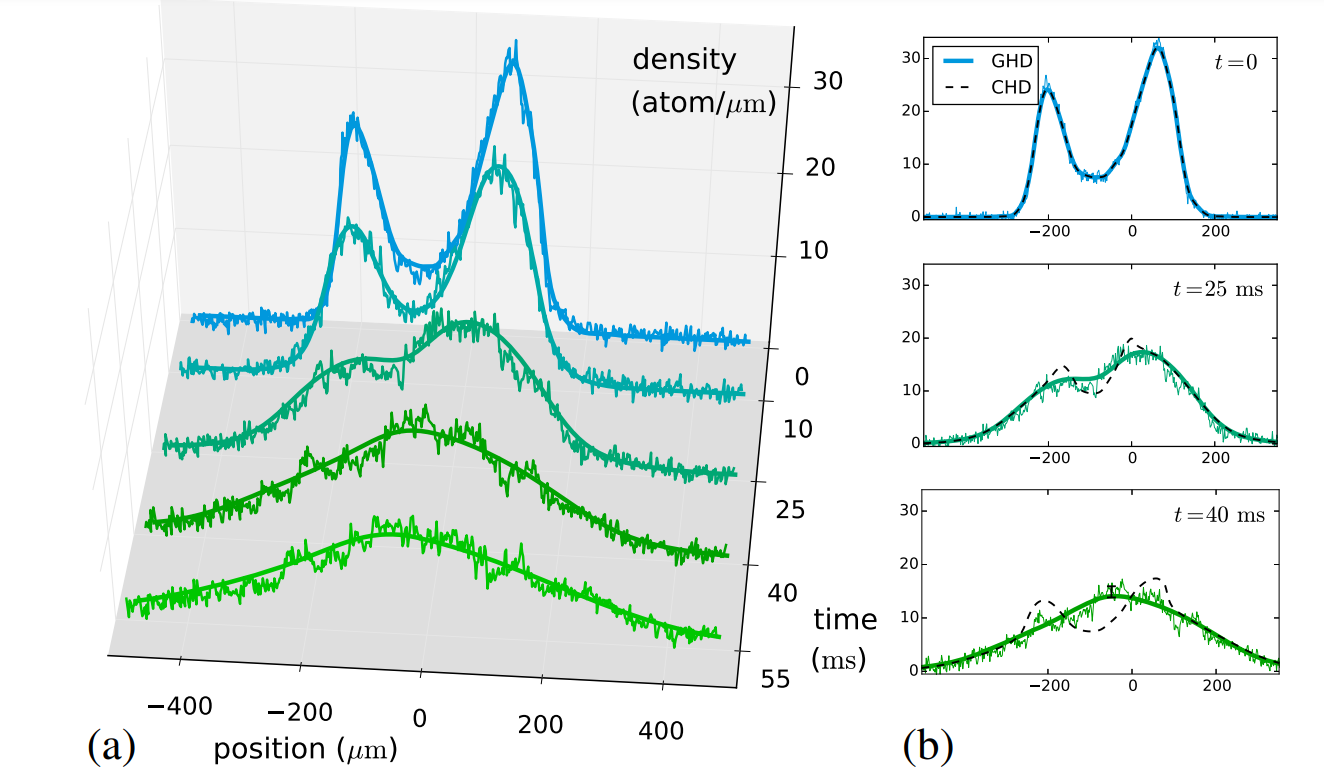}
	\caption[Experimental verification of GHD]{Experimental verification of GHD in cold bosonic atoms (described by the Lieb-Liniger model): a cloud of ultracold atoms is trapped in a quasi-one dimensional trap and initialized in a non-equilibrium state at $t=0$. a) The evolution of the distribution of the particle density is observed (noisy line) and compared to GHD (solid line). b) GHD is a much better description compared to conventional hydrodynamics (CHD), which even differs qualitatively. This shows that in this experiment, integrability is important. This figure was taken from~\cite{PhysRevLett.122.090601}, with the permission of the authors.}
	\label{fig:intro_GHD}
\end{figure}

The aim of the research of this PhD was to understand how and why hydrodynamics emerges in integrable systems, in the hope that this will pave the way to also understand hydrodynamics of non-integrable systems. Specifically, we tried to gain better understanding of quantum models. While their hydrodynamic equations are similar to classical models, the underlying associated loss of information (or entropy increase) is due to dephasing; a process which does not exists in classical models. One of the simplest quantum integrable models is the (repulsive) Lieb-Liniger model~\cite{PhysRev.130.1605,jscaux}, which will be our reference model throughout this thesis. As a model of cold bosonic atoms, it is also an experimentally relevant integrable model~\cite{Bouchoule_2022}. In particular, GHD has been verified in cold atom experiments~\cite{PhysRevLett.122.090601,Bouchoule_2022,doi:10.1126/science.abf0147}, see \cref{fig:intro_GHD}.

The purpose of this thesis is not to give an exhaustive overview of integrable models, but rather to discuss new developments in simple settings. The work presented here is mostly based on simple ideas and a lot of intuition, which led to new insights into and new tools to study integrable models. We hope we can bring across how a cascade of simple ideas, intertwined with the occasional technical computation, leads to variety of interesting results. After a swift but self-contained introduction to hydrodynamics, integrable models and GHD, we describe (attempts to) an \textit{alternative derivation of the GHD} in Lieb-Liniger in \cref{sec:LL} via an emergent classical integrable model, called \scbm s. Although some understanding is still missing, this new derivation already gives interesting deeper insights into the dynamics of quantum integrable models. In \cref{sec:scbm} we then analyze \textit{\scbm s} as standalone classical models, deriving their thermodynamics and their GHD. These models also allow for the first time to construct classical integrable models with arbitrary scattering shifts\footnote{So far all other known integrable models had fairly restricted scattering shifts.}. In \cref{sec:fixedpoint} we derive a ``\textit{space-time quadrature}'' of the (Euler) GHD equation. This powerful object not only gives rise to an efficient numerical algorithm, but also gives a mathematical handle: on the example of the Lieb-Liniger model we show that solutions to its GHD equation always exist, are unique and that they do not form discontinuities (shocks). The final \cref{sec:diff} studies the diffusive correction to GHD. We establish in hard rods that Euler GHD is accurate beyond the diffusive correction on each individual configuration, i.e. the diffusive correction vanishes (in some intrinsic sense). Assuming that this is true in any integrable model, by averaging over the intial state we obtain a non-trivial \textit{diffusive correction to GHD}. This agrees with the usual Kubo-diffusion formula result only if the state is locally in thermal equilibrium. Over time, however, long range correlations develop in the system, leading to a failure of Kubo-diffusion (or at least some naive application of it). In fact, the diffusive correction is invariant under time-reversal (implying that it is entropy conserving), showcasing that in integrable models the intrinsic noise is indeed deterministic (and not true random noise). Since this implies that diffusion cannot be the cause of thermalization, we discuss an alternative mode of thermalization based on coarse-graining.

%!TEX root = thesis.tex

\chapter{Prerequisites}
\label{sec:pre}
%\ifpdf
%\graphicspath{{Chapter1/Figs/Raster/}{Chapter1/Figs/PDF/}{Chapter1/Figs/}}
%\else
%\graphicspath{{Chapter1/Figs/Vector/}{Chapter1/Figs/}}
%\fi

\section{The hydrodynamic approximation}
\label{sec:pre_hydro}
The hydrodynamic approximation is a way to approximate the large-scale dynamics of a physical system using simplified equations. Concretely, it studies the evolution of the distribution of conserved quantities in the system. In the following we consider a 1D physical system (classical or quantum) with \underline{local} interactions that is translation invariant (hence no boundary or any external potential). For simplicity of notation, we will carry out the phenomenological derivation for a quantum system\footnote{It can easily be adapted to a classical system by considering operators like $\vu{Q}_n$ as functions on phase-space instead.}. The phenomenological derivations in this section will closely follow~\cite{BenGHD}.

Assume that this system has extensive conserved quantities (or charges) $\vu{Q}_n$. Extensive here means that we can write $\vu{Q}_n=\int\dd{x}\vu{q}_n(x)$ with (quasi-)local densities $\vu{q}_n(x)$ (see \cref{rem:pre_hydro_quasilocal}). In a typical physical Galilean invariant many-body system (such as water) there are three such extensive conserved quantities: particle number, momentum and energy.

Hydrodynamics emerges in the Euler scaling limit $N\sim L \sim T\to \infty$, where $N$ is the number of particles\footnote{If there is no meaningful particle number in the system, then $N \sim \expval{\vu{Q}_n}$ measures the amount of charge in the system} and $L$ and $T$ are the length and time scales of observation (which are large compared to the microscopic scales). This limit is chosen in a way that densities $\expval{\vu{q}_n(x)}\sim N/L$ and velocities (of sound waves) $v\sim L/T$ are $\order{1}$. This is a very natural regime for many real-world scenarios (see also  \cref{rem:pre_hydro_phys_correct,rem:pre_hydro_math_correct}). For instance, if we observe a river, then a) we observe a large amount of particles $N\sim 10^{23}$, b) our space (meters) and time scale of observations (seconds) are much larger than the microscopic scales (a water molecule is on the order of nanometers) and c) on our observation scales, we observe a finite density and finite velocity of water.

\begin{remark}\label{rem:pre_hydro_quasilocal}
	It is crucial that the conserved quantities have local densities. For instance, any quantum spin $\tfrac{1}{2}$ model defined on $L$ sites formally has $2^L$ conserved quantities (projectors on the eigenvectors of the Hamiltonians). However, most of them do not have a local densities, hence are irrelevant for hydrodynamics. To be relevant for hydrodynamics, the densities need either to be supported on a finite region (local), or decay sufficiently fast (quasi-local)~\cite{Ilievski_2016,Doyon2022,Ampelogiannis2024}. It is not well established precisely how fast they must decay; however there exists a recently developed abstract definition~\cite{Doyon2022,Ampelogiannis2024}. 
\end{remark}

\subsection{Euler hydrodynamics}
\label{sec:pre_hydro_euler}
In the Euler scaling limit $N\sim L \sim T\to \infty$, upon discarding any subleading terms, we obtain Euler hydrodynamics: the cornerstone of any hydrodynamic theory. 

\begin{figure}[!h]
	\centering
	\includegraphics{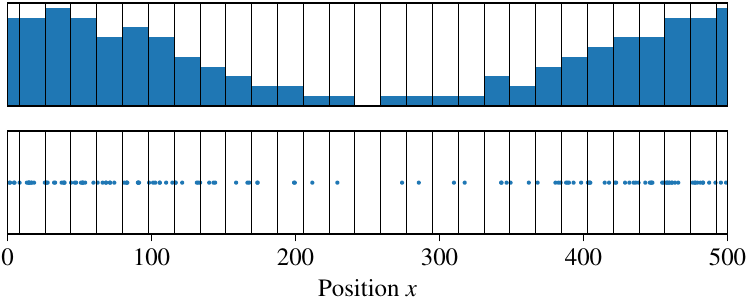}
	\caption[Schematic of coarse-graining]{Schematic of coarse-graining: a large system of size $L=500$ is divided into many fluid cells (vertical lines), which are still sufficiently big to contain many particles (bottom). Since we cannot determine the configuration inside the fluid cell, it is natural to treat it as a maximum entropy state. Averaging over the fluid cells we obtain a macrosocopic density distribution (top), which approaches a smooth distribution as $L\to \infty$.}
	\label{fig:prerequisites_LES}
\end{figure}

The first ingredient into hydrodynamics is thermalization: if we start the evolution from an arbitrary state, we expect the system to thermalize\footnote{Thermalization in the context of this thesis will mean relaxation towards a maximum entropy state (GGE). Only if the only conserved quantities are particle number, momentum and energy, then maximum entropy states are thermal states in the usual sense. Otherwise, GGE states can be non-thermal.} towards a maximum entropy state $\sim e^{-\sum_n \beta_n \vu{Q}_n}$, also called generalized Gibbs ensemble (GGE). Here the $\beta_n$ are generalized inverse temperatures that act as Lagrange parameters fixing the maximum entropy state. But since the interactions are only local and our system is very large, thermalization can only happen in a local region (growing in time). Imagine dividing space into mesoscopically sized cells of size $1 \ll \ell \ll L$, called fluid cells, which are large compared to the microscopics, but small compared to our observation length scale (see \cref{fig:prerequisites_LES}). Since we are observing the system on long time scales, each of these fluid cells will be thermalized internally. However, different fluid cells will thermalize to different GGE states, as they will have not interacted yet. Therefore, a natural guess for the state in hydrodynamics is a local equilibrium state
\begin{align}
	\vu{\rho}\ind{LES} \sim e^{-\sum_\alpha \sum_n \beta_{n,\alpha} \vu{Q}_{n,\alpha}},\label{equ:pre_hydro_LES_CG}
\end{align}
where $\alpha\in\mathbb{Z}$ labels the fluid cell, $\beta_{n,\alpha}$ determine the GGE state in fluid cell $\alpha$ and $\vu{Q}_{n,\alpha} = \int_\alpha\dd{x} \vu{q}_n(x)$ is the total charge in this cell. In the quantum (classical) setting \eqref{equ:pre_hydro_LES_CG} is a density matrix (probability distribution) on the same Hilbert space (phase space) on which the quantum (classical) theory is defined. Furthermore, since thermalization would quickly smoothen out any discontinuities, these generalized inverse temperatures $\beta_{i,\alpha}$ have to slowly vary across the system. Hence, it makes sense to assume that $\beta_{n,\alpha}= \beta_n(x_\alpha/L)$ is described by a large scale smooth function ($x_\alpha = \alpha \ell/L$ is the center of cell $\alpha$). In the thermodynamic limit the $x_\alpha$ become dense, hence we can replace the sum $\tfrac{\ell}{L}\sum_\alpha \to \int\dd{x}$ by an integral. Therefore, in the thermodynamic limit \eqref{equ:pre_hydro_LES_CG} becomes
\begin{align}
	\vu{\rho}\ind{LES} \sim e^{-\int\dd{x} \sum_n \beta_n(x/L) \vu{q}_{n}(x)}.\label{equ:pre_hydro_LES}
\end{align}

The assumption of hydrodynamics is now that the state is described at all times via \eqref{equ:pre_hydro_LES}\footnote{In fact this assumption is too naive, see \cref{sec:LL_why}.}. This is of course an approximation: even if we start from a state like \eqref{equ:pre_hydro_LES}, time evolution will generate a more complicated state. However, since we only study the system at long times, we expect that all additional details will be lost due to thermalization. Hence, at all times we should be able to describe our system using the local generalized temperatures $\beta_n(t,x/L)$.

From now on we will work in macroscopic coordinates $x \to Lx$. To determine hydrodynamics, we need to find an evolution equation for $\beta_n(t,x)$. First, let us note that $\qty{\beta_n(t,x)}$ is in one-to-one correspondence with the set of charges $\qty{\expval{\vu{q}_n(t,x)}}$ in this fluid cell. Since fluid cells are large their expected average charge densities correspond to their thermal expectation values in the thermodynamic limit:
\begin{align}
	q_n(t,x)= \expval{\vu{q}_n(t,x)} = \expval{\vu{q}_n}_{\qty{\beta_n(t,x)}} := \lim_{\ell\to \infty} \tfrac{1}{\ell} \frac{\Tr_\ell \vu{Q}_n e^{-\sum_m\beta_m(t,x) \vu{Q}_m}}{\Tr_\ell e^{-\sum_m\beta_m(t,x) \vu{Q}_m}}.\label{equ:pre_hydro_expval_q}
\end{align}
Here, $\Tr_\ell$ denotes the trace/sum\footnote{In a classical system $\Tr_\ell$ should be interpreted as a phase-space integral.} over a system of size $\ell$ with, say, periodic boundary conditions (the choice of boundary conditions should be irrelevant in the thermodynamic limit).  

\begin{remark}
	To clarify, the LHS of \eqref{equ:pre_hydro_expval_q} is an expectation value of the inhomogeneous state at a fixed point $t,x$. We set it equal to the RHS, the expectation value in an infinite homogeneous state with constant $\beta_n$. This is justified because (compared to the microscopic scale) each fluid cell can be seen as an infinitely big system. These homogeneous averages then vary slowly on an even larger scale described by $\beta_n(t,x)$.
\end{remark}

Equation \eqref{equ:pre_hydro_expval_q} can be viewed as a map $\qty{\beta_n} \mapsto \qty{q_n}$. One can show that this map is one-to-one\footnote{This follows from the convexity of the free energy.}. Hence, in order to understand the evolution of $\beta_n(t,x)$, it is sufficient to understand the evolution of $\expval{\vu{q}_n(t,x)}$. As $\vu{q}_n(t,x)$ is the density of a conserved quantity it satisfies a continuity equation
\begin{align}
	\partial_t\vu{q}_n(t,x) + \partial_x \vu{j}_n(t,x) &= 0,\label{equ:pre_hydro_micro_continuity}
\end{align} 
where $\vu{j}_n$ is the associated current. Let us now average this over \eqref{equ:pre_hydro_LES} to obtain
\begin{align}
	\partial_t q_n(t,x) + \partial_x j_n(t,x) = 0,\label{equ:pre_hydro_macro_continuity}
\end{align}
where $j_n(t,x) = \expval{\vu{j}_n(t,x)} = \expval{\vu{j}_n}_{\qty{\beta_n(t,x)}}$. Note that, similarly to $q_n$, the expectation value $j_n$ is also given as a thermodynamic expectation value in the local GGE. Since both $q_n$ and $j_n$ only depend on $\qty{\beta_n}$, we can write
\begin{align}
	\sum_m \pdv{q_n}{\beta_m} \partial_t \beta_m +  \pdv{j_n}{\beta_m} \partial_x \beta_m &= 0.\label{equ:pre_hydro_beta_eq}
\end{align}
This is an evolution equation for $\beta_n(t,x)$. Alternatively, we can invert the relation between $\qty{\beta_n} \mapsto \qty{q_n}$ to view $j_n = j_n[\qty{q_m}]$ as a function of the averaged charge densities. This gives
\begin{align}
	\partial_t q_n(t,x) + \pdv{j_n}{q_m}\partial_x q_n(t,x) &= 0.\label{equ:pre_hydro_q_eq_A}
\end{align}
Here $ \pdv{j_n}{q_m} = \sum_k  \pdv{j_n}{\beta_k} \pdv{\beta_k}{q_m}$. The coupled continuity equations \eqref{equ:pre_hydro_macro_continuity}, or alternatively \eqref{equ:pre_hydro_q_eq_A}, are called the Euler hydrodynamic equations. For a (Galilei invariant) system with the three usual conservation laws (particle number, momentum and energy) these equations are indeed equivalent to the Euler equations of hydrodynamics~\cite[Ex 2.3]{BenGHD}. 

\begin{remark}
	Going from \eqref{equ:pre_hydro_micro_continuity} to \eqref{equ:pre_hydro_macro_continuity} might seem trivial, but this is a drastic simplification: \eqref{equ:pre_hydro_micro_continuity} is a microscopic operator equation, which we cannot solve because it is not closed. On the other hand \eqref{equ:pre_hydro_macro_continuity} is just a collection of PDEs, which is closed(!). Solving \eqref{equ:pre_hydro_macro_continuity} numerically or analytically is thus significantly simpler compared to solving \eqref{equ:pre_hydro_micro_continuity}. This is what makes hydrodynamics so powerful.
\end{remark}

\begin{remark}\label{rem:pre_hydro_phys_correct}
	In a practical situation, e.g.\ an experiment, one cannot meaningfully establish that densities and velocities are of $\order{1}$ (because they are not dimensionless). Instead, it is only important that the length and time scales of observation $L$ and $T$ are much larger than the microscopic length and time scales and that $N \gg 1$ (this is typically quantified by the Knudsen number, which should ideally be small, see e.g.\ ~\cite{sone2002kinetic}). From this perspective, Euler hydrodynamics will approximate the system using only ballistic transport. For instance, it neglects diffusive transport and also other effects, such as thermal fluctuations and quantum fluctuations. If all these additional effects are indeed neglible in the specific setup, then Euler hydrodynamics will give accurate predictions. %In this sense, hydrodynamics is neither correct nor incorrect.
\end{remark}

\begin{remark}\label{rem:pre_hydro_math_correct}
	There is a different perspective on hydrodynamics, which is as a mathematical scaling limit, where $N \sim L \sim T \to \infty$ are sent to infinite while keeping their ratios fixed. We are going to employ this mathematical viewpoint throughout the thesis. To be more precise, a typical mathematical setup is as follows: First, fix functions $\beta_n(x)$. Given them, we have a specific family of initial states \eqref{equ:pre_hydro_LES} parameterized by $L$. Note that in these states are by construction finite density states, i.e.\ $N \sim L$ as $L\to \infty$. Starting from this state we can compute (in principle) $\expval{\vu{q}_n(Lt,Lx)}$ for any macroscopic time $t$ for each $L$. This way, we also ensure $T \sim L$. Next, we can (in principle) compute the limit $\lim_{L\to \infty} \expval{\vu{q}_n(Lt,Lx)}$. We then say that hydrodynamics is correct if $\lim_{L\to \infty} \expval{\vu{q}_n(Lt,Lx)}$ satisfies \eqref{equ:pre_hydro_macro_continuity}, otherwise it is incorrect. This way it is possible to prove the emergence of hydrodynamics rigorously, as it was for instance done in hard rods~\cite{Boldrighini1983}. Note that it is important to first take $L\to \infty$ before taking any $x$ or $t$ derivatives. If we state in the thesis that a hydrodynamic equation like \eqref{equ:pre_hydro_macro_continuity} is correct or incorrect it is meant in the sense of this mathematical limit. Note that in most cases we cannot evaluate the mathematical limit analytically. Nonetheless, we can compute it numerically. This is used, for instance, in chapter \ref{sec:diff} where we will show that the previous proposal for the leading order correction to GHD (diffusive correction) is incorrect, see \cref{fig:diff_diff_numerics}.
\end{remark}

Hydrodynamics is an immense simplification, at least in theory. In practice, we need to be able to compute thermal expectation values in the given system. In general, it is not known how to do this (and there are only very few exceptions). Still, hydrodynamics is useful as an effective description, similar to low energy field theory.
\begin{remark}
	While exact expressions for thermodynamic quantities are rare, for many systems there exist approximations such as the viral expansion~\cite{Masters_2008} or tabulated expressions as for hard spheres~\cite{C9CP00903E}. But even if no theory is available, for real systems, one can simply measure them in a lab. For instance, the densities of various liquids and gases are tabulated (for instance~\cite{nistfluid}). This way, even though we are still not able to exactly compute thermal properties of (say) water, we were able to predict its flow centuries ago.  
\end{remark}

As a consequence, it is also very hard to check whether hydrodynamics (or its many assumptions) are correct. In fact, very little is known about the physics behind hydrodynamics and quite often hydrodynamic modelling remains an educated guess (more so if they go beyond \eqref{equ:pre_hydro_macro_continuity}). One of the few exceptions where one can compute thermal expectation values are integrable models. This is why it is so interesting to study their hydrodynamics. In fact, GHD has contributed significantly to gaining deeper understanding of the hydrodynamic approximation.

Even though we cannot compute thermal expectation values, the general theory of thermodynamics still gives interesting insights into the equations. For instance, one can show that $\pdv{q_n}{\beta_m} = \pdv{q_m}{\beta_n}$ and $\pdv{j_n}{\beta_m} = \pdv{j_m}{\beta_n}$, i.e.\ they are symmetric matrices~\cite{BenGHD,10.21468/SciPostPhys.6.6.068}. It follows that there exist functions $f(\qty{\beta_n})$ and $g(\qty{\beta_n})$
\begin{align}
	q_n &= \fdv{f}{\beta_n}, & j_n &= \fdv{g}{\beta_n},\label{equ:pre_hydro_fg}
\end{align}
called the free energy density and free energy flux respectively. The free energy $f$ is explicitly given by the usual thermodynamic expression
\begin{align}
	f(\qty{\beta_n}) = - \lim_{\ell \to \infty} \tfrac{1}{\ell}\log \Tr_{\ell} e^{-\sum_n \beta_n \vu{Q}_n(\qty{\beta_n})}.\label{equ:pre_hydro_f_def}
\end{align}
Unfortunately, no such explicit expression is known for $g$. The existence of $f$ and $g$ has a further interesting consequence. Consider the entropy density given by $s = \sum_i \beta_iq_i - f$ and observe that it satisfies the following continuity equation
\begin{align}
	\partial_t s = \sum_{n,m} \beta_n \pdv{q_n}{\beta_m} \partial_t\beta_m = - \sum_{n,m} \beta_n \pdv{j_n}{\beta_m} \partial_x\beta_m = - \partial_x \qty(\sum_n \beta_n j_n - g).\label{equ:pre_hydro_entropy_eq}
\end{align}
Thus, the Euler equation conserves the total entropy $S= \int\dd{x}s(x)$.

\subsection{Limitation of Euler hydrodynamics}
\label{sec:pre_hydro_shock}
\begin{figure}[!h]
	\centering
	\includegraphics{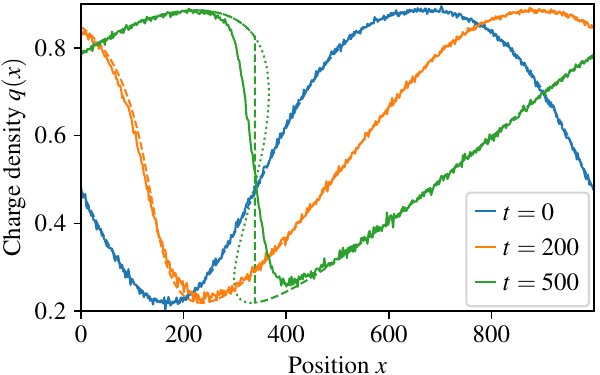}
	\caption[Shock formation]{Shock formation in a non-integrable one dimensional model with a single charge (in a periodic box of size $L=1000$): we see that the hydrodynamic prediction (dashed line) fits well with the profile obtained from microscopic simulations before the shock (orange). After shock formation (green), hydrodynamics is still accurate away from the shock, but it fails to capture the (in this case super-) diffusive broadening close to the shock (the dotted line is \eqref{equ:pre_hydro_shock_characteristics_sol}). Upon increasing $L$, the super-diffusive region becomes thinner, and the profile (slowly) approaches the hydrodynamic one as $L\to\infty$. These simulations were done in the model $(d,\sigma)=(3,996)$ described in~\cite{kim2025circuitssimpleplatformemergence}. Under a reparametrization, the hydrodynamics of this model is equivalent to Burgers' equation \eqref{equ:pre_hydro_burgers}.}
	\label{fig:pre_hydro_shock}
\end{figure}

Beside the problem of determining the thermal expectation values, Euler hydrodynamics also has a much deeper conceptual problem: hydrodynamic equations like \eqref{equ:pre_hydro_macro_continuity} tend to predict their own breakdown by developing gradient catastrophes~\cite{10.1093/oso/9780198507000.001.0001,John1985,LIU197992}. The canonical example is Burgers' equation, which is the hydrodynamic equation for a single charge with current $j = \tfrac{1}{2}q^2$:
\begin{align}
	\partial_t q + q\partial_x q &= 0.\label{equ:pre_hydro_burgers}
\end{align}
Burgers' equation can be solved exactly by the method of characteristics. Imagine that we have already found the solution $q(t,x)$ and follow a particle $x(t)$ with velocity $\dv{t}x(t) = q(t,x(t))$ starting at $x(0)=x^0$. Now observe
\begin{align}
	\dv{t} q(t,x(t)) = \partial_t q(t,x(t)) + q(t,x(t))\partial_x q(t,x(t)) = 0.\label{equ:pre_hydro_shock_characteristics}
\end{align} 
Hence, $q(t,x(t)) = q(0,x_0)$ and $x(t)=x_0+q(0,x_0)t$. This means that by going over all $x_0$, the graph of $q(t,x)$ is given by
\begin{align}
	\qty{(x,q(t,x))} = \qty{(x_0 + q(0,x_0)t,q(0,x_0))}.\label{equ:pre_hydro_shock_characteristics_sol}
\end{align}
However, this can only be a solution as long as there is exactly one $x_0$ for each $x$, otherwise the solution becomes multivalued. For short times $f(x_0) = x_0 + q(0,x_0)t$ is monotone increasing
\begin{align}
	\partial_{x_0} f(x_0) = 1 + \partial_{x_0} q(0,x_0) t,\label{equ:pre_hydro_shock_deriv}
\end{align}
but after time $t\ind{shock} = -1/\inf_{x_0} (\partial_{x_0} q(0,x_0))$ it fails to be invertible. Therefore, \eqref{equ:pre_hydro_shock_characteristics_sol} can no longer be the solution. Instead, the system develops a discontinuity (called a shock) in the profile, see \cref{fig:pre_hydro_shock}. Such shocks are known to quite generally form in most one-dimensional hydrodynamic equations, unless in special cases such as linearly degenerate systems~\cite{10.1093/oso/9780198507000.001.0001}.
Mathematically, a shock can be viewed as a weak solution to \eqref{equ:pre_hydro_burgers}. A weak solution is a way to make sense of a non-differentiable solution to a PDE (as opposed to a strong solution which is differentiable). For a hydrodynamic PDE such as \eqref{equ:pre_hydro_macro_continuity}, a weak solution has to satisfy the integrated conservation law:
\begin{align}
	\int_{x_1}^{x_2}\dd{x} \qty(q_n(t_2,x)-q_n(t_1,x)) + \int_{t_1}^{t_2}\dd{t} \qty(j_n(t,x_2)-j_n(t,x_1)) &= 0\label{equ:pre_hydro_shock_weak}
\end{align}
for any $x_1,x_2,t_1,t_2\in \mathbb{R}$. If $q_n(t,x)$ is differentiable, the above is equivalent to \eqref{equ:pre_hydro_macro_continuity}. However, if $q_n(t,x)$ is not differentiable, \eqref{equ:pre_hydro_macro_continuity} is ill-posed, but \eqref{equ:pre_hydro_shock_weak} makes sense as long as $q_n(t,x)$ is integrable. 

The problem with weak solutions, as opposed to strong solutions, is that they are not unique. Therefore, it is a non-trivial task to pick the correct one, see e.g.~\cite{10.1093/oso/9780198507000.001.0001}.

But this is only the mathematical side of the problem. Physically, the presence of a shock seriously challenges our assumptions behind hydrodynamics, especially that $\beta_n(t,x)$ is slowly varying. At the point of the shock, how are we supposed to define the fluid cell and average over a homogeneous state inside the fluid cell? To understand this one has to zoom into the shock, which turns out not to be an actual sharp discontinuity, but is smoothened out by higher order effects, like diffusion (i.e. transport, where the displacement grows as $\order{\sqrt{t}}$) or superdiffusion (i.e. transport, where the displacement grows faster than $\order{\sqrt{t}}$, but less than ballistic $\order{t}$). Therefore, after the shock forms we need to take into account these higher order effects for an accurate physical description. 

We would like to mention one important consequence: effects like diffusion increase entropy. Usually these are suppressed as $\order{1/L}$, but at the shock they become important and lead to an $\order{1}$ entropy increase rate. Indeed, for weak solutions the derivation \eqref{equ:pre_hydro_entropy_eq} fails at the shock (simply because it is not differentiable), and therefore a shock can produce entropy.

This shows that understanding a hydrodynamic equation on a mathematical level, whether it has differentiable strong or weak solutions, can lead to important physical insights. In GHD, shocks are absents and therefore Euler GHD is a reliable theory for all times. This had been conjectured in the community and was finally solved in my PhD (see \cref{sec:fixedpoint}).

\begin{remark}
	An intuitive picture about shocks is wave breaking: imagine a water wave on the ocean approaching the shore. Due to friction with the ground the upper parts of the wave are faster than the lower parts. Eventually the upper parts are in front of the lower parts and the wave breaks. During the breaking of a wave the continuous hydrodynamic description breaks down for some time. Shocks are similar, only that in one dimension the faster parts cannot get ahead of the slower parts and start to pile up.
\end{remark}

In higher dimensions, other types of gradient catastrophes can occur. The most well-known is turbulence~\cite{Frisch_1995,10.1093/acprof:oso/9780199689385.001.0001,ZHOU20101,ZHOU20211,Galtier_2022}, where the solution to the hydrodynamic equation develops a cascade of smaller and smaller vortices\footnote{In the terminology of fluid dynamics: turbulence occurs at high Reynolds numbers $\mathrm{Re}$. The Euler equation is the limit $\mathrm{Re} \to \infty$, hence it is prone to turbulence.}. This means that in addition to a diverging gradient, the direction of the gradient also starts to fluctuate widely. Despite many investigations, our understanding of turbulence remains significantly less developed compared to shocks; in particular no general theoretical frameworks exist.

\subsection{Ballistic macroscopic fluctuation theory (BMFT)}
\label{sec:pre_hydro_bmft}
Up to this point, we have discussed the traditional theory of hydrodynamics, which is a theory for averages of charge densities. However, hydrodynamic reasoning can be applied to other observables as well. One interesting development that emerged recently is ballistic macroscopic fluctuation theory (BMFT)~\cite{10.21468/SciPostPhys.15.4.136}. It is motivated by a similar existing theory, macroscopic fluctuation theory, for diffusive systems~\cite{RevModPhys.87.593}. Note that integrable systems have been crucial in developing this theory.

The idea behind BMFT is to apply hydrodynamics not just on averages, but also on fluctuations. Hydrodynamics allows us how to predict $\expval{q_n(t,x)}$ starting from $\expval{q_n(0,x)}$. But suppose that we are interested in predicting $\expval{f[\qty{q_n(t,\cdot)}]}$, where $f[\cdot]$ can be an arbitrary functional. How could we evaluate such an expression?

To study fluctuations we now slightly change notation: in \cref{sec:pre_hydro}, $q_n(t,x)$ denoted the expectation value $\expval{\vu{q}_n(t,x)}$. We will now denote by $q_n(t,x)$ a fluctuation, which is not necessarily equal to its expectation value $\expval{q_n(t,x)} := \expval{\vu{q}_n(t,x)}$, see \ref{rem:pre_hydro_BMFT_precise} for more details.

Imagine we observe a fluctuation $q_n(t,x) \neq \expval{q_n(t,x)}$ differing from the thermal average in a fluid cell. Since we know the total charge $q_n(t,x)\ell$ in this cell, our best guess for the state in this cell is the maximum entropy state with average charge $q_n(t,x)\ell$. But by the construction of Euler hydrodynamics, the current in this state is precisely given by $j_n[\qty{q_m(t,x)}]$. Hence, we expect that Euler hydrodynamics will correctly describe the evolution of $q_n(t,x)$, even though it technically differs from $\expval{q_n(t,x)}$. n this way, we can also apply hydrodynamics to states with much more complicated statistics compared to \eqref{equ:pre_hydro_LES}.

This discussion leads to what we will call the BMFT principle:
\begin{definition}[BMFT principle]\label{def:pre_hydro_BMFT}
	As $L \to \infty$, (almost) any large scale coarse-grained configuration $\qty{q_n(t,x)}$ will individually evolve according the Euler equation, up to an error decaying as $L \to \infty$.
\end{definition}

In practice this means that the quantity $\expval{f[\qty{q_n(t,\cdot)}]}$ can be computed as follows
\begin{align}
	\expval{f[\qty{q_n(t,\cdot)}]} = \expval{f[\mathcal{K}_t[\qty{q_n(0,\cdot)}]]},\label{equ:pre_hydro_BMFT_practice}
\end{align}
where $\mathcal{K}_t[\qty{q_n(0,\cdot)}]$ is the solution to \eqref{equ:pre_hydro_macro_continuity} starting from $q_n(0,x)$. This means that if we view the initial state $\expval{\ldots}$ (which might be of the form \eqref{equ:pre_hydro_LES} or not) as a probability distribution over $\qty{q_n(0,\cdot)}$, then each sample from this probability distribution will individually evolve according to the Euler equation.

Typically, thermodynamic initial states like \eqref{equ:pre_hydro_LES} satisfy a large deviation principle, i.e. as $L \to \infty$
\begin{align}
	\mathbb{P}[\qty{q_n(0,\cdot)}] \sim e^{-LI[\qty{q_n(0,\cdot)}]}\label{equ:pre_hydro_BMFT_LD}
\end{align}
for some rate function $I[\qty{q_n(0,\cdot)}]$ (see appendix \ref{app:LD} for an overview of large deviation theory), which implies that to leading order $q_n(0,\cdot)$ is Gaussian with
\begin{align}
	\expval{q_n(0,x)} &\sim \order{1}, & \mathrm{Cov}[q_n(0,x),q_n(0,y)] &\sim \order{1/L}.\label{equ:pre_hydro_BMFT_scaling}
\end{align}
In an local equilibrium state \eqref{equ:pre_hydro_LES}, different fluid cells are independent and thus
\begin{align}
	\mathrm{Cov}[q_n(0,x),q_n(0,y)] = \tfrac{1}{L}\delta(x-y) C\upd{GGE}_{nm}[\qty{\beta_n(0,x)}],\label{equ:pre_hydro_BMFT_cov_LES}
\end{align}
where $C\upd{GGE}_{nm}[\qty{\beta_n(0,x)}] = L\pdv{q_n}{\beta_m}$ are the local GGE correlations.

Since the covariance is subleading as $L \to \infty$, to leading order averages are simply solutions to the Euler equation
\begin{align}
	\expval{q_n(t,x)} = \expval{\mathcal{K}_t[\qty{q_n(0,\cdot)}} =\mathcal{K}_t[\qty{\expval{q_n(0,\cdot)}}],\label{equ:pre_hydro_BMFT_expval} 
\end{align}
which agrees with \eqref{equ:pre_hydro_macro_continuity}. However, due to the small randomness the initial state will not be precisely $\expval{q_n(0,x)}$, but instead be $q_n(0,x) = \expval{q_n(0,x)}+ \delta q_n(0,x)$, where $\delta q_n(0,x) \sim \order{1/\sqrt{L}}$ is a random fluctuation. Expanding the hydrodynamic equation to first order we find that this perturbation evolves as
\begin{align}
	\partial_t \delta q_n(t,x) + \partial_x \qty[\sum_m \pdv{j_n}{q_m}\eval_{q_m = \expval{q_m(t,x)}} \delta q_m] &= 0.\label{equ:pre_hydro_BMFT_linEuler}
\end{align}

This is the linearized Euler equation describing the evolution of perturbations over the background $\expval{q_m(t,x)}$. An important special case is when the perturbation is on top of a homogeneous GGE state, i.e. $\expval{q_m(t,x)} = \mathrm{const}$. In this case the matrix $\pdv{j_n}{q_m}$ is constant in space and time. Such an equation can be solved by diagonalizing $\pdv{j_n}{q_m}$. The eigenvalues of $\pdv{j_n}{q_m}$ are the sound velocities of the system and describe the speed at which (small) waves propagate through the otherwise homogeneous system.

If the initial fluctuations are random, we can describe their correlations at other times by $\allowbreak\mathrm{Cov}[q_n(t,x),q_m(s,y)] = \expval{\delta q_n(t,x)\delta q_m(s,y)}$, which satisfies the linearized Euler equation in both components~\cite{10.21468/SciPostPhys.15.4.136,10.21468/SciPostPhys.5.5.054}
\begin{align}
	\partial_t \expval{\delta q_n(t,x)\delta q_m(s,y)} + \partial_x \qty[\sum_k \pdv{j_n}{q_k}\eval_{q_k = \expval{q_k(t,x)}} \expval{\delta q_k(t,x)\delta q_m(s,y)}] &= 0,\label{equ:pre_hydro_BMFT_corr_1}\\
	\partial_s \expval{\delta q_n(t,x)\delta q_m(s,y)} + \partial_y \qty[\sum_k \pdv{j_m}{q_k}\eval_{q_k = \expval{q_k(s,y)}} \expval{\delta q_n(t,x)\delta q_k(s,y)}] &= 0.\label{equ:pre_hydro_BMFT_corr_2}
\end{align}
An important special case is to study the equal time correlations $\expval{\delta q_n(t,x)\delta q_m(t,y)}$. It was shown in~\cite{PhysRevLett.131.027101,10.21468/SciPostPhys.15.4.136} that those in general are long range correlated with $\expval{\delta q_n(t,x)\delta q_m(t,y)} = \order{1/L}$ even at far distant points. Furthermore, they confirmed it with numerical simulations in hard rods~\cite{PhysRevLett.131.027101}. 
\begin{remark}
	Importantly, this means that at time $t$ the state cannot be given by a local equilibrium state like \eqref{equ:pre_hydro_LES}! Even though this is only a subleading correction, it still violates (to some extent) the idea of thermalization, which is a fundamental assumption we used to derive Euler hydrodynamics. This indicates that hydrodynamics should also emerge under weaker assumptions.
\end{remark}

\begin{figure}
	\centering
	\includegraphics{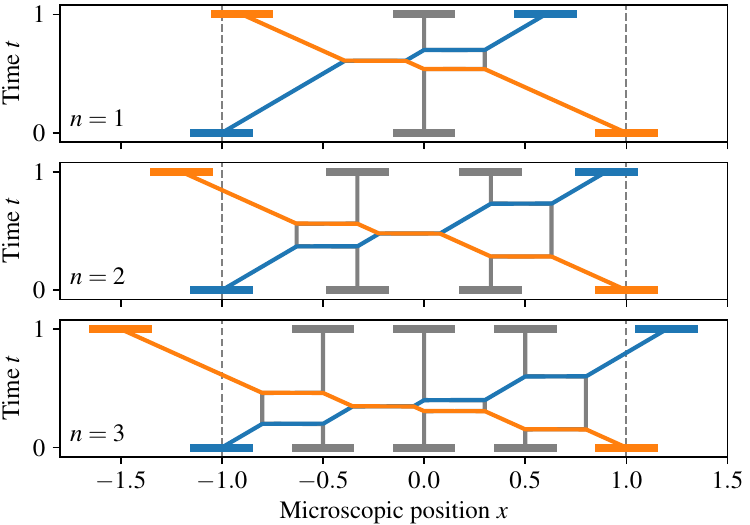}
	\caption[Microscopic origin of long range correlations]{Microscopic origin of long range correlations in hard rods (see \cref{sec:pre_int_HR}): two hard rods (blue and orange) travel through the same region in which the number of particles $n$ (gray) is fluctuating. If there are more particles in this region, both particles travel further otherwise less. This way the final position of particles becomes correlated, which is observed as non-trivial density correlations on large scales. This figure was reproduced from~\cite{hübner2025diffusivehydrodynamicshardrods}.}
	%\put(-15,0){Microscopic}
	\label{fig:pre_hydro_BMFT_LL_micro}
\end{figure}

\begin{figure}
	\centering
	\includegraphics{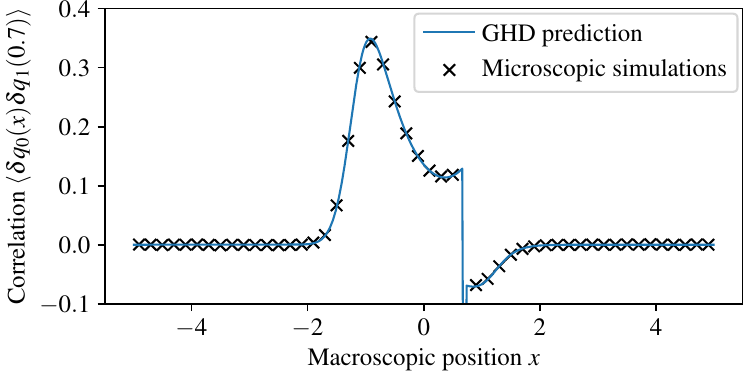}
	\caption[Long range correlations in hard rods]{Long range correlations in hard rods: the correlation between $x$ and $y=0.7$ of the first two conserved charges after time-evolution from a local equilibrium state. The correlations are a smooth function, except at $x=y=0.7$, where there is $\delta(x-y)$ peak corresponding to the local GGE correlations and additionally a jump. While long range correlations were established in~\cite{PhysRevLett.131.027101}, the existence of this jump is a new finding (see \cref{res:fixedpoint_corr_jump}). This figure was reproduced from~\cite{PhysRevLett.134.187101}.}
	\label{fig:pre_hydro_BMFT_LL_macro}
\end{figure}

This discussion is an instance where we can see that not only hydrodynamics can go much beyond the traditional setting, but also that understanding the physics behind it is very important. The BMFT principle is a powerful tool and is now well established in integrable systems (see, for instance,~\cite{PhysRevE.111.024141,horvath2024countingstatisticsquantumquenches}). In fact, we will establish a much stronger version of BMFT in hard rods (see \cref{res:diff_BMFT_diff}). In hard rods, BMFT also applies even to states which are even microscopically far from thermal equilibrium (and do not thermalize locally during the hydrodynamic evolution). This suggests that hydrodynamics may be a dynamical self-averaging phenomenon, independent of thermalization, and that the phenomenological derivation in \cref{sec:pre_hydro} is just a special case.

\begin{remark}\label{rem:pre_hydro_BMFT_precise}
	Let us try to explain what we mean by $q_n(t,x)$ more precisely. In BMFT we want to study correlation functions, i.e.\  expectation values like $\expval{\vu{q}_{n_1}(t_1,x_1)\vu{q}_{n_2}(t_2,x_2)\ldots \vu{q}_{n_k}(t_k,x_k)}$ in the Euler scaling limit. The assumption of BMFT is that one can write such expectation values as $L\to \infty$ as
	\begin{align}
		\expval{\vu{q}_{n_1}(t_1,x_1)\vu{q}_{n_2}(t_2,x_2)\ldots \vu{q}_{n_k}(t_k,x_k)} &\to \int\mathrm{D}[q_{n}(0,\cdot)] e^{-L\LD{I}[\qty{q_{n}(0,\cdot)}]} q_{n_1}(t_1,x_1)q_{n_2}(t_2,x_2)\ldots q_{n_k}(t_k,x_k).\label{equ:pre_hydro_BMFT_rem_precise}
	\end{align}
	The RHS is an average over a probability distribution $e^{-L\LD{I}[\qty{q_{n}(0,\cdot)}]}$, where $q_n(0,x)$ is the integration variable. Here, $q_n(t,x)$ is given as the solution to \eqref{equ:pre_hydro_macro_continuity} starting from $q_n(0,x)$. Note that \eqref{equ:pre_hydro_BMFT_rem_precise} is a more precise way to state the rather intuitive \cref{def:pre_hydro_BMFT}. For quantum systems, note that expression on the RHS is invariant under swapping two $q_{n}(t,x)$, while the LHS is not in a quantum system. Therefore, the RHS can only make sense in the limit $L\to\infty$ where one can neglect the commutator of (large scale averages of) charge densities. In this limit the large deviation rate function $\LD{I}[\qty{q_{n}(0,\cdot)}]$ can be obtained as the Legendre transform (see \cref{app:LD}) of the scaled cumulant generating function $\LD{F}[\qty{\lambda_n(x)}] = \lim_{L\to \infty} \tfrac{1}{L}\log \expval{e^{L\int\dd{x}\sum_n \lambda_n(x)\vu{q}_n(0,x)}}$, which can (in principle) be computed from the microscopic theory.
\end{remark}

\subsection{Diffusive hydrodynamics}
\label{sec:pre_hydro_diff}
Section \ref{sec:pre_hydro_shock} made it clear that Euler hydrodynamics is only the beginning: to understand what happens at its breakdown, we need to find higher order corrections to it. These are described by the gradient expansion~\cite{Kovtun_2012}. Recall that on the Euler scale $\expval{j_n} = j_n[\qty{\expval{q_m}}]$ is only a function of the charge densities. The idea of the gradient expansion is that in general, the currents $j = j[\qty{\expval{q_m}},\qty{\partial_x \expval{q_m}}, \qty{\partial_x^2 \expval{q_m}}, \ldots]$ depend on all derivatives of $\expval{q_m}$ as well. However, since we study large systems, higher order derivatives are suppressed by factors of $1/L$. The leading correction term is thus given by taking into account $\tfrac{1}{L}\partial_x \expval{q_n}$, leading to
\begin{align}
	\expval{j_n} = j_n[\qty{\expval{q_m}},\qty{\tfrac{1}{L}\partial_x \expval{q_m}}] = j_{\mathrm{E},n}[\qty{\expval{q_m}}] - \tfrac{1}{2L} \sum_m D_{nm}[\qty{\expval{q_n}}] \partial_x \expval{q_m} + \order{1/L^2}.\label{equ:pre_hydro_NS_curr} 
\end{align}
Here $j_{\mathrm{E},n}[\qty{\expval{q_n}}]$ is the Euler current as in \eqref{equ:pre_hydro_macro_continuity} and  $D_{nm}[\qty{\expval{q_n}}]$ is the diffusion matrix. This leads to the (Navier-Stokes like) diffusive hydrodynamics equation
\begin{align}
	\partial_t \expval{q_n} + \partial_x j_{\mathrm{E},n} &= \tfrac{1}{2L} \partial_x \qty[\sum_m D_{nm} \partial_x \expval{q_m}].\label{equ:pre_hydro_NS}
\end{align}
The diffusion matrix is model dependent. As for the Euler current, we would like to extract it from the GGE data. This is usually done based on correlation functions going back to an idea of Kubo~\cite{spohn2012large}. If \eqref{equ:pre_hydro_NS} holds it should also hold for small fluctuations $q_n(t,x) = \expval{q_n}_{\qty{\beta_n}} + \delta q(t,x)$ on top of a uniform GGE state with $\qty{\beta_n}$. Such fluctuations are intrinsic to the GGE and thus can be characterized by the two-point function $\expval{\delta q_n(t,x)\delta q_m(0,0)}_{\qty{\beta_n}}$, which analogously to \eqref{equ:pre_hydro_BMFT_corr_1} should satisfy
\begin{multline}
	\partial_t \expval{\delta q_n(t,x)\delta q_m(0,0)}_{\qty{\beta_n}} + \partial_x \qty[\sum_k \pdv{j_{\mathrm{E},n}}{q_k} \expval{\delta q_k(t,x)\delta q_m(0,0)}_{\qty{\beta_n}}]\\
	= \tfrac{1}{2L} \partial_x \qty[\sum_k D_{nk} \partial_x \expval{\delta q_k(t,x)\delta q_m(0,0)}_{\qty{\beta_n}}].\label{equ:pre_hydro_NS_corr}
\end{multline}

In case of a single conserved quantity, this is a Fokker-Planck equation for the distribution of a random particle with velocity $v=\fdv{j\ind{E}}{q}$ starting at $x=0$ with noise $\sim \sqrt{D/L}$. Its expected position $\mathbb{E}[x(t)] = vt$ and its variance $\mathrm{Var}[x]\sim D t/L$ both grow linearly in time. Generalizing this, we expect that in the long time limit $t\to \infty$~\cite{10.21468/SciPostPhys.6.4.049}
\begin{align}
	\int\dd{x}x^2\expval{\delta q_n(t,x)\delta q_m(0,0)}_{\qty{\beta_n}} =  \mathcal{D}_{nm}(\qty{\beta_n})t^2 + \tfrac{1}{L}L_{nm}(\qty{\beta_n})t + \order{1}\label{equ:pre_hydro_NS_kubo},
\end{align}
where the Drude weight $\mathcal{D}_{nm}(\qty{\beta_n})$ encodes the ballistic movement and the Onsager matrix $L_{nm}(\qty{\beta_n})$ the diffusive spreading respectively. Assuming for simplicity that $\pdv{j_{\mathrm{E},n}}{q_k}=0$, we indeed find using \eqref{equ:pre_hydro_NS_corr}:
\begin{align}
	\int\dd{x}x^2\expval{\delta q_n(t,x)\delta q_m(0,0)}_{\qty{\beta_n}} &= \frac{1}{2L}\int_0^t\dd{s}\int\dd{x}\sum_k D_{nk} x^2\partial_x^2 \expval{\delta q_k(t,x)\delta q_m(0,0)}_{\qty{\beta_n}}\\
	&=\frac{t}{L}\sum_k D_{nk} \int\dd{x} \expval{\delta q_k(0,x)\delta q_m(0,0)}_{\qty{\beta_n}}\\
	&=\frac{t}{L}\sum_k D_{nk} C\upd{GGE}_{km}[\qty{\beta_n}].
\end{align}
Here we used $\int\dd{x}\expval{\delta q_k(t,x)\delta q_m(0,0)} = \expval{\delta Q_k(x)\delta q_m(0,0)}$ is constant in time. Comparing with \eqref{equ:pre_hydro_NS_kubo} we can read off $\sum_k D_{nk} C\upd{GGE}_{km} = L_{nm}$. One can show that $L_{nm}$ is a positive definite matrix and as a consequence \eqref{equ:pre_hydro_NS} increases entropy (see ~\cite[Sec 2.4]{BenGHD}). Since entropy increase signals thermalization, this a posteriori provides a justification for the assumptions in \cref{sec:pre_hydro}. 

As for the Euler equation for a Galilei invariant system with the usual three conservation laws (particle number, momentum, energy) the momentum equation reduces to the well known (compressible) Navier-Stokes equation of hydrodynamics~\cite[Ex 2.9]{BenGHD}. 

\begin{remark}
	Compared to Euler hydrodynamics, diffusive hydrodynamics is even less understood, both mathematically\footnote{This is the essence of the Millenium prize problem on Navier-Stokes.} and physically. We would like to emphasize again that while hydrodynamics works in practice, it is still only an educated guess. Note that non-integrable 1D systems typically show super-diffusive behavior~\cite{Spohn2014}, and thus \eqref{equ:pre_hydro_NS} will not apply. In 2D systems, there might be logarithmic corrections to diffusion~\cite{Krug2018}. 
\end{remark}

In integrable systems it was indeed believed that the diffusive GHD was given by \eqref{equ:pre_hydro_NS}. Furthermore, \eqref{equ:pre_hydro_NS} is rigorously proven in hard rods for short times. However,  \eqref{equ:pre_hydro_NS} still cannot capture the correct dynamics (see \cref{sec:diff}). On the level of correlation functions \eqref{equ:pre_hydro_NS_corr} and \eqref{equ:pre_hydro_NS_kubo} still hold, however, it is not correct to plug the same result into the true out-of-equilibrium setup \eqref{equ:pre_hydro_NS}.

\begin{remark}\label{rem:pre_hydro_boltzmann}
	Recently, a highly acclaimed rigorous proof~\cite{deng2025hilbertssixthproblemderivation} of Navier-Stokes in hard spheres based on the Boltzmann equation appeared. The idea of using the Boltzmann equation dates back to Hilbert, i.e.\ Hilbert's 6'th problem. We would like to note that the Boltzmann equation only applies to dilute systems (only two-particle scattering will matter). In \eqref{equ:pre_hydro_NS}, on the other hand, we are interested in non-dilute systems, which is a different (and more intricate) scaling limit.
\end{remark}

\section{Integrable models}
\label{sec:pre_int}
There is no generally accepted definition of what an integrable model in many-body physics is. Generally, it is used if one can solve them exactly (in some field-dependent sense). Examples relevant in the context of GHD include
\begin{itemize}
	\item Classical mechanics models such as hard rods (see \cref{sec:pre_int_HR}) or the Toda chain~\cite{PhysRevB.9.1921}
	\item Classical integrable PDE's such as the Korteweg-De Vries equation equation~\cite{eilenberger2012solitons} or the non-linear Schrödinger equation~\cite{Zakharov1974}
	\item Quantum models such as the Lieb-Liniger model (see \cref{sec:pre_int_LL}) or quantum Calogero-Moser-Sutherland models~\cite{hallnäs2024calogeromosersutherlandsystems}
	\item Quantum chains such as the spin $1/2$ XXZ chain~\cite[Part III]{samaj_Bajnok_2013} or the Hubbard chain~\cite{Essler_Frahm_Göhmann_Klümper_Korepin_2005}
	\item Quantum field theories such as sine-Gordon~\cite[Part V]{samaj_Bajnok_2013} or affine Toda field theory~\cite{Korff2000}
	\item Celluar automata such as rule 54~\cite{Buca_2021} and its quantum version~\cite{PhysRevLett.126.160602}.
\end{itemize}
In addition to these also conformal field theories~\cite{DiFrancesco:1997nk}, stochastic models like the TASEP (totally asymmetric simple exclusion process)~\cite{Rost1981,spohn2012large} and statistical mechanics models like the Ising model~\cite{GesualdoDelfino_2004} or 8-vertex model~\cite[Chap 5]{lavis1999statistical} are integrable.

Since different integrable models are formulated with different mathematical descriptions across fields, it is hard to clearly define integrability. A common property of integrable models across fields is that they possess an infinite number of conserved quantities with local densities. This is commonly believed to imply factorized scattering based on an argument by Parke~\cite{PARKE1980166}. Factorized scattering means that multi-particle scattering events can be decomposed into two-particle scattering events. This simplification is what enables us to solve models exactly.

Despite these common features, the way explicit computations are performed varies from field to field with often model specific technicalities. In order to focus on GHD instead of peculiarities of integrable models, we decided to restrict (most parts of) our discussion to two very simple models: the classical hard rods model and the quantum Lieb-Liniger model. They are sufficiently simple to showcase the important features of integrable models in an accessible manner. We believe that most of the results can be carried over to other integrable models by making appropriate adjustments. 

\begin{remark}
	In classical Hamiltonian systems there exist the commonly accepted notion of Liouville-integrability~\cite{arutyunov2020elements}. We would like to note that if a system is Liouville-integrable, then it does not necessarily posses an infinite number of conserved quantities with local densities (see \cref{rem:scbm_local_integrability} for further discussion). As those are required for GHD we need to restrict to integrable models with such an infinite number of conserved quantities with local densities.
\end{remark}

\subsection{Hard rods model: a paradigmatic classical model}
\label{sec:pre_int_HR}
Hard rods is the hard spheres model in one dimension~\cite{spohn2012large}. Formally, its Hamiltonian is given by
\begin{align}
	H(\vec{x},\vec{\rapidity}) = \tfrac{1}{2} \sum_i \rapidity_i^2 + \sum_{i\neq j} V(x_i-x_j),\label{equ:pre_int_HR_def}
\end{align}
where $V(x) = \infty$ if $\abs{x}<d$ and otherwise $0$ and $d$ is the hard rod diameter. Particle trajectories are simple. Particles evolve like free particles $\dv{t}x_i = \rapidity_i$ until they hit another particle. During scattering both particles instantaneously exchange their momenta $\rapidity_i \leftrightarrow \rapidity_j$ (like billiard balls).

Therefore, particles might exchange their momenta, but the number of particles with each momentum $\rapidity$ is conserved. Thus, for any integer power $n$, the quantity $Q_n = \sum_i \rapidity_i^n$ is a conserved quantity with local density $q_n(x) = \sum_i \delta(x-x_i) \rapidity_i^n$. For $n=0$ this is the particle number $Q_0=N$, $Q_1$ is the total momentum and $Q_2$ is the energy. In addition to these, higher order $Q_n$ are also conserved. This is the infinite collection of conserved quantities.

Alternatively, for each $\rapidity\in \mathbb{R}$, the quantity $Q(\rapidity) = \sum_i \delta(\rapidity-\rapidity_i)$ is a conserved quantity with density $\rho(x,\rapidity) = \sum_i \delta(x-x_i)\delta(\rapidity-\rapidity_i)$. This representation will be important for GHD. 

\begin{figure}[!h]
	\centering
	\includegraphics{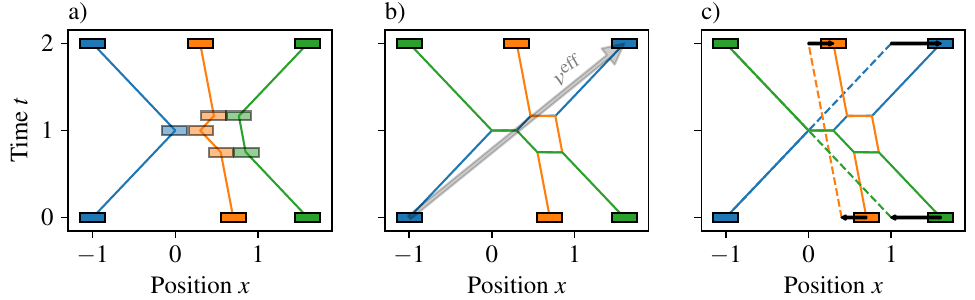}
	\caption[Dynamics of hard rods]{Dynamics of hard rods: a) physical hard rods scatter by exchanging their momenta $\rapidity$. b) Equivalently, one can consider tracer dynamics (particles exchange positions instead of momenta during scattering). Therefore, particles effectively travel a larger distance in a given time interval, which will give rise to the effective velocity in GHD. c) Using the contraction map \eqref{equ:pre_int_HR_contract} hard rods can be mapped to non-interacting particles. In these coordinates evolution is trivial. To obtain the location of hard rods at a later time one simply has to expand back to original coordinates. This figure was reproduced from~\cite{hydrowithoutaveraging}.}
	\label{fig:pre_int_HR_dynamics}
\end{figure}

In the context of GHD, it is useful to think about hard rods not in terms of physical particles but instead in terms of tracer particles (see \cref{fig:pre_int_HR_dynamics} b)). Tracer particles, instead of exchanging their momenta during scattering, exchange their positions $x_i \leftrightarrow x_j$. This means that particle $i$ will keep its momentum $\rapidity_i$ for all times $t$. When two particles meet they both jump forward by a distance $d$. Note that this representation is merely a relabeling of particles during scattering, hence has no physical effect on indistinguishable particles. 

One advantage of tracer particles is that their dynamics is known explicitly: define the contracted coordinate
\begin{align}
	\hat{x}_i = x_i - d \sum_{j\neq i} \theta(x_i-x_j),\label{equ:pre_int_HR_contract}
\end{align}
i.e. the first particle on the left is unchanged, the second particle is shifted by $-d$, the third particle by $-2d$ and so on. This ``contracts'' away the size $d$ of the particles. By investigating a scattering event it is not hard to see the evolution in contracted coordinates is free~\cite{Boldrighini1983}
\begin{align}
	\hat{x}_i(t) = \hat{x}_i+p_i t.\label{equ:pre_int_HR_contract_evol}
\end{align}

After evolving up to time $t$ in the contracted coordinates, we then need to revert to physical coordinates. This can be done by keeping the first particle on the left fixed, shifting the second particle by $d$, the third by $2d$ and so on; in formula:
\begin{align}
	x_i(t) &= \hat{x}_i(t) + d \sum_{j\neq i} \theta(\hat{x}_i(t)-\hat{x}_j(t)).\label{equ:pre_int_HR_expand}
\end{align}
Note that this can be efficiently implemented on a computer by sorting the particles in ascending $\hat{x}_i(t)$ before expanding. Note that \eqref{equ:pre_int_HR_contract}, \eqref{equ:pre_int_HR_contract_evol} and \eqref{equ:pre_int_HR_expand} provide an explicit analytical solution for the dynamics of hard rods. Having such an explicit expression is unique even among integrable models and makes hard rods one of the most important integrable models.

\subsection{Lieb-Liniger model: a paradigmatic quantum model}
\label{sec:pre_int_LL}
Integrability in quantum models is usually associated to the Bethe ansatz. We will discuss the coordinate Bethe ansatz and its thermodynamic limit, the thermodynamic Bethe ansatz (TBA), using the repulsive Lieb-Liniger model~\cite{PhysRev.130.1605} as an example. These approaches are basically an explicit guess of the eigenstates in terms of linear combinations of plane waves. We will not discuss alternative approaches, such as the algebraic Bethe ansatz~\cite{Korepin_Bogoliubov_Izergin_1993} (this -- in spirit -- defines creation and annihilation operators for Bethe states similar to the treatment of the harmonic oscillator).

The Lieb-Liniger model describes a bosonic quantum gas of $N$ particles with pairwise $\delta$ interaction
\begin{align}
	\vu{H}\ind{LL} = -\sum_{i=1}^N \partial_{x_i^2} + 2c\sum_{i<j}\delta(x_i-x_j),\label{equ:pre_int_LL_def}
\end{align}
where $c$ is the interaction strength. We now derive the coordinate Bethe Ansatz for this model, closely following~\cite{jscaux}.

First, consider the two particle problem $\vu{H}\ind{LL} = -\partial_{x_1}^2 - \partial_{x_2}^2 + 2c\delta(x_1-x_2)$. Since we are dealing with a bosonic model, the wave function is fully symmetric and thus it is sufficient to find an eigenstate for $x_1<x_2$. For $x_1<x_2$ the eigenstates are simply linear combinations of plane wave states $e^{i\rapidity_1x_1+i\rapidity_2x_2}$. The $\delta$ function implies a boundary condition at $x_1=x_2$, which can be found as follows: first express the Hamiltonian in relative coordinates $x = (x_1+x_2)/2$, $y=x_1-x_2$:
\begin{align}
	\vu{H}\ind{LL} = -\tfrac{1}{2}\partial_x^2 -2\partial_y^2+2c\delta(y).\label{equ:pre_int_LL_twoparticles}
\end{align}
Now integrate $\vu{H}\ind{LL}\psi = E\psi$ over $y$ from $0^-$ to $0^+$, yielding $\partial_y \psi(y=0^-)- \partial_y \psi(y=0^+) + c \psi(y=0) = 0$. Expressing this in the old coordinates and making use of the symmetry of the wave function we end up with
\begin{align}
	(\partial_{x_1}-\partial_{x_2}-c)\psi\eval_{x_1=x_2^-} &= 0.\label{equ:pre_int_LL_bc}
\end{align}
This can be solved by a superposition of incoming and outgoing waves
\begin{align}
	\psi(x_1,x_2) = S_1e^{i\rapidity_1x_1+i\rapidity_2x_2} + S_2e^{i\rapidity_2x_1+i\rapidity_1x_2}.\label{equ:pre_int_LL_twoparticle_psi}
\end{align}
Inserting this into \eqref{equ:pre_int_LL_bc} we find
\begin{align}
	\frac{S_2}{S_1} &= - \frac{c+i(\rapidity_1-\rapidity_2)}{c-i(\rapidity_1-\rapidity_2)}  = -e^{i\phi(\rapidity_1-\rapidity_2)}, & \phi(\rapidity) &=2\arctan(\rapidity/c).\label{equ:pre_int_LL_twoparticle_ratio}
\end{align}
This construction can be extended to $N$ particles, with eigenstates given by the Bethe state
\begin{align}
	\psi(\vec{x}|\vec{\rapidity}) = \frac{\chi(\vec{x})}{\mathcal{N}_N} \sum_{\sigma\in \permutation_N} (-1)^\sigma e^{i\Phi(\vec{x}_{\sigma^{-1}},\vec{\rapidity})},\label{equ:pre_int_LL_Nparticle}
\end{align}
where $\chi(\vec{x})=\prod_{i<j} \sgn(x_i-x_j)$, $\mathcal{N}_N=\sqrt{(2\pi)^NN!}$, $\sigma$ runs over all permutations of $N$ elements and the Bethe phase is given by
\begin{align}
	\Phi(\vec{x},\vec{\rapidity}) = \sum_i \rapidity_i x_i + \tfrac{1}{4} \sum_{i\neq j} \sgn(x_i-x_j) \phi(\rapidity_i-\rapidity_j).\label{equ:pre_int_LL_phase}
\end{align}
The energy of this eigenstate is given by $E=\sum_i\rapidity_i^2$. Showing that \eqref{equ:pre_int_LL_phase} is an eigenstate is not too hard: restricted to the ordering $x_1 < x_2 < \ldots < x_N$, this is simply a superposition of plane waves. Furthermore, at each boundary $x_i=x_{i+1}^-$ it satisfies the correct boundary condition \eqref{equ:pre_int_LL_bc} 
by construction.

Note that \eqref{equ:pre_int_LL_Nparticle} is antisymmetric in $\rapidity$, meaning that if any $\rapidity_i=\rapidity_j$ the wave function vanishes. This is similar to the Pauli principle for fermions: we say that the Lieb-Liniger quasi-particle has fermionic statistics (as opposed to the physical particle which is bosonic).

\begin{remark}
	One can show that \eqref{equ:pre_int_LL_Nparticle} is a complete basis of $L^2\ind{sym}(\mathbb{R}^N)$, in the sense of~\cite{Gaudin_2014}
	\begin{align}
		\int\dd[N]{x}\psi^*(\vec{x}|\vec{\rapidity})\psi(\vec{x}|\vec{\rapidityp}) &= \sum_{\sigma \in \permutation_N} (-1)^\sigma \prod_i\delta(\rapidity_i-\rapidityp_{\sigma_i}),\label{equ:pre_int_LL_basis_1}\\
		\int\dd[N]{\rapidity}\psi(\vec{x}|\vec{\rapidity})\psi^*(\vec{y}|\vec{\rapidity}) &= \sum_{\sigma \in \permutation_N} (-1)^\sigma \prod_i\delta(x_i-y_{\sigma_i}).\label{equ:pre_int_LL_basis_2}
	\end{align}
\end{remark}

As in hard rods we now define the conserved charges for the Lieb-Liniger model as $Q_n = \sum_i \rapidity_i^n$. More precisely, we define the operators $\vu{Q}_n$ as
\begin{align}
	\vu{Q}_n\psi(\vec{x}|\vec{\rapidity}) = \qty[\sum_i \rapidity_i^n]\psi(\vec{x}|\vec{\rapidity}).\label{equ:pre_int_LL_Qn}
\end{align}

\begin{remark}
	One can show that each $\vu{Q}_n$ has a local density (for $n=0,1,2,\ldots$)~\cite[App 2]{doyon2023abinitioderivationgeneralised}. This is important as these charges would otherwise not be relevant for hydrodynamics. 
\end{remark}
 
\begin{remark}
	On a technical note, the rapidities (or asymptotic momenta) $\rapidity_i$ are not the physical momenta. For instance, if we would measure the kinetic energy $\expval{\sum_i (-i\partial_{x_i})^2}$ in a Bethe state, we would not obtain $\expval{\vu{Q}_2} = \sum_i \rapidity_i^2 = \expval{\vu{H}\ind{LL}}$ (which is the total energy).
\end{remark}

\subsubsection{Finite system size and Bethe quantization condition}
For hydrodynamics, we need to be able to compute thermal expectation values in states with a finite density of particles. Unfortunately, no matter how many particles we add, we are always technically in a zero density state. Hence, \eqref{equ:pre_int_LL_Nparticle} is not really useful for thermodynamics and we first need to study the system on a finite size $\ell$ (with periodic boundary conditions). Luckily, states like \eqref{equ:pre_int_LL_Nparticle} are still eigenstates if they additionally satisfy the periodic boundary condition $\psi(x_i=0) = \psi(x_i=\ell)$ for each $i$. This happens if
\begin{align}
	(-1)^{N-1}e^{i\rapidity_i\ell + \sum_{j\neq i} \phi(\rapidity_i-\rapidity_j)} = 1,\label{equ:pre_int_LL_bethe_quant_prod}
\end{align}
which implies the Bethe quantization condition~\cite{jscaux}
\begin{align}
	\rapidity_i + \tfrac{1}{\ell}\sum_{j\neq i} \phi(\rapidity_i-\rapidity_j) = \tfrac{2\pi}{\ell} (I_i + \tfrac{1}{2} \theta(N\;\mathrm{even})),\label{equ:pre_int_LL_bethe_quant_sum}
\end{align}
where the quantum numbers $I_i$ are integers. One can show that for any $\qty{I_i}_{i=1:N}$ a unique solution $\qty{\rapidity_i}$ to \eqref{equ:pre_int_LL_bethe_quant_sum} exists\footnote{This is because $\rapidity_i$ can be obtained as the minimizer of a convex function}, that solutions $\rapidity_i \leftrightarrow \rapidity_j$ are symmetric under exchange of $I_i \leftrightarrow I_j$ and that $\rapidity_i=\rapidity_j$ if $I_i=I_j$. Hence we need to restrict to $I_i\neq I_j$ and the full set of eigenstates of the model is thus given by any integers satisfying $I_1 < I_2 < \ldots < I_n$.

\subsection{Thermodynamic Bethe ansatz (TBA)}
\label{sec:pre_int_TBA}
In the finite system we can now compute thermal properties, like the (grand canonical) partition function
\begin{align}
	Z[\beta] &= \Tr_\ell e^{-\sum_n\beta_n\vu{Q}_n}.\label{equ:pre_int_TBA_Z}
\end{align}
For the Lieb-Liniger model this thermodynamic Bethe ansatz (TBA) was first carried out in~\cite{10.1063/1.1664947}, but similar equations were later also found in quantum spin chains~\cite{10.1143/PTP.46.401,10.1143/PTP.50.1519,Takahashi_1999} and quantum field theories~\cite{ZAMOLODCHIKOV1990695}. We will follow the derivation outlined in~\cite{jscaux},  which is close to the original derivation~\cite{10.1063/1.1664947}.

Note that any Bethe state can be uniquely expressed as $\qty{m_a}_{a\in \mathbb{Z}}$, where $m_a = 1$ if there is $I_i=a$ and zero otherwise. Hence, summing over all quantum states becomes an infinite collection of sums of the form
\begin{align}
	Z[\beta] &= \qty[\prod_{a\in\mathbb{Z}}\sum_{m_a=0}^1] e^{-\sum_a\sum_n\beta_n\rapidity^n_a[\qty{m_a}]} = \qty[\prod_{a\in\mathbb{Z}}\sum_{m_a=0}^1] e^{-\sum_a m_a\beta(\rapidity_a[\qty{m_a}])},\label{equ:pre_int_TBA_Z_2}
\end{align}
where we have again defined $\beta(\rapidity) = \sum_n\beta_n\rapidity^n$. Now group $\zeta$ into cells $\set{A}_\alpha = [(\alpha-1/2)\ell\Delta \zeta,(\alpha+1/2)\ell\Delta \zeta]$ of (integer) size $1 \ll \ell\Delta \zeta \ll \ell$. For each of these cells denote the total number of particles in them by $n_\alpha = \sum_{a\in \set{A}_\alpha} m_a$, which is a number between $0\leq n_\alpha \leq \ell\Delta \zeta$. Assuming that, up to subleading terms, $\rapidity_a[\qty{m_a}] = \rapidity_\alpha[\qty{n_\alpha}]$ depends only on $n_\alpha$, we can split the sum as follows
\begin{align}
	Z[\beta] &= \qty[\prod_{\alpha\in\mathbb{Z}}\sum_{n_\alpha=0}^{\ell\Delta \zeta}] e^{-\sum_\alpha n_\alpha\beta(\rapidity_\alpha[\qty{n_\alpha}])} \prod_{\alpha\in\mathbb{Z}}\qty[\qty(\prod_{a\in \set{A}_\alpha}\sum_{m_a=0}^1) \delta_{\sum_{a\in \set{A}_\alpha} m_a=n_\alpha}]. \label{equ:pre_int_TBA_Z_3}
\end{align}
The last bracket is simply given by 
\begin{align}
	{\ell \Delta \zeta \choose n_\alpha} \sim \exp(-\ell \Delta \zeta \qty[\tfrac{n_\alpha}{\ell\Delta \zeta}\log \tfrac{n_\alpha}{\ell\Delta \zeta} + (1-\tfrac{n_\alpha}{\ell\Delta \zeta})\log (1-\tfrac{n_\alpha}{\ell\Delta \zeta})]) =: \exp(-\ell \Delta \zeta \gamma(\tfrac{n_\alpha}{\ell\Delta \zeta})).\label{equ:pre_int_TBA_combinatorial_factor}
\end{align}
Now we need to understand how $\rapidity_\alpha$ will depend on $\qty{n_\alpha}$. From \eqref{equ:pre_int_LL_bethe_quant_sum} we can infer
\begin{align}
	\rapidity_\alpha + \tfrac{1}{\ell}\sum_\beta n_\beta \phi(\rapidity_\alpha-\rapidity_\beta) = 2\pi \alpha \Delta \zeta + \order{1/\ell}. \label{equ:pre_int_TBA_eq_discrete}
\end{align}
Let us define $\zeta= \alpha \Delta\zeta$, $n(\zeta) = n_\alpha/(\ell \Delta \zeta)$ and $p(\zeta) = p_\alpha$. Then \eqref{equ:pre_int_LL_bethe_quant_sum} becomes
\begin{align}
	2\pi \zeta = \rapidity(\zeta) + \Delta\zeta \sum_\beta n(\zeta_\beta) \phi(\rapidity(\zeta)-\rapidity(\zeta_\beta)) \to \rapidity(\zeta) + \int\dd{\zeta'} n(\zeta') \phi(\rapidity(\zeta)-\rapidity(\zeta')).\label{equ:pre_int_TBA_eq_cont} 
\end{align}
In the last step we took the continuum limit $\Delta \zeta \to 0$. Alternatively, we can write $\zeta(\rapidity)$ and take a derivative w.r.t to $\rapidity$ find that $\dv{\zeta}{\rapidity}$ satisfies
\begin{align}
	1+ \int\dd{\rapidityp} \varphi(\rapidity-\rapidityp) n(\zeta(\rapidityp)) \dv{\zeta(\rapidityp)}{\rapidity} = 2\pi \dv{\zeta(\rapidity)}{\rapidity}.  \label{equ:pre_int_TBA_eq_deriv}
\end{align}
Here $\varphi(\rapidity) = \phi'(\rapidity)$. We will abuse notation and write $n(\rapidity) = n(\zeta(\rapidity))$. Equations like \eqref{equ:pre_int_TBA_eq_deriv} typically appear in quantum integrable models. In fact, for any function $f(\rapidity)$ one defines $f\upd{dr}(\rapidity)$ as the solution to the dressing equation
\begin{align}
	f\upd{dr}(\rapidity) &= f(\rapidity) + \int\tfrac{\dd{\rapidityp}}{2\pi} \varphi(\rapidity-\rapidityp) n(\rapidityp) f\upd{dr}(\rapidityp).\label{equ:pre_int_TBA_dressing}
\end{align}
Comparing with \eqref{equ:pre_int_TBA_eq_deriv}, we identify $\dv{\zeta(\rapidity)}{\rapidity} = \tfrac{1}{2\pi}1\upd{dr}$. The function $n(\rapidity)$ is the occupation function, measuring the proportion of levels occupied in the cell $\alpha$ (it is by construction, bounded by $0\leq n(\rapidity)\leq 1$). The quantity $\rho\ind{s}(\rapidity) = \dv{\zeta(\rapidity)}{\rapidity}$ is interpreted as the number of available states $\dd{\zeta}$ in a small rapidity interval $\dd{\rapidity}$. Their product is the quasi-particle density
\begin{align}
	\rho(\rapidity) = n(\rapidity)\dv{\zeta}{\rapidity} = \int\dd{\zeta} n(\zeta) \delta(\rapidity - \rapidity(\zeta)). \label{equ:pre_int_TBA_rho}
\end{align}
It measures the number of quasi-particles in a small $\dd{\rapidity}$ and can be used to compute conserved quantities as $\tfrac{1}{\ell}Q_n = \int\dd{\rapidity}\rho(\rapidity)\rapidity^n$.

When we take the continuum limit $\Delta \zeta \to 0$ of \eqref{equ:pre_int_TBA_Z_3}, the sum $\prod_{\alpha} \sum_{n_\alpha=0}^{\ell \Delta\zeta}$ approaches a functional integral $\int_0^1\mathrm{D}[n(\zeta)]$, where for each $\zeta\in \mathbb{R}$ the value $n(\zeta)$ is integrated from $0$ to $1$. We finally find
\begin{align}
	Z[\beta] &= \int_0^1\mathrm{D}[n(\zeta)] e^{-\ell \qty[\int\dd{\zeta} \gamma(n(\zeta)) + \int\dd{\zeta} n(\zeta)\rapidity(\zeta)]}\\
	&= \int_0^1\mathrm{D}[n(\zeta)] e^{-\ell \qty[\int\tfrac{\dd{\rapidity}}{2\pi} 1\upd{dr}(\rapidity) \gamma(n(\rapidity)) + \int\dd{\rapidity} n(\rapidity)\rapidity(\rapidity)]} = \int_0^1\mathrm{D}[n(\zeta)]  e^{- \ell f[n(\rapidity)]}.\label{equ:pre_int_TBA_pathintegral}
\end{align}
As $\ell \to \infty$ this integral is dominated by the minimum of $f[n(\rapidity)]$, which we study in the next section.

\section{Generalized hydrodynamics (GHD)}
\label{sec:pre_GHD}
Now that we have seen how GGE states can be treated in the Lieb-Liniger model, we can carry out the procedure to find the hydrodynamic equations as described in \cref{sec:pre_hydro}. This will lead us to generalized hydrodynamics (GHD). GHD (in quantum models) was originally developed in parallel by two groups~\cite{PhysRevX.6.041065,PhysRevLett.117.207201}. The phenomenological derivation given below follows~\cite{PhysRevX.6.041065,BenGHD}. The other phenomenological derivation in~\cite{PhysRevLett.117.207201} is based on a phenomenological kinetic theory picture. At first this may seem like a ``less solid'' derivation, but we would like to remind the reader that hydrodynamics is still also just a phenomenological guess. In fact, we will discuss in the beginning of \cref{sec:LL}, that since the derivation of hydrodynamics works in any model it also cannot describe its model specific origins. Therefore, alternative derivations can be way more insightful. We will demonstrate this by giving such an alternative model specific derivation for hard rods. This is also the main motivation for trying to find a new derivation of quantum models in \cref{sec:LL}.

\begin{remark}\label{rem:pre_GHD_integrable_PDE}
	Related to this discussion, GHD had already been established independently in integrable PDEs in 2003~\cite{EL2003374} using a model specific approach. This was later connected to the hydrodynamic approach in~\cite{Bonnemain_2022}. Intriguingly, in integrable PDEs there is no a priori notion of ``sum over all states'' (as required to define GGE states)~\cite{Babelon_Bernard_Talon_2003}. Hence, the same hydrodynamic equation has to emerge for any kind of ``sum over all states'' notion. This again highlights that there must be more to hydrodynamics than what is suggested by the phenomenological derivation in \cref{sec:pre_hydro}.
\end{remark}
  
\subsection{GHD from TBA}
\label{sec:pre_GHD_TBA}
In the Lieb-Liniger model we found that the free energy at $\beta(\rapidity)$ is given by
\begin{align}
	f[\beta] = \inf_{n} f[n,\beta] = \inf_{n} \int\tfrac{\dd{\rapidity}}{2\pi} 1\upd{dr}(\rapidity)\qty[\gamma(n(\rapidity)) + n(\rapidity)\beta(\rapidity)]. \label{equ:pre_int_GHD_TBA_free_energy} 
\end{align}
It turns out that similar expressions can be found for integrable models of any kind, classical or quantum~\cite{BenGHD,Tongeren_2016}. The details of the model enter only two places. First, the function $\gamma(n)$ describes the entropy dependent on the statistics of the quasi-particle (which may be different from the statistics of the physical particle): for instance, classical particles are described by $\gamma(n) = n\log n - n$, fermions by $\gamma(n) = n\log n+(1-n)\log(1-n)$ and bosons by $\gamma(n) = n\log n - (1+n)\log(1+n)$. Second, the scattering shift $\varphi(\rapidity-\rapidityp)$ describes the scattering of two particles and enters the dressing equation \eqref{equ:pre_int_TBA_dressing}. We will assume that $\varphi(\rapidity)=\varphi(-\rapidity)$ is symmetric and we will define the integral operator $\vu{T}f(\rapidity) = \int\tfrac{\dd{\rapidityp}}{2\pi} \varphi(\rapidity-\rapidityp) f(\rapidityp)$. Using this, we can write $f\upd{dr} = (\vb{1}-\vu{T}n)^{-1} f$. Similarly, we define the transposed dressing operation as $f\upd{drT} = (\vb{1}-n\vu{T})^{-1} f$. We outline basic properties of these operations in appendix \ref{app:dr}, which will be heavily used in the following computation.

The minimum of a functional like \eqref{equ:pre_int_GHD_TBA_free_energy} is obtained when its variation under any perturbation $\delta n(\rapidity)$ vanishes
\begin{align}
	\delta f[n,\beta] &= \int\tfrac{\dd{\rapidity}}{2\pi} [\vu{T}\delta n 1\upd{dr}]\upd{dr} \qty[\gamma + n\beta] + (\gamma' + \beta)\delta n = \int\tfrac{\dd{\rapidity}}{2\pi} \delta n \qty(\vu{T}\qty[\gamma + n\beta]\upd{drT} + \gamma' + \beta) = 0.\label{equ:pre_int_GHD_TBA_free_energy_delta} 
\end{align}
Since this has to vanish for all $\delta n$, we obtain
\begin{align}
	0&= \vu{T}\qty[\gamma + n\beta]\upd{drT} + \gamma' + \beta = \qty[\vu{T}\gamma + \vu{T}n\beta]\upd{dr} + \gamma' + \beta.\label{equ:pre_int_GHD_TBA_free_energy_delta_2} 
\end{align}
Applying $\vb{1}-\vu{T}n$ cancels the dressing and we finally find:
\begin{align}
	0 &= \beta+\gamma' + \vu{T}(\gamma-n\gamma') =: \beta - \varepsilon + \vu{T}F(\varepsilon).\label{equ:pre_int_GHD_TBA_free_energy_delta_3} 
\end{align}
Here, to connect with the usual notation in the literature, we introduced $\varepsilon(\rapidity) = \gamma'(n(\rapidity))$ (interpreted as energy of an excitation with rapidity $\rapidity$) and $F(\varepsilon(\rapidity)) = \gamma-n\gamma'$. Equation \eqref{equ:pre_int_GHD_TBA_free_energy_delta_3} has to be solved for $n$. If a solution is found, it can be inserted into \eqref{equ:pre_int_GHD_TBA_free_energy}
\begin{align}
	f[\beta] = \int\tfrac{\dd{\rapidity}}{2\pi} 1\upd{dr}(\rapidity)\qty[\gamma + n\beta] = \int\tfrac{\dd{\rapidity}}{2\pi} \qty[\gamma + n\beta]\upd{drT} =  \int\tfrac{\dd{\rapidity}}{2\pi} \gamma + n\beta + n\vu{T}\qty[\gamma + n\beta]\upd{drT}.\label{equ:pre_int_GHD_TBA_free_energy_at_min} 
\end{align}
Using \eqref{equ:pre_int_GHD_TBA_free_energy_delta_2}, this can be simplified to
\begin{align}
	f[\beta] =  \int\tfrac{\dd{\rapidity}}{2\pi} \gamma(n(\rapidity))-n(\rapidity)\gamma'(\rapidity) = \int\tfrac{\dd{\rapidity}}{2\pi} F(\varepsilon(\rapidity)).\label{equ:pre_int_GHD_TBA_free_energy_at_min_2} 
\end{align}

This provides us with the free energy density $f[\beta]$, from which we can find $q_n$ by taking derivatives w.r.t $\beta$. We still need to compute the free energy flux $g[\beta]$. Unfortunately, while for $f[\beta]$ there exists an explicit expression, no such expression for $g[\beta]$ is available. Hence, we need to do an educated guess: in a relativistic theory there exists crossing symmetry, roughly telling us that exchanging $x \leftrightarrow t$ is implemented by $\rapidity \leftrightarrow E$ and $q \leftrightarrow j$. Expressing \eqref{equ:pre_int_GHD_TBA_free_energy_at_min_2} in terms of $\zeta$, i.e.
\begin{align}
	f[\beta] =  \int\tfrac{\dd{\zeta}}{2\pi} \dv{\rapidity}{\zeta}\; (\gamma(n(\zeta))-n(\zeta)\gamma'(\zeta)),\label{equ:pre_int_GHD_TBA_free_energy_summary} 
\end{align}
we see that momentum only appears in $\dv{\rapidity}{\zeta}$. Thus, it makes sense to replace $\dv{\rapidity}{\zeta} \to \dv{E}{\zeta} = \dv{E(\rapidity)}{\rapidity} \dv{\rapidity}{\zeta} =: v(\rapidity) \dv{\rapidity}{\zeta}$.
Hence, one proposes the free energy flux $g[\beta]$ to be
\begin{align}
	g[\beta] =  \int\tfrac{\dd{\zeta}}{2\pi} v(\rapidity)\dv{\rapidity}{\zeta}( \gamma(n(\zeta))-n(\zeta)\gamma'(\zeta)) = \int\tfrac{\dd{\rapidity}}{2\pi} v(\rapidity) \qty( \gamma(n(\rapidity))-n(\rapidity)\gamma'(\rapidity)).\label{equ:pre_int_GHD_TBA_free_energy_flux} 
\end{align}

\begin{remark}
	This is of course just a guess. It can be derived in systems where at least one current is itself a conserved quantity (most known systems have this property\footnote{A typical example of this is the total momentum $\sum_i \rapidity_i$, which is also the current of the total particle number $N$.}.)~\cite{10.21468/SciPostPhys.9.3.040}. In addition, there exists many model specific derivations, for instance~\cite{PhysRevLett.117.207201,EL2003374,Borsi_2021,10.1063/1.1692711,doyon2023abinitioderivationgeneralised,PhysRevLett.132.251602}. 
\end{remark}

Now let us take a functional derivative of \eqref{equ:pre_int_GHD_TBA_free_energy_at_min_2} and \eqref{equ:pre_int_GHD_TBA_free_energy_flux} w.r.t $\beta(\rapidity)$ to find
\begin{align}
	\rho(\rapidity) &:= -\int\tfrac{\dd{\rapidityp}}{2\pi} n(\rapidityp) \gamma''(\rapidityp)) \fdv{n(\rapidityp)}{\beta(\rapidity)}, & j(\rapidity) &:= -\int\tfrac{\dd{\rapidityp}}{2\pi} v(\rapidityp) n(\rapidityp) \gamma''(\rapidityp)) \fdv{n(\rapidityp)}{\beta(\rapidity)}.\label{equ:pre_int_GHD_TBA_rhoj_delta} 
\end{align}
Perturbing \eqref{equ:pre_int_GHD_TBA_free_energy_delta_3} under a small change $\delta \beta$, we find that the corresponding $\delta n$ satisfies
\begin{align}
	\gamma''(n)\delta n = -(\delta \beta)\upd{dr},\label{equ:pre_int_GHD_TBA_random_eq} 
\end{align}
hence $\gamma''(n(\rapidity))\fdv{n(\rapidity)}{\beta(\rapidityp)} = - [\delta(\rapidity-\rapidityp)]\upd{dr}$ (where the dressing applies on $\rapidity$). From this we find using \eqref{equ:pre_int_GHD_TBA_rhoj_delta}
\begin{align}
	\rho(\rapidity) &= \tfrac{1}{2\pi} 1\upd{dr}(\rapidity) n(\rapidity), & j(\rapidity) &= \tfrac{1}{2\pi} v\upd{dr}(\rapidity) n(\rapidity).\label{equ:pre_int_GHD_TBA_rhoj} 
\end{align}

This expresses $\rho(\rapidity)$ and $j(\rapidity)$ in terms of the occupation function $n(\rapidity)$. For hydrodynamics, we would like to be able to compute $j(\rapidity)$ directly for a given $\rho(\rapidity)$. Note that by the definition of the dressing $j(\rapidity)$ satisfies $j = \tfrac{1}{2\pi} nv + n\vu{T}j$. Multiplying this by $1\upd{dr}(\rapidity)$ and using
\begin{align}
	1\upd{dr} = 1 + \vu{T}n1\upd{dr} = 1+2\pi \vu{T}\rho\label{equ:pre_int_GHD_TBA_onedr} 
\end{align}
we find $\qty(1 + 2\pi \vu{T})j(\rapidity) = \rho(\rapidity) v(\rapidity) + 2\pi \rho(\rapidity)\vu{T}j(\rapidity)$ or equivalently
\begin{align}
	j = \rho v + 2\pi (\rho\vu{T}j - j \vu{T}\rho ).\label{equ:pre_int_GHD_TBA_j_derivation} 
\end{align}
This is a self-consistency equation for $j(\rapidity)$ given only $\rho(\rapidity)$. This equation is typically expressed in terms of the effective velocity $v\upd{eff}(\rapidity) = j(\rapidity)/\rho(\rapidity) = v\upd{dr}(\rapidity)/1\upd{dr}(\rapidity)$ 
\begin{align}
	v\upd{eff}(\rapidity) &= v(\rapidity) +  \int\dd{\rapidityp} \varphi(\rapidity-\rapidityp) \rho(\rapidityp) (v\upd{eff}(\rapidityp)-v\upd{eff}(\rapidity)).\label{equ:pre_int_GHD_TBA_veff} 
\end{align}
This is the effective velocity equation of GHD and in many ways the most important equation of GHD in general. In terms of the effective velocity we can write the Euler equation \eqref{equ:pre_hydro_macro_continuity} as
\begin{align}
	\partial_t \rho(t,x,\rapidity) + \partial_x (v\upd{eff}(t,x,\rapidity)\rho(t,x,\rapidity)) &= 0,\label{equ:pre_int_GHD_TBA_GHD} 
\end{align} 
where $v\upd{eff}(t,x,\rapidity)$ satisfies \eqref{equ:pre_int_GHD_TBA_veff} at each space-time point $t,x$. Note that \eqref{equ:pre_int_GHD_TBA_GHD} is just an equation for a phase-space density of particles $\rho(t,x,\rapidity)$ moving with velocity $v\upd{eff}(t,x,\rapidity)$, explaining the name effective velocity. 

\begin{remark}
	As for hard rods we have now treated the $\rapidity \in \mathbb{R}$ as the `index' labeling the conserved quantity.
\end{remark}

\begin{remark}
	Here, we only discussed the case of a single species of quasi-particles. In many models there exist more quasi-particles. This means that the index of the conserved quantity is given by $\rapidity \to (a,\rapidity)$, where $a$ labels the species of the particle. This can easily be accounted for in the above derivation: quantities like the quasi-particle density $\rho(t,x,\rapidity) \to \rho_a(t,x,\rapidity)$ receive an additional index $a$ and one has to replace $\int\dd{\rapidity} \to \sum_a \int\dd{\rapidity}$. Furthermore, in some models it is more natural to use a different index $\lambda$, instead of $\rapidity = P(\lambda)$, to label the conserved quantities. This can also be accounted for easily by replacing $1\upd{dr} \to {P'}\upd{dr}$ and $v\upd{dr} \to {E'}\upd{dr}$. In fact this can be seen as a ``gauge'' freedom of GHD~\cite{Bonnemain_2022}
\end{remark}

\begin{remark}
	Using large deviation theory (see appendix \ref{app:LD}), one can compute the following GGE correlation functions from \eqref{equ:pre_int_GHD_TBA_free_energy}~\cite{10.21468/SciPostPhys.5.5.054,SciPostPhys.3.6.039}
	\begin{align}
		C\ind{GGE}(\rapidity,\rapidityp) &= L\expval{\delta\rho(\rapidity)\delta\rho(\rapidityp)} =  \int\frac{\dd{\rapidity_0}}{2\pi} \qty[\delta(\rapidity-\rapidity_0)]\upd{drT}\qty[\delta(\rapidityp-\rapidity_0)]\upd{drT} \frac{1\upd{dr}(\rapidity_0)}{\gamma''(n(\rapidity_0))}, \label{equ:pre_int_GHD_TBA_corr}
	\end{align}
	where the $\cdot\upd{drT}$ act on $\rapidity$ and $\rapidityp$ respectively.
\end{remark}

\subsection{Derivation in hard rods}
\label{sec:pre_GHD_HR}

Before diving deeper into GHD we would like to present an alternative derivation in hard rods based on the explicit solution (\ref{equ:pre_int_HR_contract}-\ref{equ:pre_int_HR_expand}), which will highlight the meaning of \eqref{equ:pre_int_GHD_TBA_veff}~\cite{Boldrighini1983,PhysRevLett.120.045301}. 

Note that we will work in macroscopic coordinates $x\to Lx$, $t \to Lt$ (and $\hat{x} \to L\hat{x}$), which is equivalent to rescaling $d \to d/L$. Note that since $\hat{x}_i(t) < \hat{x}_j(t)$ is equivalent to $x_i(t) < x_j(t)$, we can write \eqref{equ:pre_int_HR_expand} as
\begin{align}
	x_i(t) &= \hat{x}_i + \rapidity_it + \tfrac{d}{L} \sum_{j\neq i} \theta(x_i(t)-x_j(t)).\label{equ:pre_int_GHD_HR_selfconsits} 
\end{align}
Taking a time derivative, we find
\begin{align}
	\dot{x}_i(t) &= \rapidity_i + \tfrac{d}{L} \sum_{j\neq i} \delta(x_i(t)-x_j(t)) (\dot{x}_i(t)-\dot{x}_j(t)).\label{equ:pre_int_GHD_HR_particles_timederiv} 
\end{align}
In the continuum limit we assume that $\rho(t,x,\rapidity) = \tfrac{1}{L}\sum_i \delta(x-x_i(t))\delta(\rapidity-\rapidity_i)$ can be replaced by a continuous distribution. Defining a function $v\upd{eff}(t,x,\rapidity)$ such that $\dot{x}_i(t) = v\upd{eff}(t,x_i(t),\rapidity_i)$, we find
\begin{align}
	v\upd{eff}(t,x,\rapidity) &= \rapidity + d \int\dd{\rapidityp} \rho(t,x,\rapidityp) (v\upd{eff}(t,x,\rapidity)-v\upd{eff}(t,x,\rapidityp)).\label{equ:pre_int_GHD_HR_veff} 
\end{align}
This is the effective velocity equation \eqref{equ:pre_int_GHD_TBA_veff} for a model with $\varphi(\rapidity) = -d$ and $v(\rapidity) = \rapidity$.

\begin{remark}
	This quick derivation is far from a rigorous proof. The arguments can be made more solid by doing a continuum limit of the explicit solution, similar to what we will do in \cref{sec:scbm_GHD}. With this strategy the GHD of hard rods was rigorously proven in 1983~\cite{Boldrighini1983}, long before present day GHD was developed. To my knowledge, this was the first ever rigorous proof of the Euler equation in any interacting many-body system.  
\end{remark}

From this derivation we learn that the effective velocity is indeed the observed velocity of hard rods (also see \cref{fig:pre_int_HR_dynamics} b)). The idea is that in addition to the bare velocity $\rapidity$ the jumps from hitting other particles average to an additional velocity contribution. The contribution from the jumps is given by the individual jump size $d$ multiplied by the frequency of jumps. This frequency is proportional to the particle density (a higher density will imply more scattering events) and to the relative velocity of particles (faster particles will hit more particles). Hence, we obtain the consistency equation \eqref{equ:pre_int_GHD_HR_veff}. 

Note that in hard rods it is possible to find the explicit solution to \eqref{equ:pre_int_GHD_HR_veff}
\begin{align}
	v\upd{eff}(\rapidity) = \frac{\rapidity - d \int\dd{\rapidityp}\rapidityp \rho(\rapidityp)}{1-d\int\dd{\rapidityp} \rho(\rapidityp)}.\label{equ:pre_int_GHD_HR_veff_explicit} 
\end{align}
We can see that the interaction has two effects on the velocity: first, the velocity is shifted by $-d \int\dd{\rapidityp}\rapidityp\rho(\rapidityp)$ depending on the average speed of the other particles. Second, the denominator (since it is smaller than $1$) amplifies the velocity depending on the total particle density $\bar{\rho} = \int\dd{\rapidityp} \rho(\rapidityp)$. Note that $1-d\bar{\rho} = (L-dN)/L$ is the ratio of available space for hard rods (or ratio of length in contracted vs physical space). Since particles ``skip'' a spatial interval of size $d$ every time they hit another particle, they appear faster.

\subsection{General properties of GHD}
\label{sec:pre_GHD_GHD}
\subsubsection{Interpretation of the effective velocity}
\label{sec:pre_GHD_GHD_veff}
\begin{figure}[!h]
	\centering
	\includegraphics{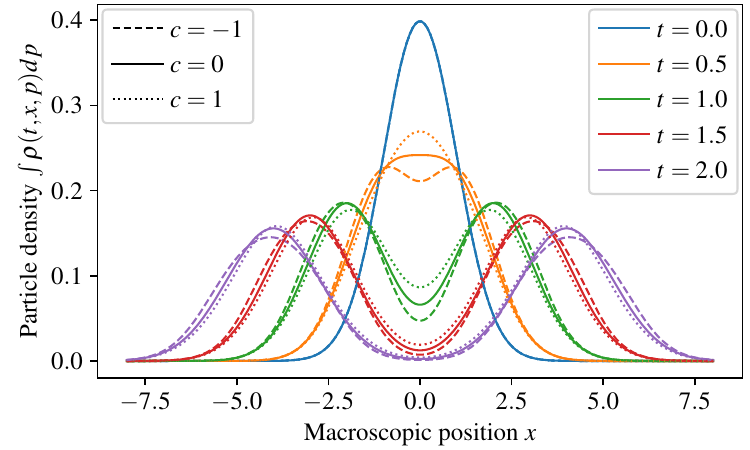}
	\caption[Simulation of GHD]{Simulation of GHD with Lieb-Liniger phase shift for different $c=-1,0,1$ starting from $\rho^0(x,\rapidity) = \tfrac{5A}{4\pi}e^{-x^2/2}\qty(e^{-25(\rapidity-1)^2/2} + e^{-25(\rapidity+1)^2/2})$, where $A=1$. In agreement with our interpretation of the effective velocity we find that for $c<0$ ($c>0$) the density spreads faster (slower) compared to the non-interacting case $c=0$. Note that the effect is strongest in high density regions, while the tails are unaffected by the interaction (in these low density regions evolution is effectively non-interacting). The simulations were done using IFluid~\cite{10.21468/SciPostPhys.8.3.041}. Also see \cref{fig:scbm_GHD} for a simulation with stronger interaction.}
	\label{fig:pre_GHD_numerics}
\end{figure}
Based on our understanding in hard rods we can now interpret \eqref{equ:pre_int_GHD_TBA_veff} in a general model as follows~\cite{PhysRevLett.120.045301}: a quasi-particle with rapidity $\rapidity$ has a bare velocity $v(\rapidity)$. When it scatters with another particle with rapidity $\rapidityp$ it slows down or speeds up, such that after scattering its position is effectively shifted by $\varphi(\rapidity-\rapidityp)$ (forward if $\varphi(\rapidity-\rapidityp) < 0$ and backward if $\varphi(\rapidity-\rapidityp) > 0$). We will see this explicitly later in \cref{fig:scbm_twoparticle}. Furthermore, we interpret $1\upd{dr}(\rapidity) = 1 + \int\dd{\rapidityp}\varphi(\rapidity-\rapidityp)\rho(\rapidityp)$ as available space for a particle with rapidity $\rapidity$.

\begin{remark}\label{rem:pre_GHD_flea_gas}
	This picture can be implemented numerically: evolve particles with their velocity $v(\rapidity)$ and instantaneously shift particle positions whenever they meet. This algorithm is known as the flea gas algorithm~\cite{PhysRevLett.120.045301} of GHD. It works in practice, but it is not an integrable model. The integrable models we introduce in \cref{sec:scbm} can be seen as an improved version of it. 
\end{remark}

\subsubsection{Transport form of the GHD equation}
\label{sec:pre_GHD_GHD_n}
In analogy to the conservation form of GHD equation \eqref{equ:pre_int_GHD_TBA_GHD} for $\rho$, there is also an equation for the occupation function $n$~\cite{PhysRevX.6.041065}: using \eqref{equ:pre_int_GHD_TBA_rhoj} and \eqref{equ:app_dr_delta} we find
\begin{align}
	0 &= \partial_t(1\upd{dr} n) + \partial_x(v\upd{dr} n) = n(\partial_t 1\upd{dr} + \partial_x v\upd{dr}) + 1\upd{dr} \partial_tn + v\upd{dr}\partial_x n\\
	&=  n\qty[\vu{T}(1\upd{dr}\partial_t n + v\upd{dr}\partial_x n)]\upd{dr} + 1\upd{dr} \partial_tn + v\upd{dr}\partial_x n.\label{equ:pre_GHD_n_deriv} 
\end{align}
We see that \eqref{equ:pre_GHD_n_deriv} is satisfied if $1\upd{dr} \partial_tn + v\upd{dr}\partial_x n = 0$, or equivalently
\begin{align}
	\partial_tn(t,x,\rapidity) + v\upd{eff}(t,x,\rapidity)\partial_x n(t,x,\rapidity) = 0.\label{equ:pre_GHD_n} 
\end{align}
This is a transport equation (note that $v\upd{eff}$ is now outside the derivative), which has the following interesting property: imagine that we already know $v\upd{eff}(t,x,\rapidity)$. Consider the GHD characteristics, i.e. particle trajectories $\dot{x}(t) = v\upd{eff}(t,x(t),\rapidity)$ traveling along this velocity field. Now evaluate $n(t,x(t),\rapidity)$ on this trajectory and observe
\begin{align}
	\dv{t} n(t,x(t),\rapidity) = \partial_t n(t,x(t),\rapidity) + \partial_x n(t,x(t),\rapidity)v\upd{eff}(t,x(t),\rapidity) = 0.\label{equ:pre_GHD_characteristic_derivation} 
\end{align}
Hence $n(t,x(t),\rapidity) = n(0,x(0),\rapidity)$ is constant along a GHD characteristic. Recall that $n(t,x,\rapidity)$ describes the average occupation of quantum numbers around rapidity $\rapidity$ at position $x$. We see that GHD never changes this occupation and only moves it to a different location. In particular, if in a fermionic model like Lieb-Liniger the initial state satisfies the physical constraint $0\leq n(t=0,x,\rapidity) \leq 1$, it will satisfy it for all times (this is an important consistency check!).

We would like to point out that from the derivation of \eqref{equ:pre_GHD_n} it follows that
\begin{align}
	\partial_t 1\upd{dr} + \partial_x v\upd{dr} &= 0,\label{equ:pre_GHD_GHD_for_dressing} 
\end{align}
which is a useful formula in GHD.

\subsubsection{Conservation laws}
\label{sec:pre_GHD_GHD_entropies}
By integrating the GHD equation over space we immediately find that for each $\rapidity\in \mathbb{R}$, $Q(\rapidity) = \int\dd{x} \rho(x,\rapidity)$ is indeed conserved. This corresponds to the microscopic conservation laws.

In addition, GHD also conserves quantities of the kind $B(t) = \int\dd{x}\dd{\rapidity} b(n(t,x,\rapidity))1\upd{dr}(t,x,\rapidity)$~\cite{10.21468/SciPostPhys.6.6.070}:
\begin{align}
	\dv{t} B(t) &= \int\dd{x}\dd{\rapidity} (b'(n)\partial_t n 1\upd{dr} + b(n)\partial_t 1\upd{dr}) = -\int\dd{x}\dd{\rapidity} (b'(n)v\upd{eff}\partial_x n 1\upd{dr} + b(n)\partial_x v\upd{dr})\\
	&= -\int\dd{x}\dd{\rapidity} \partial_x (b(n)v\upd{dr}) = 0. \label{equ:pre_GHD_entropies_deriv} 
\end{align}
What are these additional quantities? Note that for $b(n) = \gamma(n)$, the quantity $B$ is the entropy of the GGE \eqref{equ:pre_int_GHD_TBA_free_energy}. We already know from general hydrodnamics that it has to conserve the entropy. However, note that Euler GHD \eqref{equ:pre_int_GHD_TBA_GHD} does not depend on the quasi-particle statistics. Since, in principle, a quantum and a classical particle model could have the same GHD equation (we will construct a classical model with the same GHD as Lieb-Liniger in \cref{sec:scbm}),  \eqref{equ:pre_int_GHD_TBA_GHD} must conserve all possible entropies.

\subsubsection{Contracted coordinates}
\label{sec:pre_GHD_GHD_hatx}
Generalizing the definition of contracted space in hard rods \eqref{equ:pre_int_HR_contract}, one defines the contracted space coordinate $\hat{x} = \hat{X}(t,x,\rapidity)$~\cite{DOYON2018570} in GHD as
\begin{align}
	\hat{X}(t,x,\rapidity) &= x + \int\dd{y}\dd{\rapidityp} \theta(x-y)\varphi(\rapidity-\rapidityp)\rho(t,y,\rapidityp).\label{equ:pre_GHD_hatX_def} 
\end{align}
Its derivative is given by $\partial_x \hat{X}(t,x,\rapidity) = 1\upd{dr}(t,x,\rapidity)$. Hence, space is locally rescaled by $\dd{\hat{x}} = 1\upd{dr}\dd{x}$, which agrees with the interpretation of $1\upd{dr}$ in \cref{sec:pre_GHD_GHD_veff}. Similarly, the time derivative of \eqref{equ:pre_GHD_hatX_def} is
\begin{align}
	\partial_t \hat{X}(t,x,\rapidity) &= -\int\dd{y}\dd{\rapidityp} \theta(x-y)\varphi(\rapidity-\rapidityp)\partial_x (v\upd{eff}(t,x,\rapidityp) \rho(t,y,\rapidityp))\\
	&= -\int\dd{\rapidityp} \varphi(\rapidity-\rapidityp)v\upd{eff}(t,x,\rapidityp) \rho(t,y,\rapidityp) = - \vu{T}nv\upd{dr} = v(\rapidity) - v\upd{dr}(t,x,\rapidity). \label{equ:pre_GHD_hatX_def_timederiv} 
\end{align}

One can now define the quasi-particle density in reduced coordinates $\hat{\rho}(t,\hat{x},\rapidity)$ as the push-forward $\hat{\rho}(t,\cdot,\rapidity) = \hat{X}(t,\cdot,\rapidity)_*\rho(t,\cdot,\rapidity)$ (we introduce the push forward of measures, one of my favorite mathematical notions, in appendix \ref{app:push}). Push forward means that for any observable $\hat{\testfunction}(\hat{x},\rapidity)$ we have
\begin{align}
	\expval{\hat{\rho},\hat{\testfunction}} := \int\dd{\hat{x}}\dd{\rapidity} \hat{\rho}(t,\hat{x},\rapidity)\hat{\testfunction}(\hat{x},\rapidity) = \int\dd{x}\dd{\rapidity}\rho(t,x,\rapidity) \hat{\testfunction}(\hat{X}(t,x,\rapidity),\rapidity)\label{equ:pre_GHD_hatX_weak}  
\end{align}
Taking the time derivative we find
\begin{align}
	\dv{t}\expval{\hat{\rho},\hat{\testfunction}} &= \int\dd{x}\dd{\rapidity}-\partial_x(v\upd{eff}(t,x,\rapidity)\rho(t,x,\rapidity)) \hat{\testfunction}(\hat{X}(t,x,\rapidity),\rapidity)+ \rho(t,x,\rapidity)\nonumber\\
	&\alignshift\times\partial_{\hat{x}} \hat{\testfunction}(\hat{X}(t,x,\rapidity),\rapidity) (v(\rapidity) - v\upd{dr}(t,x,\rapidity))\\
	&= \int\dd{x}\dd{\rapidity}\rho(t,x,\rapidity) \qty[v\upd{eff}(t,x,\rapidity) 1\upd{dr}(t,x,\rapidity) + v(\rapidity) - v\upd{dr}(t,x,\rapidity)] \partial_{\hat{x}} \hat{\testfunction}(\hat{X}(t,x,\rapidity),\rapidity)\\
	&= \int\dd{x}\dd{\rapidity}v(\rapidity)\rho(t,x,\rapidity) \partial_{\hat{x}} \hat{\testfunction}(\hat{X}(t,x,\rapidity),\rapidity) = \expval{\hat{\rho},v(\rapidity)\partial_{\hat{x}}\hat{\testfunction}}.
\end{align}
Since this must hold for any $\hat{\testfunction}(\hat{x},\rapidity)$ we conclude
\begin{align}
	\partial_t \hat{\rho}(t,\hat{x},\rapidity) + v(\rapidity)\partial_{\hat{x}} \hat{\rho}(t,\hat{x},\rapidity) &= 0,\label{equ:pre_GHD_hatX_GHD_eq} 
\end{align}
which is the GHD equation for non-interacting particles. This can be solved explicitly by $\hat{\rho}(t,\hat{x},\rapidity) = \hat{\rho}(0,\hat{x}-v(\rapidity),\rapidity)$. Using \eqref{equ:app_push_local_formula} we can write\sloppy $\rho(t,x,\rapidity) = \hat{\rho}(t,\hat{X}(t,x,\rapidity),\rapidity)\dv{\hat{X}(t,x,\rapidity)}{x} = \hat{\rho}(t,\hat{X}(t,x,\rapidity),\rapidity)1\upd{dr}(t,x,\rapidity)$ and in combination with \eqref{equ:pre_int_GHD_TBA_rhoj} identify
\begin{align}
	\tfrac{1}{2\pi} n(t,x,\rapidity) = \hat{\rho}(t,\hat{X}(t,x,\rapidity),\rapidity).\label{equ:pre_GHD_hatX_n_relation} 
\end{align}

The contracted coordinate is a powerful mapping and will appear throughout the thesis. Starting from an initial $\rho(t=0,x,\rapidity)$ we can compute $\hat{\rho}(t=0,\hat{x},\rapidity)$ and from it $\hat{\rho}(t,\hat{x},\rapidity)$. Unfortunately, unlike hard rods, in general we do not know how to map $\hat{x}$ back to physical coordinates $x$, which we need in order to determine $\rho(t,x,\rapidity)$. This is because \eqref{equ:pre_GHD_hatX_def} depends on $\rho(t,x,\rapidity)$ itself. Instead, it has to be solved self-consistently: for a given $\hat{\rho}(t,\hat{x},\rapidity)$ find a $\rho(t,x,\rapidity)$, s.t. $\hat{\rho}(t,\cdot,\rapidity) = \hat{X}(t,\cdot,\rapidity)_*\rho(t,\cdot,\rapidity)$. Such a self-consistency equation at time $t$ is called a quadrature. 
 
This quadrature can be used to find numerical solutions to the GHD equation. Compared to usual finite element simulations of PDEs like the GHD equation, solving a quadrature numerically might be more involved, but it scales better for long times $t$: finite-element simulations, where one discretizes time into small steps $\Delta t$, accumulate errors with each step. The longer the simulation time $t$, the smaller $\Delta t$ has to be for an accurate result. Therefore, large $t$ require many simulation steps. On the other hand the complexity for solving a quadrature does not depend on $t$. Hence, even if solving a quadrature takes much longer than doing a single step in a finite element simulation, for sufficiently large simulation times $t$, solving a quadrature is more efficient. This idea has been developed and demonstrated in~\cite{DOYON2018570}. One result in this thesis is an upgraded version of this quadrature that in addition also gives many important mathematical insights (see \cref{sec:fixedpoint}). 

\subsubsection{Conjectured absence of shocks}
\label{sec:pre_GHD_GHD_noshock}
As we discussed in \cref{sec:pre_hydro_shock}, Euler hydrodynamics typically predicts its own breakdown by developing shocks (in one dimension).

It had been conjectured that the GHD equation does not produce shocks. This is due to multiple observations. First, GHD formally satisfies the linear degeneracy condition (this is explained and demonstrated in~\cite[Eq. (176)]{BenGHD}), which (in the case of a finite number of conserved charges) is known to prevent shock formation~\cite{10.1093/oso/9780198507000.001.0001}. Roughly speaking, linear degeneracy means that hydrodynamic modes do not self-interact. For a single conservation law, the only such option is $j(q) = vq$, where $v\in \mathbb{R}$ is a constant (this simply describes an evolution by a constant velocity); for a finite number of conservation laws the condition is more complicated. Unfortunately, this condition has so far only been studied for a finite number of conservation laws, but not for the infinite case (furthermore, in GHD, the ``index'' of the conserved quantity is the continuous variable $\rapidity$).

The second observation, made in~\cite{El2011,Pavlov2012}, is that finite component GHD, i.e. studying states of the form $\rho(t,x,\rapidity) = \sum_i\rho_i(t,x)\delta(\rapidity-\rapidity_i)$ reduces to a finite set of hydrodynamic equations. These equations are not only linearly degenerate, but in addition also have a ``semi-Hamiltonian'' structure, for which solutions in terms of a quadrature is known~\cite{MR1086085}. There had been attempts to try to generalize this solution to full GHD~\cite{PhysRevLett.119.220604}, but (up to our work) was only possible for special initial conditions. 

The last observation is more phenomenological: in the Lieb-Liniger model at zero temperature the local states are given by zero entropy states $n(x,\rapidity) = 0,1$, i.e. a (collection of) Fermi-seas. Since a Fermi-sea is only non-zero in an interval $[\rapidity_1,\rapidity_2]$, the hydrodynamic equations reduce to a set of two coupled hydrodynamic equations (since there are only two parameters left). The resulting equation develops shocks. In~\cite{PhysRevLett.119.195301,PhysRevResearch.6.013328} it was observed that GHD does not have a shock, instead it produces another Fermi-sea. 

Due to these observations it was conjectured that GHD should not produce shocks. This conjecture was solved in my PhD, as outlined in \cref{sec:fixedpoint}. 

\subsubsection{Diffusive GHD}
\label{sec:pre_GHD_GHD_diff}
Going beyond Euler hydrodynamics, the Navier-Stokes like diffusive correction (as in \cref{sec:pre_hydro_diff}) to the GHD equation is given by
\begin{align}
	\partial_t \rho + \partial_x(v\upd{eff}\rho) &= \tfrac{1}{2L} \partial_x \qty[\vu{D} \partial_x \rho].\label{equ:pre_GHD_diff_GHD} 
\end{align}
Note that the diffusion matrix is an integral operator with kernel $\vu{D}(\rapidity,\rapidityp)$. Its explicit formula has been derived in~\cite{10.21468/SciPostPhys.6.4.049,PhysRevLett.121.160603,10.21468/SciPostPhys.9.5.075,Durnin_2021}:
\begin{align}
	\vu{D} = (\vb{1}-n\vu{T})^{-1}1\upd{dr}\vu{\tilde{D}} \tfrac{1}{1\upd{dr}}(\vb{1}-n\vu{T}),  \label{equ:pre_GHD_diff_D1} 
\end{align}
where
\begin{align}
	1\upd{dr}(\rapidity)^2\vu{\tilde{D}}(\rapidity,\rapidityp) &= w(\rapidity)\delta(\rapidity-\rapidityp) - W(\rapidity,\rapidityp)\\
	W(\rapidity,\rapidityp) &= \tfrac{1\upd{dr}(\rapidity)}{2\pi\gamma''(n(\rapidity))} \qty[\varphi\upd{dr}(\rapidity,\rapidityp)]^2 \abs{v\upd{eff}(\rapidity)-v\upd{eff}(\rapidityp)}\label{equ:pre_GHD_diff_D2_W}\\
	w(\rapidityp) &= \int\dd{\rapidity} W(\rapidity,\rapidityp).\label{equ:pre_GHD_diff_D2_w} 
\end{align}
Here, $\varphi\upd{dr}(\rapidity,\rapidityp) = \qty[\varphi(\cdot - \rapidityp)]\upd{dr}(\rapidity,\rapidityp)$ is the dressing of $\varphi(\rapidity-\rapidityp)$ in $\rapidity$. An in-depth discussion of the properties of \eqref{equ:pre_GHD_diff_GHD}, in particular that it increases entropy can be found in~\cite{10.21468/SciPostPhys.6.4.049}.

We would like to note that diffusive GHD \eqref{equ:pre_GHD_diff_GHD}, unlike Euler GHD, depends on the particle statistics (due to the $\gamma''$ in \eqref{equ:pre_GHD_diff_D2_W}). Hence, the evolution of $\rho(t,x,\rapidity)$ of a classical and a quantum model with the same Euler GHD will differ by $\order{1/L}$.

This equation has been rigorously proven in hard rods if the state at time $t$ is local equilibrium state~\cite{Boldrighini1997}\footnote{The proof is restricted to states with a constant particle density.}. One result of this thesis (discussed in \cref{sec:diff}) is that \eqref{equ:pre_GHD_diff_GHD} is not the correct equation on the diffusive scale. This is due to the long range correlations discussed in \cref{sec:pre_hydro_bmft} (meaning that the state is not a local equilibrium state).

\begin{remark}
	As GHD does not produce shocks, Euler hydrodynamics is sufficient to describe the system at very large scales. The solution to \eqref{equ:pre_GHD_diff_GHD} is remains $\order{1/L}$ close to the Euler solution. Therefore, it is extremely hard to observe (and check) \eqref{equ:pre_GHD_diff_GHD} in experimental and numerical setups.  
\end{remark}
\begin{remark}
	Diffusive GHD still leads to an $\order{1}$ effect on the very long diffusive time-scale $t \to L^2 t$, where it was believed to describe thermalization. In light of the new diffusive GHD equation, we will argue in \cref{sec:diff_entropy} that thermalization should occur earlier.  
\end{remark}

\subsection{Further developments in GHD}
\label{sec:pre_GHD_further}
What was mentioned so far is only the tip of the iceberg for the developments of GHD. When first introduced, GHD solved a very hard problem: predicting dynamics in quantum integrable models (at least on large scales). This was crucial to connect with experiments such as the one presented in \cref{fig:intro_GHD}. But more importantly it introduced a way of thinking about integrable models, which revolutionized the field of integrability. In the following we would like to give an overview of some further developments. 

\subsubsection{GHD with external potential}
\label{sec:pre_GHD_further_potential}
If one applies an external potential to an integrable model, integrability and translation invariance (two crucial ingredients for GHD) are both broken. However, if the external potential is slowly varying, then we can assume it to be constant in each fluid cell and thus the system should still thermalize locally to a GGE. Following this idea, one obtains the following GHD equation in an external potential~\cite{10.21468/SciPostPhys.2.2.014}
\begin{align}
	\partial_t \rho(t,x,\rapidity) + \partial_x ((\partial_\rapidity E)\upd{eff}(t,x,\rapidity) \rho(t,x,\rapidity)) - \partial_\rapidity ((\partial_x E)\upd{eff}(t,x,\rapidity) \rho(t,x,\rapidity)) &= 0,\label{equ:pre_GHD_further_external} 
\end{align}
where $E(x,\rapidity)$ describes the position dependent single particle energy and $f\upd{eff}$ is the solution to \eqref{equ:pre_int_GHD_TBA_veff} with $v$ replaced by $f$. This equation explicitly does not conserve most of the conserved quantities of the integrable model, but interestingly still conserves all entropies \eqref{equ:pre_GHD_entropies_deriv}~\cite{10.21468/SciPostPhys.6.6.070}.

In a similar spirit, GHD has been extended to include slow changes in the interaction parameter (i.e. $c \to c(t,x)$ in \eqref{equ:pre_int_LL_def})~\cite{PhysRevLett.123.130602}, external noise~\cite{PhysRevB.102.161110} and particle losses~\cite{10.21468/SciPostPhys.9.4.044}.

\subsubsection{Impurities}
\label{sec:pre_GHD_further_impurity}
If an external potential is very localized, i.e. it is an impurity, then the GHD equation holds everywhere except at the location of the impurity. The impurity can be taken into account as a boundary condition for the GHD equation at this point (relating the incoming to the outgoing state). Those have been found in integrable impurities~\cite{PhysRevLett.131.156303} and for external potentials of a mesoscopic size~\cite{Hübner_2024}.

\subsubsection{Quantum GHD}
\label{sec:pre_GHD_further_QGHD}
Since GHD describes finite temperature states, quantum effects are strongly suppressed. To observe them, one has to also perform a low-temperature limit in addition to the hydrodynamic limit. In this regime quantum fluctuations should be transported on top of the evolving GHD background. Quantum GHD~\cite{PhysRevLett.124.140603} gives a time-dependent quadratic Hamiltonian for the evolution of such fluctuations. Quantum GHD is a good starting point to understand the evolution quantum properties such as entanglement in integrable models, e.g.~\cite{Ruggiero_2022,Scopa_2022,Capizzi_2023}.

\subsubsection{Experimental verifications of GHD}
\label{sec:pre_GHD_further_experiments}
GHD and related results have been verified experimentally in a number of different systems. This includes cold atom systems like in \cref{fig:intro_GHD} modeling the Lieb-Liniger model~\cite{PhysRevLett.122.090601,PhysRevLett.126.090602,doi:10.1126/science.abf0147,PhysRevX.12.041032}. GHD has also been studied in the presence of a trapping potential, like in the famous quantum Newton cradle experiment~\cite{Kinoshita2006,10.21468/SciPostPhys.6.6.070}. Furthermore, GHD of integrable PDEs is studied in optical fibres~\cite{PhysRevResearch.5.L042002} and water tanks~\cite{PhysRevE.109.034207}.
 
\subsubsection{Numerical implementations of GHD}
\label{sec:pre_GHD_further_numerics}
The GHD equation is still an incredibly complicated PDE. The canonical way to solve such a PDE is via a finite element method (discretizing $t, x$ and $\rapidity$). However, in each time step of the simulation one has to solve the effective velocity equation \eqref{equ:pre_int_GHD_TBA_veff} at every point, which is time intensive. Therefore, it is crucial to minimize the number of the space and time steps, by using well-designed stable and accurate algorithms. From my experience the most efficient (by far) implementation up to date is the publicly available package IFluid~\cite{10.21468/SciPostPhys.8.3.041,MOLLER2023112431}, which, among others, implements 4th order accurate BSL (backwards semi-Lagrangian) schemes for the GHD equation \eqref{equ:pre_int_GHD_TBA_GHD} and also for its extensions like \eqref{equ:pre_GHD_diff_GHD} and \eqref{equ:pre_GHD_further_external}.

As a byproduct of the work carried out in thesis, we uncover alternative numerical methods: as we discuss in \cref{sec:fixedpoint}, one can solve \eqref{equ:pre_int_GHD_TBA_GHD} by solving the fixed point problem \eqref{equ:fixedpoint_fixedpoint} numerically. This directly gives the solution of the GHD equation at a given space time point $t,x$, so it may be more efficient for long time simulations. However, this is only possible without the presence of an external potential. Similarly, GHD can also be solved by finding the stationary point of an action, see \cref{sec:scbm_app_optimization}.

Another strategy is to simulate GHD via a classical particle model specifically designed in such a way that its large scale dynamics is given by the correct GHD equation. One example of this is the flea gas algorithm (see \cref{rem:pre_GHD_flea_gas}), or the \scbm s we introduce in \cref{sec:scbm}. We demonstrate this in \cref{fig:scbm_GHD}. While particle simulations are very efficient, the outcome has to be averaged over many samples. Still, depending on the scenario and on the observable of interest, this approach might converge faster than finite-element simulations.

%!TEX root = thesis.tex

\chapter{Ab initio derivation of GHD in quantum models}
\label{sec:LL}
This chapter concerns a new derivation of GHD in the Lieb-Liniger model, which will also be applicable to any other integrable model solved via the coordinate Bethe ansatz. Unfortunately, it is not complete. Still, it provides an alternative way to think about GHD in quantum models, which will also be the basis for \cref{sec:scbm}. 

\begin{remark}
	I found the simple derivation leading to the classical particle model \eqref{equ:LL_SP_QP_def}, from which GHD follows directly, early on in my PhD. This shows that the wave function behaves in an intuitive GHD-like manner. However, this is GHD on the wrong object! GHD is supposed to describe expectation values of charge densities, which are non-trivial to obtain from the wave function. Together with my supervisor, I published the preprint~\cite{doyon2023abinitioderivationgeneralised}, in which GHD was derived using some (unverified) assumptions. Reinvestigating the problem at some later point, I got uncertain of these assumptions. It is an ongoing discussion whether or not these assumptions hold, and there are many other proposals to derive GHD. Below, I give an overview over my favorite line of attempt, which will lead us extremely close to the GHD equation.
\end{remark}

\section{Why do we need another derivation?}
\label{sec:LL_why}
In the same way that the derivation of GHD in hard rods (\cref{sec:pre_GHD_HR}) is more intuitive (and significantly simpler) than deriving GHD phenomenologically using TBA (\cref{sec:pre_GHD_TBA}), the hope of this alternative ``ab initio'' derivation is to gain a better understanding of what is actually happening in the quantum model from the perspective of the microscopic wave function. ``Ab initio'' here means that we  will start from the exact solution to the microscopic dynamics and try to derive GHD.

In general, the phenomenological derivation of hydrodynamics presented in \cref{sec:pre_hydro} is somewhat unphysical. This is because, thermalization in many-body systems does not actually mean that the local state of the system is given by \eqref{equ:pre_hydro_LES}, but only that it appears as such when measuring local observable. Instead, in a quantum system the state could be given by a pure state, that locally looks like an eigenstate of the system. Then, assuming the (generalized) eigenstate thermalization hypothesis\footnote{This states that expectation values of local observables in eigenstates are close to their GGE values.}~\cite{DAlessio03052016}, measuring any local observable will give a value close to a GGE value. This means on the level of one-point functions we will not be able to distinguish such pure states from local equilibrium states \eqref{equ:pre_hydro_LES}.

The phenomenological derivation in \cref{sec:pre_hydro} cannot provide any understanding of what is happening there. For Euler hydrodynamics this might not be important, but further understanding might enable to apply hydrodynamic ideas to other problems. 

In addition to this, the standard phenomenological derivation is also demanding: we want to study a large scale problem on the infinite system. However, we now split space into fluid cells: in order to compute thermal states in them, we need to first consider finite $\ell$, which gives rise to the complicated Bethe quantization conditions\footnote{Not to mention the fact that we assume periodic boundary conditions for these fluid cells, which is clearly unphysical: fluid cells are connected to the neighboring fluid cells, not to themselves.}. Next, we need to do the $\ell \to \infty$ limit of those to achieve the TBA. The interaction, encoded in the phase shift $\phi(\rapidity-\rapidityp)$, determines the ratio of quantum numbers per rapidity interval, leading to the interaction term in the effective velocity equation \eqref{equ:pre_int_GHD_TBA_veff}.

But all of this does not exist on the infinite system. If we have $N$ particles, we can put them into a finite region of size $L$, with arbitrarily high density\footnote{The only reason why we cannot study GGE states on the infinite line is because GGE states are uniform. Hence their average density is always $N/\infty = 0$.}. If we never use fluid cells then we do not need the Bethe quantization condition (any set of rapidities gives rise to an eigenstate). So how does the interaction come into play here? How does GHD emerge? 

This discussion also shows that a derivation starting from the microscopic wave function might not only be more physical, but also might be significantly simpler. For instance, our attempt at a new derivation does not require to assume the string hypothesis needed to study models with strings. We explain this in \cref{sec:LL_attractive} on the example of the attractive Lieb-Liniger model.

\begin{remark}
	Another potential strategy to derive the GHD equation is to use a perturbative series around the free fermion point~\cite{PhysRevLett.128.190401}.
\end{remark}

\section{Stationary phase approximation of the Bethe phase}
\label{sec:LL_SP}
Our starting point is that any $N$-particle wavefunction $\Psi(t,\vec{x})$ can be decomposed into eigenstates \eqref{equ:pre_int_LL_Nparticle}, whose time evolution is trivial
\begin{align}
	\Psi(t,\vec{x}) &= \int\dd[N]{\rapidity} A(\vec{\rapidity}) \psi(\vec{x}|\vec{\rapidity}) e^{-i\sum_i \rapidity_i^2 t}.\label{equ:LL_SP_psi_starting_point}
\end{align}
Here $A(\vec{\rapidity})$ can in principle be an arbitrary square-integrable function\footnote{Only the anti-symmetric part of $A(\vec{\rapidity})$ will matter.}, but since we are interested in hydrodynamics it should describe a large scale distribution of particles. An intuitive choice would be for instance a product of the form
\begin{align}
	A(\vec{\rapidity}) &= \prod_{i=1}^N \unsymA_i\qty(\tfrac{\rapidity_i-\rapidity^0_i}{\Delta \rapidity_i}) e^{-i\sum_i L \rapidity_i \hat{x}_i^0}.\label{equ:LL_SP_psi_amplitude}
\end{align}
Here each particle is described by a wave packet of the form $\unsymA_i$ (say a Gaussian) centered around rapidity $\rapidity_i^0$ and location controlled by $\hat{x}_i^0$ (as we will see $\hat{x}_i^0$ is not directly the location of a wave packet in physical space, see \eqref{equ:LL_SP_QP_def} for their relation). The width of the wave packet in rapidity space is controlled by $\Delta \rapidity_i$, which by the uncertainty principle should correspond to a width in space $\Delta x_i \sim 1/\Delta \rapidity_i$. Since we are interested in the Euler scaling limit where $L \sim T \to \infty$, let us change to macroscopic coordinates $\vec{x}\to L\vec{x}$, $t\to Lt$:
\begin{align}
	\Psi(Lt,L\vec{x}) &=\frac{\chi(\vec{x})}{\mathcal{N}_N} \sum_{\sigma\in \permutation_N} (-1)^\sigma \int\dd[N]{\rapidity} e^{-iLS_t(\vec{x}_{\sigma^{-1}},\vec{\hat{x}}^0,\vec{\rapidity})} \prod_i\unsymA_i\qty(\tfrac{\rapidity_i-\rapidity^0_i}{\Delta \rapidity_i}), \label{equ:LL_SP_psi_macroscopic}
\end{align}
where the total phase is given by
\begin{align}
	S_t(\vec{x},\vec{\hat{x}}^0,\vec{\rapidity}) &= \sum_i \rapidity_i (x_i-\hat{x}_i^0) +\tfrac{1}{4L}\sum_{i\neq j}\sgn(x_i-x_j) \phi(\rapidity_i-\rapidity_j)- \sum_i\rapidity_i^2t.\label{equ:LL_SP_phase}
\end{align}
The fast-oscillating phase in \eqref{equ:LL_SP_psi_macroscopic} suggests to approximate it via the stationary phase approximation, i.e.\ the dominating part of the $\rapidity$ integral should come from the point where
\begin{align}
	0 &= \partial_{\rapidity_i} S_t(\vec{x},\vec{\hat{x}}^0,\vec{\rapidity}) = x_i-\hat{x}_i^0 + \tfrac{1}{2L} \sum_{j\neq i}\sgn(x_i-x_j) \varphi(\rapidity_i-\rapidity_j)- 2\rapidity_i t.\label{equ:LL_SP_phase_SP}
\end{align}
In appendix \ref{app:convexity_of_bethe_phase} we show that the Hessian $\vb{H}_{ij}(t,\vec{x},\vec{\hat{x}}^0,\vec{\rapidity}) = \partial_{\rapidity_i}\partial_{\rapidity_j} S_t(\vec{x},\vec{\hat{x}}^0,\vec{\rapidity})$ is always negative definite if $t>t\ind{c}=\tfrac{N}{L}\sup_\rapidity\abs{\varphi'(\rapidity)}$, hence $S_t(\vec{x},\vec{\hat{x}}^0,\vec{\rapidity})$ is a strictly concave function in $\vec{\rapidity}$. This implies that at least for $t>t\ind{c}$ a solution to \eqref{equ:LL_SP_phase_SP} always exists and is unique. Denoting the solution to \eqref{equ:LL_SP_phase_SP} by $\vec{\rapidity}(t,\vec{x},\vec{\hat{x}}^0)$, we can write the wave function approximately as
\begin{align}
	\Psi(Lt,L\vec{x}) &\sim\frac{\chi(\vec{x})}{\mathcal{N}_N} \sum_{\sigma\in \permutation_N} (-1)^\sigma \frac{e^{-iLS_t(\vec{x}_{\sigma^{-1}},\vec{\hat{x}}^0,\vec{\rapidity}(t,\vec{x}_{\sigma^{-1}},\vec{\hat{x}}^0))}}{\sqrt{\det \vb{H}(t,\vec{x}_{\sigma^{-1}},\vec{\hat{x}}^0,\vec{\rapidity}(t,\vec{x}_{\sigma^{-1}},\vec{\hat{x}}^0))}} \prod_i\unsymA_i\qty(\tfrac{\rapidity_i(t,\vec{x}_{\sigma^{-1}},\vec{\hat{x}}^0)-\rapidity^0_i}{\Delta \rapidity_i}).\label{equ:LL_SP_SP_WF}
\end{align}

\subsection{Gas of wave packets and quasi-particle trajectories}
\label{sec:LL_SP_QP}
\begin{figure}[!h]
	\centering
	\includegraphics{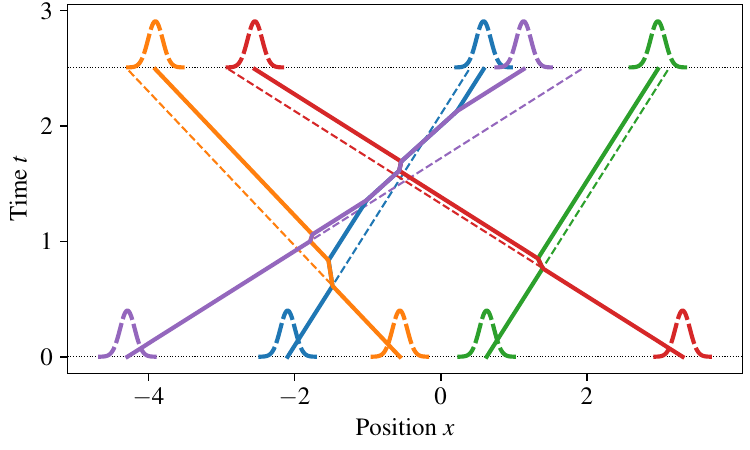}
	\caption[Evolution of Lieb-Liniger wave packets]{Evolution of Lieb-Liniger wave packets: the position of Lieb-Liniger wave packets (sketched as Gaussian bumps), follow trajectories (solid lines) given as solutions to \eqref{equ:LL_SP_QP_def}. During scattering particles `stick together' for a short amount of time, giving rise to an effective positions shift $\varphi(\rapidity-\rapidityp)$ compared to their non-interacting evolution (dashed lines). This figure was reproduced from~\cite{doyon2023abinitioderivationgeneralised}.}
	\label{fig:LL_scbm}
\end{figure}

Now consider the case where $1/L \ll \Delta \rapidity_i \ll 1$ is small. With this choice, both the spreading of the wave function in rapidity and macroscopic position $\Delta x/L \sim 1/\Delta \rapidity_i \ll 1$ is small (hence we effectively deal with a classical particle). This means that the wave function will de facto vanish for all $\vec{x}$ unless $\vec{\rapidity}(t,\vec{x}_{\sigma^{-1}},\vec{\hat{x}}^0) \approx \vec{\rapidity}^0$ for some permutation $\sigma$. We therefore interpret the inverse $\vec{x}(t,\vec{\rapidity},\vec{\hat{x}}^0)$ to $\vec{\rapidity}(t,\vec{x},\vec{\hat{x}}^0)$ as the trajectories of the individual wave packet $x_i(t) = x_i(t,\vec{\rapidity}^0,\vec{\hat{x}}^0)$ with rapidity $\rapidity^0_i$ and initial location controlled by $\hat{x}^0_i$. We call this the \textit{gas of wave packets}, depicted in \cref{fig:LL_scbm}, which will correspond to the quasi-particles in GHD.

From \eqref{equ:LL_SP_phase_SP}, the trajectories satisfy the following equation
\begin{align}
	\hat{x}_i^0 + 2\rapidity_i t &= x_i(t,\vec{\rapidity},\vec{\hat{x}}^0) + \tfrac{1}{2L} \sum_{i\neq j} \sgn(x_i(t,\vec{\rapidity},\vec{\hat{x}}^0)-x_j(t,\vec{\rapidity},\vec{\hat{x}}^0))\varphi(\rapidity_i-\rapidity_j).\label{equ:LL_SP_QP_def}
\end{align} 
Note that, due to the jump of the $\sgn(x)$ function a solution does not necessarily exists. However, \eqref{equ:LL_SP_QP_def} can formally be viewed as the minimization condition of the (strictly) convex function
\begin{align}
	\mcl{A}_t(\vec{x},\vec{\rapidity},\vec{\hat{x}}^0) &= \tfrac{1}{2} \sum_i \qty(x_i - \hat{x}_i^0 - 2\rapidity_i t)^2 + \tfrac{1}{4L} \sum_{i\neq j} \abs{x_i-x_j}\varphi(\rapidity_i-\rapidity_j).\label{equ:LL_SP_QP_action}
\end{align}
This function is convex since $\varphi(\rapidity) > 0 $. Therefore, a unique minimizer always exists\footnote{The minimizer of a (strictly) convex function always exists, even if it is not differentiable.}, which we can use as a definition for $\vec{x}(t,\vec{\rapidity},\vec{\hat{x}}^0)$. Alternatively, we can regularize the sign function by a smooth function, say $\sgn_\alpha(x) = \tanh(x/\alpha)$ for $\alpha\to 0$. The latter will allow us to properly define this particle model as a classical integrable model, which we discuss in detail in \cref{sec:scbm}. Here, we will stick with the straightforward definition as minimizer of \eqref{equ:LL_SP_QP_action}.

Solutions to this equation are depicted in \cref{fig:LL_scbm}. We can observe that particles, unless they scatter, simply follow straight trajectories with velocity $2\rapidity_i$. During scattering, two (or more) particles stick together for a short time and evolve as a composite particle. This phenomenon, and the associated non-differentiability of the trajectories, originates from \eqref{equ:LL_SP_QP_action}. If particles are close, it is beneficial to put two particles at the same $x_i=x_j$ in order to minimize \eqref{equ:LL_SP_QP_action}. If instead one chooses to smoothens the $\sgn(x)$ functions, the trajectories will be a smooth approximation of this behavior. 

By construction, particles $i$ and $j$ scatter for precisely the time such that after scattering their trajectories are displaced in space by $\tfrac{1}{L}\varphi(\rapidity_i-\rapidity_j)$ (see discussion in \cref{sec:scbm_scbm_twoparticle}). This is precisely the behavior we expect from the GHD quasi-particles.

Doing a similar analysis as for hard rods (see \cref{sec:pre_GHD_HR}), it is easy to see that $L \to \infty$ leads to the GHD equation (we will do a more careful derivation in \cref{sec:scbm_GHD}). Indeed, taking a time-derivative of \eqref{equ:LL_SP_QP_def} and denoting $v\upd{eff}(t,x_i(t,\vec{\rapidity},\vec{\hat{x}}^0),\rapidity_i) = \dot{x}_i(t,\vec{\rapidity},\vec{\hat{x}}^0)$ and $\rho(t,x,\rapidity) = \tfrac{1}{L}\sum_i \delta(x-x_i(t,\vec{\rapidity},\vec{\hat{x}}^0))\delta(\rapidity-\rapidity_i)$ we find
\begin{align}
	2\rapidity &= v\upd{eff}(t,x,\rapidity) + \int\dd{\rapidityp} \varphi(\rapidity-\rapidityp) \rho(t,x,\rapidityp) (v\upd{eff}(t,x,\rapidity)-v\upd{eff}(t,x,\rapidityp)),\label{equ:LL_SP_QP_GHD_quick} 
\end{align}
which is just the effective velocity equation \eqref{equ:pre_int_GHD_TBA_veff} of GHD with the Lieb-Liniger scattering phase $\varphi(\rapidity) = \phi'(\rapidity) = \tfrac{2c}{c^2+\rapidity^2}$ (note that the bare velocity of Lieb-Liniger is $v(\rapidity) = 2\rapidity$, since the usual $1/2$ is missing from its kinetic term in \eqref{equ:pre_int_LL_def}). 

\begin{result}
	The large scale locations of localized wave-packets in the Lieb-Liniger model evolve like the quasi-particles of its GHD\footnote{These wave packets behave very similar to solitons observed in integrable PDEs.~\cite{eilenberger2012solitons}}. This is the microscopic origin of the quasi-particle interpretation of GHD in quantum models.\label{res:modulation_localized_wavepackets}
\end{result}

We choose the specific setting \eqref{equ:LL_SP_psi_amplitude} as it is pedagogical, but the idea generalizes to other settings: GHD should emerge via a stationary phase approximation that confines the evolution along the classical trajectories $x_i(t,\vec{\rapidity}^0,\vec{\hat{x}}^0)$. This classical particle model (which is indeed integrable, see \cref{sec:scbm_canonical_trafo}), acts as an intermediate mesoscopic classical theory, which in the large scale limit will give rise to GHD.

For instance, in this setting we assumed that $\Delta \rapidity_i$ is small, so that it was justified to consider only a single configuration $\rapidity_i \to \rapidity_i^0$. But this is not necessary: if $\Delta \rapidity_i$ is not small, then the initial state can be seen as a quantum superposition of classical configurations. For each $\vec{\rapidity}$, each of these configurations will independently evolve according to the classical evolution $\vec{x}(t,\vec{\rapidity},\vec{\hat{x}}^0)$. At time $t$ the wave function is then given by a quantum super position of these trajectories weighted with their initial amplitude $A(\vec{p})$. This is similar to the emergence of classical mechanics from quantum mechanics in the path integral picture: a stationary phase approximation of the quantum phase confines quantum trajectories to the classical trajectory. However, since the wave function at initial time is a superposition of many classical configurations, the final wave function will still be a quantum superposition of the evolved configurations.

\begin{result}
	Extending \cref{res:modulation_localized_wavepackets}, on large scales, the wave function of the Lieb-Liniger model is given by a quantum superposition of classical configurations, each of which evolves like the quasi-particles in GHD.
\end{result}

\begin{remark}
	In light of the emergence of a classical particle model, it seems tempting to think of this particle model as a semi-classical limit of the Lieb-Liniger model. Even though we will indeed later call these models semi-classical Bethe models, we would like to stress that they are not: the limit we are taking here is not $\hbar \to 0$ (in fact $\hbar = 1$ is constant throughout the computation), but instead a large scale limit $L\sim T\sim N \to \infty$. For instance, if $N$ is kept finite, while $L\to \infty$, then the scattering shifts would be negligible as they are $\order{1/L}$. It is only due to $N\sim L$ that they add to a finite contribution.
\end{remark}

Before concluding this derivation we would like to point out an important result that might easily be overlooked: equation \eqref{equ:LL_SP_SP_WF} gives a concrete approximation for the wave function at time $t$. It is not clear how good this approximation is, whether it can be controlled in some norm and under what assumptions. It is obtained via a fairly uncontrolled stationary phase approximation, where the number of integrals $N\sim L$ increases as well (also note the quickly increasing permutation sum we have fully ignored so far). But even if its precision is unclear, it is a formula that can provide intuition to understand dynamics of the full wave function, also to situations where GHD might not apply! It is a prediction for the microscopic theory that can (in principle) be checked or falsified. The phenomenological derivation of GHD in \cref{sec:pre_GHD_TBA} cannot provide that, due to the various reasons laid out in \cref{sec:LL_why}. 

\begin{result}
	Equation \eqref{equ:LL_SP_SP_WF} is a prediction of the dynamics of the microscopic wave function in the GHD regime. Accuracy and justification of this approximation are unclear.
\end{result}

\begin{openproblem}
	To what extend and under what conditions is \eqref{equ:LL_SP_SP_WF} justified. What are the correction terms? 
\end{openproblem}

\section{Two attempts on deriving GHD}
The findings in the previous section are interesting, but they are not yet GHD. GHD is not a statement about the evolution of the wave function, but instead a prediction about the evolution of observables. It is actually quite interesting that GHD also predicts the dynamics of wave packets, because the density of wave packets is not the quasi-particle density of GHD. This is due to the sum of permutations, which suppresses wave packets if they come too close.

We are going to present two approaches to derive GHD from the wave function, each avoiding the treatment of permutations from their own perspective. As we will see, they both get quite close, but their failure also highlights the need to properly deal with permutations.

\begin{remark}
	This second part of the chapter is going to be significantly more technical than the first part. It is also not relevant for the remaining parts of the thesis. Before continuing here, it might also be interesting to first take a look at \cref{sec:scbm}, where we study \eqref{equ:LL_SP_QP_def} in more detail.
\end{remark}

\subsection{Definition of the quasi-particle density}

For these computations we first need to define the (large scale) quasi-particle density $\quasiparticleop(x,\rapidity)$. We will define it in a distributional sense as follows: for any one-particle observable $\testfunction(x,\rapidity)$ define the operator $\expval{\quasiparticleop, \testfunction} = \int\dd{x}\dd{\rapidity} \quasiparticleop(x,\rapidity) \testfunction(x,\rapidity)$  acting on a wave function as
\begin{align}
	\expval{\quasiparticleop, \testfunction} \psi(\vec{x}|\vec{\rapidity}) &= \frac{\chi(\vec{x})}{\mathcal{N}_N}\sum_{\sigma\in \permutation_N} (-1)^\sigma \qty[\frac{1}{L}\sum_i \testfunction\qty(\tfrac{x_{\sigma^{-1}_i}}{L},\rapidity_i)] e^{i\Phi(\vec{x}_{\sigma^{-1}}|\vec{\rapidity})}.\label{equ:LL_deriv_cq_def}
\end{align}
Why do we choose this definition? First observe that if $\testfunction(x,\rapidity) = \testfunction(\rapidity)$ is independent of $x$, this simply measures the total charge associated with $\testfunction(\rapidity)$ divided by the macroscopic length scale
\begin{align}
	\expval{\quasiparticleop, \testfunction}\eval_{\testfunction(x,\rapidity) = \testfunction(\rapidity)} \psi(\vec{x}|\vec{\rapidity}) &= \qty[\frac{1}{L}\sum_i \testfunction\qty(\rapidity_i)] \psi(\vec{x}|\vec{\rapidity}).\label{equ:LL_deriv_cq_def_total_charge}
\end{align}
For instance, if $\testfunction(\rapidity) = 1$, it measures the total particle density, if $\testfunction(\rapidity) = \rapidity$ the total momentum density and so on. Similarly, if we choose $\testfunction(x,\rapidity) = \theta(a<x<b)\testfunction(\rapidity)$, \eqref{equ:LL_deriv_cq_def} will measure the total density of charge $\testfunction(\rapidity)$ in the interval $(a,b)$. Therefore, formally if $\testfunction(x,\rapidity) = \delta(x-x')\delta(\rapidity-\rapidity')$, then $\expval{\quasiparticleop, \testfunction} = \quasiparticleop(x,\rapidity)$ measures the density of particles with rapidity $\rapidity'$ at $x'$, which is precisely what the quasi-particle density should measure.

\begin{remark}
	The operator $\expval{\quasiparticleop, \testfunction}$ might not necessarily be Hermitian. This can be solved by replacing $\expval{\quasiparticleop, \testfunction} \to \tfrac{1}{2}\qty(\expval{\quasiparticleop, \testfunction} + \expval{\quasiparticleop, \testfunction}^\dagger)$, which will not affect our computations.  
\end{remark}

\begin{remark}
	As it turns out, the microscopic version, i.e.\ setting $L=1$ in \eqref{equ:LL_deriv_cq_def}, the choice $\testfunction(x,\rapidity) = \delta(x-x')\rapidity^n$ for $n\in \mathbb{N}$ indeed defines a local density of the conserved quantity $\sum_i \rapidity_i^n$~\cite{doyon2023abinitioderivationgeneralised}. 
\end{remark}

\subsection{Attempt 1: Neglecting all non-trivial permutations}\label{sec:LL_deriv_1}
For the first attempt we assume that $\Delta\rapidity_i\ll 1$ is small, meaning that we can set $\rapidity_i \to \rapidity_i^0$. To measure an observable we need to sandwich $\expval{\vu{\rho},\testfunction}$ between a bra and a ket state. Applying the stationary phase approximation on both of these states we find that they are only non-zero if $\vec{x}_{\sigma^{-1}\ind{bra/ket}} = \vec{x}(t,\vec{\rapidity}^0,\vec{\hat{x}}^0)$. Here $\sigma\ind{bra}$ and $\sigma\ind{ket}$ are the permutations in the bra and in the ket state respectively. This implies that $\vec{x}_{\sigma^{-1}\ind{bra}} = \vec{x}_{\sigma^{-1}\ind{ket}}$, which, unless some $x_i = x_j$ implies that $\sigma\ind{bra} = \sigma\ind{ket} = \sigma$. We will now assume that only such permutations contribute. The remaining sum over $\sigma$ can be absorbed into the integral $\vec{x} \to \vec{x}_\sigma$ leading to:

\begin{align}
	\expval{\expval{\quasiparticleop, \testfunction}} &\sim \int\dd[N]{x} \frac{1}{\det \vb{H}(t,\vec{x},\vec{\hat{x}}^0,\vec{\rapidity}(t,\vec{x},\vec{\hat{x}}^0))} \prod_i\unsymA_i\qty(\tfrac{\rapidity_i(t,\vec{x},\vec{\hat{x}}^0)-\rapidity^0_i}{\Delta \rapidity_i})^2\qty[\tfrac{1}{L}\sum_i \testfunction(x_{i},\rapidity_i(t,\vec{x},\vec{\hat{x}}^0))].
\end{align}
Since $\Delta\rapidity_i \to 0$ we know that this integral is only nonzero around $\vec{x} = \vec{x}(t,\vec{\rapidity}^0,\vec{\hat{x}}^0)$ and thus
\begin{align}
	\expval{\expval{\quasiparticleop, \testfunction}} &\to \tfrac{1}{L}\sum_i \testfunction(x_i(t,\vec{\rapidity}^0,\vec{\hat{x}}^0),\rapidity_i^0).\label{equ:LL_deriv_1_final}
\end{align}
Here the prefactor was fixed using $\expval{\expval{\quasiparticleop, 1}} = N/L$. Expression \eqref{equ:LL_deriv_1_final} is precisely what to expect from GHD: for each particle $i$, we evaluate $\testfunction(x_i(t),\rapidity_i^0)$ at the location of the classical particle $x_i(t)$. We have already seen in \eqref{equ:LL_SP_QP_GHD_quick} that those trajectories satisfy the effective velocity equation, hence $\expval{\expval{\quasiparticleop, \testfunction}} \to \int\dd{x}\dd{\rapidity}\rho(t,x,\rapidity)\testfunction(x,\rapidity)$, where $\rho(t,x,\rapidity)$ satisfies the GHD equation with initial condition $\rho^0(x,\rapidity) = \tfrac{1}{L} \sum_i \delta(x-x_i(0,\vec{\rapidity}^0,\vec{\hat{x}}^0))\delta(\rapidity-\rapidity_i)$. 

This is a simple straightforward derivation of GHD at least in this specific setting. However, there is an obvious problem: it treats wave packets as classical particles, but it does not take their fermionic nature into account. Since we are free to choose the $\hat{x}^0_i$ and $\rapidity_i^0$, there is nothing preventing us from packing as many particles as we want into a single place. This seems somewhat unphysical. 

As we will discuss later, the problem with this approach is that instead of $\vec{x}_{\sigma^{-1}\ind{bra}} = \vec{x}_{\sigma^{-1}\ind{ket}}$, we only actually know that $\vec{x}_{\sigma^{-1}\ind{bra}} - \vec{x}_{\sigma^{-1}\ind{ket}} = \order{\Delta \rapidity}$. Therefore, the assumption that only $\sigma\ind{bra} = \sigma\ind{ket}$ contribute to the permutation sum is likely not justified. The number of particles in an interval of size $\Delta \rapidity$ is proportional to $L\Delta\rapidity$, hence we would need to sum over all permutations $\order{(L\Delta\rapidity)!}$ of these closeby particles.

\subsection{Attempt 2: \wignerqpd}
\label{sec:LL_deriv_2}
The last attempt applies to the particular ``gas of wave packets'' initial states. In the following, we would like to present another attempt that applies to much more general settings. The idea leading to this approach was to avoid dealing with the sum over permutations altogether. If we choose a fully antisymmetric amplitude $A(\vec{p})$ in \eqref{equ:LL_SP_psi_starting_point}, the sum over permutations would never appear throughout the computation. However, it is non-trivial to write down meaningful amplitudes. For GHD each particle needs two pieces of information: $\hat{x}_i^0$ and $\rapidity_i$. The modulus of $A(\vec{p})$ can naturally be viewed as describing the distribution of $\rapidity_i$, but the $\hat{x}_i^0$ is contained in the (fast-oscillating) phase. How can we extract them and make sure they are distributed over a large system? What we need is a phase-space description of a wave function. 

Luckily, such a description exists in terms of the well known \wignerqpd~(see for instance ~\cite{VITatarskii_1983}):
\begin{align}
	\hat{W}_N(\vec{\hat{x}}^0,\vec{\rapidity}) &= \tfrac{1}{\pi^N} \int\dd[N]{\rapidity'}A(\vec{\rapidity}-\vec{\rapidity}')A(\vec{\rapidity}+\vec{\rapidity}')^* e^{-2i\vec{\rapidity}'\vec{\hat{x}}^0}.\label{equ:LL_deriv_2_wigner_def}
\end{align}
This function has the features of a phase-space distribution. Its marginals are given by 
\begin{align}
	\int\dd[N]{\hat{x}^0}\hat{W}_N(\vec{\hat{x}}^0,\vec{\rapidity}) &= \abs{A(\vec{\rapidity})}^2, & 
	\int\dd[N]{\rapidity}\hat{W}_N(\vec{\hat{x}}^0,\vec{\rapidity}) &= \abs{\tilde{A}(\vec{\hat{x}}^0)}^2,\label{equ:LL_deriv_2_wigner_marginals}
\end{align}
where $\tilde{A}(\vec{\hat{x}}^0)$ is the wave function in position space, i.e. the Fourier transform of $A(\vec{\rapidity})$. This in particular implies $	\int\dd[N]{\hat{x}^0}\dd[N]{\rapidity}\hat{W}_N(\vec{\hat{x}}^0,\vec{\rapidity}) = 1$. Due to these properties, in standard textbooks on quantum mechanics, $\hat{W}_N(\vec{\hat{x}}^0,\vec{\rapidity})$ is often interpreted as the classical phase-space distribution that the quantum state corresponds to. However, there are two important caveats~\cite{VITatarskii_1983}. First, unlike a classical phase-space distribution (which is a probability distribution) $\hat{W}_N(\vec{\hat{x}}^0,\vec{\rapidity})$ can become negative\footnote{However, one can show that convoluting the Wigner function with a Gaussian obeying the Heisenberg uncertainty principle is always positive~\cite{VITatarskii_1983}. Intuitively, this means negative values can only appear in regions of size $\order{1}$, or $\order{\hbar}$ in physical units.}. Second, in the presence of a potential the evolution equation becomes significantly more complicated (and non-local) compared to the classical case. Still, it is often used as a convenient tool to study the semi-classical approximation of quantum mechanics. 

For us, these technicalities are not important. We only use the quasi-probability distribution to describe our initial state. We can explicitly write the expectation value
\begin{align}
	\expval{\expval{\quasiparticleop, \testfunction}} &= \int\frac{\dd[N]{x}\dd[N]{\hat{x}^0}\dd[N]{\rapidity}\dd[N]{\rapidity'}}{\mathcal{N}_N^2} \hat{W}_N(\vec{\hat{x}}^0,\tfrac{\vec{\rapidity}+\vec{\rapidity}'}{2})e^{i\vec{\hat{x}}^0(\vec{\rapidity}'-\vec{\rapidity}) + i\Phi(\vec{x},\vec{\rapidity}) - i\Phi(\vec{x},\vec{\rapidity}') -i (\vec{\rapidity}^2-\vec{\rapidity}'^2)t}\qty[\tfrac{1}{L}\sum_i \testfunction(x_i/L,\rapidity_i)].\label{equ:LL_deriv_2_obs_basic}
\end{align}

In this expression it is obvious that we packed all details about the initial state (including the antisymmetrizing sum over permutations) into $\hat{W}_N(\vec{\hat{x}}^0,\vec{\rapidity})$. In order to now take the limit $N,L \to \infty$ we need to allow $\hat{x}^0_i \sim \order{L}$, hence we are going to consider large scale initial states
\begin{align}
	\hat{W}_{N}(\vec{\hat{x}}^0,\vec{\rapidity}) = \tfrac{1}{L^N} \hat{W}_{N,L}(\vec{\hat{x}}^0/L,\vec{\rapidity}),\label{equ:LL_deriv_2_wigner_scaling}
\end{align}
where $\hat{W}_{N,L}(\vec{\hat{x}}^0,\vec{\rapidity})$ is a function varying on $\order{1}$ in $\hat{x}^0$. Rescaling $x$ and $\hat{x}^0$ by $L$ and also looking at large times $t\to Lt$ we find
\begin{align}
	\expval{\expval{\quasiparticleop, \testfunction}} &\sim\int\dd[N]{x}\dd[N]{\hat{x}^0}\dd[N]{\rapidity}\dd[N]{\rapidity'} \hat{W}_{N,L}(\vec{\hat{x}}^0,\tfrac{\vec{\rapidity}+\vec{\rapidity}'}{2})e^{iL\vec{\hat{x}}^0(\vec{\rapidity}'-\vec{\rapidity}) + i\Phi(L\vec{x},\vec{\rapidity}) - i\Phi(L\vec{x},\vec{\rapidity}') -iL (\vec{\rapidity}^2-\vec{\rapidity}'^2)t}\qty[\tfrac{1}{L}\sum_i \testfunction(x_i,\rapidity_i)]\nonumber\\
	&=\int\dd[N]{x}\dd[N]{\hat{x}^0}\dd[N]{\rapidity}\dd[N]{\rapidity'} \hat{W}_{N,L}(\vec{\hat{x}}^0,\tfrac{\vec{\rapidity}+\vec{\rapidity}'}{2})e^{iL(S_t(\vec{x},\vec{\hat{x}}^0,\vec{\rapidity})-S_t(\vec{x},\vec{\hat{x}}^0,\vec{\rapidity}'))}\qty[\tfrac{1}{L}\sum_i \testfunction(x_i,\rapidity_i)].\label{equ:LL_deriv_2_obs_scaling}
\end{align}
Before we start analyzing this expression, let us
discuss the initial state.
\subsubsection{Discussion of admissible initial states}

First, let us note that at this point we can fully forget about the amplitude $A(\vec{\rapidity})$ and only describe our state using $\hat{W}_{N,L}(\vec{\hat{x}}^0,\vec{\rapidity})$. In fact, we do not even need to require that the initial state is pure. If the initial state is mixed, one can just straightforwardly use the \wignerqpd~of the mixed state\footnote{The \wignerqpd~for a mixed state described by a density matrix $\quasiparticleop$ is simply obtained by replacing $A(\vec{\rapidity}-\vec{\rapidity}')A(\vec{\rapidity}+\vec{\rapidity}')^* \to \bra{\vec{\rapidity}-\vec{\rapidity}'}\quasiparticleop\ket{\vec{\rapidity}+\vec{\rapidity}'}$ in \eqref{equ:LL_deriv_2_wigner_def}.}. On the other hand, we cannot put any $\hat{W}_{N,L}(\vec{\hat{x}}^0,\vec{\rapidity})$ there, because then the corresponding density matrix would not necessarily be fully anti-symmetric. 

Ideally, we would initialize the system in a local equilibrium state \eqref{equ:pre_hydro_LES}, but it is not clear how to compute these states. We can loosen this assumption and try to consider state satisfying the large deviation scaling \eqref{equ:pre_hydro_BMFT_scaling}. But even more importantly, the state should locally look somewhat close to a GGE. How can we do this?

Observe that we can view the Bethe wave function \eqref{equ:pre_int_LL_Nparticle} as a unitary map $\vu{U}$ with $\bra{\vec{x}}\vu{U}\ket{\vec{\rapidity}} = \psi(\vec{x}|\vec{\rapidity})$. This map maps a free fermionic wave function, i.e.\ a fully antisymmetric amplitude, onto a Lieb-Liniger wave function, i.e.\ satisfying the correct Lieb-Liniger boundary conditions whenever $x_i=x_j$. The \wignerqpd~defined in \eqref{equ:LL_deriv_2_wigner_def} is then the \wignerqpd~of this free fermionic wave function. Hence, we interpret $\hat{W}_{N,L}(\vec{\hat{x}}^0,\vec{\rapidity})$ as the quantum state in what we will later call ``contracted'' non-interacting system. 

Similar to the equivalent map for classical models \eqref{equ:scbm_map_def}, this map is a non-local map. In particular, it does not map local equilibrium states onto local equilibrium states. Instead, it introduces additional long range correlations (see \cref{sec:fixedpoint_corr_corr}). However, this is only an inconvenience: still, for classical systems, any ``admissible'' state (see \cref{rem:LL_wigner_admissble}) of the non-interacting system gives rise to an ``admissible'' state for the interacting system. Thus, we expect the same to hold for the quantum map:

\begin{conjecture}
	Any $\hat{W}_{N,L}(\vec{\hat{x}}^0,\vec{\rapidity})$ describing an ``admissible'' large scale initial state for free fermions, should, under the mapping $\vu{U}$, give rise to an ``admissible'' initial state for the interacting system (and vice versa).
\end{conjecture}

\begin{remark}\label{rem:LL_wigner_admissble}
	We leave the meaning of ``admissible'' vague on purpose: it should be a somewhat physical state satisfying the large deviation scaling \eqref{equ:pre_hydro_BMFT_scaling} and should locally be sufficiently close to a GGE.
\end{remark}

Assuming that this conjecture is true, we can bypass the problem of choosing a precise state: we simply choose the \wignerqpd~of any state that is ``admissible'' for free fermions.

\begin{remark}
	States that certainly are ``admissible'' are local equilibrium states of free fermions, which are also the Gibbs states of free fermions in an external potential. Those can be computed by solving the corresponding single particle problem, and have been studied quite extensively (see for instance~\cite{Dean_2019}). This also includes marginals of \wignerqpd~for free fermionic local equilibrium states. Unfortunately, we are not aware of any studies/explicit expression of the full $N$ particle \wignerqpd~of such states.
\end{remark} 

\subsubsection{Large scale analysis}
Now that we have discussed ``admissible'' initial states, we can continue to perform the stationary phase approximation of \eqref{equ:LL_deriv_2_obs_scaling}. For our discussion later, we will use a slightly different, but equivalent way (see appendix \ref{app:SP_alter}). The main assumption in this attempt is that $\hat{W}_{N,L}(\vec{\hat{x}}^0,\vec{\rapidity})$ is smooth in $\vec{\rapidity}$ as $L \to \infty$.

First, let us denote the Fourier transform $\tilde{\hat{W}}_{N,L}(\vec{p},\vec{p}') = \int\dd[N]{\hat{x}^0}\hat{W}_{N,L}(\hat{x}^0,\vec{p}')e^{-i\vec{p}\vec{\hat{x}}^0}$ and use it to write
\begin{align}
	\expval{\expval{\quasiparticleop, \testfunction}} &\sim\int\dd[N]{x}\dd[N]{\rapidity}\dd[N]{\rapidity'} \tilde{\hat{W}}_{N,L}(L(\vec{\rapidity}'-\vec{\rapidity}),\tfrac{\vec{\rapidity}+\vec{\rapidity}'}{2})e^{i\Phi_{Lt}(L\vec{x},\vec{\rapidity}) - i\Phi_{Lt}(L\vec{x},\vec{\rapidity}')}\qty[\tfrac{1}{L}\sum_i \testfunction(x_i,\rapidity_i)]\label{equ:LL_deriv_2_deriv_1}\\
	&\sim\int\dd[N]{x}\dd[N]{\rapidity}\dd[N]{q} \tilde{\hat{W}}_{N,L}(\vec{q},\vec{\rapidity}+\tfrac{1}{2L}\vec{q})e^{i\Phi_{Lt}(L\vec{x},\vec{\rapidity}) - i\Phi_t(L\vec{x},\vec{\rapidity}+\vec{q}/L)}\qty[\tfrac{1}{L}\sum_i \testfunction(x_i,\rapidity_i)]\label{equ:LL_deriv_2_deriv_2}\\
	&= \int\dd[N]{x}\dd[N]{\rapidity}\dd[N]{q} \tilde{\hat{W}}_{N,L}(\vec{q},\vec{\rapidity}+\order{1/L})e^{-i\grad_{\rapidity}\Phi_{Lt}(L\vec{x},\vec{\rapidity})\vec{q}/L + \order{1/L}}\qty[\tfrac{1}{L}\sum_i \testfunction(x_i,\rapidity_i)]\label{equ:LL_deriv_2_deriv_3}\\
	&\approx\int\dd[N]{x}\dd[N]{\rapidity} \hat{W}_{N,L}(\grad_{\rapidity}\Phi_{Lt}(L\vec{x},\vec{\rapidity})/L,\vec{\rapidity})\qty[\tfrac{1}{L}\sum_i \testfunction(x_i,\rapidity_i)]\label{equ:LL_deriv_2_deriv_4}.
\end{align}
Here we defined $\vec{\rapidity}'=\vec{\rapidity}+\vec{q}/L$, $\Phi_{t}(\vec{x},\vec{\rapidity}) = \Phi(\vec{x},\vec{\rapidity}) - \vec{\rapidity}^2t$ and discarded terms of subleading order. We will now denote
\begin{align}
	\hat{x}^0_i(t,\vec{x},\vec{\rapidity}) &= \partial_{\rapidity_i}\Phi_{Lt}(L\vec{x},\vec{\rapidity})/L = x_i + \tfrac{1}{2L} \sum_{j\neq i}\sgn(x_i-x_j) \varphi(\rapidity_i-\rapidity_j)- 2\rapidity_i t, \label{equ:LL_deriv_2_hatx_def}
\end{align}
which gives the initial location of the quasi-particles in the contracted coordinates. It is also the inverse function of $\vec{x}(t,\vec{\hat{x}}^0,\vec{\rapidity})$ and $\vec{\rapidity}(t,\vec{x},\vec{\hat{x}}^0)$ in $\vec{\hat{x}}^0$. Using this we finally arrive at
\begin{align}
	\expval{\expval{\quasiparticleop, \testfunction}} &\sim \int\dd[N]{x}\dd[N]{\rapidity} \hat{W}_{N,L}(\vec{\hat{x}}^0(t,\vec{x},\vec{\rapidity}),\vec{\rapidity})\qty[\tfrac{1}{L}\sum_i \testfunction(x_i,\rapidity_i)].\label{equ:LL_deriv_2_final_x}
\end{align}
We can also do a change of coordinates $x \to \hat{x}^0$:
\begin{align}
	\expval{\expval{\quasiparticleop, \testfunction}} &\sim \int\tfrac{\dd[N]{\hat{x}^0}\dd[N]{\rapidity}}{\det \partial_{x_i}\hat{x}^0_j} \hat{W}_{N,L}(\vec{\hat{x}}^0,\vec{\rapidity})\qty[\tfrac{1}{L}\sum_i \testfunction(x_i(t,\vec{\hat{x}}^0,\vec{\rapidity}),\rapidity_i)].\label{equ:LL_deriv_2_final_hatx}
\end{align}
Let us try to understand this expression. If the determinant from the change of coordinates was not there, we would have
\begin{align}
	\expval{\expval{\quasiparticleop, \testfunction}} &= \int\dd[N]{\hat{x}^0}\dd[N]{\rapidity} \hat{W}_{N,L}(\vec{\hat{x}}^0,\vec{\rapidity})\qty[\tfrac{1}{L}\sum_i \testfunction(x_i(t,\vec{\hat{x}}^0,\vec{\rapidity}),\rapidity_i)].\label{equ:LL_deriv_2_final_hatx_nodet}
\end{align}
This is precisely how we would compute the expectation value of a one-particle observable $\testfunction(x,\rapidity)$ for a classical dynamical system, where the initial configuration of particles is randomly choosen according to the probability distribution $\hat{W}_{N,L}(\vec{\hat{x}}^0,\vec{\rapidity})$, and the time evolution map is given by $\vec{x}(t,\vec{\hat{x}}^0,\vec{\rapidity})$. If \eqref{equ:LL_deriv_2_final_hatx_nodet} was true, it would be simple to derive GHD from it, following the same steps that we will later do for the classical model (see \cref{sec:scbm_GHD}).

But, unfortunately, there is the determinant. The Jacobian is explicitly given by
\begin{align}
	\partial_{x_j}\hat{x}^0_i(t,\vec{x},\vec{\rapidity}) &= \delta_{ij} + \tfrac{1}{L} \sum_{j\neq i}\delta(x_i-x_j) \varphi(\rapidity_i-\rapidity_j),\label{equ:LL_deriv_2_final_jacobian}
\end{align}
which is either trivial or singular, whenever $x_i=x_j$. One can now either regularize the $\delta$ function, or write \eqref{equ:LL_deriv_2_final_hatx} instead as\footnote{This is the mathematical proper way of changing coordinates $x\to \hat{x}^0$ in the presence of jumps.}
\begin{align}
	\expval{\expval{\quasiparticleop, \testfunction}} &\sim \int_{\msf{\Sigma}_t}\dd[N]{\hat{x}^0}\dd[N]{\rapidity} \hat{W}_{N,L}(\vec{\hat{x}}^0,\vec{\rapidity})\qty[\tfrac{1}{L}\sum_i \testfunction(x_i(t,\vec{\hat{x}}^0,\vec{\rapidity}),\rapidity_i)],\label{equ:LL_deriv_2_final_hatx_proper}
\end{align}
where $\msf{\Sigma}_t$ is the image of $(\vec{x},\vec{\rapidity}) \mapsto \vec{\hat{x}}^0(t,\vec{x},\vec{\rapidity})$. This image is not $\mathbb{R}^{2N}$ due to the jumps in \eqref{equ:LL_deriv_2_hatx_def}. The time-evolution of this set is simple
\begin{align}
	\msf{\Sigma}_t = \qty{(\vec{\hat{x}}^0+2\vec{\rapidity}t,\vec{\rapidity})|(\vec{\hat{x}}^0,\vec{\rapidity}) \in \msf{\Sigma}_0}.\label{equ:LL_deriv_2_final_Sigma}
\end{align}
If indeed $\msf{\Sigma}_0 = \mathbb{R}^{2N}$, then $\msf{\Sigma}_t=\msf{\Sigma}_0 = \mathbb{R}^{2N}$ and we would find \eqref{equ:LL_deriv_2_final_hatx_nodet}. It might be that the integral over  $\mathbb{R}^{2N}\backslash\msf{\Sigma}_t$ is negligible as $N,L\to \infty$. After all, the individual jumps in \eqref{equ:LL_deriv_2_hatx_def}, decay as $1/L$, meaning that as $L\to \infty$ the function becomes effectively continuous. However, at the same time the number of jumps grows as $N^2$, meaning that the total volume of $\mathbb{R}^{2N}\backslash\msf{\Sigma}_t \sim N^2/L$ grows. In general, we thus believe that \eqref{equ:LL_deriv_2_final_x} is not accurate as $L\to \infty$, and that the assumptions leading to \eqref{equ:LL_deriv_2_final_x} were too crude. We discuss these approximations in more detail in the next section.

\begin{result}
	Even though both attempt like do not produce asymptotically correct formulas for $L\to \infty$, they still show how GHD emerges from quantum mechanics via the classical particle model.
\end{result}

\section{Discussion of the assumptions}

\subsection{Violation of unitarity}
In our opinion, the clearest argument to immediately recognize that \eqref{equ:LL_deriv_2_final_x} cannot be correct is the following: we definitely know that if we choose $\testfunction(x,\rapidity)=1$, then $\expval{\expval{\quasiparticleop, \testfunction}} = N/L$ for all times $t$. This can be computed from the exact microscopic expression \eqref{equ:LL_deriv_2_obs_basic} based on
\begin{align}
	\int\dd[N]{\hat{x}^0}\dd[N]{\rapidity} \hat{W}_{N,L}(\vec{\hat{x}}^0,\vec{\rapidity}) = 1.
\end{align}
In our final result \eqref{equ:LL_deriv_2_final_hatx_proper} we instead find
\begin{align}
	\expval{\expval{\quasiparticleop, 1}} &\sim \tfrac{N}{L}\int_{\Sigma_t}\dd[N]{\hat{x}^0}\dd[N]{\rapidity} \hat{W}_{N,L}(\vec{\hat{x}}^0,\vec{\rapidity}).\label{equ:LL_wrong_unitarity_wigner}
\end{align}
As $\Sigma_t$ is time-dependent this will also likely be a time-dependent number (although we cannot prove that it is).  Therefore, the assumptions used to derive \eqref{equ:LL_deriv_2_final_hatx_proper} are likely not justified. 
\begin{remark}
	For convenience, we have not explicitly written out the prefactor hidden in $\sim$ in \eqref{equ:LL_deriv_2_final_hatx_proper}. Typically, one would say that one has to fix this prefactor by properly normalizing \eqref{equ:LL_deriv_2_final_hatx_proper}, such that $\expval{\expval{\quasiparticleop, 1}} = N/L$. However, here, by going through the computation (\ref{equ:LL_deriv_2_deriv_1}-\ref{equ:LL_deriv_2_deriv_4}) again, one can compute the prefactor explicitly: it only depends on $N$ and $L$, but not on any details of the state. In particular, it also does not depend on $t$.
\end{remark}

This means that the determinant term in \eqref{equ:LL_deriv_2_final_hatx} or the $\Sigma_t$ in \eqref{equ:LL_deriv_2_final_hatx_proper} should not appear. In fact, expression \eqref{equ:LL_deriv_2_final_hatx_nodet} is more reasonable as it also gives $\expval{\expval{\quasiparticleop, 1}} = 1$.

This is very peculiar. Our approximation violates unitarity, or to be more precise the conservation of total probability, that is fundamental in quantum mechanics. At what point have we lost this in our approximation scheme?

Recall that we absorbed the sum over permutations into the amplitude $A$ and later into the \wignerqpd. The claim of this section is: we loose the correct normalization, as soon as we forget that this sum is still in there. To see this, note that Bethe wave functions are only an orthonormal basis when the sum over permutations is included. Furthermore, this only happens if the scattering phase $\phi(\rapidity)$ is precisely given by the Lieb-Liniger one. A generic $\phi(\rapidity)$ will not lead to an orthogonal basis. Any derivation of GHD has to make use of these facts somehow, otherwise normalization is not guaranteed.
	
To illustrate where our approximation goes wrong, it is instructive to study the following example. Due to the orthogonormality of Bethe states, we know that for any fully antisymmetric $A(\vec{\rapidity})$ we have
\begin{align}
	\int\dd[N]{x}\dd[N]{\rapidity'} \psi(\vec{x}|\rapidity)^*\psi(\vec{x}|\rapidity')A(\vec{\rapidity}')e^{i(\vec{\rapidity}^2-\vec{\rapidity}'^2)t} &= N!A(\vec{\rapidity}).\label{equ:LL_wrong_unitarity_L2}
\end{align}
At large times $t\to Lt$, we can explicitly write this as follows
\begin{align}
	A(\vec{\rapidity}) \sim \int\dd[N]{x}\dd[N]{\rapidity'} e^{i\Phi_{Lt}(L\vec{x},\vec{\rapidity}')-\Phi_{Lt}(L\vec{x},\vec{\rapidity})}A(\vec{\rapidity}')
\end{align}
and now do a saddle point approximation as before over $x$ and $\rapidity'$. The saddle point condition over $x$ implies $\rapidity_i=\rapidity_i'$ and the one over $\rapidity$ implies
\begin{align}
	0 &= x_i + \tfrac{1}{2L} \sum_{j\neq i}\sgn(x_i-x_j) \varphi(\rapidity_i-\rapidity_j)- 2\rapidity_i t.\label{equ:LL_wrong_unitarity_L2_hatx}
\end{align}
Hence we find $A(\vec{\rapidity}) \sim \frac{1}{\det \partial_{x_i}\hat{x}_j^0}A(\vec{\rapidity})$,
where the determinant turns out to be the same determinant as before! This clearly indicates that the determinant must be an artifact of the saddle point approximation.

One might conclude from this that such determinants are negligible. But we do not think that this is the case. Note that, in this example taking the saddle point approximation is clearly not justified. This can easily be seen by comparing \eqref{equ:LL_wrong_unitarity_L2_hatx} to \eqref{equ:LL_SP_QP_def}. Equation \eqref{equ:LL_wrong_unitarity_L2_hatx} describes the configuration where all $\hat{x}^0_i = 0$, whose solution is also $x_i=0$ for all $i$. Already due to the fact that the Lieb-Liniger model is translation invariant the point $\vec{x}=0$ cannot have any physical meaning. 

The reason why the saddle point does not work here, is because the integral over $x$ is over an unbounded region of $\mathbb{R}^N$. For the $\rapidity$ integral there is the $A(\vec{\rapidity})$ which effectively restricts the integral to some finite region. For $x$ there is no such amplitude $A(\vec{x})$ and hence the integral is over an unbounded region. Importantly, the stationary phase approximation is only justified for integrals $\int\dd[N]{x}f(x)e^{iLg(x)}$, where the amplitude $f(x)$ decays sufficiently fast at $\abs{x}\to \infty$.

\subsection{Is the stationary phase approximation justified?}
At this point the reader might wonder whether the stationary phase approximation in \eqref{equ:LL_SP_SP_WF} is justified at all. The stationary phase approximation (see appendix \ref{app:SP}), as a mathematical statement, is an asymptotic expansion of integrals of the form $\int\dd[N]{x}f(\vec{x})e^{iLg(\vec{x})}$ for fixed $N$ as $L\to \infty$. Is it even valid if $N\sim L\to \infty$?

To answer this, we discuss in appendix \ref{app:SP} that the stationary phase approximation of $\int\dd[N]{x}f(\vec{x})e^{iLg(\vec{x})}$ at a stationary point $\vec{x}_0$ is valid if:
\begin{enumerate}
	\item For $\vec{x}=\vec{x}_0 + \order{1/\sqrt{L\vb{H}_g(\vec{x}_0)}}$ the function $f(\vec{x})$ is constant.
	\item For $\vec{x}=\vec{x}_0 + \order{1/\sqrt{L\vb{H}_g(\vec{x}_0)}}$ the function $g(\vec{x})$ is well approximated by its second order Taylor polynomial.
\end{enumerate}

Here, $\vb{H}_g(\vec{x}_0)$ is the Hessian of $g(\vec{x})$. Applied to our setting this means that we need that conditions 1 and $2$ must be satisfied in a neighborhood $\delta \rapidity_i,\delta x_i\sim 1/\sqrt{L}$ around the optimal $\vec{x},\vec{\rapidity}$. 

\subsubsection{Discussion of the phase}
\label{sec:LL_wrong_SP_phase}
For the fast-oscillating phase we need to understand whether its quadratic Taylor polynomial is a good approximation in this region. At first this might not seem to be the case because of the many jumps in the phase \eqref{equ:LL_SP_phase}. However, the typical inter particle distance is $1/L$ and we are integrating over a region $\delta x \sim 1/\sqrt{L}$. Hence, we can expect that the phase self-averages in this region.

For instance, the phase seen by an individual $i$ particle in \eqref{equ:LL_SP_phase} is given by
\begin{align}
	S_t(\vec{x},\vec{\rapidity}) &= \rapidity_i (x_i-\hat{x}_i^0) +\tfrac{1}{2L}\sum_{j\neq i}\sgn(x_i-x_j) \phi(\rapidity_i-\rapidity_j)- \rapidity_i^2 t + \mathrm{const}(x_i)\\
	&\approx \rapidity_i (x_i-\hat{x}_i^0) +\tfrac{1}{2}\int\dd{x'}\dd{\rapidity'}\rho(x',\rapidity')\sgn(x_i-x') \phi(\rapidity_i-\rapidity')- \rapidity_i^2 t + \mathrm{const}(x_i).\label{equ:LL_wrong_SP_single_particle}
\end{align}

This is the phase seen by particle $i$ in the region of integration $\delta \rapidity_i, \delta x_i\sim 1/\sqrt{L}$. Note that this is a smooth function and hence the stationary phase approximation should be justified. In fact, the stationary phase approximation should be carried out over \eqref{equ:LL_wrong_SP_single_particle} instead of the microscopic phase \eqref{equ:LL_SP_phase}. This is also the reason why we can neglect the fact that the phase is not actually smooth in $x$.

Hence, we conclude that at the level of the phase, the stationary phase approximation is justified.

\subsubsection{The curse of the permutations}
After understanding the phase, we now need to understand whether or not the amplitude, i.e. what we called $f$ in \eqref{equ:app_SP_basic} can be treated as constant over $\delta \rapidity_i,\delta x_i\sim 1/\sqrt{L}$.

We will now argue that this is not the case and the reason for this failure is the sum over permutations which so far only had a spectating role.

First, let us note that if an arbitrary $A\upd{bare}(\vec{p})$ is antisymmetrized, i.e. $A(\vec{p})=\sum_\sigma (-1)^\sigma A\upd{bare}(\vec{p}_\sigma)$, $A(\vec{p}) = 0$, whenever some $p_i=p_j$. This means that even if $A\upd{bare}(\vec{p})$ is a smooth slowly varying function, if we vary a single $p_i$ in a range $[a,b]$, the function will have at least as many zeros as the number of other particles such that $\rapidity_j \in [a,b]$. In our case $\rapidity_i$ fluctuates with $\delta \rapidity_i \sim 1/\sqrt{L}$ and in such an interval we expect $\sim \sqrt{L}$ particles as $L \to \infty$. Hence, the amplitude will certainly not be almost constant over the region of integration. Instead, it will actually become increasingly rougher.

In the two attempts of a derivation presented earlier, we choose two different strategies to deal with the sum over permutations. In the first attempt (\cref{sec:LL_deriv_1}), we kept the sum over permutations explicit and assumed that the amplitude $A(\vec{p})$ is smooth. In this way, for each individual permutation, the stationary phase approximation is justified. The problem arises once we take the sum over all permutations: this sum has $N!$ terms, each term is only exact up to $\sim 1/L$, implying that the error could increase with $L$. Luckily, as we observed in \cref{sec:LL_deriv_1}, the fast oscillating phase suppresses any permutation where $\abs{x_{\sigma_i}- x_i} \gg 1/\sqrt{L}$ are not close. Therefore, only permutations that act ``locally'' will be relevant. If we imagine to divide space into $\sim\sqrt{L}$ cells of size $\sim 1/\sqrt{L}$, we will have $\sim \sqrt{L}$ particles in each. This means we can give a rough estimate of the number of ``local'' permutations as $\sim (\sqrt{L}!)^{\sqrt{L}}$, which is much less than $N!$. Still, it is much larger than the error of each individual term.

In the Wigner function attempt (\cref{sec:LL_deriv_2}), we tried to avoid the sum over permutations and `hid' it inside the initial state already from the start. In our large scale analysis, we actually only do two approximations, both in \eqref{equ:LL_deriv_2_deriv_3}. First, we neglect subleading terms in the fast oscillating phase. As discussed in \cref{sec:LL_wrong_SP_phase}, we believe that such approximations on the phase are justified. In the second approximation we replace
\begin{align}
	\tilde{\hat{W}}_{N,L}(\vec{q},\vec{\rapidity}+\order{1/L}) \to \tilde{\hat{W}}_{N,L}(\vec{q},\vec{\rapidity}).\label{equ:LL_wrong_wigner_replacement}
\end{align}
From \eqref{equ:LL_deriv_2_wigner_def} one can derive
\begin{align}
	\tilde{\hat{W}}_{N,L}(\vec{q},\vec{\rapidity}) &= A(\vec{\rapidity}+\vec{q})A(\vec{\rapidity}-\vec{q})^*.\label{equ:LL_wrong_wigner_A}
\end{align}

As discussed above, in an interval of size $\sim 1/L$, we expect that the fully antisymmetric amplitude $A(\vec{\rapidity})$ has $\order{1}$ zeros. Hence, $\tilde{\hat{W}}_{N,L}(\vec{q},\vec{\rapidity})$ is not constant in the region $\vec{\rapidity} + \order{1/L}$ and the approximation \eqref{equ:LL_deriv_2_deriv_3} is not justified.

\begin{result}
	The two attempts to derive the GHD equation highlight that the local statistics of the particles is important and has to be incorporated into the derivation. 
\end{result}

\begin{remark}
	It might seem strange that in \eqref{equ:app_SP_basic} we need that the function is constant over an interval of size $\delta \rapidity \sim 1/L$, instead of the $\delta x,\delta \rapidity \sim 1/\sqrt{L}$. We discuss in appendix \ref{app:SP_alter} how both scalings are related: the upshot is that we only need that $\delta x\delta \rapidity \sim 1/L$, hence if $\delta x \sim 1$, then $\delta \rapidity \sim 1/L$ is also allowed. 
\end{remark}

\subsection{How could a proper derivation look like?}
Even though we are not able to go further at this point, understanding the failure of \eqref{equ:LL_deriv_2_deriv_3} is conceptually important: we need to understand the local statistics. If our reasoning is correct, we do not need to understand the local statistics of the Lieb-Liniger model, but it will be sufficient to understand the much simpler statistics of free fermions and apply it on $	\hat{W}_{N,L}(\vec{\hat{x}}^0,\vec{\rapidity})$. Obviously, it is not feasible to do this for the precise microscopic $\hat{W}_{N,L}(\vec{\hat{x}}^0,\vec{\rapidity})$. But, since in any macroscopic or mesoscopic region there are many microscopic regions, we can expect that the effect of the microscopic statistics will average. Even more, it is not unreasonable to believe that such averaged local statistics will be universal\footnote{Similar to how many random matrices ensembles have the same local statistics as the standard GUE ensemble~\cite{10.1093/oxfordhb/9780198744191.013.6}.}. If the local statistics are universal, it means one can compute them in much simpler states, for instance in (free fermion) GGE states. Since such states have been studied extensively, and we are convinced that it will be possible to compute the required quantities. The problem here is more to identify which quantities are important and how one would use them to actually compute \eqref{equ:LL_deriv_2_deriv_3}. Note that such knowledge would be useful not just for the Lieb-Liniger model, but for any quantum integrable model with fermionic quasi-particles.

While this strategy might help to uncover clear mathematical understanding of the emergence of GHD, it is also fairly technical. While studying the classical models in \cref{sec:scbm}, we will be able to do similar computations in a large deviation setting. This does not require us to understand the local statistics (which is a Poisson point process). Can we do something similar in a quantum model? Recall that the stationary phase approximation effectively approximates the function by a Gaussian. We can interpret this Gaussian as an effective quadratic theory $\tfrac{1}{2}\sum_{ij}\partial_{x_i}\partial_{x_j}f(x_0)x_ix_j$ describing the fluctuations around the optimal point. The natural analogue of this for fermionic excitations should be a quadratic theory $\tfrac{1}{2}\sum_{ij}\partial_{x_i}\partial_{x_j}f(x_0)\vu{c}_i^\dagger \vu{c}_j$ for some effective fermionic quasi-particles created by $\vu{c}_i^\dagger$, however it is not clear where and how such a theory would emerge.

\section{Attractive Lieb-Liniger}
\label{sec:LL_attractive}

In this section we would like to explain, on the example of the attractive Lieb-Liniger model, how the derivation of \cref{sec:LL_SP} can be extended to models with strings. The eigenstates for the attractive Lieb-Liniger model are still given by \eqref{equ:pre_int_LL_Nparticle}, but $\rapidity_i$ does not necessarily have to be real~\cite{10.1063/1.1704156,jscaux}. To see this, restrict to $x_1<x_2<\ldots< x_N$ and write the eigenstate as 
\begin{align}
		\psi(\vec{x}|\vec{\rapidity}) \sim \sum_{\sigma\in \permutation_N} (-1)^\sigma e^{i\sum_i\rapidity_{\sigma_i}x_i} \prod_{i<j} B(\rapidity_{\sigma_j}-\rapidity_{\sigma_i}),\label{equ:LL_attractive_Bethe}
\end{align}
where $B(\rapidity) = e^{\tfrac{i}{2}\phi(\rapidity)}\sqrt{c^2+\rapidity^2} = c+i\rapidity$. Now let us take $x_1 \to -\infty$. To avoid blowup, for each permutation we either have $\Im \rapidity_{\sigma_1} \leq 0$ or $\prod_{i<j} B(\rapidity_{\sigma_j}-\rapidity_{\sigma_i}) = 0$. This implies that if any $\Im \rapidity_i > 0$, there has to be another particle $j$ such that $B(\rapidity_j-\rapidity_i) = 0$, implying $\rapidity_j = \rapidity_i + ic$. Similarly, by sending $x_N \to \infty$ we find that if $\Im \rapidity_i < 0$, then there must be another $\rapidity_j = \rapidity_i - ic$. This is only possible for $c<0$, where additional solutions of the form~\cite{10.1063/1.1704156,jscaux}
\begin{align}
	\rapidity_{a,k} &= \rapidity_a + ik\abs{c} & & k=-(l-1)/2, -(l-1)/2 +1 ,\ldots, (l-1)/2,\label{equ:LL_attractive_string_rapidities}
\end{align}
called strings of length $l = 1, 2, 3, \ldots$, are present. Here, $\rapidity_a\in \mathbb{R}$ is the center of the string. Strings of length $l \geq 2$ are bound states of particles and are interpreted as new types of quasi-particles with asymptotic momentum $P_a = \sum_{k = -(l-1)/2}^{(l-1)/2} \rapidity_{a,k} = l \rapidity_a$ and energy $E_l = \sum_{k = -(l-1)/2}^{(l-1)/2} \rapidity_{a,k}^2 = l \rapidity_a^2 - \abs{c}^2 \sum_{k = -(l-1)/2}^{(l-1)/2} k^2$~\cite{10.1063/1.1704156,jscaux} (see also the experiment~\cite{horvath2025observingbethestringsattractive}). 

\begin{remark}
	Dealing with strings becomes difficult if we restrict to a finite system. There we do not need them to decay exponentially as $x\to \pm\infty$. Instead, they need to satisfy the Bethe quantization condition \eqref{equ:pre_int_LL_bethe_quant_prod}, which is a completely different condition~\cite{jscaux}. Hence, strings will not have the form \eqref{equ:LL_attractive_string_rapidities}. The precise form is not known in general, but typically one assumes that in the thermodynamic limit $\ell\to\infty$ strings have the form \eqref{equ:LL_attractive_string_rapidities}. This is known as the string hypothesis~\cite[Chap 6]{arutyunov2020elements}. It makes sense: if $\ell$ is much larger than the size of the string $\sim \abs{c}$, then the exponential decay is a solution also in the periodic box, up to an exponential small $e^{-\ell/\abs{c}}$ error.
\end{remark}

\begin{remark}
	Note that while the string hypothesis is often applicable in many situations, exceptions are also known. For instance, in many models there exist states whose Bethe roots are either incompatible with the string hypothesis~\cite{ISLER1993209,PhysRevB.60.7271,TakehisaFujita_2003} or which converge only very slowly to them~\cite{DEVEGA1985439,KAROWSKI1988473,GRANET201896}. These violations of the string hypothesis have to be taken into account, e.g.\ in the theory of critical spin chains~\cite{DEVEGA1985439,KAROWSKI1988473,GRANET201896}. Therefore, avoiding the string hypothesis might help to apply GHD to more general situations.
\end{remark}

The analysis we are doing here is on the infinite system and hence does not require Bethe quantization and thus also no string hypothesis. 

Let us label particles as $i \to (l,a,k)$, where $l$ is the string length, $a=1, \ldots, N_l$ is its label and $k=-(l-1)/2, \ldots, (l-1)/2$ is the $k$ particle in $a$. We will denote the individual rapidities as $\rapidity_{l,a,k} = \rapidity_{l,a} + ik\abs{c}$, the string center by $\rapidity_{l,a}$ and the string energy and momentum by $E_l(\rapidity_{l,a})$ and $P_l(\rapidity_{l,a})$. With this, the general wave function is given by
\begin{align}
	\Psi(t,\vec{x}) &= \qty[\prod_{l=1}^\infty\int\dd[N_l]{\rapidity_{l,\cdot}}] A(\vec{\rapidity}) \psi(\vec{x}|\vec{\rapidity}) e^{-i\sum_{l,a,k} \rapidity_{l,a,k}^2 t}.\label{equ:LL_attractive_WF}
\end{align}
Making again the choice $A(\vec{\rapidity}) = \prod_{l=1}^\infty\prod_{a=1}^{N_l}  \unsymA_{l,a}\qty(\tfrac{\rapidity_{l,a}-\rapidity^0_{l,a}}{\Delta \rapidity_{l,a}}) e^{-iL\sum_{l,a} \rapidity_{l,a} \hat{x}_{l,a}^0}$, evaluating the wavefunction at large scales $x\to Lx, t\to Lt$ we find analogously to \eqref{equ:LL_SP_phase} the following fast oscillating phase
\begin{align}
	S_t(\vec{x},\vec{\rapidity}) &= \sum_{l,a,k} \rapidity_{l,a,k} (x_{l,a,k}-\hat{x}_{l,a}^0) +\tfrac{1}{4L}\sum_{(l,a,k)\neq (l',a',k')}\sgn(x_{l,a,k}-x_{l',a',k'}) \phi(\rapidity_{l,a,k}-\rapidity_{l',a',k})- \sum_{l,a,k}\rapidity_{l,a,k}^2 t.\label{equ:LL_attractive_phase_basic}
\end{align}
Before we take the saddle point of this phase, let us observe that due the exponential decay of the wave function of a string\footnote{This is $\order{1}$ on the microscopic scale.}, the wave function will only be non-zero if (in macroscopic coordinates) all string components are at the same location $x_{l,a,k} = x_{l,a}$. Hence, the fast oscillating phase becomes
\begin{align}
	S_t(\vec{x},\vec{\rapidity}) &= \sum_{l,a} P_l(\lambda_{l,a}) (x_{l,a}-\hat{x}_{l,a}^0) +\tfrac{1}{4L}\sum_{(l,a)\neq (l',a')}\sgn(x_{l,a}-x_{l',a'}) \phi_{ll'}(\rapidity_{l,a}-\rapidity_{l',a'})- \sum_{l,a}E_l(\rapidity_{l,a}) t,\label{equ:LL_attractive_phase_string}
\end{align}
where $\varphi_{ll'}(\rapidity) = \sum_{kk'}\phi(\rapidity +i(k-k')\abs{c})$ is the scattering phase of two strings. These are precisely the properties of strings found also in the TBA formalism. Therefore, an analogous derivation as the one leading to \eqref{equ:LL_SP_QP_def} leads to the GHD of the attractive Lieb-Liniger model (which has been obtained from the TBA here~\cite{Koch_2022}). 

\section{Conclusion}
In this chapter, we studied the large scale dynamics of the Lieb-Liniger model starting from its wave function. A stationary phase approximation revealed an emergent classical particle model, which gives rise to the GHD equation (and will be studied in \cref{sec:scbm}). While obtaining this classical model is easy, showing the emergence of the GHD equation on the level of expectation values is complicated: we proposed two possible strategies, both of which are likely using too crude assumptions. The main problem seems to be the sum over permutations which interferes with our approximation schemes.

Nevertheless, this new derivation is in many ways more straightforward and more physical compared to the phenomenological derivation from the TBA in \cref{sec:pre_GHD_TBA} and gives additional physical insights into the origin of GHD in quantum models. For instance, in models with strings one does not need to assume the string hypothesis (as demonstrated on the attractive Lieb-Liniger model in \cref{sec:LL_attractive}).

%!TEX root = thesis.tex

\chapter{Semi-classical Bethe models: A new family of integrable models}
\label{sec:scbm}
In \cref{sec:LL} we identified classical particle trajectories, describing the evolution of the positions of the quantum wave-packets as solutions to
\begin{align}
	\hat{x}_i^0 + v(\rapidity_i)t &= x_i(t) + \tfrac{1}{2} \sum_{j\neq i} \sgn_\alpha(x_i(t)-x_j(t))\varphi(\rapidity_i-\rapidity_j).\label{equ:scbm_self_consistency_bare_time_evol} 
\end{align}
From \cref{fig:LL_scbm} it is natural to interpret these solutions as particle trajectories. A natural question arises: can these trajectories be understood to originate from some classical particle model? Is this model integrable, and if so, what is its GHD?

This chapter is mostly based on two publications~\cite{PhysRevLett.132.251602,doyon2023generalisedtbartdeformationsclassicalfree}. We will see that these models are indeed integrable models and that their GHD coincides with the one of the quantum model. While we were not able to derive the GHD of the quantum model, we can use well-established techniques to derive the thermodynamics and GHD of these classical models. 

An interesting observation is that these models make sense for any phase shift $\varphi(\rapidity-\rapidityp)$ (unlike the quantum model which requires $\varphi(\rapidity-\rapidityp)$ to be of Lieb-Liniger type). Hence, they are a family of classical integrable models with arbitrary phase shift. This is a novelty in integrability, as other families of integrable models often have very restricted shapes of phase-shifts.

\begin{remark}
	Unlike \eqref{equ:LL_SP_QP_def}, we will now work with a smooth regularization of the sign function (this simplifies the mathematical treatment). We assume that $\sgn_\alpha(x)$ is a smooth monotone increasing function with $\sgn_\alpha(x) \to \pm \infty$ as $x\to \pm \infty$, for instance $\sgn_\alpha(x) = \tfrac{x}{x^2+\alpha^2}$ or $\sgn_\alpha(x) = \tanh(x/\alpha)$, but many more are possible. We interpret the parameter $\alpha$ as the interaction range.\label{rem:scbm_sgn_regularization}
\end{remark}

\section{Construction of the model via canonical transformations}
\label{sec:scbm_canonical_trafo}
Consider the $N$ particle ``contracted'' classical phase space $(\vec{\hat{x}},\vec{\rapidity}) \in \mathbb{R}^{2N}$ with Poisson bracket $\qty{\hat{x}_i,\rapidity_j} = \delta_{ij}$ and a non-interacting Hamiltonian
\begin{align}
	H(\vec{\hat{x}},\vec{\rapidity}) = \sum_{i=1}^N E(\rapidity_i),\label{equ:scbm_free_Hamiltonian}
\end{align}
where the single-particle energy $E(\rapidity)$ is such that $E'(\rapidity) = v(\rapidity)$. Its trajectories are trivial $\hat{x}_i(t) = \hat{x}_i^0 + v(\rapidity_i) t$. Now apply a canonical transformation~\cite[Table 9.1]{goldstein2002classical} to ``real'' coordinates $(\vec{x}, \vec{\scbmmomentum})$ with generating function
\begin{align}
	\Phi(\vec{x},\vec{\rapidity}) = \sum_{i=1}^N x_i\rapidity_i + \tfrac{1}{4} \sum_{j\neq i}\sgn_\alpha(x_i-x_j)\phi(\rapidity_i-\rapidity_j). \label{equ:scbm_generating_function}
\end{align}
The corresponding equations are (recall $\varphi(\rapidity) = \phi'(\rapidity)$)
\begin{align}
	\hat{x}_i = \partial_{\rapidity_i} \Phi &= x_i+\tfrac{1}{2}\sum_{j\neq i}\sgn_\alpha(x_i-x_j)\varphi(\rapidity_i-\rapidity_j) =: \scbmmap_{\vec{\rapidity}}(\vec{x})_i\label{equ:scbm_map_def}\\
	\scbmmomentum_i &= \rapidity_i+\tfrac{1}{2}\sum_{j\neq i}\partial_x\sgn_\alpha(x_i-x_j)\phi(\rapidity_i-\rapidity_j).
\end{align}
Here we introduced the map $\scbmmap_{\vec{\rapidity}}: \vec{x} \mapsto \vec{\hat{x}}$, whose properties we will study in detail later. For now note that $\hat{x}^0 = \scbmmap_{\vec{\rapidity}}(\vec{\hat{x}}^0)$ is satisfied at initial time, but also at any later time $\scbmmap_{\vec{\rapidity}}(\vec{x}(t)) = \vec{\hat{x}}(t) = \vec{\hat{x}}^0 + v(\vec{\rapidity})t$, which is precisely \eqref{equ:scbm_self_consistency_bare_time_evol}. Using this transformation we can view $\hat{x}_i$ and $\rapidity_i$ as functions of $\vec{x}$ and $\vec{\scbmmomentum}$, denoted by $\hat{x}_i[\vec{x},\vec{\scbmmomentum}]$ and $\rapidity_i[\vec{x},\vec{\scbmmomentum}]$ respectively.

\begin{result}
	The semi-classical Bethe models can be constructed via the canonical transformation generated by \eqref{equ:scbm_generating_function}. 
\end{result}

By construction, these models can be solved by quadrature, i.e.\ if we can compute $\scbmmap_{\vec{\rapidity}}^{-1}$ we can find the solution at any time as $\vec{x}(t) = \scbmmap_{\vec{\rapidity}}^{-1}(\vec{\hat{x}}^0 + v(\vec{\rapidity})t) = \scbmmap_{\vec{\rapidity}}^{-1}(\scbmmap_{\vec{\rapidity}}(\vec{x}^0) + v(\vec{\rapidity})t)$. Furthermore, by construction $\vu{Q}_a[\vec{x},\vec{\scbmmomentum}] = \sum_{i=1}^N \rapidity_i[\vec{x},\vec{\scbmmomentum}]^a$ are an infinite family of conserved quantities, with local densities
\begin{align}
	\vu{q}_a(x)[\vec{x},\vec{\scbmmomentum}] &= \sum_{i=1}^N \delta(x-x_i) \rapidity_i[\vec{x},\vec{\scbmmomentum}]^a.\label{equ:scbm_local_density}
\end{align}
Note that $\vu{Q}_2[\vec{x},\vec{\scbmmomentum}]$ is the Hamiltonian expressed in the new coordinates. Hence, we conclude that they are integrable models.

\begin{result}
	Semi-classical Bethe models are integrable models. This was rigorously established in~\cite{doyon2023generalisedtbartdeformationsclassicalfree} for models satisfying $\varphi(\rapidity) \geq 0$. 
\end{result}

\begin{remark}\label{rem:scbm_local_integrability}
	Any model obtained via a canonical transformation from a non-interacting model is solvable by inverting the canonical transformation. Also, it always has infinitely many conservation laws and is Liouville-integrable. However, generically the densities of these conserved quantities are not (quasi-)local on $(x,\scbmmomentum)$ space. Thus, establishing the integrability (in the sense we use throughout the thesis) of our models in~\cite{doyon2023generalisedtbartdeformationsclassicalfree} required showing locality, which was a non-trivial task.
\end{remark}

In \cref{sec:LL} equation \eqref{equ:scbm_self_consistency_bare_time_evol} was derived from a quantum model with a specific $\varphi$. However, nothing stops us from simply plugging any $\varphi(\rapidity)$ into \eqref{equ:scbm_generating_function} and to obtain a classical integrable model with an arbitrary scattering shift. This is an interesting result since up to this point only integrable models with very specific scattering shifts were available.   

\begin{result}
	For any scattering shift there exist a classical integrable model. For $\varphi(\rapidity) \geq 0$ this is rigorously shown for a large family in~\cite{doyon2023generalisedtbartdeformationsclassicalfree}. For negative $\varphi(\rapidity) < 0$ a formal construction was not yet possible, but intuitively should exist (see \cref{sec:scbm_negphi}).
\end{result}

\section{Properties of semi-classical Bethe models}

\subsection{Two particle scattering}
\label{sec:scbm_scbm_twoparticle}
\begin{figure}[!h]
	\centering
	\includegraphics{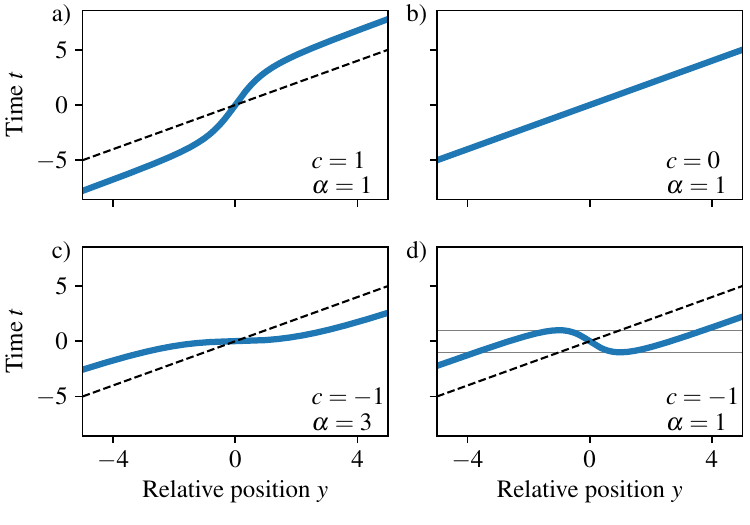}
	\caption[Two particle scattering in \scbm s]{Two particle scattering in \scbm s~in relative coordinates $y=x_1-x_2$ as function of time $t$ for Lieb-Liniger shift \eqref{equ:pre_int_LL_twoparticle_ratio}: a) For positive $\varphi(\rapidity) > 0$, then scattering is an effective time-delay or a negative position shift compared to the b) non-interacting evolution $\varphi(\rapidity) = 0$. c) For negative $\varphi(\rapidity) < 0$ particles speed up during scattering, leading to a positive position shift. d) However for strong negative $\varphi(\rapidity) < 0$, it can happen that trajectories ``bend backwards in time'', i.e.\ become multivalued during scattering. This figure was reproduced from~\cite{PhysRevLett.132.251602}.}
	\label{fig:scbm_twoparticle}
\end{figure}

To understand the evolution of the model it is instructive to first study the two particle case (see \cref{fig:scbm_twoparticle}). For two particles the relative coordinate $y(t) = x_1(t)-x_2(t)$ decouples from the center of mass $(x_1(t)+x_2(t))/2$:
\begin{align}
	\hat{y}^0+(v(\rapidity_1)-v(\rapidity_2))t &= y(t) + \sgn_\alpha(y(t))\varphi(\rapidity_1-\rapidity_2)\label{equ:scbm_twoparticle_relative}.
\end{align}
Here $\hat{y}^0 = \hat{x}^0_1-\hat{x}^0_2$. We can easily solve this for $t$, giving us a relation $t(y)$, which has to be inverted. Since $\sgn_\alpha(y)$ is a bounded function we know that $y(t)\to \hat{y}^0+(v(\rapidity_1)-v(\rapidity_2))t \mp \varphi(\rapidity_1-\rapidity_2)$ for long times $t\to \pm \infty$ (assuming $v(\rapidity_1)>v(\rapidity_2)$). This means that the trajectories before and after scattering are precisely shifted by $\varphi(\rapidity_1-\rapidity_2)$. The scattering occurs when the LHS of \eqref{equ:scbm_twoparticle_relative} is around $0$ and $t(y)$ will interpolate between both trajectories smoothly. If it is invertible then $y(t)$ will do the same. In general for positive $\varphi(\rapidity_1-\rapidity_2) > 0$ particles slow down and for negative $\varphi(\rapidity_1-\rapidity_2) < 0$ they accelerate during scattering. However, if the scattering shift becomes too negative, trajectories ``bend backwards in time'', i.e. $t(y)$ is not invertible anymore, and thus trajectories become multivalued during scattering (see \cref{sec:scbm_negphi}). 

For the moment let us exclude this case. We can extend the two-particle observations to more particles: particles will follow straight lines according to their individual constant (bare) velocity $v(\rapidity_i)$. When two particles scatter, these trajectories get shifted by the scattering shift. This behavior generalizes to one of tracer particles in hard rods.

\begin{result}
	The particles of semi-classical Bethe models evolve like tracer particles.
\end{result}

We will see that this is very convenient for deriving GHD and has applications as effective models for solitons, see \cref{sec:scbm_app_other}.

\subsection{Convex optimization problem for positive scattering shift}
We have seen that inverting $\scbmmap_{\vec{\rapidity}}$ is crucial for obtaining the trajectories of particles. Note that one can obtain the solution to \eqref{equ:scbm_self_consistency_bare_time_evol} as the stationary point of the following action
\begin{align}
	\scbmaction(\vec{x}|\vec{\hat{x}}) = \tfrac{1}{2}\sum_{i} (x_i-\hat{x}_i)^2 + \tfrac{1}{4} \sum_{i\neq j}\mathrm{abs}_\alpha(x_i-x_j)\varphi(\rapidity_i-\rapidity_j).\label{equ:scbm_action}
\end{align}
Here $\mathrm{abs}_\alpha(x) = \int_0^x\dd{y} \sgn_\alpha(y)$ is a regularization of the absolute value function. In case $\varphi(\rapidity)\geq 0$ action \eqref{equ:scbm_action} is (strictly) convex, meaning that there exist always a unique solution $\vec{x}$ for any configuration in ``contracted space'' $\vec{\hat{x}}$.

\begin{result}
	For positive scattering shift, there always exists a unique solution $\vec{x}$, which is also smooth in $\vec{\hat{x}}$. This in particular implies that trajectories $\vec{x}(t)$ are smooth in time.
\end{result}
This is an extremely strong result and not only tells us that these semi-classical Bethe models are well-defined models, but also enables us to compute their trajectories efficiently using convex optimization solvers.

\subsection{What about negative scattering shift?}
\label{sec:scbm_negphi}
So what can we say about negative phase-shift and the presence of multivalued trajectories? Note that no matter how small the scattering shift, if sufficiently many particles scatter simultaneously there will always be a situation where multivalued trajectories occur. This seems to indicate that a construction is not possible: note that the existence of multiple solutions implies that $\scbmmap_{\vec{\rapidity}}$ (and also the canonical transformation \eqref{equ:scbm_generating_function}) is not invertible. Thus, formally the model is not well-defined globally on phase-space $\mathbb{R}^{2N}$.

\begin{figure}[!h]
	\centering
	\includegraphics{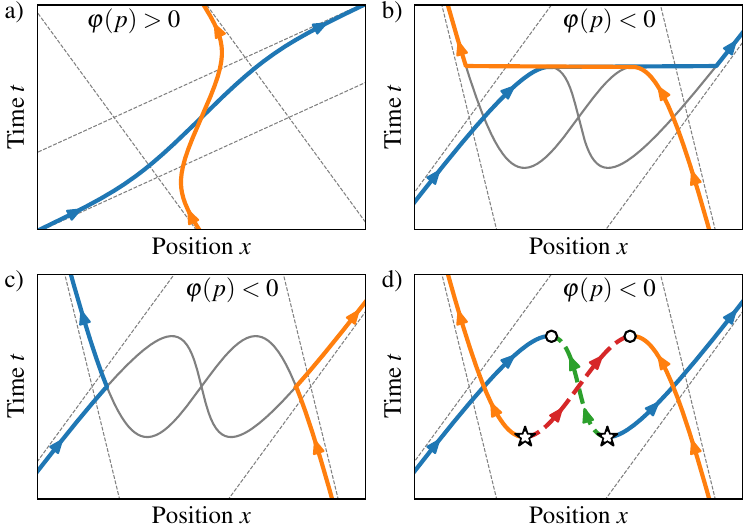}
	\caption[Interpretations of multivalued trajectories]{Interpretations of multivalued trajectories: a) If $\varphi(\rapidity) > 0$ multivalued trajectories are not possible. We propose 3 different interpretations of multivalued trajectories: b) particles jump at some point during scattering (like tracer hard rods, see \cref{fig:pre_int_HR_dynamics} b), c) particles exchange their rapidities during scattering (like physical hard rods, see \cref{fig:pre_int_HR_dynamics} a)) or c) that the backwards in time part of a trajectory should be interpreted as an anti-particle (dashed lines). A particle/ anti-particle pair is created at the point where a trajectory become multivalued (stars). The anti-particle eventually annihilates with the incoming particle (circles), allowing the newly created particle to escape to infinity (with an effectively shifted position by $\varphi(\rapidity)$). This figure was reproduced from~\cite{PhysRevLett.132.251602}.}
	\label{fig:scbm_interpretations}
\end{figure}

Can we resolve this unsatisfying result? We did not develop a full solution, but the following ideas seem to be promising. Note that on each individual solution branch during scattering the canonical map is well-defined. Therefore, the ``only'' thing to do is to make sure that these branches are glued together in a way that preserves the symplectic form of the phase-space. 

In~\cite{PhysRevLett.132.251602} we proposed three different intuitive schemes to choose a specific branch during scattering, depicted in \cref{fig:scbm_interpretations}:
\begin{itemize}
	\item Ignore additional solutions and let particles jump, similar to hard rods tracer particles
	\item Similar to physical hard rods, exchange particles during scattering
	\item Inspired by Feynman's picture of anti-particles, interpret solutions running backwards in time as anti-particles.
\end{itemize}
All of these solutions work in practice, however, formal constructions beyond the intuitive pictures are still lacking.

\begin{openproblem}
	For scattering shifts that are not exclusively positive, construct the \scbm~globally on phase-space. 
\end{openproblem}

\section{Generalized hydrodynamics}
\label{sec:scbm_GHD}
The GHD of these models can easily be obtained using the quick argument presented in \eqref{equ:LL_SP_QP_GHD_quick}. However, here we would like to give a more robust derivation of the GHD equation:

First, recall that we obtain the GHD equation in the Euler scaling limit $L,T,N \to \infty$. Let us set $T=L$ and rescale $x\to Lx, \hat{x} \to L\hat{x}$ and $t\to Lt$. Equation \eqref{equ:scbm_self_consistency_bare_time_evol} becomes
\begin{align}
	\hat{x}_i^0 + v(\rapidity_i)t &= x_i(t) + \tfrac{1}{2L} \sum_{j\neq i} \sgn_\alpha(L(x_i(t)-x_j(t)))\varphi(\rapidity_i-\rapidity_j)\\
	&\to x_i(t) + \tfrac{1}{2L} \sum_{j\neq i} \sgn(x_i(t)-x_j(t))\varphi(\rapidity_i-\rapidity_j).\label{equ:scbm_ghd_self_consistency_bare_time_evol} 
\end{align}
Here we used that as $L \to \infty$ for any finite $x$, $\sgn_\alpha(Lx) \to \sgn(x)$, meaning the regularization $\alpha$ does not play an important role on these large scales. Now note that if two particles $i$ and $j$ have equal $\rapidity_i=\rapidity_j$ and $\hat{x}_i^0 > \hat{x}_j^0$ then $x_i(t) > x_j(t)$. If $\varphi(\rapidity)\geq 0$ this holds even microscopically, otherwise it might be broken during scattering, depending on the precise way to deal with multivalued trajectories. However, by $\hat{x}_i^0 > \hat{x}_j^0$ we mean that $\hat{x}_i^0$ is macroscopically apart from $\hat{x}_j^0$, hence both particles will never scatter (see \cref{rem:scbm_ghd_macroscopic_scattering}).

We can therefore assume that $x_i(t) \to \ghdcharacteristichatx(t,\hat{x}_i^0,\rapidity_i)$ where $\ghdcharacteristichatx(t,\hat{x}^0,\rapidity)$ is an increasing function in $\hat{x}^0$. We can thus write
\begin{align}
	\hat{x}^0 + v(\rapidity)t &= \ghdcharacteristichatx(t,\hat{x}^0,\rapidity) + \tfrac{1}{2L} \sum_{j}\sgn(\ghdcharacteristichatx(t,\hat{x}^0,\rapidity)-\ghdcharacteristichatx(t,\hat{x}^0_j,\rapidity_j))\varphi(\rapidity-\rapidity_j)\\
	&= \ghdcharacteristichatx(t,\hat{x}^0,\rapidity) + \tfrac{1}{2}\int\dd{\hat{y}^0}\dd{\rapidity'}\hat{\rho}^0(\hat{y}^0,\rapidity')\sgn(\ghdcharacteristichatx(t,\hat{x}^0,\rapidity)-\ghdcharacteristichatx(t,\hat{y}^0,\rapidity'))\varphi(\rapidity-\rapidity'),\label{equ:scbm_ghd_self_consistency_characteristic}
\end{align} 
where we defined the quasi-particle density in contracted coordinates and in physical coordinates as
\begin{align}
	\hat{\rho}^0(\hat{x}^0,\rapidity) &= \tfrac{1}{L} \sum_i \delta(\hat{x}^0-\hat{x}^0_i)\delta(\rapidity-\rapidity_i), & \rho(t,x,\rapidity) &= \tfrac{1}{L} \sum_i \delta(x-x_i(t))\delta(\rapidity-\rapidity_i).\label{equ:scbm_ghd_rho_def} 
\end{align}
Note that the contracted quasi-particle density $\hat{\rho}^0(\hat{x}^0,\rapidity)$ is related to the physical quasi-particle density via (for any observable $\testfunction(x,\rapidity)$)
\begin{multline}
	\int\dd{x}\dd{\rapidity} \rho(t,x,\rapidity) \testfunction(x,\rapidity) = \tfrac{1}{L} \sum_i \testfunction(x_i(t),\rapidity_i)\\
	\approx \tfrac{1}{L} \sum_i \testfunction(\ghdcharacteristichatx(t,\hat{x}^0_i,\rapidity_i),\rapidity_i) = \int\dd{\hat{x}^0}\dd{\rapidity}\hat{\rho}^0(\hat{x}^0,\rapidity)\testfunction(\ghdcharacteristichatx(t,\hat{x}^0,\rapidity),\rapidity).
\end{multline}
In other words, as $L\to \infty$, $\rho(t,\cdot,\rapidity) = \ghdcharacteristichatx(t,\cdot,\rapidity)_*\hat\rho^0(\cdot,\rapidity)$ is push-forward of $\hat\rho^0(\hat{x}^0,\rapidity)$ via $\ghdcharacteristichatx(t,\hat{x}^0,\rapidity)$ (see appendix \ref{app:push}), or equivalently $\hat{\rho}^0(\hat{x}^0,\rapidity) = \rho(t,\ghdcharacteristichatx(t,\hat{x}^0,\rapidity),\rapidity)\partial_{\hat{x}^0}\ghdcharacteristichatx(t,\hat{x}^0,\rapidity)$.

Now assume that $\hat{\rho}^0(\hat{x}^0,\rapidity)$ approaches a continuous function as $L\to\infty$ (in a weak limit sense). Then obtaining $\ghdcharacteristichatx(t,\hat{x}^0,\rapidity)$ as solution to \eqref{equ:scbm_ghd_self_consistency_characteristic} is a well-defined problem. In case $\varphi(\rapidity)\geq 0$ this solution can again be obtained as the minimizer of a convex action (see \cref{sec:scbm_app_optimization}), hence always exists and is unique. Let us take the time-derivative of \eqref{equ:scbm_ghd_self_consistency_characteristic}
\begin{align}
	v(p) &= \dv{t}\ghdcharacteristichatx(t,\hat{x}^0,\rapidity) + \int\dd{\hat{y}^0}\dd{\rapidity'}\hat{\rho}^0(\hat{y}^0,\rapidity')\delta(\ghdcharacteristichatx(t,\hat{x}^0,\rapidity)-\ghdcharacteristichatx(t,\hat{y}^0,\rapidity'))\varphi(\rapidity-\rapidity')\nonumber\\
	&\alignshift\times\qty(\dv{t}\ghdcharacteristichatx(t,\hat{x}^0,\rapidity)-\dv{t}\ghdcharacteristichatx(t,\hat{y}^0,\rapidity'))\\
	&= \dv{t}\ghdcharacteristichatx(t,\hat{x}^0,\rapidity) + \int\dd{\rapidity'}\rho(t,Z(t,\hat{x}^0,\rapidity),\rapidity')\varphi(\rapidity-\rapidity')\nonumber\\
	&\alignshift\times\qty(\dv{t}\ghdcharacteristichatx(t,\hat{x}^0,\rapidity)-\dv{t}\ghdcharacteristichatx(t,Z^{-1}(t,\ghdcharacteristichatx(t,\hat{x}^0,\rapidity),\rapidity'),\rapidity'))\label{equ:scbm_ghd_veff_tmp},
\end{align}
where $\ghdcharacteristichatx^{-1}$ is the inverse function of $\ghdcharacteristichatx$ in $\hat{x}^0$. Now identify $v\upd{eff}(t,x,\rapidity) = \dv{t}\ghdcharacteristichatx(t,\ghdcharacteristichatx^{-1}(t,x,\rapidity),\rapidity)$ and observe that \eqref{equ:scbm_ghd_veff_tmp} becomes
\begin{align}
	v\upd{eff}(t,x,\rapidity) &= v(\rapidity) + \int\dd{\rapidity'}\rho(t,x,\rapidity')\varphi(\rapidity-\rapidity')\qty(v\upd{eff}(t,x,\rapidity')-v\upd{eff}(t,x,\rapidity)),
\end{align}
which is precisely the effective velocity equation \eqref{equ:pre_int_GHD_TBA_veff} of GHD. As a last step we need to derive an equation for $\rho$. The simplest way is to integrate $\rho$ against a test function $\testfunction(x,\rapidity)$ and use properties of the push-forward (see appendix \ref{app:push}):
\begin{align}
	&\int\dd{x}\dd{\rapidity} \partial_t \rho(t,x,\rapidity)\testfunction(x,\rapidity) = \dv{t} \int\dd{\hat{x}^0}\dd{\rapidity} \hat{\rho}^0(\hat{x}^0,\rapidity)\testfunction(\ghdcharacteristichatx(\hat{x}^0,\rapidity),\rapidity)\\
	&= \int\dd{\hat{x}^0}\dd{\rapidity} \hat{\rho}^0(\hat{x}^0,\rapidity)\partial_x\testfunction(\ghdcharacteristichatx(\hat{x}^0,\rapidity),\rapidity)\dv{t} \ghdcharacteristichatx(\hat{x}^0,\rapidity)\\
	&= \int\dd{\hat{x}^0}\dd{\rapidity} \hat{\rho}^0(\hat{x}^0,\rapidity)\partial_x\testfunction(\ghdcharacteristichatx(\hat{x}^0,\rapidity),\rapidity)v\upd{eff}(t,\ghdcharacteristichatx(\hat{x}^0,\rapidity),\rapidity)\\
	&= \int\dd{x}\dd{\rapidity} \rho(x,\rapidity)\partial_x\testfunction(x,\rapidity)v\upd{eff}(t,x,\rapidity) = -\int\dd{x}\dd{\rapidity} \partial_x(v\upd{eff}(t,x,p)\rho(x,\rapidity)\testfunction(x,\rapidity).
\end{align}
Since this has to hold for any $\testfunction(x,\rapidity)$ we conclude the GHD equation
\begin{align}
	\partial_t \rho(t,x,\rapidity) + \partial_x(v\upd{eff}(t,x,\rapidity)\rho(t,x,\rapidity)) = 0.
\end{align}

\begin{figure}[!h]
	\centering
	\includegraphics{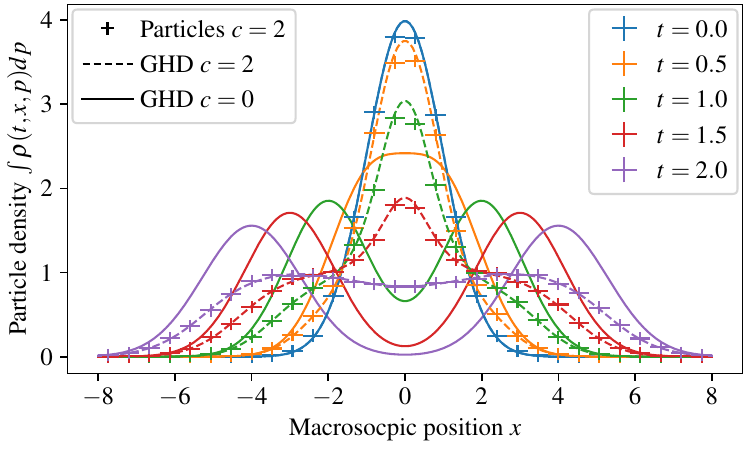}
	\caption[GHD of \scbm s]{GHD of \scbm~with Lieb-Liniger phase shift: the GHD simulation (dashed lines) agrees with the particle simulations. The non-interacting evolution (solid lines) is given as comparision. The initial state is the same as in \cref{fig:pre_GHD_numerics}, but with $A=10$ and $c=2$. Note that this state has occupation function $n>1$ and thus is not a physical state for the quantum model. The classical model does not have this restriction. The GHD simulations were done in IFluid~\cite{10.21468/SciPostPhys.8.3.041}. The particle distributions where obtained numerically by convex optimization of \eqref{equ:scbm_action}. Note that the size of the $+$ represents the numerical error: the particle distribution was averaged over a small  $\Delta x$ ($x$ elongation of $+$) and averaged over $100$ randomly generated initial configurations (the standard deviation of this is the $y$ elongation of $+$), each having $\approx 3000$ particles. This figure was reproduced from~\cite{PhysRevLett.132.251602}.}
	\label{fig:scbm_GHD}
\end{figure}

\begin{result}
	In the Euler scaling limit semi-classical Bethe models satisfy the GHD equation. 
\end{result}

\begin{remark}\label{rem:scbm_ghd_macroscopic_scattering}
	The derivation makes many assumptions, most importantly that $x_i(t)>x_j(t)$ if $\rapidity_i=\rapidity_j$ and $\hat{x}_i^0>\hat{x}_j^0$. We already discussed that this might be broken microscopically during scattering if negative $\varphi(\rapidity)$ are allowed. But since scattering is microscopic, this will not affect macroscopic dynamics. However, in principle one could imagine situations where a macroscopic amount of particles scatter simultaneously. In such cases GHD would break down (see \cref{sec:fixedpoint_blowup}).
\end{remark}

Interestingly, this derivation does not require any thermodynamics. In fact, it does not even require averaging over the initial state. Therefore, if $\hat{\rho}^0(\hat{x}^0,\rapidity)$ or equivalently $\rho(0,x,\rapidity)$ of \eqref{equ:scbm_ghd_rho_def} indeed approach a continuous function, then $\rho(t,x,\rapidity)$ will satisfy the GHD equation. 

\begin{result}
	Semi-classical Bethe models satisfy GHD not just on average, but independently on (almost) every individual configuration as $L\to \infty$. In other words they satisfy the BMFT principle (see \cref{def:pre_hydro_BMFT}).
\end{result}

\section{Thermodynamics}
\label{sec:scbm_thermo}
Similar to hard rods, we do not need the thermodynamic properties of these systems to derive GHD. However, it is still interesting to compute them: in fact, we are able to compute the free energy even in the presence of an external field (which is possible only in very few other models, like hard rods). The full microscopic derivation is given in \cite{doyon2023generalisedtbartdeformationsclassicalfree}, here we would like to present a simpler derivation in the large system size limit.

Since thermodynamics does not depend on time, in this section we will set $t=0$. To set the stage let us define the two distributions
\begin{align}
	\hat{\rho}(\hat{x},\rapidity) & = \tfrac{1}{L} \sum_i \delta(\hat{x}-\hat{x}_i)\delta(\rapidity-\rapidity_i), & \rho(x,\rapidity) &= \tfrac{1}{L} \sum_i \delta(x-x_i)\delta(\rapidity-\rapidity_i),
\end{align}
which we assume to become continuous functions as $L \to \infty$ ($x$ and $\hat{x}$ are macroscopic positions). The map \eqref{equ:scbm_map_def} between $x$ and $\hat{x}$ can be written in this limit as
\begin{align}
	\hat{X}(x,\rapidity) &= x + \tfrac{1}{2}\int\dd{y}\dd{\rapidity'}\rho(y,\rapidity') \sgn(x-y)\varphi(\rapidity-\rapidity').
\end{align}
We can use this to formally relate $\hat{\rho}(\cdot,\rapidity) = \hat{X}(\cdot,\rapidity)_* \rho(\cdot,\rapidity)$.

We want to compute the partition function of our model with external fields, whose Hamiltonian can be written as $H\ind{inhom} = \int\dd{x}\sum_{n=0}^\infty \beta_n(x) \vu{q}_n(x)$ (note that this is equivalent to a local equilibrium state \eqref{equ:pre_hydro_LES}). For instance, the usual canonical ensemble can be obtained by setting $\beta_2(x)=\beta$, meaning that $H = \beta\vu{Q}_2$. Note that $x$ here is a macroscopic coordinate, meaning that the $\beta_a$ vary on a macroscopic scale. Note that we can write $H\ind{inhom}=\sum_{i=1}^N \beta(x_i,\rapidity_i)$ with $\beta(x,\rapidity) = \sum_{n=0}^\infty \beta_n(x)\rapidity^n$.

We will compute the grand-canonical partition function
\begin{align}
	Z = \sum_{N=0}^\infty \frac{1}{N!} \int\dd[N]{x}\dd[N]{\pi} e^{-H\ind{inhom}} = \sum_{N=0}^\infty \tfrac{1}{N!} \int\dd[N]{\hat{x}}\dd[N]{\rapidity} e^{-\sum_{i=1}^N \beta(x_i(\vec{\hat{x}},\vec{\rapidity}),\rapidity_i)}.
\end{align}
Note that we are using classical particle statistics. It is well known that (in the sense of large deviation theory) the measure $\sum_{N=0}^\infty \tfrac{1}{N!} \int\dd[N]{\hat{x}^0}\dd[N]{\rapidity} \to \int D[\hat{\rho}] e^{L\mathcal{S}[\hat{\rho}]}$ can be viewed as a measure on the space of measures $\hat{\rho}(\hat{x},\rapidity)$ with entropy $\mathcal{S}[\hat{\rho}] = -\int\dd{\hat{x}}\dd{\rapidity} \hat{\rho}(\hat{x},\rapidity)\log \hat{\rho}(\hat{x},\rapidity)- \hat{\rho}(\hat{x},\rapidity)$ (this follows from Sanov's theorem~\cite{TOUCHETTE20091})\footnote{This can be understood as follows: $\hat{\rho}(\hat{x},\rapidity)$ is a macro-state, which represents a large number $e^{L\mathcal{S}[\hat{\rho}]}$ of micro-states $\vec{\hat{x}},\vec{\rapidity}$.}. Thus we can write:
\begin{align}
	Z = \int D[\hat{\rho}] e^{L\mathcal{S}[\hat{\rho}]- L \int\dd{\hat{x}}\dd{\rapidity}\hat{\rho}(\hat{x},\rapidity)\beta(X(\hat{x},\rapidity),\rapidity)} = \int D[\hat{\rho}] e^{-L\hat{\mathcal{F}}[\hat{\rho}]}.\label{equ:scbm_thermo_Z_final}
\end{align}
Here $X(\hat{x},\rapidity)$ is the inverse function of $\hat{X}(x,\rapidity)$ in $x$. We are now going to change the integration over $D[\hat{\rho}]$ into $D[\rho]$ and write $Z = \int D[\rho] e^{-L\mathcal{F}[\rho]}$. In the sense of large deviation theory we can neglect the change of measure determinant and write the free energy as (recall the properties of the push-forward, appendix \ref{app:push})
\begin{align}
	\mathcal{F}[\rho] &= \hat{\mathcal{F}}[\hat{\rho}] = \int\dd{x}\dd{\rapidity} \rho(x,\rapidity) \log \frac{\rho(x,\rapidity)}{1\upd{dr}(x,\rapidity)} +\rho(x,\rapidity) + \rho(x,\rapidity) \beta(x,\rapidity).\label{equ:scbm_thermo_F}
\end{align}
Here $1\upd{dr}(x,\rapidity) = \dv{\hat{X}(x,\rapidity)}{x} = 1 + \int\dd{\rapidity'}\rho(x,\rapidity')\varphi(\rapidity-\rapidity')$ coincides with $1\upd{dr}$ as in \eqref{equ:pre_int_GHD_TBA_onedr}. Note that \eqref{equ:scbm_thermo_F} is the integral over a local free energy that has the form \eqref{equ:pre_int_GHD_TBA_free_energy} with classical particle statistics\footnote{Up to a factor of $2\pi$ that can be absorbed by a rescaling.} $\gamma(n)=n\log n - n$. Furthermore, these equations are independent for different $x$. The local $\rho(x,\rapidity)$ can then be found by optimizing \eqref{equ:scbm_thermo_F} for each $x$ independently as in \cref{sec:pre_GHD_TBA}. This means that the local density approximation (LDA) holds. 

\begin{result}
	The full thermodynamic large deviation rate function of semi-classical Bethe models can be computed as $L \to \infty$. They satisfy the local density approximation (LDA). This is rigorously proven in~\cite{doyon2023generalisedtbartdeformationsclassicalfree}, which also contains a result for finite $L$.
\end{result}

\section{Applications of semi-classical Bethe models}

\subsection{Monte-Carlo simulation of GHD}\label{sec:scbm_app_mc}
The GHD equation is a complicated non-linear PDE. Beyond the usual space-time discretization techniques, there exists also other numerical solution approaches. One of them is to simulate particle dynamics that give rise to the GHD equation and average over many simulations. In many PDE's this Monte-Carlo method converges much faster than usual simulations. Particle models are even more important when going beyond standard GHD, for instance for studying the GHD with external fields or diffusive corrections (see the recent work~\cite{urilyon2025simulatinggeneralisedfluidsinteracting} using \scbm s for this purpose). 

Our semi-classical Bethe models can be constructed for a given GHD equation and then be used to simulate them (see \cref{fig:scbm_GHD}). This can be seen as an upgrade to the flea gas algorithm (see \cref{rem:pre_GHD_flea_gas}) as they are proper integrable models.

Compared to the Flea gas, semi-classical Bethe models have a great advantage for numerical simulations: by solving the quadrature \eqref{equ:scbm_ghd_self_consistency_bare_time_evol} we can find the solution directly at any time $t$, which for large times is far more efficient than the flea gas (its computation time scales at least linearly in time). For positive $\varphi(\rapidity)$ one can alternatively apply a efficient convex optimization algorithms on \eqref{equ:scbm_action}.

\begin{result}
	One can use semi-classical Bethe models to Monte-Carlo simulate GHD equations. The computational effort is constant in time due to the existence of a quadrature. 
\end{result} 

So far we discussed the simulation of Euler GHD. Can we also use semi-classical Bethe models to simulate diffusive GHD, say of the quantum Lieb-Liniger model? The naive answer is no: diffusive GHD depends on the particle statistics, which is fermionic in the quantum case, but classical in semi-classical Bethe models. We believe however, that there is a workaround, see \cref{conj:diff_simulation_fermionic}.

\begin{remark}
	As discussed in \cref{sec:scbm_negphi}, so far we are not able to define \scbm s with negative scattering shift microscopically as a Hamiltonian evolution on a phase space due to multivalued trajectories. For the purpose of simulating GHD however, this does not matter: we expect that all trajectories will be microscopically close. Hence, for a large scale simulation it is sufficient to simply pick any of those.
\end{remark}

\subsection{GHD as an optimization problem}\label{sec:scbm_app_optimization}
Semi-classical Bethe models can also be used to obtain another rewriting of GHD. We can take the hydrodynamic limit of the action \eqref{equ:scbm_action} and obtain an action formulation for GHD. There are many ways of doing this, one possible option is to derive an action for the GHD trajectory:
\begin{align}
	\scbmaction_t[\ghdcharacteristichatx|\hat{\rho}] &= \tfrac{1}{2} \int\dd{\hat{x}}\dd{\rapidity}\hat{\rho}(\hat{x},\rapidity) (\ghdcharacteristichatx(\hat{x},\rapidity)-\hat{x}-v(\rapidity)t)^2 + \tfrac{1}{4}\int\dd{\hat{x}}\dd{\hat{y}}\dd{\rapidity}\dd{\rapidity}'\abs{\ghdcharacteristichatx(\hat{x},\rapidity)-\ghdcharacteristichatx(\hat{y},\rapidity')} \varphi(\rapidity-\rapidity').\label{equ:scbm_GHD_optim}
\end{align}
A stationary point of this action at any time $t$ is $Z(t,\hat{x},\rapidity)$ from \eqref{equ:scbm_ghd_self_consistency_characteristic}, from which the solution to the GHD equation can be constructed. This, and further formulations will be published in~\cite{ghdoptimization}. 

\begin{result}
	The solution to the GHD equation can be obtained as a stationary point of the action \eqref{equ:scbm_GHD_optim}.
\end{result}

For positive $\varphi(\rapidity)$ this action is again convex, meaning that it has a unique minimizer and thus a solution to the GHD equation always exists and is \sout{unique} (see \cref{rem:scbm_convexghd_uniqueness2}).

\begin{result}
	For $\varphi(\rapidity) \geq 0$, the GHD equation has a \sout{unique} solution for all times $t$.\label{res:scbm_convexghd_positive}
\end{result}

\begin{remark}
	While intuitively \cref{res:scbm_convexghd_positive} seems to be clear, mathematically speaking it is lacking a precise definition of what is a solution. There are many different notions of solutions (e.g. strong or weak solutions) and one might have to specify how to deal with discontinuities. Existence and uniqueness properties depend strongly on these choices. While a solution exists we do not know whether it will differentiable or even continuous.\label{rem:scbm_convexghd_uniqueness} 
\end{remark}

\begin{remark}
	Why have we crossed out uniqueness? This is related to \cref{rem:scbm_convexghd_uniqueness}: uniqueness here means uniqueness of a certain type of mathematical solution. However, there might be more physical solutions. In fact, we will give an example of this in \cref{sec:fixedpoint_blowup}.\label{rem:scbm_convexghd_uniqueness2}  
\end{remark}

These remarks show that existence and uniqueness of a PDE like the GHD equation is a delicate matter. We will study this question in \cref{sec:fixedpoint} in great detail using a different tool. 

\subsection{Other applications: $T\bar{T}$ transformations and soliton finding}
\label{sec:scbm_app_other}
We would like to mention two further applications, which are not too relevant for the scope of this thesis, but are relevant to other communities.

The first is that the semi-classical Bethe models explicitly implement (generalized) $T\bar{T}$ deformations, which are perturbations of integrable models conserving integrability. They are important tools to construct new integrable models, especially in QFT~\cite{Jiang_2021}. However, while they can be defined implicitly, no explicit construction was known. Semi-classical Bethe models solve this problem in classical physics, it is however not clear how to translate this to quantum physics. Note that in particular the emergence of anti-particles as in \cref{sec:scbm_negphi} is a natural concept in QFT and thus it would be interesting to formulate a similar construction for quantum systems. Note that finding a similar construction (or showing that none exists) will be particularly interesting in QFT as it is commonly believed that only a very restrictive set of phase-shifts are allowed~\cite{mussardo2009statistical}.

The other application is about soliton finding. This is a problem in all integrable models, but in particular in integrable PDE's where the GHD quasi-particles are solitons (wave-packets that do not change shape). During scattering these solitons effectively are shifted by the scattering shift, similar to what happens in semi-classical Bethe models. However, during the interaction of two (or more) solitons the wave-packets lose their shape and oscillate widely, so it is very hard to determine the position of a soliton, even approximately. By comparing with the evolution of a corresponding semi-classical Bethe model one can find approximate position of these solitons (for instance this has recently been shown in the Toda lattice~\cite{aggarwal2025asymptoticscatteringrelationtoda,aggarwal2025effectivevelocitiestodalattice}.

\section{Conclusion}
Based on the emergent classical dynamics identified in \cref{sec:LL}, in this chapter we introduced a new family of integrable models, called \scbm s. They are convenient models to study GHD in the sense that their particles behave like quasi-particles in GHD. For negative scattering shift, the trajectories become multivalued during scattering, making these models ill-defined (for now). We discussed possible interpretations of this, including the creation of particle-antiparticle pairs. Leveraging the simple structure of these models we derived their GHD and their thermodynamics and discussed possible applications.

%!TEX root = thesis.tex

\chapter{The GHD equation: Existence, Uniqueness and Absence of Shock formation}
\label{sec:fixedpoint}

In \cref{sec:LL} we identified \scbm s while studying quantum integrable models. In \cref{sec:scbm} we showed that their large scale dynamics is given by GHD. Thus, it is now a natural next step to study the GHD equation itself. In particular, we will recast the GHD equation into a powerful fixed point problem for its solution. This chapter is based on two publications~\cite{hübner2024newquadraturegeneralizedhydrodynamics,hübner2024existenceuniquenesssolutionsgeneralized}.

\section{Space-time quadrature of the GHD equation}
Since we are going to use slightly different notation as in previous chapters, let us reintroduce it from scratch. Our starting point is the GHD equation $\partial_t \rho(t,x,\rapidity) + \partial_x (v\upd{eff}(t,x,\rapidity)\rho(t,x,\rapidity)) = 0$. For now assume that $\rho(t,x\to-\infty,\rapidity) \to 0$. Define the contracted coordinate $\hat{x}$ as
\begin{align}
	\hat{X}(t,x,\rapidity) &= x + \int_{-\infty}^x\dd{y}\dd{\rapidity'} \rho(t,y,\rapidity')\varphi(\rapidity-\rapidity')\label{equ:fixedpoint_hatX_def}.
\end{align}
We have seen in \cref{sec:pre_GHD_GHD_hatx} that the GHD equation in contracted coordinates $\hat{\rho}(t,\cdot,\rapidity) = \hat{X}(t,\cdot,\rapidity)_* \rho(t,\cdot,\rapidity)$ is non-interacting:
\begin{align}
	\partial_t \hat{\rho}(t,\hat{x},\rapidity) + v(\rapidity)\partial_x\hat{\rho}(t,\hat{x},\rapidity) = 0\label{equ:fixedpoint_hatx_GHD_def}.
\end{align}
The crucial step now is to consider the effect of this coordinate change on the height fields
\begin{align}
	\heightfield(t,x,\rapidity) &= \int_{-\infty}^x\dd{y}\rho(t,y,\rapidity), & \hat{\heightfield}(t,\hat{x},\rapidity) &= \int_{-\infty}^{\hat{x}}\dd{\hat{y}}\hat{\rho}(t,\hat{y},\rapidity).\label{equ:fixedpoint_heightfield}
\end{align}
It is easy to see that similarly to $\hat{\rho}(t,\hat{x},\rapidity)$, $\hat{\heightfield}(t,\hat{x},\rapidity) = \hat{\heightfield}^0(\hat{x}-v(\rapidity)t,\rapidity)$ also evolves trivially. Note that it follows from the definition of the push-forward that $\heightfield(t,x,\rapidity) = \hat{\heightfield}(t,\hat{X}(t,x,\rapidity),\rapidity)$.Therefore, 
\begin{align}
	\heightfield(t,x,\rapidity) = F_{t,x}[\heightfield(t,x,\cdot)] &= \hat{\heightfield}(t,x + \int\dd{\rapidity'}\varphi(\rapidity-\rapidity')\heightfield(t,x,\rapidity'),\rapidity)\\
	&= \hat{\heightfield}^0(x-v(\rapidity)t + \int\dd{\rapidity'}\varphi(\rapidity-\rapidity')\heightfield(t,x,\rapidity'),\rapidity)\label{equ:fixedpoint_fixedpoint}.
\end{align}
This is a functional fixed-point equation for $\heightfield(t,x,\rapidity)$. Such equations can be solved for instance by iteration. Once a solution $\heightfield(t,x,\rapidity)$ is found, the solution $\rho(t,x,\rapidity)$ can be computed from it, for instance by taking a derivative in $x$ (there are also other explicit formulas~\cite{hübner2024newquadraturegeneralizedhydrodynamics}). Note that $\hat{\heightfield}^0(\hat{x},\rapidity)$ can be obtained from the initial data. Most importantly, note that the fixed-point equation decouples for different space-time points $(t,x)$. This means that one can compute the solution of the GHD equation directly at any space-time point, without needing to consider any other point. This is a drastic simplification of the GHD equation (which has the three degrees of freedom $t,x,\rapidity$) onto an equation with one remaining degree of freedom $\rapidity$; useful for both numerical simulations and analytical considerations.

\begin{result}
	The GHD equation has a space-time quadrature, i.e.\, its solution can be found by solving a fixed-point equation of a function in $\rapidity$ only ($t$ and $x$ appear as external parameters).
\end{result} 

\section{Application to repulsive Lieb-Liniger model}
The new fixed-point approach is an incredibly powerful as we will now demonstrate on the example of the (repulsive) Lieb-Liniger model. 

For that we will need the following two features of the Lieb-Liniger model. First, recall $\varphi(\rapidity) = \tfrac{2c}{c^2+\rapidity^2} \geq 0$ from which one can explicitly compute $\int\dd{\rapidity}\varphi(\rapidity) = 2\pi$. Second, recall that the occupation function $0 \leq n < 1$ is bounded (we exclude the boundary case $n = 1$). From this and \eqref{equ:pre_GHD_hatX_n_relation} it follows that $0\leq \hat{\rho}< \tfrac{1}{2\pi}$. This bound implies that its height field $\abs{\hat{\heightfield}^0(\hat{x},\rapidity)-\hat{\heightfield}^0(\hat{y},\rapidity)} < \tfrac{\abs{\hat{x}-\hat{y}}}{2\pi}$ is a Lipschitz function. 

Now fix $t$ and $x$ and observe that in the supremum norm $\norm{f} = \sup_\rapidity \abs{f(p)}$ we have that for any height fields $\heightfield_1(\rapidity)$ and $\heightfield_2(\rapidity)$:
\begin{align}
	\norm{F_{t,x}[\heightfield_1]-F_{t,x}[\heightfield_2]}_\infty &\leq \tfrac{1}{2\pi} \norm{\int\dd{\rapidity'}\varphi(\rapidity-\rapidity')(\heightfield_1(\rapidity')-\heightfield_2(\rapidity'))}_\infty \\
	&< \tfrac{1}{2\pi} \qty(\int\dd{\rapidity'}\varphi(\rapidity-\rapidity'))\norm{\heightfield_1-\heightfield_2}_\infty = \norm{\heightfield_1-\heightfield_2}_\infty.
\end{align}
Mathematically speaking this means that the functional fixed-point map $F_{t,x}$ is a contracting map in the Banach space of bounded functions\footnote{Note that $\Phi(t,x,\rapidity)$ is not necessarily a bounded function as $x \to \infty$. However, here we study the problem for fixed $t$ and $x$. Therefore, we only require that $\Phi(t,x,\rapidity)$ is a bounded function in $\rapidity$.}. The Banach fixed-point theorem states that a contracting map on a Banach space always has a unique fixed point.

This immediately implies that a unique $\heightfield(t,x,\rapidity)$ exists for all times $t$ and all positions $x$. The same is then obviously true for its derivative $\rho(t,x,\rapidity)$.

\begin{result}
	The GHD equation of the repulsive Lieb-Liniger model always has a unique solution. 
\end{result}

But we can go even further. Let us differentiate \eqref{equ:fixedpoint_fixedpoint} w.r.t $x$:
\begin{align}
	\rho(t,x,\rapidity) &= \hat{\rho}^0\qty(x-v(\rapidity)t + \int\dd{\rapidity'}\varphi(\rapidity-\rapidity')\heightfield(t,x,\rapidity'),\rapidity) \qty(1+\int\dd{\rapidity'}\varphi(\rapidity-\rapidity')\rho(t,x,\rapidity')).
\end{align}
Using
\begin{align}
	n(t,x,\rapidity) = 2\pi \hat{\rho}^0(x-v(\rapidity)t + \int\dd{\rapidity'}\varphi(\rapidity-\rapidity')\heightfield(t,x,\rapidity'),\rapidity)\label{equ:fixedpoint_smooth_n}
\end{align}
we observe
\begin{align}
	\rho(t,x,\rapidity) &= \tfrac{1}{2\pi}n(t,x,\rapidity) + n(t,x,\rapidity)\int\tfrac{\dd{\rapidity'}}{2\pi}\varphi(\rapidity-\rapidity')  \rho(t,x,\rapidity')
\end{align}
or in other words $\rho(t,x,\rapidity) = \qty(\tfrac{1}{2\pi}n(t,x,\rapidity))\upd{drT}$.

Now assume that the initial $\rho(t=0,x,\rapidity)$ is smooth and thus also $\hat{\rho}^0(\hat{x},\rapidity)$ is a smooth function. Observe the following:
Since the transposed dressing equation has a unique solution (see below \cref{thm:fixedpoint_existsol}) on the space of bounded functions, we know that $\rho(t,x,\rapidity)$ is bounded. But because of this, its height field $\heightfield(t,x,\rapidity)$ is differentiable. Then \eqref{equ:fixedpoint_smooth_n} in turn implies that $n(t,x,\rapidity)$ is differentiable, implying  $\rho(t,x,\rapidity)$ is differentiable as well. This then implies that $\heightfield(t,x,\rapidity)$ is twice differentiable and so on. By iterating this argument\footnote{These arguments overlook many mathematical details. The actual proof requires much more sophisticated analysis.} we conclude that $\rho(t,x,\rapidity)$ is a smooth function in $x$ (and similarly also in $t$ and in $\rapidity$).

\begin{result}
	The solution to the GHD equation of the repulsive Lieb-Liniger model always remains smooth for all times if it is initially smooth. In particular, shock formation is absent.
\end{result}
This finally solved the conjecture (see \cref{sec:pre_GHD_GHD_noshock}) that solutions to the GHD equation do not develop shocks (at least in this model).

\begin{remark}\label{rem:fixedpoint_jacobian_drT}
	The appearance of the transposed dressing equation has deeper reasons. Without going into details, one can view the transposed dressing operation as the Jacobian of the fixed point equation \eqref{equ:fixedpoint_fixedpoint}. In a nutshell this means that the fixed point equation will have a unique solution if the dressing equation has one as well (for any $n(t,x,\rapidity)$) and vice versa.
\end{remark}

Another conclusion from the above construction is that for any function $b(n,\rapidity)$ the quantity $B(t) = \int\dd{x}\dd{\rapidity} b(n(t,x,\rapidity))1\upd{dr}(t,x,\rapidity)$ is conserved in time: using $1\upd{dr}(t,x,\rapidity) = \partial_x \hat{X}(t,x,\rapidity)$ we have
\begin{align}
	B(t) = \int\dd{x}\dd{\rapidity} b(2\pi \hat{\rho}^0(\hat{X}(t,x,\rapidity),\rapidity)\partial_x \hat{X}(t,x,\rapidity) = \tfrac{1}{2\pi} \int\dd{\hat{x}}\dd{\rapidity}b(2\pi\hat{\rho}^0(\hat{x},\rapidity),\rapidity).
\end{align}
This includes, as special cases, conserved quantities and any entropy.
\begin{result}\label{res:fixedpoint_entropy}
	The solution to the GHD equation conserves all conserved quantities and any entropy (see \cref{sec:pre_GHD_GHD_entropies}).
\end{result}

\begin{remark}
	It is easy to derive more fixed-point equations of various other quantities (see~\cite{hübner2024newquadraturegeneralizedhydrodynamics,hübner2024existenceuniquenesssolutionsgeneralized} and~\cite[App A]{10.21468/SciPostPhys.18.4.135} for details). Depending on the precise situation one might be more suitable than the others.
\end{remark}

\begin{remark}
	The Banach fixed point theorem has also been applied to establish existence and uniqueness of solutions to the TBA equations \eqref{equ:pre_int_GHD_TBA_free_energy_delta_3}, see~\cite{FRING1999579}.
\end{remark}

\section{Mathematical statements}
Due to its importance we turned these arguments into a proper mathematical proof~\cite{hübner2024existenceuniquenesssolutionsgeneralized}. We would like to state the exact theorems here. 

We assume that $\varphi(\rapidity)$ and $n^0(x,\rapidity) \geq 0$ are measurable functions and satisfy
\begin{align}
	\sup_\rapidity \int\dd{\rapidityp}\abs{\varphi(\rapidity-\rapidityp)}\sup_{x} n^0(x,\rapidityp) < C,
\end{align}
where $C=2\pi$ if $\varphi(\rapidity-\rapidityp) \geq 0$ or $C=\pi$ otherwise. In the Lieb-Liniger model this includes all states with $n^0(x,\rapidity)<1$, but also some states where $n^0(x,\rapidity)$ reaches its maximal value $1$ (in particular it includes all known physically relevant states in Lieb-Liniger). In addition, we also need that the initial state decays sufficiently fast in $\rapidity$:
\begin{align}
	\sup_{x,\rapidity} \abs{v(\rapidity)n^0(x,\rapidity)} < \infty.
\end{align}

\begin{theorem}\label{thm:fixedpoint_existsol}
	Under the above assumptions the GHD equation has a weak essentially unique solution satisfying for all $t_1,t_2,x_1,x_2,\rapidity \in \mathbb{R}$:
	\begin{align}
		\int_{x_1}^{x_2}\dd{x} \rho(t_2,x,\rapidity) - \rho(t_1,x,\rapidity) + \int_{t_1}^{t_2}\dd{t} j(t,x_2,\rapidity) - j(t,x_1,\rapidity) &= 0.\label{equ:fixedpoint_math_weak_GHD}
	\end{align}
	Furthermore,
	\begin{itemize}
		\item if additionally both $n^0(x,\rapidity)$ and $v(\rapidity)n^0(x,\rapidity)$ are continuous in $x$ (pointwise in $\rapidity$), then there exists a unique continuous solution (in $t,x$ pointwise in $\rapidity$).
		\item if additionally, both $n^0(x,\rapidity)$ and $v(\rapidity)n^0(x,\rapidity)$ are $r\geq 1$ times continuously differentiable in $t,x$ (pointwise in $\rapidity$) and $\sup_{x,\rapidity} \abs{(1+\abs{v(p)})^{a+1}\partial_x^a n^0(x,\rapidity)} < \infty$ for $1 \leq a \leq r$, then the solution will be a strong $r$ times continuously differentiable (in $t,x$ pointwise in $\rapidity$) solution to the GHD equation.
	\end{itemize}
\end{theorem}

Note that initial states do not need to decay at $x\to \pm \infty$ and can have discontinuities (in which case we can only find a weak solution \eqref{equ:fixedpoint_math_weak_GHD} to the GHD equation.

As part of the proof of \cref{thm:fixedpoint_existsol} we showed that the dressing equation \eqref{equ:pre_int_TBA_dressing} and the effective velocity equation \eqref{equ:pre_int_GHD_TBA_veff} are well-defined and have a unique solution in the space of bounded functions.

\begin{remark}
	These results only apply to the GHD equation on the infinite line. However, one can easily extend it to the GHD equation of a periodic box by considering periodic initial states.
\end{remark}

\subsection{What about other models?}
\Cref{thm:fixedpoint_existsol} gives a strong result for a large family of models and demonstrates that the hydrodynamic approximation of integrable models (aka GHD) is much different from the hydrodynamics of non-integrable models, for instance due to the absence of shock formation.

The results are for now restricted to sufficiently small scattering phase shift or equivalently sufficiently low density. If we go beyond these restrictions our workhorse, the Banach fixed-point theorem, does not apply anymore. This, however, does not mean that the fixed-points are non-unique.

To study a different model, here are some options:
\begin{itemize}
	\item Try to find a different Banach space on which the fixed-point equation is contracting. If this is possible, most of the above results can easily be extended.
	\item Try to use a different fixed-point theorem other than the Banach fixed-point theorem. 
	\item In case the fixed point equation seems not to be contracting in any space\footnote{Any easy way of checking this is to perform the fixed point iteration numerically. If the solution blows up, then the fixed point equation cannot be contracting in any space.}, try to restructure the fixed-point equation. For instance, consider the strategy used in~\cite[App A]{10.21468/SciPostPhys.18.4.135} for the (trotterized) XXZ chain.
\end{itemize}

On the other hand it is also easy to establish whether it is possible to have singular behavior of the GHD equation in a certain model: if one finds any (physical) $n$ such that there exists a non-zero solution to $f= n\vu{T} f$, this means that $\vb{1}-n\vu{T}$ cannot be inverted, hence the transposed dressing operation is not well-defined. Since by \cref{rem:fixedpoint_jacobian_drT} the transposed dressing equation is the Jacobian of the fixed-point equation \eqref{equ:fixedpoint_fixedpoint}, we expect this to signal a singularity in its solution. Take such an $n$ and evolve it back in time, which gives by construction an initial state that will develop some singularity in finite time. Thus, the absence of singularities in the GHD equation is strongly linked to the well-definedness of the (transposed) dressing operation of the model.

We will discuss an example in the next section, however, we would like to mention that such singularities are probably going to be very different from shocks. The reason for this is that we expect them to be short-lived on the infinite line (with $\rho(t,x,\rapidity) \to 0$ as $x\to\pm \infty$): as $t \to \infty$ the solution to the GHD equation will spread out further and eventually end up in a low density state. In this low density state the solution to the GHD equation will be unique and thus has to coincide with the solution from the fixed-point equation. However, from \cref{res:fixedpoint_entropy} we know that the solution to the fixed-point equation has the same entropy as at initial time (note that \eqref{equ:fixedpoint_fixedpoint} does not require the existence of a solution at intermediate times!). Therefore, entropy cannot increase irreversibly, as it happens after shock formation.

\begin{conjecture}\label{conj:fixedpoint_shock_entropy}
	Any singularity in the solution of a GHD equation does not lead to irreversible dynamics. In particular, there is no persistent entropy increase.
\end{conjecture}  

\section{Macroscopic scattering events in hard rods with negative length}\label{sec:fixedpoint_blowup}
In this section we will discuss one mechanism (and the only one known to us) via which the GHD equation can develop gradient catastrophes. It appears when too many particles collide simultaneously, leading to a (universality breaking) macroscopic scattering event. Unlike shocks, they are not characterized by a discontinuity, but instead by a blowup developing in finite time.

\begin{figure}[!h]
	\centering
	\includegraphics{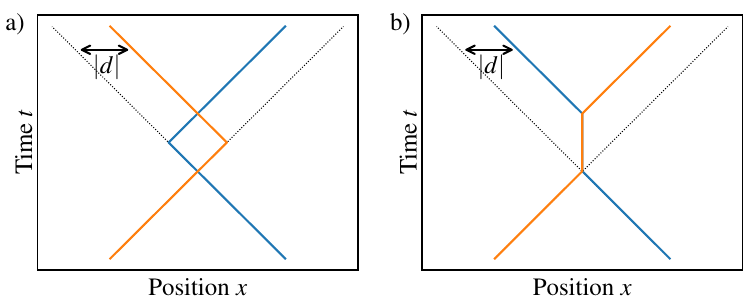}
	\caption[Hard rods with negative length]{Two particle scattering in different implementations of hard rods with negative length $d<0$: a) Similar to physical hard rods, but particles go through each other a distance $d$ before scattering. b) Implementation as \scbm: particles stick together during scattering, until an effective position shift $d$ is reached.}
	\label{fig:fixedpoint_shock_twoparticle}
\end{figure}

As instructive example, consider hard rods with negative length $-\abs{d}$, i.e.\ $v(\rapidity) = \rapidity$ and $\varphi(\rapidity) = -d \geq 0$. To demonstrate the breaking of universality we will use two different microscopic implementations of this model, depicted in \cref{fig:fixedpoint_shock_twoparticle}. One of them is based on physical hard rods~\cite{MR4772242,ferrari2024diffusivefluctuationsgeneralized}, where particle penetrate each other for a short time (the trajectories can be found using (\ref{equ:pre_int_HR_contract}-\ref{equ:pre_int_HR_expand}), but with negative $d$). The other implementation is a \scbm~with trajectories given by \eqref{equ:scbm_self_consistency_bare_time_evol} with $\varphi(\rapidity) = -d>0$. Note that apart from regions in which particle scatter, the trajectories of both implementations exactly coincide. 

To gain intuition, let us study the transposed dressing equation first. In hard rods it is explicitly given by
\begin{align}
	f\upd{drT}(\rapidity) = f(\rapidity) -\tfrac{d}{2\pi} n \int\dd{\rapidity}f\upd{drT}(\rapidity).
\end{align}
By integrating both sides over $\rapidity$ one can easily find
\begin{align}
	f\upd{drT}(\rapidity) &= f(\rapidity) - \tfrac{dn(\rapidity)}{2\pi+d\bar{n}}\bar{f},
\end{align}
where we denoted $\bar{g} = \int\dd{\rapidity}g(\rapidity)$. For positive length $d > 0$ (normal hard rods) this is always a finite number. However, for negative length $d < 0$, this is ill-defined whenever $\bar{n} = 2\pi/\abs{d}$.

In order to see the effect in GHD, we want to start from a state where $\bar{n} < \tfrac{2\pi}{\abs{d}}$ (a low density state) and evolve to a state where $\bar{n} > \tfrac{2\pi}{\abs{d}}$ (a high density state), thereby crossing the singular $\bar{n} = 2\pi/\abs{d}$ line.

In a nutshell, we simply want to collide as many particles in the same region as possible. It is easy to find a suitable initial state: simply choose an $n(x,\rapidity)$ where $\bar{n} > \tfrac{2\pi}{\abs{d}}$ and evolve backwards until $\bar{n} < \tfrac{2\pi}{\abs{d}}$ (this will always happen for sufficiently long times as particle separate eventually). 

\begin{figure}[!h]
	\centering
	\includegraphics{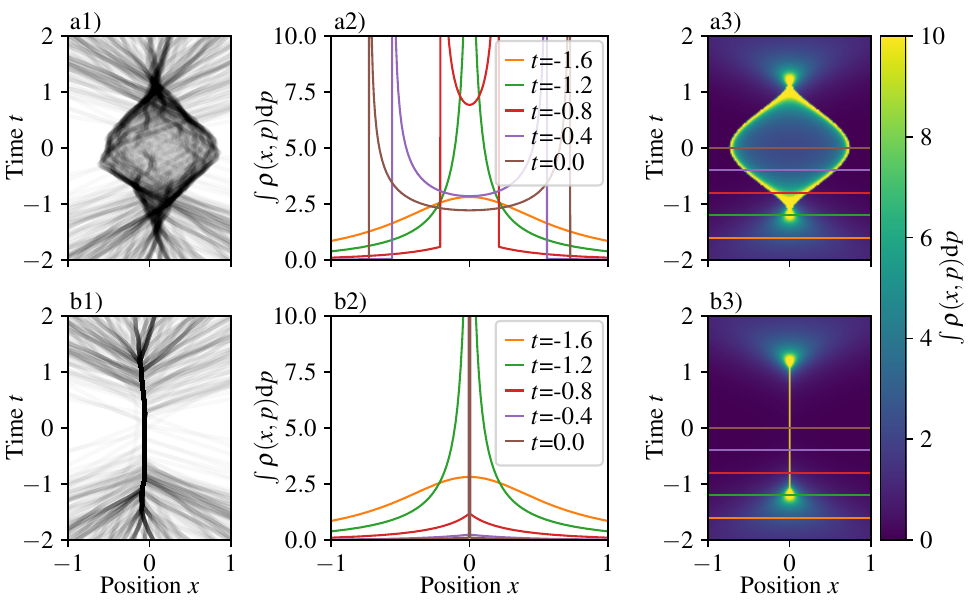}
	\caption[Macroscopic scattering event in hard rods with negative length]{Macroscopic scattering event in hard rods with negative length $d=-1$ for $\hat{\rho}^0(\hat{x},\rapidity) = \tfrac{5}{2\pi}e^{-\hat{x}^2/2}\qty(e^{-(\rapidity-1)^2/2} + e^{-(\rapidity+1)^2/2})$, realized either as hard rods a1)-a3) or as \scbm~ b1)-b3): a1)/b1) Trajectories of microscopic particles, a2)/a3) particle density profiles and a3)/b3) heatmap of the particle density (lines correspond to the profiles of a2)/b2)). Before and after the microscopic scattering event both descriptions agree, but throughout it their behavior is fundamentally different: hard rods a1)-a3) penetrate each other for a short time, while in the \scbm~b1)-b3) they accumulate in one point. In either case the particle density diverges during the macroscopic scattering event.}
	\label{fig:fixedpoint_shock}
\end{figure}

Equivalently, instead of using $n(x,\rapidity)$ we can characterize the initial state by $\hat{\rho}^0(\hat{x},\rapidity)$. For \cref{fig:fixedpoint_shock} we use\footnote{The conventions for $\hat{x}$ in chapters \ref{sec:scbm} and \ref{sec:fixedpoint} differ by a constant shift. Here we use the one of  \cref{sec:scbm}, since conveniently the resulting $\rho(t,x,\rapidity)$ is symmetric in space and time.}
\begin{align}
	\hat{\rho}^0(\hat{x},\rapidity) = \tfrac{5}{2\pi}e^{-\hat{x}^2/2}\qty(e^{-(\rapidity-1)^2/2} + e^{-(\rapidity+1)^2/2}),
\end{align}  
which for $d=-1$ is above the threshold. In \cref{fig:fixedpoint_shock} we not only observe a divergent density, but also what happens inside the singular region. Interestingly, both realizations of the microscopic model differ (see \cref{fig:fixedpoint_shock} a1) and b1)), thus breaking the universality of hydrodynamics. In fact, what we observe is a macroscopic scattering event: usually in GHD, the precise trajectories during $n$ particle scattering are not observable, allowing for a universal description in terms of the scattering phases only. However, in this case a macroscopic number of particles $n$ scatters simultaneously, meaning that the microscopic implementation of scattering becomes important. For usual hard rods with negative length particles penetrate each other for a short time before scattering, leading to \cref{fig:fixedpoint_shock} a1). In \scbm s particles stick together during scattering, implying that a macroscopic amount of them is located in a microscopic region of space (see \cref{fig:fixedpoint_shock} b1)).

\begin{result}
	In hard rods with negative length, a failure of invertibility of the dressing/transposed dressing equation signals a macroscopic scattering event: the model specific implementation of scattering becomes visible on the macroscopic scale, implying a breakdown of hydrodynamic universality.
\end{result}

\begin{remark}
	I expect the same to be true in any model with $\varphi(\rapidity) \geq 0$. For instance, I tried colliding many particles in the Toda model~\cite{10.1063/1.5096892} (in the Toda model particles are also allowed to penetrate each other for a short time), leading to similar behavior as in \cref{fig:fixedpoint_shock} a1).
\end{remark}

Studying the same scenario using GHD we observe a blow up of the quasi-particle density (in finite Euler time). After that, there cannot be a unique solution. Interestingly, two GHD algorithms presented in this thesis capture these different microscopic behaviors. Naturally, the convex optimization algorithm \eqref{equ:scbm_GHD_optim} optimizes the action \eqref{equ:scbm_GHD_optim} by placing many particles at the same position, i.e. the quasi-particle density contains a singular term of the form $\rho(t,x,\rapidity) = \delta(x) \ldots$, leading to \cref{fig:fixedpoint_shock} b3). On the other hand, the solution to the fixed-point problem \eqref{equ:fixedpoint_fixedpoint} agrees with the behavior in \cref{fig:fixedpoint_shock} a1) in the following sense: during the macrosocopic scattering event \eqref{equ:fixedpoint_fixedpoint} has multiple solutions, each giving rise to a possible $\rho(t,x,\rapidity)$. The $\rho(t,x,\rapidity)$ shown in \cref{fig:fixedpoint_shock} a3) is then the sum of all of these contributions\footnote{The intuition for doing this comes from the push-forward, which in case of multivaluedness is also a sum of all individual contributions (see \cref{rem:app_push_multivalued}). In particular, this scheme conserves $\int\dd{x}\rho(t,x,\rapidity)$ for each $\rapidity$.}.  

\begin{result}
	On the level of GHD the onset of a macroscopic scattering event is signaled by a divergent quasi-particle density. During it, there is no unique solution to the GHD equation. Instead, different solution schemes give rise to different solutions. 
\end{result}

\begin{remark}
	The diverging quasi-particle density is more a failure of the mathematical description in terms of a quasi-particle density, rather than a new physical effect. In particular, unlike shocks, the duration of macroscopic scattering events is finite, and after its completion the solution to the GHD equation is again unique (in particular there is no entropy increase, see \cref{conj:fixedpoint_shock_entropy}). However, in an actual physical system the blow up will be problematic: integrable models are typically an approximation of more complicated systems with non-integrable couplings. It is very likely that a divergent quasi-particle density will activate those additional terms, leading to break-down of integrability and thus also of GHD. 
\end{remark}

\section{Linearized GHD equation and evolution of correlation functions}
\subsection{Linearized GHD equation}
In addition to solving the GHD equation one can also use the fixed-point formalism to find an explicit formula for the evolution of correlation functions, which is related to the solution of the linearized Euler equation.

Consider a known solution to the GHD equation $\rho(t,x,\rapidity)$ starting from $\rho^0(x,\rapidity)$. How will the solution $\rho(t,x,\rapidity) \to \rho(t,x,\rapidity) + \delta\rho(t,x,\rapidity)$  change if we perturb the initial state $\rho^0(x,\rapidity) \to \rho^0(x,\rapidity) + \delta\rho^0(x,\rapidity)$.

Perturbing \eqref{equ:fixedpoint_fixedpoint} we find
\begin{align}
	\delta\heightfield(t,x,\rapidity) &= 2\pi\hat{\rho}^0(\hat{u}(t,x,\rapidity),\rapidity) \vu{T}\delta \heightfield(t,x,\rapidityp) + \delta\hat{\heightfield}^0(\hat{u}(t,x,\rapidity),\rapidity),\label{equ:fixedpoint_corr_deltaheightfield_deriv}
\end{align}
where $\hat{u}(t,x,\rapidity) = x-v(\rapidity)t+2\pi\vu{T}\heightfield(t,x,\rapidityp)$ is the initial location (in contracted coordinates) of the GHD characteristic ending at $(x,\rapidity)$ at time $t$. Using appendix \ref{app:dr} this implies $\delta\heightfield(t,x,\rapidity) = (\vb{1}-n(t,x,\cdot)\vu{T})^{-1}\delta\hat{\heightfield}^0(\hat{u}(t,x,\rapidityp),\rapidityp)$. Evaluating this at time $t=0$ we find a similar relation for $\delta\hat{\heightfield}^0(\hat{x},\rapidity)$ in terms of $\delta\heightfield^0(x,\rapidity)$. Finally, this gives the following formula
\begin{align}
	\delta\heightfield(t,x,\rapidity) &= (\vb{1}-n_t\vu{T})^{-1} \insertionop_{u(t,x,\cdot)} (\vb{1}-n_0\vu{T}) \delta\heightfield^0(x,\rapidity).\label{equ:fixedpoint_corr_deltaheightfield}
\end{align}
Here $\insertionop_{y}$ is the operator evaluating a function at $x = y$, $u(t,x,\cdot) = X(\hat{u}(t,x,\rapidity),\rapidity)$ is the initial location of the GHD characterisitic in physical coordinates and $n_t$ is a shorthand for $n(t,x,\rapidity)$. This formula implements the three steps of the evolution of an integrable model (contraction, time-evolution and expansion) on the level of perturbations. Starting from a localized perturbation $\delta\rho^0(x,\rapidity) = \delta(x-x_0)\delta(\rapidity-\rapidity_0)$ we have $\delta\heightfield^0(x,\rapidity) = \theta(x-x_0)\delta(\rapidity-\rapidity_0)$ and thus by taking a spatial derivative of \eqref{equ:fixedpoint_corr_deltaheightfield} we find the solution kernel to the linearized GHD equation
\begin{align}
	\linghdkernel(t,x,\rapidity|x^0,\rapidity^0) &= \tfrac{1}{2}\partial_x \qty[\sgn(x-\ghdcharacteristic(t,x^0,\rapidity))\qty(\delta(\rapidity-\rapidity^0) - \tfrac{1}{2\pi}n(t,x,\rapidity)\varphi(\rapidity-\rapidity^0))]\upd{drT}.\label{equ:fixedpoint_corr_linGHD_kernel}
\end{align}
Here $\ghdcharacteristic(t,x,\rapidity)$ is the GHD characteristic, i.e.\ the endopint of a particle with momentum $\rapidity$ starting at $x$ as function of time $t$:

\begin{result}
	The linearized GHD equation has the explicit solution $\delta\rho(t,x,\rapidity) = \int\dd{x^0}{\rapidity^0} \linghdkernel(t,x,\rapidity|x^0,\rapidity^0) \delta\rho^0(x^0,\rapidity^0)$, where $\linghdkernel(t,x,\rapidity|x^0,\rapidity^0)$ is given by \eqref{equ:fixedpoint_corr_linGHD_kernel}. This agrees with an earlier result~\cite[Sec 3.2]{10.21468/SciPostPhys.5.5.054}.
\end{result}
\begin{figure}[!h]
	\centering
	\includegraphics{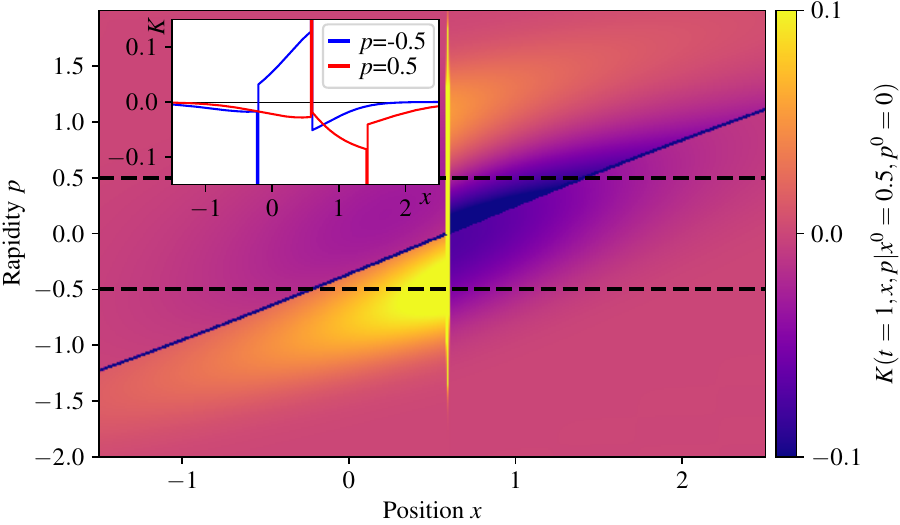}
	\caption[Solution kernel to the linearized GHD equation]{Solution kernel to the linearized GHD equation of the Lieb-Liniger model starting from $x^0=0.5$ and $\rapidity^0=0$ at time $t=1$ on top of a background, which evolved from  $\rho^0(x,\rapidity) = e^{-(x^2+\rapidity^2)/2}/4$ at $t=0$. One can clearly identify the two rays of $\delta$ peaks present in \eqref{equ:fixedpoint_corr_linGHD_singular}. Their crossing point is the location of the GHD quasi-particle that started at $(x^0,\rapidity^0)$ at $t=0$. One ray is at the same location as this particle, the other is at locations of quasi-particles with other rapidities $\rapidity\neq\rapidity^0$ that were also excited at $x^0$ at $t=0$. Additionally, there is a smooth background (coming from the mean-field interaction of particles), which jumps at the location of the $\delta$ peaks (also see the inset, which shows cuts at fixed $\rapidity=\pm 0.5$). This behavior translates to the one of correlation functions, see \cref{fig:pre_hydro_BMFT_LL_macro}. This figure was reproduced from~\cite{hübner2024newquadraturegeneralizedhydrodynamics}.}
	\label{fig:fixedpoint_corr_linGHD}
\end{figure}
It is instructive to extract the singular part of this contribution 
\begin{align}
	\linghdkernel(t,x,\rapidity|x^0,\rapidity^0) &= \delta(x-Y(t,x^0,\rapidity^0))\qty[\delta(\rapidity-\rapidity^0)]\upd{drT}\nonumber\\
	&\alignshift-\delta(x-Y(t,x^0,\rapidity))\tfrac{1}{2\pi}n(t,x,\rapidity)\varphi(\rapidity,\rapidity^0)\nonumber\\
	&\alignshift-\tfrac{1}{2} \sgn(x-Y(t,x^0,\rapidity^0))\tfrac{1}{2\pi}\partial_x n(t,x,\rapidity)\varphi(\rapidity-\rapidity^0))\nonumber\\
	&\alignshift+\tfrac{1}{2}\sgn(x-Y(t,x^0,\rapidity^0))\qty[\partial_xn(t,x,\rapidity)\vu{T}\qty[\delta(\rapidity-\rapidity^0)]\upd{drT}]\upd{drT} +\mathrm{(continuous)}.\label{equ:fixedpoint_corr_linGHD_singular}
\end{align}
This structure is shown in \cref{fig:fixedpoint_corr_linGHD}.

\subsection{Correlation function}
\label{sec:fixedpoint_corr_corr}
The solution to the linearized Euler equation can directly be used to study the evolution of two-point correlation functions. Since both components evolve independently according the linearized Euler equation we have:
\begin{align}
	\expval{\delta \rho(t,x,\rapidity)\delta \rho(s,y,q)} &= \int\dd{x^0}\dd{\rapidity^0}\dd{y^0}\dd{q^0} \linghdkernel(t,x,\rapidity|x^0,\rapidity^0)\linghdkernel(s,y,q|y^0,q^0)\expval{\delta\rho(0,x^0,\rapidity^0)\delta \rho(0,y^,q^0)}.\label{equ:fixedpoint_corr_kernel}
\end{align}
As mentioned in \cref{sec:pre_hydro_bmft}, a particularly interesting case is the equal time correlation function $t=s$, where we observe long-range correlations. This can be understood best by looking at the evolution of the height fields, which is given by
\begin{align}
	\expval{\delta\heightfield(t,x,\rapidity)\delta\heightfield(t,y,\rapidityp)} &= (\vb{1}-n_t\vu{T})^{-1}_1(\vb{1}-n_t\vu{T})^{-1}_2 \insertionop_{1,u(t,x,\cdot)} \insertionop_{2,u(t,x,\cdot)}(\vb{1}-n_0\vu{T})_1(\vb{1}-n_0\vu{T})_2 \expval{\delta\heightfield^0\delta\heightfield^0}.\label{equ:fixedpoint_corr_corr_heightfield}
\end{align}
Here the indices $1$ and $2$ refer to whether we apply the operator to the first $(x,\rapidity)$ or the second entry $(y,\rapidityp)$ of the correlation function. As discussed above the operator $(\vb{1}-n_0\vu{T})$ implements the transformation to contracted coordinates, followed by a time-evolution and then the operator $(\vb{1}-n_t\vu{T})^{-1}$ transforms the result back to physical coordinates. 

It is interesting to study their evolution starting from a local equilibrium state \eqref{equ:pre_hydro_LES}, which has spatially uncorrelated GGE correlations $L\expval{\delta\rho^0(x,\rapidity)\delta\rho^0(y,\rapidityp)} = \delta(x-y)C\ind{GGE}^0(x,\rapidity,\rapidityp)$, see \eqref{equ:pre_int_GHD_TBA_corr}, or in terms of the height fields
\begin{align}
	L\expval{\delta\heightfield^0(x,\rapidity)\delta\heightfield^0(y,\rapidityp)} &= D\ind{GGE}^0(x\wedge y,\rapidity,\rapidityp),\label{equ:fixedpoint_corr_EV_heightfield_init}
\end{align}
where $x\wedge y = \min(x,y)$ and $D\ind{GGE}^0(x,\rapidity,\rapidityp) = \int_{-\infty}^x \dd{x} C\ind{GGE}^0(x,\rapidity,\rapidityp)$. Note that $\expval{\delta\heightfield^0(x,\rapidity)\delta\heightfield^0(y,\rapidityp)}$ is a continuous function. The map to contracted coordinates is given by
\begin{align}
	\expval{\delta\hat{\heightfield}^0(\hat{x},\rapidity)\delta\hat{\heightfield}^0(\hat{y},\rapidity)} &= \insertionop_{1,X(\hat{x},\rapidity)}\insertionop_{2,X(\hat{y},\rapidityp)}(\vb{1}-n_0\vu{T})_1(\vb{1}-n_0\vu{T})_2 \expval{\delta\heightfield^0(x,\rapidity)\delta\heightfield^0(y,\rapidityp)}.\label{equ:fixedpoint_corr_corr_hatheightfield_init}
\end{align}
Taking two derivatives we obtain
\begin{align}
	L\expval{\delta\hat{\rho}^0(\hat{x},\rapidity)\delta\hat{\rho}^0(\hat{y},\rapidity)} &= \insertionop_{1,X(\hat{x},\rapidity)}\insertionop_{2,X(\hat{y},\rapidityp)}\tfrac{1}{1\upd{dr}_11\upd{dr}_2}(\vb{1}-n_0\vu{T})_1(\vb{1}-n_0\vu{T})_2 C\ind{GGE}^0(x,\rapidity,\rapidityp)\delta(x-y)\\
	&= \tfrac{1}{2\pi\gamma''(2\pi \hat{\rho}^0(\hat{x},\rapidity))}\delta(\hat{x}-\hat{y})\delta(\rapidity-\rapidityp) + \ldots =: \delta(\hat{x}-\hat{y})\hat{C}\ind{GGE}(\hat{x},\rapidity,\rapidityp) + \ldots,\label{equ:fixedpoint_corr_corr_hatrho_init}
\end{align}
where the remaining terms will only contribute a continuous background. Note that this singular part is invariant under non-interacting time-evolution
\begin{align}
	L\expval{\delta\hat{\rho}(t,\hat{x},\rapidity)\delta\hat{\rho}(t,\hat{y},\rapidity)}	&= \tfrac{1}{2\pi\gamma''(2\pi \hat{\rho}(t,\hat{x},\rapidity))}\delta(\hat{x}-\hat{y})\delta(\rapidity-\rapidityp) + \ldots.\label{equ:fixedpoint_corr_corr_hatrho_t}
\end{align}
Therefore, when expanding into physical coordinates at time $t$ we will again find a $\delta$ peak with the correct local GGE correlations. However, in this process we need to take two derivatives of
\begin{align}
	L\expval{\delta \heightfield(t,x,\rapidity)\delta \heightfield(t,y,\rapidityp)} = (\vb{1}-n_t\vu{T})_1^{-1}(\vb{1}-n_t\vu{T})_2^{-1}\hat{D}\ind{GGE}(t,\hat{X}(t,x\wedge y,\rapidity),\rapidity,\rapidityp) + \ldots,\label{equ:fixedpoint_corr_corr_heightfield_t}
\end{align}
which can also act on the dressings. Thus we find
\begin{align}
	L\expval{\delta \rho(t,x,\rapidity)\delta \rho(t,y,\rapidityp)}&\eval_{x=y} = \delta(x-y) C\ind{GGE}(t,x,\rapidity,\rapidityp)\nonumber\\
	&+ \qty(\qty[(\vb{1}-n_t\vu{T})^{-1}\partial_xn_{t}\vu{T}]_1 - \qty[(\vb{1}-n_t\vu{T})^{-1}\partial_xn_{t}\vu{T}]_2) C\ind{GGE}(t,y,\rapidity,\rapidityp) \sgn(x-y)\nonumber\\
	&+L\expval{\delta \rho(t,x,\rapidity)\delta \rho(t,x,\rapidityp)}_\epsilon,\label{equ:fixedpoint_corr_corr_rho_t} 
\end{align}
where $\expval{\delta \rho(t,x,\rapidity)\delta \rho(t,x,\rapidityp)}_\epsilon$ represents a continuous contribution at $x=y$. Formally, we define it as follows:
\begin{align}
	\expval{\delta \rho(t,x,\rapidity)\delta \rho(t,x,\rapidityp)}_\epsilon &= \lim_{\epsilon\to 0}\sum_{\sigma=-1,1} \expval{\delta \rho(t,x+\sigma\epsilon,\rapidity)\delta \rho(t,x-\sigma\epsilon,\rapidityp)}.
\end{align}
To conclude, at $x=y$ we find a continuous background, but even more intriguingly the correlations also have a jump at this point (see \cref{fig:pre_hydro_BMFT_LL_macro}). Both of these observations mean that the system is locally out-of-equilibrium.

What is so special about the jump? Integrable models like Lieb-Liniger are PT-symmetric. It follows from PT symmetry that, if the state is PT symmetric infinitesimally close to a point $x$, then the correlations must be symmetric around this point\footnote{At time $t=0$ PT symmetry ($x\to -x$ and $t\to -t$) only swaps $x\to-x$, hence correlations are symmetric in $x$}, hence there cannot be a jump. We conclude that the evolved state breaks local PT symmetry. Furthermore, the initial state \eqref{equ:pre_hydro_LES} is PT symmetric. Therefore, the dynamics spontaneously breaks PT symmetry locally. This breaking does not build up slowly in time, but a finite jump immediately appears at arbitrarily small Euler time $t$\footnote{This is does not mean that the jump actually appears immediately. It only means that the jump appears on a shorter time scale than the Euler time scale.}. Note that the jump is proportional to $\partial_x n(t,x,\rapidity)$, hence it is not present in homogeneous states, i.e.\ in equilibrium GGE states (GGE are time-invariant by construction). Therefore, the jump and the local PT symmetry breaking is an out-of-equilibrium effect. 

\begin{result}\label{res:fixedpoint_corr_jump}
	The equal time two-point correlation functions spontaneously break PT symmetry locally by instantly developing a jump at $x=y$ of order $1/L$.  
\end{result}

\begin{remark}
	One may argue that this $1/L$ PT breaking is not too surprising since the initial state does not satisfy global PT symmetry, but only local PT symmetry asymptotically as $L\to\infty$. Thus some finite size corrections in $L$ are naturally expected. However, I still want to include it in the list of results, as it is in drastic contrast to the local equilibrium assumption usually done in hydrodynamics. In fact, for me personally this observation has probably been the most influential of my whole PhD.
\end{remark}

It will also be interesting for us to have the continuous part of the long-range correlations at $x=y$ and $t \to 0^+$. This can be computed by taking derivatives of \eqref{equ:fixedpoint_corr_corr_heightfield}:
\begin{align}
	L\expval{\delta \rho(0^+,x,\rapidity)\delta \rho(0^+,x,\rapidityp)}_\epsilon &= \frac{1}{2}(\vb{1}-n_t\vu{T})^{-1}_1(\vb{1}-n_t\vu{T})^{-1}_2\sgn(v\upd{eff}(x,\rapidity)-v\upd{eff}(x,\rapidityp))\nonumber\\
	&\alignshift\times\qty[\partial_x n_1\vu{T}_1 (\vb{1}-n\vu{T})_2 C\ind{GGE}(x,\rapidity,\rapidityp) - \partial_x n_2\vu{T}_2 (\vb{1}-n\vu{T})_1 C\ind{GGE}(x,\rapidity,\rapidityp)],\label{equ:fixedpoint_corr_corr_rho_t0plus}
\end{align}
which can alternatively also be expressed in terms of correlations of $\delta n = \tfrac{2\pi}{1\upd{dr}}(\vb{1}-n\vu{T})\delta \rho$:
\begin{align}
	&\expval{\delta n(0^+,x,\rapidity)\delta n(0^+,x,\rapidityp)}_\epsilon = \frac{(2\pi)^2}{2}\sgn(v\upd{eff}(x,\rapidity)-v\upd{eff}(x,\rapidityp))\nonumber\\
	&\alignshift\times\qty[\bra{\rapidity}\vu{T}_1 (\vb{1}-n\vu{T})_1^{-1}\ket{\rapidityp}\frac{\partial_x n(x,\rapidity)}{2\pi \gamma''(n(x,\rapidityp))1\upd{dr}(x,\rapidity)} - \bra{\rapidityp}\vu{T}_1 (\vb{1}-n\vu{T})_1^{-1}\ket{\rapidity} \frac{\partial_x n(x,\rapidityp)}{2\pi \gamma''(n(x,\rapidity))1\upd{dr}(x,\rapidityp)}].\label{equ:fixedpoint_corr_corr_n_t0plus}
\end{align}

\section{Conclusion}
In this chapter we introduced a new powerful technique to analyze the GHD equation: by recasting it into a fixed-point problem, where space and time appear only as external variables, we can directly study its solution at any space-time point. On the example of the Lieb-Liniger model we established the global existence and uniqueness of solutions as well as absence of shock formation in GHD, thereby solving a community conjecture.

However, this does not mean, that global solutions exist in all models. On the example of hard rods with negative length we showed that colliding too many particles, leads to a finite time blow up of the solution and a subsequent macroscopic scattering event. There the underlying particle scattering mechanism becomes apparent on a macroscopic scale, hence breaking the universality of GHD.

The fixed-point method is also a convenient tool to study the evolution of initial state fluctuations on top of the hydrodynamic background. We use our theory to establish that the long range correlations at $x=y$ are non-trivial: in addition to a continuous background they also have a local PT breaking discontinuity.

%!TEX root = thesis.tex

\chapter{The effect of long range correlations on diffusive GHD}
\label{sec:diff}
In this chapter we go beyond Euler hydrodynamics and study its diffusive correction. We will finally conclude that the previously established Navier-Stokes like diffusive GHD equation \eqref{equ:pre_GHD_diff_GHD} is incorrect. To be more precise, \eqref{equ:pre_GHD_diff_GHD} only applies at local equilibrium states (where it was also rigorously proven in hard rods), but is affected at later times by the long range correlations.

\section{Why might diffusive GHD be wrong?}
The Navier-Stokes like diffusive equation was well established in the community and used in many works, e.g.~\cite{10.21468/SciPostPhys.6.4.049,PhysRevLett.121.160603,10.21468/SciPostPhys.9.5.075,Durnin_2021,MOLLER2023112431,PhysRevB.98.220303,PhysRevLett.122.127202,Bastianello_2021,DeNardis_2022,PhysRevLett.125.240604,Doyon_2017,PhysRevResearch.6.023083,Bulchandani_2024,PhysRevLett.134.010405,10.21468/SciPostPhysCore.7.2.025,10.21468/SciPostPhysCore.1.1.002}. It was believed to be the reason for thermalization of integrable models (in particular in integrability breaking external potentials). However, I slowly started to question this theory due to multiple independent inconsistencies:
\begin{enumerate}
	\item Long range correlations: As found in \cref{sec:fixedpoint_corr_corr}, long range correlations have a non-trivial local behavior at $x=y$. As we discuss in the next section this has the same order of magnitude as the diffusion term in \eqref{equ:pre_GHD_diff_GHD}.
	\item Wrong thermalization timescale: I was doing (heavy) numerical simulations to understand the numerically observed (and still unexplained) failure of thermalization of hard rods in a harmonic trap~\cite{PhysRevLett.120.164101,PhysRevE.108.064130}. As I increased the system size $L$, the scaling of the thermalization time did not scale diffusively ($t\sim L^2$), but much slower.
	\item Dependence on particle statistics: While Euler GHD does not depend on the particle statistics, \eqref{equ:pre_GHD_diff_GHD} does. The particle statistics emerges from the notion of ``sum over all states'', required to define GGE states. As mentioned in \cref{rem:pre_GHD_integrable_PDE}, in integrable PDEs there is no unambiguous way of defining such measures. Using different measures might lead to different particle statistics, and thus potentially to different diffusive equations \eqref{equ:pre_GHD_diff_GHD}.
\end{enumerate}

In addition to this, \eqref{equ:pre_GHD_diff_GHD} had never been properly checked in numerical simulations with the required accuracy\footnote{Recall that diffusion is a small effect that is suppressed as $1/L$.}. These ideas let to a careful investigation of the diffusive correction in hard rods~\cite{hübner2025diffusivehydrodynamicshardrods}. Then, based on this, a general theory was proposed in~\cite{PhysRevLett.134.187101}. The derivation presented here based on ``Hydrodynamics without averaging'' was developed later~\cite{hydrowithoutaveraging}, but is also very insightful on its own.

\section{Why might long range correlations affect diffusive hydrodynamics?}
We will start by giving a simple argument to illustrate why long range correlations should affect the diffusive equation. This is simply based on the fact that these fluctuations have magnitude $\order{1/L}$ and thus they are of the same order as the diffusive correction to GHD.

To see that the long range correlations indeed give contributions of the same order, let us for the moment ignore diffusion and only take into account fluctuations. We can write the quasi-particle density as $\rho(t,x,\rapidity) = \expval{\rho(t,x,\rapidity)} + \delta \rho(t,x,\rapidity)$, where the fluctuations are small $\delta \rho(t,x,\rapidity) \sim 1/\sqrt{L}$. By averaging the GHD equation we can now derive an equation for $\expval{\rho(t,x,\rapidity)}$
\begin{align}
	\partial_t \expval{\rho(t,x,\rapidity)} &= -\partial_x\expval{ j[\rho(t,x,\cdot)](\rapidity)} = -\partial_x\expval{ j[\expval{\rho(t,x,\cdot)} + \delta \rho(t,x,\cdot)](\rapidity)}\\
	&= -\partial_x\Big[j[\expval{\rho(t,x,\cdot)}] + \int\dd{\rapidity'} \fdv{j[\expval{\rho(t,x,\cdot)}]}{\rho(\rapidity')}\underbrace{\expval{\delta \rho(t,x,\rapidity')}}_{=0}\nonumber\\
	&\alignshift+ \tfrac{1}{2}\int\dd{\rapidity'}\dd{\rapidity''} \fdv[2]{j[\expval{\rho(t,x,\cdot)}]}{\rho(\rapidity')\rho(\rapidity'')}\expval{\delta \rho(t,x,\rapidity')\delta \rho(t,x,\rapidity'')}\Big] + \order{1/\sqrt{L^3}}.\label{equ:diff_LR_basic}
\end{align}
Note that $\expval{\delta \rho(t,x,\rapidity')\delta \rho(t,x,\rapidity'')} = \expval{\rho(t,x,\rapidity')\rho(t,x,\rapidity'')}\upd{c}$ is of order $1/L$. However, $\expval{\rho(t,x,\rapidity')\rho(t,x,\rapidity'')}\upd{c}$ is evaluated at the same point $x$, where it is actually singular: it has a $\delta$ function, a jump and a regular part. We do not know how to do deal with these singularities, but at least the non-singular part of the long range correlations at $x=y$ should contribute. 

The diffusive correction to GHD \eqref{equ:pre_GHD_diff_GHD} was obtained by assuming that the state is in local equilibrium~\cite{Boldrighini1997,10.21468/SciPostPhys.6.4.049}. Therefore, one reasonable way to proceed would be to repeat these derivations, but in a state which includes long range correlations. Unfortunately, it is not clear how the state will actually look like beyond the one-point and two-point function we know from GHD\footnote{Also those are only large scale averages of the precise microscopic quantities.}. For instance, following the derivation \eqref{equ:diff_LR_basic}, we need to know exactly how the singularities (which are not actual microscopic singularities) of the two-point function look on the microscopic scale.

Even if we could solve this problem, it would still be a very cumbersome task to derive the equation. In the following we would like to present an alternative approach that also gives much deeper insights into the physics of GHD.

\begin{remark}
	The idea that a diffusive order correction might arise from the correlations in the system as in \eqref{equ:diff_LR_basic} had been proposed earlier under the name ``diffusion from convection''~\cite{10.21468/SciPostPhys.9.5.075}.
\end{remark}

\section{Hydrodynamics without averaging}
Hydrodynamics is usually described in terms of averages $\expval{\rho(t,x,\rapidity)}$ over, for instance, a local equilibrium state. However, from the BMFT perspective (\cref{sec:pre_hydro_bmft}) each individual configuration will evolve via the Euler GHD equation. The only randomness comes from the initial state, which can be viewed as a probability distribution of these configurations. In other words, Euler GHD is self-averaging (almost surely w.r.t. to the initial state). In general, we have seen that the fluctuations of $\rho(t,x,\rapidity)$ are of order $1/\sqrt{L}$. This is sufficient to show BMFT, since these fluctuations are subleading compared to the Euler scale quantities. Therefore, for now, there is no reason to assume that it might be exact on order $1/L$, which is required to make statements about diffusion (that indeed Euler scale BMFT is accurate beyond the diffusive scale $1/L$ will be the big result of this section). At first, fluctuations of the order $1/\sqrt{L}$ might seem to be a clear-cut indication that the BMFT can be at most accurate up to $\order{1/\sqrt{L}}$. However, note that the initial state already has fluctuations of order $1/\sqrt{L}$ build in. So even if Euler scale GHD were microscopically exact, we would still observe fluctuations of the order $1/\sqrt{L}$ coming solely from the initial state. We can only say that the inaccuracy of GHD is at most $\order{1/\sqrt{L}}$, as otherwise it would dominate the fluctuations. 

\begin{remark}
	In this view the additional new term in \eqref{equ:diff_LR_basic} comes only from the initial state fluctuations. It is thus only an apparent, but not an intrinsic correction to Euler GHD. On the other hand, we expected the RHS of \eqref{equ:pre_GHD_diff_GHD} to be such an intrinsic correction. The aim of this section is to identify true intrinsic diffusive corrections to GHD (we will see that there are none).
\end{remark}

To disentangle the inaccuracy of GHD (which we will call intrinsic noise) from the initial state fluctuations (initial state noise) we would like to introduce a new paradigm to think about GHD (and hydrodynamics more generally): hydrodynamics without averaging. The idea is to completely abandon any initial state noise: the initial state is a fixed deterministic configuration. We will then compute the error of the solution to the GHD equation compared to the exact evolution of this individual state. Then we will try to understand how this error scales as $N\sim L\to \infty$. This is not well-defined: the error will of course depend on the specific state. Even worse, there is no way to take a limit $L\to \infty$ as we can choose completely different states for each $N$. This is a massive mathematical advantage of considering local equilibrium states: it defines a family of initial states depending only on a single parameter $L$, with an unambiguous limit $L\to \infty$. Still, we will see that we can indeed meaningfully estimate the scaling of the error. 

If the state and its the evolution are deterministic, how can any randomness ever emerge? Also, GHD is fundamentally collective theory. For instance, the effective velocity only appears due to a large number of scattering events. How can we describe this on a deterministic state, which we know microscopically exact? The idea is to assume that one can only observe the system up to some scale. All details beyond this scale contribute to an intrinsic noise, of which only self-averaged behavior is observable on the larger scales. For instance, the velocities of particles with self-average to the effective velocity and its fluctuations should be well-described by white noise or similar (we never study the nature of the noise, but this would be an interesting topic of future research).

Even though a mathematically less clear description, ``hydrodynamics without averaging'' has many advantages. Once established, one can easily average it over any initial state. This might be a local equilibrium state, but it can also be a state with long range correlations or a completely exotic state. This averaging will then introduce state-dependent additional terms into the evolution equation, similar to \eqref{equ:diff_LR_basic}. Hence, in different physical situations one obtains different corrections to Euler GHD. This is the true power of this approach.

\begin{remark}
	While having a fixed deterministic configuration is natural in classical systems, it does not really make sense in quantum systems. Here, the corresponding states would be pure quantum states. However, observables like charge densities are always fluctuating in such states, unless they are an eigenstate of the observable. Hence, the initial state would have to be a slowly modulated state that is locally an (approximate) eigenstate of all charge densities\footnote{The charge densities do not need to commute microscopically, however the commutator of their mesoscopic scale averages is suppressed as $1/L$. Hence if we choose the pure state to be an eigenstate of mesocopic averages of charge densities, then their fluctuations should also be suppressed as $L \to \infty$.}.
\end{remark}

\begin{openproblem}
	Extend ``hydrodynamics without averaging'' to quantum systems.
\end{openproblem} 

\section{Derivation in hard rods}
\label{sec:diff_HR}
Given an initial configuration of $N$ hard rods $\qty{x_i,\rapidity_i}$, their positions at time $t$ are given by
\begin{align}
	x_i(t) &= \hat{x}_i(t) + \tfrac{d}{L} \sum_{j\neq i} \theta(\hat{x}_i(t)-\hat{x}_j(t)),\label{equ:diff_HR_particle_xt}\\
	\hat{x}_i(t) &= \hat{x}_i+p_i t\label{equ:diff_HR_particle_hatxt},\\
	\hat{x}_i &= x_i - \tfrac{d}{L} \sum_{j\neq i} \theta(x_i-x_j).\label{equ:diff_HR_particle_hatx}
\end{align}
Compared to \eqref{equ:pre_int_HR_contract}, here we already used macroscopic coordinates $x\to Lx, t\to Lt ,\hat{x} \to L\hat{x}$. Our aim is to compare this exact evolution to the GHD prediction. For that, we need to choose a good quantifier of the accuracy. A natural idea is to go via observables: imagine a hard rods experiment, where one initializes the system in a specific state, measures this state once to determine an initial coarse grained $\rho\ind{CG}(x,\rapidity)$. Then one evolves the system up to time $t$ and measures a single observable $\expval{\rho\ind{micro}(t),\testfunction}=\tfrac{1}{L}\sum_i\testfunction(x_i(t),\rapidity_i)$. In parallel, one numerically solves the GHD equation starting from $\rho\ind{CG}(x,\rapidity)$ and then computes $\expval{\rho\ind{CG}(t),\testfunction} = \int\dd{x}\dd{\rapidity} \rho(t,x,\rapidity)\testfunction(x,\rapidity)$. The question we are asking is how precisely does a GHD solution predict the value of this observable.

This GHD prediction of such observables is given by:
\begin{align}
	\expval{\rho\ind{CG}(t),\testfunction} &= \int\dd{x}\dd{\rapidity} \rho\ind{CG}(x,\rapidity) \testfunction(\ghdcharacteristic(t,x,\rapidity),\rapidity),\label{equ:diff_HR_GHD_obs}\\
	\ghdcharacteristic(t,x,\rapidity) &= \hat{X}(x) +\rapidity t + d \int\dd{y}\dd{\rapidityp} \rho\ind{CG}(y,\rapidityp) \theta(\hat{X}(x)-\hat{X}(y) + (\rapidity-\rapidityp) t) - \tfrac{d}{2L},\label{equ:diff_HR_GHD_Xt}\\
	\hat{X}(x) &= x - d \int\dd{y}\dd{\rapidityp} \rho\ind{CG}(y,\rapidityp) \theta(x-y) + \tfrac{d}{2L}.\label{equ:diff_HR_GHD_hatX}
\end{align}

As for the microscopic evolution, this can be viewed in 3 steps: 1. transform to contracted coordinates, 2. non-interacting evolution, 3. transforming back to physical coordinates. We will treat each of these steps individually. Note that in \eqref{equ:diff_HR_GHD_hatX}, compared to \eqref{equ:pre_GHD_hatX_def}, we added the constant shift $\tfrac{d}{2L}$ to the definition of $\hat{X}(x)$. This shift simply cancels for the GHD evolution, but we will see that the $\tfrac{d}{2L}$ naturally emerges in \cref{sec:diff_HR_contract}.

\subsection{Fluid cell averaging}
\label{sec:diff_HR_obs}
Before we can study any dynamics we need to explain how to obtain a coarse grained $\rho\ind{CG}(x,\rapidity)$ that corresponds to our initial data. Here we choose to do this by fluid cell averaging. Choose $1/L \ll \Delta x, \Delta \rapidity \ll 1$ and divide space into cells
\begin{align}
	\set{A}_\alpha &= \qty[x_\alpha-\Delta x/2,x_\alpha +\Delta x/2], & \set{B}_\beta &= \qty[\rapidity_\beta-\Delta \rapidity/2,\rapidity_\beta+\Delta \rapidity/2], & \set{C}_{\alpha,\beta} = A_\alpha \times B_\beta,\label{equ:diff_HR_cells_def}
\end{align}
where $\alpha,\beta\in \mathbb{Z}$ label the cells and $x_\alpha=\alpha\Delta x, \rapidity_\beta=\beta\Delta\rapidity$ are the centers of the cells. Now let us denote by $n_{\alpha,\beta}$ the number of particles in $\set{C}_{\alpha,\beta}$ and by
\begin{align}
	\rho_{\alpha,\beta} &= \frac{n_{\alpha,\beta}}{L\Delta x\Delta \rapidity} =  \frac{1}{L\Delta x\Delta \rapidity} \sum_i \theta((x_i,\rapidity_i)\in \set{C}_{\alpha,\beta})\label{equ:diff_HR_rho_cg}
\end{align}
the density of particles in cell $\set{C}_{\alpha,\beta}$ (note that we expect $\rho_{\alpha,\beta}$ to be of order $1$). The coarse grained $\rho\ind{CG}(x,\rapidity)$ is now the piecewise constant function
\begin{align}
	\rho\ind{CG}(x,\rapidity) &= \sum_{\alpha,\beta} \theta((x,\rapidity)\in \set{C}_{\alpha,\beta}) \rho_{\alpha,\beta}.\label{equ:diff_HR_rho_cg_cont}
\end{align}

At this point we can ask, already at time $t=0$: what is the error of using $\rho\ind{CG}(x,\rapidity)$ on computing an observable $\testfunction(x,\rapidity)$ instead of the exact $\rho\ind{micro}(x,\rapidity)$? To answer this let us write $x_i = x_\alpha + y_i \Delta x$ and $\rapidity_i = \rapidity_\beta + \rapidityp_i \Delta \rapidity$, where $-1/2 < y_i,\rapidityp_i<1/2$ denote the positions of particles inside a cell.

Hence, we can write (by a abuse of notation we write $i \in \set{C}_{\alpha,\beta}$ if $(x_i,\rapidity_i)\in \set{C}_{\alpha,\beta}$)
\begin{align}
	\expval{\rho\ind{micro},\testfunction} &= \tfrac{1}{L} \sum_{\alpha,\beta} \sum_{i\in \set{C}_{\alpha,\beta}} \testfunction(x_\alpha + y_i \Delta x, \rapidity_\beta + \rapidityp_i \Delta \rapidity)\\
	&= \tfrac{1}{L} \sum_{\alpha,\beta} n_{\alpha,\beta} \Big[\testfunction(x_\alpha, \rapidity_\beta) + \partial_x \testfunction(x_\alpha, \rapidity_\beta) \qty[y]_{\alpha,\beta}\Delta x + \partial_\rapidity \testfunction(x_\alpha, \rapidity_\beta) \qty[\rapidityp]_{\alpha,\beta}\Delta \rapidity\nonumber\\
	&\hspace{2.3cm}+ \tfrac{1}{2}  \partial_x^2 \testfunction(x_\alpha, \rapidity_\beta) \qty[y^2]_{\alpha,\beta}\Delta x^2 +  \partial_x\partial_\rapidity \testfunction(x_\alpha, \rapidity_\beta) \qty[y\rapidityp]_{\alpha,\beta}\Delta x\Delta \rapidity\nonumber\\
	&\hspace{2.3cm}+ \tfrac{1}{2}  \partial_\rapidity^2 \testfunction(x_\alpha, \rapidity_\beta) \qty[\rapidityp^2]_{\alpha,\beta}\Delta \rapidity^2 + \order{\Delta x^3}\Big].\label{equ:diff_HR_obs_micro}
\end{align}
Here we introduced $\qty[f]_{\alpha,\beta} = \tfrac{1}{n_{\alpha,\beta}} \sum_{i\in \set{C}_{\alpha,\beta}} f_i$ to represent averages over a cell. Similarly, we can also compute
\begin{align}
	\expval{\rho\ind{CG},\testfunction} &= \sum_{\alpha,\beta} \rho_{\alpha,\beta} \Delta x\Delta\rapidity \int_{-1/2}^{1/2}\dd{y}\dd{\rapidityp}  \testfunction(x_\alpha+y\Delta x,\rapidity_\beta+\rapidityp\Delta\rapidity)\\
	&= \sum_{\alpha,\beta} \rho_{\alpha,\beta} \Delta x\Delta\rapidity \qty[\testfunction(x_\alpha,\rapidity_\beta) + \tfrac{1}{24} \partial_x^2\testfunction(x_\alpha,\rapidity_\beta) \Delta x^2 + \tfrac{1}{24} \partial_\rapidity^2\testfunction(x_\alpha,\rapidity_\beta) \Delta \rapidity^2 + \order{\Delta x^3}]\label{equ:diff_HR_obs_cg}.
\end{align}
Hence, the difference is given by
\begin{align}
	\expval{\rho\ind{CG},\testfunction}-\expval{\rho\ind{micro},\testfunction} &=\Delta x\Delta\rapidity \sum_{\alpha,\beta} \rho_{\alpha,\beta} \Big[-\partial_x \testfunction(x_\alpha, \rapidity_\beta) \qty[y]_{\alpha,\beta}\Delta x - \partial_\rapidity \testfunction(x_\alpha, \rapidity_\beta) \qty[\rapidityp]_{\alpha,\beta}\Delta \rapidity\nonumber\\
	&\alignshift+ \tfrac{1}{2}(\tfrac{1}{12}-\qty[y^2]_{\alpha,\beta}) \partial_x^2\testfunction(x_\alpha, \rapidity_\beta) \Delta x^2 -\partial_x\partial_\rapidity \testfunction(x_\alpha, \rapidity_\beta) \qty[y\rapidityp]_{\alpha,\beta}\Delta x\Delta \rapidity\nonumber\\
	&\alignshift+  \tfrac{1}{2}(\tfrac{1}{12}-\qty[\rapidityp^2]_{\alpha,\beta}) \partial_\rapidity^2\testfunction(x_\alpha, \rapidity_\beta) \Delta \rapidity^2 + \order{\Delta x^3}\Big]\label{equ:diff_HR_obs_diff}.
\end{align}

In the worst case, $\qty[y]_{\alpha,\beta}$ and $\qty[\rapidityp]_{\alpha,\beta}$ can be of order $\order{1}$. Then the error would be
\begin{align}
	\expval{\rho\ind{CG},\testfunction}-\expval{\rho\ind{micro},\testfunction} &= \order{\Delta x}.
\end{align}

\begin{remark}
	If $\testfunction(x, \rapidity)$ is Lipschitz, i.e. $\abs{\partial_x \testfunction(x, \rapidity)} < C$ and $\abs{\partial_\rapidity \testfunction(x, \rapidity)} < C$, then the first term is strictly bounded by $\tfrac{1}{2}C(\Delta x+\Delta p)N/L$. Note that this shows that coarse-graining, as described above, is asymptotically exact as $L\to \infty$.
\end{remark}

However, this error estimate is not realistic. Intuitively, the $y_i$ and $q_i$ should be distributed uniformly in each fluid cell. In fact, any  $\qty[f]_{\alpha,\beta}$ should self average. To account for this, assume that each $y_i$ is independently distributed as
\begin{align}
	f(y) = 1+ \Delta x\frac{\partial_x\rho(x_\alpha,\rapidity_\beta)}{\rho_{\alpha,\beta}} y\label{equ:diff_HR_local_avg_def}
\end{align}
and similarly for $\rapidityp_i$. Here, we took into account that the density is not fully homogeneous across the cell. With these assumptions we find
\begin{align}
	\mathbb{E}[\avg{y}_{\alpha,\beta}]&= \tfrac{\Delta x}{12}\tfrac{\partial_x\rho(x_\alpha,\rapidity_\beta)}{\rho_{\alpha,\beta}} \sim \Delta x, & \mathrm{Var}[\avg{y}_{\alpha,\beta}] &= \tfrac{1}{12n_{\alpha,\beta}} \sim \tfrac{1}{L\Delta x\Delta \rapidity}.\label{equ:diff_HR_obs_local_def}
\end{align}
Under these averages we determine
\begin{align}
	\mathbb{E}&\qty[	\expval{\rho\ind{CG},\testfunction}-\expval{\rho\ind{micro},\testfunction}]\nonumber\\
	&= \tfrac{\Delta x\Delta \rapidity}{12} \sum_{\alpha,\beta} \Big[-\partial_x\phi(x_\alpha,\rapidity_\beta)\partial_x\rho(x_\alpha,\rapidity_\beta) \Delta x^2 -\partial_\rapidity\testfunction(x_\alpha,p_\beta)\partial_\rapidity\rho(x_\alpha,\rapidity_\beta) \Delta \rapidity^2 + \order{\Delta x^3}\Big]\\
	&\to -\tfrac{1}{12}\int\dd{x}\dd{\rapidity}\partial_x\testfunction(x,\rapidity)\partial_x\rho(x,\rapidity)\Delta x^2+\partial_\rapidity\testfunction(x,\rapidity)\partial_\rapidity\rho(x,\rapidity)\Delta \rapidity^2 + \order{\Delta x^3}\label{equ:diff_HR_obs_diff_mean}
\end{align}
and
\begin{align}
	\mathrm{Var}\qty[\expval{\rho\ind{CG},\testfunction}-\expval{\rho\ind{micro},\testfunction}] &= \tfrac{1}{12}\tfrac{\Delta x\Delta \rapidity}{L}\sum_{\alpha,\beta} \rho_{\alpha,\beta}\Big[\partial_x\testfunction(x_\alpha,\rapidity_\beta)^2 \Delta x^2 +\partial_\rapidity\testfunction(x_\alpha,\rapidity_\beta)^2 \Delta \rapidity^2 + \order{\Delta x^3}\Big]\\
	&\to \tfrac{1}{12L} \int\dd{x}\dd{\rapidity}\rho(x,\rapidity)\Big[\partial_x\testfunction(x,\rapidity)^2\Delta x^2+\partial_p\testfunction(x,\rapidity)^2\Delta \rapidity^2\Big] + \order{\Delta x^3/L}.\label{equ:diff_HR_obs_diff_var}
\end{align}
We can interpret $\expval{\rho\ind{CG},\testfunction}-\expval{\rho\ind{micro},\testfunction}$ as the emergent noise we observe due to coarse-graining. This noise has expectation value $\sim \Delta x^2$ and standard deviation $\sim \Delta x/\sqrt{L}$. The deviation we will observe when measuring $\testfunction(x,\rapidity)$ is then given by the dominant term:
\begin{align}
	\expval{\rho\ind{CG},\testfunction}-\expval{\rho\ind{micro},\testfunction} \sim \max(\Delta x^2,\Delta x/\sqrt{L}).\label{equ:diff_HR_obs_diff_scale}
\end{align}
This means that if we scale $\Delta x,\Delta \rapidity \sim L^{\mu-1}$, where $0<\mu<1$, we find that the scaling of the error $\expval{\rho\ind{CG},\testfunction}-\expval{\rho\ind{micro},\testfunction} \sim L^{-\nu}$ is
\begin{align}
	\nu = \begin{cases}
		2-2\mu & \mu > 1/2\\
		\tfrac{3}{2}-\mu & \mu < 1/2.
	\end{cases}\label{equ:diff_HR_obs_diff_mu_nu}
\end{align} 

We give a plot of this exponent in \cref{fig:diff_HR_scaling}, together with numerical results. We see that there is some kind of ``phase-transition'' here: if cells are large $\mu > 1/2$, then the error is dominated by a systematic error $\sim \Delta x^2$, coming from the fact that $\avg{y}_{\alpha,\beta}\sim \partial_x\rho(x,\rapidity)\Delta x$ is shifted due to slope of the density. If $\mu <1/2$, on the other hand, the error is dominated by statistical fluctuations $\abs{\avg{y}_{\alpha,\beta}} \sim 1/\sqrt{n_{\alpha,\beta}}$, because particles are distributed all over the fluid cell.

At this point the reader might wonder why we do ``hydrodynamics without averaging'', but still average over some probability distribution \eqref{equ:diff_HR_local_avg_def} to obtain these results. The answer is that we only use this probability distribution to estimate the magnitude of terms. After all, \eqref{equ:diff_HR_local_avg_def} is not physical, because it also allows hard rods to overlap (which is forbidden). In particular, \eqref{equ:diff_HR_obs_diff_mean} and \eqref{equ:diff_HR_obs_diff_var} are not the actual errors. The only robust results are the scalings which we expect to be universal. Discarding all next-to-leading order terms in \eqref{equ:diff_HR_obs_diff} we thus arrive at the following result:
\begin{result}\label{res:diff_HR_obs_scaling}
	The error of measuring an observable after coarse graining an individual configuration is given by
	\begin{align}
		\expval{\rho\ind{CG},\testfunction}-\expval{\rho\ind{micro},\testfunction} &=-\Delta x\Delta\rapidity \sum_{\alpha,\beta} \rho_{\alpha,\beta} \Big[\partial_x \testfunction(x_\alpha, \rapidity_\beta) \qty[y]_{\alpha,\beta}\Delta x + \partial_\rapidity \testfunction(x_\alpha, \rapidity_\beta) \qty[\rapidityp]_{\alpha,\beta}\Delta \rapidity + \ldots\Big],\label{equ:diff_HR_obs_diff_final}
	\end{align}
	which in a ``generic'' state decays as $\expval{\rho\ind{CG},\testfunction}-\expval{\rho\ind{micro},\testfunction} \sim \max(\Delta x^2,\Delta x/\sqrt{L})$. We cannot give a clear definition of ``generic'', but intuitively it means that locations of particles are spread evenly inside each cell.
\end{result}

This result is fully deterministic, but we have discarded subleading terms. Recall that the idea of ``hydrodynamics without averaging'' is that we can later on average over a probability measure. In this regard \eqref{equ:diff_HR_obs_diff_final} is quite powerful: it pinpoints which pieces of information, namely $\qty[y]_{\alpha,\beta}$ and $\qty[q]_{\alpha,\beta}$, which are necessary (and sufficient) to compute the error of coarse graining. The precise statistics of the error will then depend on the precise state, but only through $\qty[y]_{\alpha,\beta}$ and $\qty[q]_{\alpha,\beta}$. Furthermore, we need only either their expectation values or variances, depending on whether $\Delta x \gg \sqrt{L}$ or $\Delta x \ll \sqrt{L}$.

\begin{remark}
	\Cref{res:diff_HR_obs_scaling} only applies if $\testfunction(x,\rapidity)$ is smooth (or at least differentiable). If $\testfunction(x,\rapidity)$ has jumps then the scalings will be different.
\end{remark}

\begin{remark}\label{rem:diff_HR_obs_smooth}
	There are other ways to obtain a continuous distribution from a collection of particles. For instance, one could smoothen the location of each individual particle as
	\begin{align}
		\rho\ind{smooth}(x,\rapidity) = \tfrac{1}{L\Delta x\Delta \rapidity} \sum_i \eta(\tfrac{x-x_i}{\Delta x},\tfrac{\rapidity-\rapidity_i}{\Delta \rapidity}),
	\end{align}
	where $\eta(y,\rapidityp)$ is an arbitrary smoothing kernel (for instance a Gaussian). The error of this approximation can be analyzed similarly and leads to different scalings (in fact, for this scheme the error is smaller). However, towards smaller $\Delta x$ the ``phase transition'' of the error, where the dominant part of the error changes from a systematic to a statistical error, still occurs. Hence, unsurprisingly, the error of the approximation depends strongly on the approximation itself.
\end{remark}

\subsection{Contracting space}
\label{sec:diff_HR_contract}
After we understood the effect of fluid cell averaging, we can now study the first step in the hard rods evolution: transforming to contracted coordinates. This map is microscopically given by
\begin{align}
	x_i \to \hat{x}_i = x_i - \tfrac{d}{L}\sum_{j\neq i}\theta(x_i-x_j).\label{equ:diff_HR_contract_micro_hatx}
\end{align}

We will now compare the value of an observable $\expval{\hat{\rho}\ind{micro},\hat{\testfunction}} = \tfrac{1}{L} \sum_i \testfunction(\hat{x}_i,\rapidity_i)$ to the one obtained with the coarse grained density $\expval{\hat{\rho}\ind{CG},\hat{\testfunction}} = \int\dd{x}\dd{\rapidity}\rho\ind{CG}(x,\rapidity)\testfunction(\hat{X}(x),\rapidity)$. For a continuous density the map from $x\to \hat{x}$ is given by \eqref{equ:diff_HR_GHD_hatX}
\begin{align}
	\hat{X}(x) &= x - d\int\dd{x'}\dd{\rapidity'}\rho\ind{CG}(x',\rapidity')\theta(x-x')+\tfrac{d}{2L}\\
	&= x_\alpha + y \Delta x - d\Delta x\Delta \rapidity \sum_{\alpha',\beta'} \rho_{\alpha',\beta'} \int_{-1/2}^{1/2}\dd{y'}\dd{\rapidityp'} \theta(x_\alpha-x_{\alpha'}+(y-y')\Delta x) +\tfrac{d}{2L}\\
	&= x_\alpha + y \Delta x - d\Delta x\Delta \rapidity \sum_{\alpha',\beta'} \rho_{\alpha',\beta'} \theta(x_\alpha-x_{\alpha'}) +\tfrac{d}{2L}\label{equ:diff_HR_contract_cg_hatx}
\end{align}
Here we use the convention $\theta(0) = 1/2$ (which is important for $\alpha = \alpha'$). For convenience, let us denote $\hat{X}_\alpha = x_\alpha - \tfrac{d}{L}\sum_{\alpha',\beta'}n_{\alpha',\beta'}\theta(x_\alpha-x_\alpha')$. Now we can compute
\begin{align}
	\expval{\hat{\rho}\ind{CG},\hat{\testfunction}} &= \Delta x\Delta \rapidity\sum_{\alpha,\beta} \rho_{\alpha,\beta} \qty[ \hat{\testfunction}(\hat{X}_\alpha,\rapidity_\beta) - \tfrac{d}{2L}\partial_{\hat{x}}\hat{\testfunction}(\hat{X}_\alpha,\rapidity_\beta) + \order{\Delta x^2}].\label{equ:diff_HR_contract_cg_obs}
\end{align}
Similarly, we can write
\begin{align}
	\hat{x}_i &= x_\alpha + y_i\Delta x - \tfrac{d}{L}\sum_{\alpha',\beta'}n_{\alpha',\beta'}\theta(x_\alpha-x_\alpha') -\tfrac{d}{L}\sum_{j\in \set{A}_{\alpha}}(\delta_{i\neq j}\theta(y_i-y_j) -\tfrac{1}{2})\\
	&= x_\alpha + y_i\Delta x - \tfrac{d}{L}\sum_{\alpha',\beta'}n_{\alpha',\beta'}\theta(x_\alpha-x_\alpha') -\tfrac{d}{2L}\sum_{j\in \set{A}_{\alpha}}\delta_{i\neq j}\sgn(y_i-y_j) + \tfrac{d}{2L}.
\end{align}
Thus, we find
\begin{align}
	\expval{\hat{\rho}\ind{micro},\hat{\testfunction}} &= \tfrac{1}{L} \sum_{\alpha,\beta} n_{\alpha,\beta} \Big[\hat{\testfunction}(\hat{X}_\alpha,\rapidity_\beta) +  \partial_{\hat{x}}\hat{\testfunction}(\hat{X}_\alpha,\rapidity_\beta) \avg{y}_{\alpha,\beta} + \partial_\rapidity\hat{\testfunction}(\hat{X}_\alpha,\rapidity_\beta) \avg{\rapidityp}_{\alpha,\beta}\Big]\\
	&- \tfrac{d}{2L^2} \sum_{\alpha,\beta} \partial_{\hat{x}}\hat{\testfunction}(\hat{X}_\alpha,\rapidity_\beta) \qty[\sum_{i\in \set{C}_{\alpha,\beta}}\sum_{j\in \set{A}_\alpha}\delta_{i\neq j}\sgn(y_i-y_j)-n_{\alpha,\beta}] + \order{\Delta x^2}.\label{equ:diff_HR_contract_micro_obs}
\end{align}
Therefore, the difference is given by
\begin{align}
	\expval{\hat{\rho}\ind{CG},\hat{\testfunction}}-\expval{\hat{\rho}\ind{micro},\hat{\testfunction}} &= -\tfrac{1}{L} \sum_{\alpha,\beta} n_{\alpha,\beta} \Big[ \partial_{\hat{x}}\hat{\testfunction}(\hat{X}_\alpha,\rapidity_\beta) \avg{y}_{\alpha,\beta} + \partial_\rapidity\hat{\testfunction}(\hat{X}_\alpha,\rapidity_\beta) \avg{\rapidityp}_{\alpha,\beta}\Big]\\
	&\alignshift+ \tfrac{d}{2L^2} \sum_{\alpha,\beta} \partial_{\hat{x}}\hat{\testfunction}(\hat{X}_\alpha,\rapidity_\beta) \qty[\sum_{i\in \set{C}_{\alpha,\beta}}\sum_{j\in \set{A}_\alpha}\delta_{i\neq j}\sgn(y_i-y_j)] + \order{\Delta x^2}\label{equ:diff_HR_contract_diff}
\end{align}
Note that the $\tfrac{d}{2L}$ in \eqref{equ:diff_HR_GHD_hatX} canceled a term in \eqref{equ:diff_HR_contract_diff}. We can analyze the scaling of these terms as before, which is done in appendix \ref{app:diff_err_scaling}. We find again the same result
\begin{align}
	\expval{\hat{\rho}\ind{CG},\hat{\testfunction}}-\expval{\hat{\rho}\ind{micro},\hat{\testfunction}} \sim \max(\Delta x^2,\Delta x/\sqrt{L}),\label{equ:diff_HR_contract_scaling}
\end{align}
which we verify in numerical simulations in \cref{fig:diff_HR_scaling}.

\subsection{Non-interacting time evolution}
\label{sec:diff_HR_free}
Next, we study the non-interacting time evolution. For this, let us for the moment ignore the fact that we are dealing with hard rods and just study non-interacting particles. Their trajectories are simply given by
\begin{align}
	x_i(t) = x_i+\rapidity_i t.\label{equ:diff_HR_free_micro_def}
\end{align}
If we measure an observable at time $t$ we therefore find
\begin{align}
	\expval{\rho\ind{micro}(t),\testfunction} &= \tfrac{1}{L} \sum_i \testfunction(x_i+\rapidity_i t,\rapidity_i) = \tfrac{1}{L} \sum_{\alpha,\beta} n_{\alpha,\beta} \Big[\testfunction(x_\alpha,\rapidity_\beta) + \partial_x \testfunction(x_\alpha,\rapidity_\beta) (\avg{y}_{\alpha,\beta} + \avg{\rapidityp}_{\alpha,\beta} t)\\
	&\alignshift+ \partial_\rapidity \testfunction(x_\alpha,\rapidity_\beta) \avg{\rapidityp}_{\alpha,\beta} + \order{\Delta x^2 +t^2\Delta p^2}\Big]\label{equ:diff_HR_free_micro_obs}
\end{align}
Similarly, if we coarse grain we observe
\begin{align}
	\expval{\rho\ind{CG}(t),\testfunction} &= \int\dd{x}\dd{\rapidity}\rho\ind{CG}(x,\rapidity)\testfunction(x+\rapidity t,\rapidity) = \Delta x\Delta \rapidity\sum_{\alpha,\beta} \rho_{\alpha,\beta} \testfunction(x_\alpha+\rapidity_\beta t,\rapidity_\beta) + \order{\Delta x^2 +t^2\Delta p^2}\label{equ:diff_HR_free_cg_obs}
\end{align}
Hence, the difference is simply given by
\begin{multline}
	\expval{\rho\ind{CG}(t),\testfunction}-\expval{\rho\ind{micro}(t),\testfunction}\\
	= -\tfrac{1}{L} \sum_{\alpha,\beta} n_{\alpha,\beta} \Big[\partial_x \testfunction(x_\alpha,\rapidity_\beta) (\avg{y}_{\alpha,\beta} + \avg{\rapidityp}_{\alpha,\beta} t) + \partial_\rapidity \testfunction(x_\alpha,\rapidity_\beta) \avg{\rapidityp}_{\alpha,\beta} \Big] + \order{\Delta x^2 +t^2\Delta p^2}.\label{equ:diff_HR_free_diff}
\end{multline}
As in \eqref{equ:diff_HR_obs_diff_scale}, we thus conclude
\begin{align}
	\expval{\rho\ind{CG}(t),\testfunction}-\expval{\rho\ind{micro}(t),\testfunction} &\sim \max(\Delta x^2+ t^2\Delta \rapidity^2,\sqrt{\Delta x^2+ t^2\Delta \rapidity^2}/\sqrt{L}).\label{equ:diff_HR_free_diff_scaling}
\end{align}
If $t = \order{1}$ this reduces to \eqref{equ:diff_HR_obs_diff_scale} (we choose to keep $t$ explicit here for later reference). We check this in \cref{fig:diff_HR_scaling}.

\subsection{Full time evolution}
\label{sec:diff_HR_full}
Now, we want to study the full time-evolution consisting of the four steps: 1. coarse graining, 2. contracting, 3. non-interacting time evolution and 4. expanding (which is very similar to contracting). We understand each of these steps in detail now, but unfortunately we cannot combine them. The whole derivation has to be done in one go. This can be done~\cite{hydrowithoutaveraging}, but the computation is lengthy and does not add any conceptual understanding, hence we will skip it here. One finds the same scaling (verified in \cref{fig:diff_HR_scaling}):
\begin{result}
	If the coarse grained density is evolved to time $t = \order{1}$ using the GHD equation, the value of an observable $\testfunction(x,\rapidity)$ differs from the exact microscopic result by
	\begin{align}
		\expval{\rho\ind{CG}(t),\testfunction}-\expval{\rho\ind{micro}(t),\testfunction} &=\max(\Delta x^2,\Delta x/\sqrt{L}).\label{equ:diff_HR_full_diff_scaling}
	\end{align}
\end{result}

\begin{remark}
	This result is also important for finite-element numerical simulations of the GHD equations. Even if higher order schemes are used and numerical errors are thus negligible, the result will still differ from the microscopic result by \eqref{equ:diff_HR_full_diff_scaling}. 
\end{remark}

\begin{remark}\label{rem:diff_HR_full_norm}
	The reason why we cannot combine the individual results to obtain a result for the full time evolution is that by measuring a single observable $\testfunction(x,\rapidity)$ we cannot reconstruct $\rho(x,\rapidity)$. If instead we would be able to show in some norm that $\norm{\rho\ind{CG} - \rho\ind{micro}} \sim \max(\Delta x^2,\Delta x/\sqrt{L})$, and that all operations are continuous in this norm, we would immediately find the same result for the full time evolution.
\end{remark}

\begin{openproblem}
	Identify a norm which satisfies the requirements mentioned in \cref{rem:diff_HR_full_norm}.
\end{openproblem}

Let us now conclude the big result of this chapter. We are free to choose $\Delta x$. In particular, we can choose $\Delta x\ll 1/\sqrt{L}$, in which case the error will be $\Delta x/\sqrt{L} \ll 1/L$. This is smaller than the diffusive scale!
\begin{result}\label{res:diff_BMFT_diff}
	In hard rods, there is no (intrinsic) emergent noise on the diffusive scale and there is no (intrinsic) diffusive correction to GHD. As a consequence, BMFT is also accurate on the diffusive scale.
\end{result}

This result means that if one prepares a \underline{single} large scale configuration of hard rods, then Euler GHD describes the evolution more accurately than any GHD equation with a $1/L$ correction like \eqref{equ:pre_GHD_diff_GHD}. This was quite surprising to us, since we expected the RHS of \eqref{equ:pre_GHD_diff_GHD} to emerge from intrinsic noise. In fact, the RHS of \eqref{equ:pre_GHD_diff_GHD} is only a result of fluctuations in the initial state, as we will discuss in \cref{sec:diff_diff}.

\subsection{Verification with numerical simulations}
\begin{figure}[!h]
	\centering
	\includegraphics{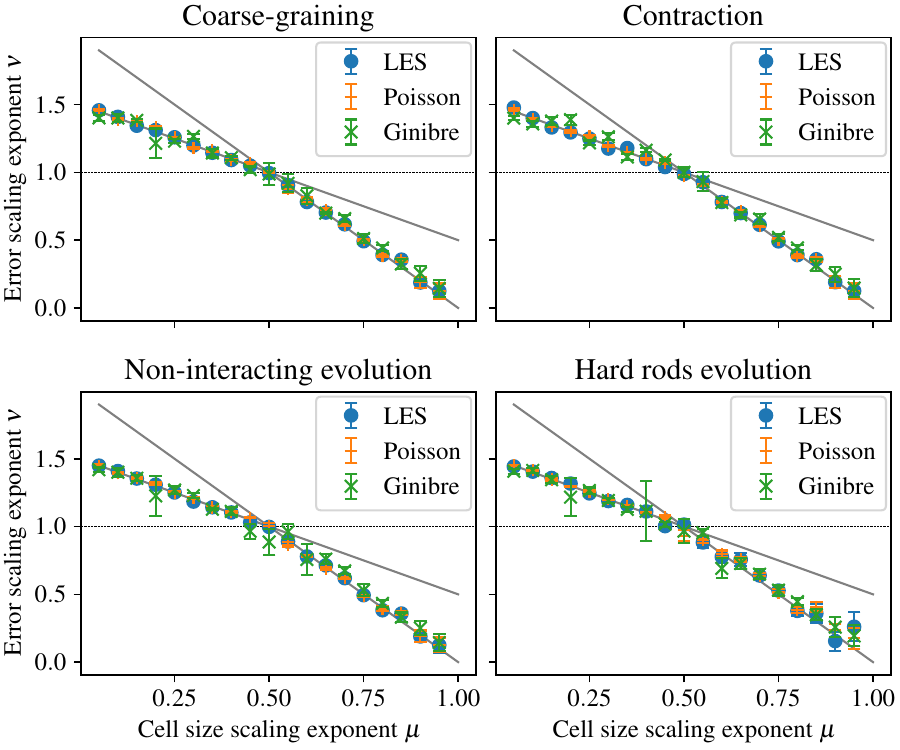}
	\caption[Error scaling]{Numerically extracted scaling of the error $\xi=L^{-\nu}$ of coarse-graining on a scale $\Delta x=\Delta\rapidity\sim L^{\mu-1}$ for coarse-graining the initial state (\cref{sec:diff_HR_obs}), contracting space (\cref{sec:diff_HR_contract}), doing non-interacting time evolution (\cref{sec:diff_HR_free}) and the full hard rods evolution (\cref{sec:diff_HR_full}). This is done for the three different ensembles described in the text: local equilibrium states (LES), Poisson point process (Poisson) and Ginibre ensemble (Ginibre). For the first two $L$ ranged from $1000$ to $10000$ and the error was averaged over $S=10000$ samples. For the Ginibre ensemble, $L$ ranged from $50$ to $500$ and the error was averaged over $S=1000$ samples. Therefore, the Ginibre ensemble results are less converged. The data agrees well with the theoretical prediction (gray lines) from \eqref{equ:diff_HR_full_diff_scaling}. Note that if $\nu > 1$ (dashed line), then the error is smaller than $1/L$. This shows that there is no (intrinsic) diffusive correction to Euler GHD in hard rods. This figure was reproduced from~\cite{hydrowithoutaveraging}.}
	\label{fig:diff_HR_scaling}
\end{figure}
Verifying these results with numerical simulations is not trivial, because as $L \to \infty$ we need to consider different initial states with different errors $\expval{\rho\ind{CQ},\testfunction}-\expval{\rho\ind{micro},\testfunction}$. To quantify the scaling of the error reliably we will average the square of the error over a suitable ensemble of states $\mathbb{E}[\ldots]$:
\begin{align}
	\xi &= \sqrt{\mathbb{E}[(\expval{\rho\ind{CQ},\testfunction}-\expval{\rho\ind{micro},\testfunction})^2]} = \sqrt{\mathbb{E}[\expval{\rho\ind{CQ},\testfunction}-\expval{\rho\ind{micro},\testfunction}]^2 + \mathrm{Var}[\expval{\rho\ind{CQ},\testfunction}-\expval{\rho\ind{micro},\testfunction}]}\label{equ:diff_HR_numerics_error_def}
\end{align}
for different $L \gg 1$ and then extract its scaling $\xi \sim L^{-\nu}$ by fitting $\xi = bL^{-\nu}$ to the data\footnote{Actually, we fit $\xi = bL^{-\nu}$ separately to the expectation value and the variance and then take the slower decay. This is to avoid finite size effects from the crossover happening at $\mu = 1/2$}. This is done for different coarse-graining scales $\Delta x=\Delta\rapidity \sim L^{\mu-1}$, where $0 < \mu < 1$ to obtain the scaling of the error as function of $\mu$.

We use three different ensembles for $\mathbb{E}[\ldots]$:
\begin{itemize}
	\item Local equilibrium state (LES): This is the canonical initial state ensemble \eqref{equ:pre_hydro_LES} with average density $\rho(x,\rapidity) = \frac{5}{2\pi}e^{-\tfrac{1}{2}(x^2+\rapidity^2)}$. Note that those can be easily generated numerically in hard rods using an efficient algorithm (see for instance~\cite{PhysRevLett.134.187101}).
	\item Poisson point process (Poisson): We generate a Poisson point process with average density $\hat{\rho}(\hat{x},\rapidity) = \frac{5}{2\pi}e^{-\tfrac{1}{2}(\hat{x}^2+\rapidity^2)}$ in contracted space and expand it to physical coordinates $\hat{x} \to x$. Note that this is a cheap way of generating a state with local GGE correlations, but additional long range correlations.
	\item Ginibre ensemble (Ginibre): To generate a sample, we first draw $N\sim \mathrm{Pois}(5L)$ and then compute the eigenvalues $z_i$ of a random $N\times N$ matrix $\vb{Z}=(\vb{X}+i\vb{Y})/\sqrt{2N}$, where both $\vb{X},\vb{Y}$ filled with i.i.d. standard Gaussians (i.e.\ a Ginibre ensemble). It is well known that as $N \to \infty$ these eigenvalues are distributed uniformly in the disc $\abs{z}<1$~\cite{byun2024progress}. We then define the initial particle configuration $(\hat{x}_i,\rapidity_i)$ in contracted space via $\hat{x}_i + i \rapidity_i = f(\abs{z_i})z_i$. Here, $f(r) = \sqrt{-2\log(1-\abs{z}^2)}/\abs{z}$ is chosen such that the average density becomes $\hat{\rho}(\hat{x},\rapidity) = \frac{5}{2\pi}e^{-\tfrac{1}{2}(\hat{x}^2+\rapidity^2)}$. We then expand this configuration to physical space.
\end{itemize}

While the first two states are somewhat physical, the third one was chosen clearly not to be. This is because of the eigenvalue repulsion of random matrices: if we would not do the stretching $z\to f(\abs{z})z$, then in the correlations in contracted space would be given by the one of the Ginibre ensemble~\cite{10.1093/imrn/rnm006} $\expval{\delta\hat{\rho}(\hat{x},\rapidity)\delta\hat{\rho}(\hat{x},\rapidityp)} \sim (\delta''(x-y)\delta(\rapidity-\rapidityp) + \delta(x-y)\delta''(\rapidity-\rapidityp))$, which is clearly fundamentally different from the one we found in hard rods $\expval{\delta\hat{\rho}(\hat{x},\rapidity)\delta\hat{\rho}(\hat{x},\rapidityp)} \sim \delta(x-y)\delta(\rapidity-\rapidityp)$, see \eqref{equ:fixedpoint_corr_corr_hatrho_init}. During time evolution, which is trivial in contracted space, those local correlations remain invariant, hence the Ginibre ensemble states will never(!) locally thermalize to a hard rods GGE state, even on the level of the singular $\delta$ correlation part (this picture is not changed by the streching $z\to f(\abs{z})z$). They are completely unphysical states and were chosen to demonstrate that GHD works fully independently of GGE states. 

The numerically extracted scalings are shown in \cref{fig:diff_HR_scaling}, which agree well with the predictions of ``hydrodynamics without averaging'' derived in the previous sections\footnote{We measure the observable $\testfunction(x,\rapidity) = \tfrac{1}{2\pi} e^{-\tfrac{1}{2}(x^2+\rapidity^2)}$, the hard rods size is $d=0.3$ and evolve to $t=1$.}.  

\begin{remark}
	The fact that GHD also applies to unphysical states like the Ginibre ensemble shows that the physical picture of hydrodynamics based on local thermalization as discussed in \cref{sec:pre_hydro} is strongly misleading (at least in integrable models): hydrodynamics does not emerge due to thermalization, but rather is a dynamical self-averaging effect. It would be interesting to try to initialize non-integrable models in unphysical states and to see whether hydrodynamics still emerges.  
\end{remark}

\begin{remark}
	While the original idea of ``hydrodynamics without averaging'' was to avoid averaging over a specific ensemble, in order to check it we need to average. In the usual formulation of hydrodynamics this averaging is essential. Instead, here it is merely a tool to extract an estimate of the scaling of the error. To show hydrodynamics in the usual formulation it is sufficient to show that $\mathbb{E}[\expval{\rho\ind{CQ},\testfunction}-\expval{\rho\ind{micro},\testfunction}]^2 \to 0$, but here we show $\mathbb{E}[(\expval{\rho\ind{CQ},\testfunction}-\expval{\rho\ind{micro},\testfunction})^2] \to 0$, which is a much stronger statement. It means that as $L\to\infty$ any configuration will satisfy the GHD equation individually (almost surely w.r.t. ensemble $\mathbb{E}[\ldots]$). Unfortunately, this does not show it for all configurations, since as $L\to\infty$ almost surely states of each ensemble only cover a very tiny subset of the full configuration set $\mathbb{R}^{2N}$. 
\end{remark}

\subsection{Are there better coarse grainings?}
The results of \cref{sec:diff_HR} imply that Euler GHD BMFT is valid on the diffusive scale, but cannot be more precise than $1/L^{3/2}$. However, already the coarse-graining of the initial state has only this precision, therefore Euler GHD based on it cannot be more precise. It might be that the GHD evolution is actually much more precise than this. 

The question arises whether there are other types of coarse-grainings that are more precise from the start and whether there is an optimal way of coarse-graining. If there are arbitrarily good coarse-grainings, one could actually check up to what order BMFT works. If Euler GHD BMFT is also arbitrarily precise, then this would imply that there are no corrections to Euler GHD on any hydrodynamic scale. On the other hand if there are arbitrarily good coarse-grainings but the error of Euler GHD decays with a fixed scaling, then this means that Euler GHD will break down at this scale due to a coarse-graining independent emergent noise. It would then be very interesting to understand the nature of this noise.

\subsection{What about other integrable models?}
The analysis done on hard rods heavily relies on the explicit formulas (\ref{equ:diff_HR_particle_xt}-\ref{equ:diff_HR_particle_hatx}) for the microscopic dynamics. Therefore, it cannot be easily generalized to other integrable models.

The \scbm s are very close to hard rods in the sense that they also have a similar (although implicit) map to contracting coordinates. It seems reasonable to assume that also this map will also be accurate beyond $1/L$. In addition, in our derivation of GHD, the only approximation we ever do, except coarse-graining, is a microscopic one in \eqref{equ:scbm_ghd_self_consistency_bare_time_evol}. Hence, it seems reasonable that the limitations of the accuracy of GHD should be dictated by the coarse-graining and thus Euler GHD should be accurate on the diffusive scale as well.

\begin{conjecture}
	In any integrable model, Euler GHD BMFT is accurate up to (including) the diffusive scale. This means that on each configuration Euler GHD has an error smaller than $1/L$.
\end{conjecture}

\section{Diffusive GHD}
\label{sec:diff_diff}
In this section we would like to explain how diffusive GHD, and in particular \eqref{equ:pre_GHD_diff_GHD} emerges. We know that \eqref{equ:pre_GHD_diff_GHD} must emerge at least in some situations since it had been proven in hard rods for $t\to 0^+$~\cite{Boldrighini1997}. The main point is: even though Euler GHD is exact on the diffusive scale on each individual configuration, the initial state fluctuates with $\delta\rho \sim 1/\sqrt{L}$. Since GHD evolution is a non-linear map, each configuration evolves slightly differently, leading to an overall shift of the average. To understand this consider the following simple problem: given a fixed map $\vec{y}=f(\vec{x})$, and a random variable $\vec{x}\in\mathbb{R}^n$, what is the statistics of $\vec{y}$. Assuming $\mathbb{E}[x_i]=\order{1}$, $\mathbb{E}[\delta x_i \delta x_j] \sim 1/L \ll 1$ and $\mathbb{E}[\delta x_{i_1}\delta x_{i_2}\ldots \delta x_{i_k}] \ll 1/L$ for $k\geq 2$, we find using $x_i=\mathbb{E}[x_i] + \delta x_i$
\begin{align}
	\mathbb{E}[y_i] &= \mathbb{E}[f_i(\mathbb{E}[\vec{x}] + \delta \vec{x})] = f_i(\mathbb{E}[\vec{x}]) +  \partial_{x_j}f_i(\mathbb{E}[\vec{x}]) \underbrace{\mathbb{E}[\delta x_j]}_{=0} + \tfrac{1}{2} \partial_{x_j}\partial_{x_k}f_i(\mathbb{E}[\vec{x}]) \mathbb{E}[\delta x_j\delta x_k] + \ldots. \label{equ:diff_diff_LD_expansion}
\end{align} 
If we are interested in terms including $1/L$, then we need to take the small $\order{1/L}$ shift due to the variance into account. This additional contribution is what will give rise to diffusive GHD. 

Before starting, let us briefly establish that the equation we will obtain certainly cannot be \eqref{equ:pre_GHD_diff_GHD}. This is because \eqref{equ:pre_GHD_diff_GHD} increases entropy while the Euler GHD equation is entropy conserving. Hence, the averaged entropy\footnote{Note that this is different from the entropy of the averaged quasi-particle density.} will be conserved. In fact since the Euler GHD equation is time reversal symmetric, the resulting equation should also be symmetric under time reversal.

By applying the reasoning of \eqref{equ:diff_diff_LD_expansion} to the Euler GHD equation, we repeat the derivation of \eqref{equ:diff_LR_basic}. Unfortunately, this is not well-defined since $\expval{\delta\rho(t,x,\rapidity)\delta\rho(t,y,\rapidityp)}$ is singular at $x=y$, see \eqref{equ:fixedpoint_corr_corr_rho_t}. However, from our discussion in \cref{sec:diff_HR} we know that the GHD equation only makes sense after coarse-graining over a fluid cell of size $1/L \ll \Delta x \ll 1$. Performing such a coarse-graining \eqref{equ:fixedpoint_corr_corr_rho_t}
\begin{align}
	\int_{x-\Delta x/2}^{x+\Delta x/2}\tfrac{\dd{x'}\dd{x''}}{\Delta x^2}\expval{\delta \rho(t,x',\rapidity)\delta \rho(t,x'',\rapidityp)} &= \tfrac{1}{L\Delta x} C\ind{GGE}(t,x,\rapidity,\rapidityp) + \expval{\delta \rho(t,x,\rapidity)\delta \rho(t,x,\rapidityp)}_\epsilon,\label{equ:diff_diff_corr_cg}
\end{align}
where $\expval{\delta \rho(t,x,\rapidity)\delta \rho(t,x,\rapidityp)}_\epsilon$ represents the (unchanged) continuous part. We see that the jump disappeared and that furthermore the local GGE correlations become dependent on the coarse-graining scale $\Delta x$. This is strange as it would not lead to a universal equation like \eqref{equ:pre_GHD_diff_GHD}. 

Before going on, it is instructive to compute the contribution coming from the regular part at $t=0^+$ after initializing the system in local equilibrium state \eqref{equ:pre_hydro_LES} at time $t=0$. For that it is convenient to expand $j = \tfrac{1}{2\pi}n v\upd{dr}$ in $n$ and not in $\rho$ (see appendix \ref{app:dr_hessj}): 
\begin{align}
	\expval{j(\rapidity)} &=  \tfrac{1}{2\pi}v\upd{dr}n + \tfrac{1}{2\pi}(\vb{1}-n\vb{T})^{-1}\qty[\expval{\delta n \delta v\upd{dr}}- v\upd{eff}\expval{\delta n\delta 1\upd{dr}}].
\end{align}
Using $\expval{\delta n \delta f\upd{dr}} = (\vb{1}-\vb{T}n)^{-1}_2\vu{T}_2f\upd{dr}_2\expval{\delta n \delta n}$ (which follows from \eqref{equ:app_dr_delta}) and inserting \eqref{equ:fixedpoint_corr_corr_n_t0plus}, we find
\begin{align}
	&(\vb{1}-n\vb{T})\expval{j(0^,x,\rapidity)}_\epsilon\nonumber\\
	&= \frac{1}{2\pi}nv + \frac{1}{2L}\qty[\int\dd{\rapidityp}\qty[(\varphi(\rapidity-\rapidityp))\upd{dr}]^2 \abs{v\upd{eff}(\rapidity)-v\upd{eff}(\rapidityp)}\qty(\tfrac{\partial_x n(x,\rapidity)}{2\pi\gamma''(n(x,\rapidityp))}\tfrac{1\upd{dr}(\rapidityp)}{1\upd{dr}(\rapidity)} - \tfrac{\partial_x n(x,\rapidityp)}{2\pi\gamma''(n(x,\rapidity))})]\\
	&= \frac{1}{2\pi}nv + \frac{1}{2L 1\upd{dr}(\rapidity)}  \qty[w(\rapidity)\partial_x n(\rapidity) - \int\dd{\rapidityp}W(\rapidity,\rapidityp)\partial_x n(\rapidityp)]
\end{align}
Here by abuse of notation $\expval{j(0^,x,\rapidity)}_\epsilon$ stands for the contribution due to $\expval{\delta n \delta n}_\epsilon$. Multiplying both sides by $(\vb{1}-n\vu{T})^{-1}$ and inserting into \eqref{equ:diff_LR_basic}, we find that the correction term is precisely the Navier-Stokes like diffusion equation \eqref{equ:pre_GHD_diff_GHD}! Interestingly, it emerges purely due to the regular part of the long range correlations (which are immediately appear at $t=0^+$), the singular part in \eqref{equ:diff_diff_corr_cg} does not seem to affect the diffusive correction.

\begin{figure}[!h]
	\centering
	\includegraphics{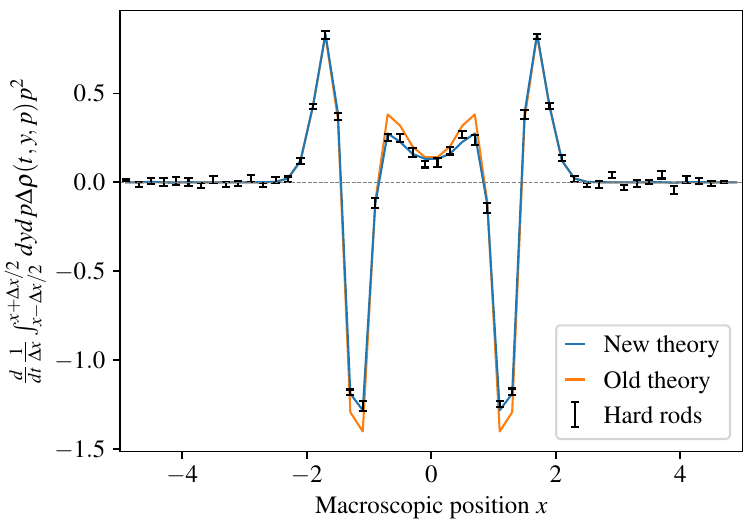}
	\caption[Numerical check of diffusive GHD]{Numerical check of diffusive GHD: we plot the time-derivative of the $1/L$ correction integrated over $\rapidity^2$ and averaged over a fluid cell of size $\Delta x = 0.2$. This is done at time $t=1$ after releasing the system from a local equilibrium state \eqref{equ:pre_hydro_LES}, with average density $\rho^0(x,\rapidity) = e^{-(\rapidity+\tanh{x})^2/2}/\sqrt{2\pi}$. While not present at $t=0$, long range correlations have developed at $t=1$ (these are plotted in \cref{fig:pre_hydro_BMFT_LL_macro}). The result obtained from microscopic hard rods simulations are compared to the old \eqref{equ:pre_GHD_diff_GHD} and the new \eqref{equ:diff_diff_diff_new} diffusive theory. The hard rods numerics are well-described by the new diffusive equation, but differs significantly from the old diffusive equation \eqref{equ:pre_GHD_diff_GHD}, proving that it cannot be valid. This plot is reproduced from~\cite{PhysRevLett.134.187101}, where also further details are given.}
	\label{fig:diff_diff_numerics}
\end{figure}

Hence, it seems natural to assume that we can neglect the singular part at any time and only the regular part contributes. This leads to a new proposal for the diffusive GHD equation
\begin{multline}
	\partial_t \expval{\rho(t,x,\rapidity)} + \partial_x \qty[v\upd{eff}(t,x,\rapidity)\expval{\rho(t,x,\rapidity)}]\\
	+ \tfrac{1}{2}\partial_x\qty[ \int\dd{\rapidityp}\dd{\rapidityp'}\frac{\delta j(t,x,\rapidity)}{\delta\rho(\rapidityp)\delta\rho(\rapidityp')}\expval{\delta\rho(t,x,\rapidityp)\delta\rho(t,x,\rapidityp')}_{\epsilon}] + \order{1/L^2}= 0.\label{equ:diff_diff_diff_new}
\end{multline}
This equation is confirmed by numerical simulations in hard rods, see \cref{fig:diff_diff_numerics}. Furthermore, in hard rods one can derive \eqref{equ:diff_diff_diff_new} from the known microscopic dynamics (\ref{equ:diff_HR_particle_xt}-\ref{equ:diff_HR_particle_hatx})~\cite{hübner2025diffusivehydrodynamicshardrods}. Hence, it is natural to conjecture \eqref{equ:diff_diff_diff_new} to hold in any model.

\begin{result}
	The equation for $\expval{\rho(t,x,\rapidity)}$, i.e. the diffusive GHD equation, is given by \eqref{equ:diff_diff_diff_new} instead of \eqref{equ:pre_GHD_diff_GHD}. Thus, integrable models on the diffusive scale are described by a set of two-coupled equations \eqref{equ:diff_diff_diff_new} and 
	\begin{align}
		\partial_t \expval{\delta\rho(t,x,\rapidity)\delta\rho(t,y,\rapidityp)} &+ \partial_x \qty[\int\dd{\rapidity'}\fdv{j(t,x,\rapidity)}{\rho(\rapidity')}\expval{\delta\rho(t,x,\rapidity')\delta\rho(t,y,\rapidityp)}]\\
		&+\partial_y \qty[\int\dd{\rapidityp'}\fdv{j(t,y,\rapidityp)}{\rho(\rapidityp')}\expval{\delta\rho(t,x,\rapidity)\delta\rho(t,y,\rapidityp')}] = 0.\label{equ:diff_diff_diff_corr}
	\end{align}
\end{result}

Furthermore, the microscopic derivation in~\cite{hübner2025diffusivehydrodynamicshardrods} teaches us that an order of limits is important: first evolve the state to $t+\Delta t$, then take $L \to \infty$ and as a last step take $\Delta t \to 0$. The derivation is long, but in a nutshell, the absence of a contribution from the $\delta$ peak can be seen as follows: the small $\Delta t$ means that the correlations are evaluated not at $x$, but at $x-v\upd{eff}(x,\rapidity)\Delta t$. This means that different momenta are evaluated at different locations, thereby avoiding the $\delta$ peak. 

Due to our findings in \cref{sec:diff_HR}, we know that we can replace the microscopic dynamics by the Euler GHD equation, without changing the diffusive behavior. Indeed, applying the expansion \eqref{equ:diff_diff_LD_expansion} on \eqref{equ:diff_HR_GHD_obs} can be used to compute \eqref{equ:diff_diff_diff_new}~\cite{hydrowithoutaveraging}. While in other integrable models there is no explicit solution to the GHD equation, we still know that the solution satisfies the self-consistency equation \eqref{equ:fixedpoint_fixedpoint}. By expanding this equation like in \eqref{equ:diff_diff_LD_expansion}, one naively obtains
\begin{multline}
	\expval{\heightfield(t,x,\rapidity)} = \expval{\hat{\heightfield}^0}(\expval{\hat{u}}(t,x,\rapidity),\rapidity)+ 2\pi\expval{\delta\partial_{\hat{x}}\hat{\heightfield}^0(\expval{\hat{u}}(t,x,\rapidity),\rapidity)(\vu{T}\delta \heightfield(t,x,\rapidity))}\\
	+ \frac{(2\pi)^2}{2} \expval{\partial_{\hat{x}}^2\hat{\heightfield}^0}(\expval{\hat{u}}(t,x,\rapidity),\rapidity)\expval{(\vu{T}\delta \heightfield(t,x,\rapidity))^2}, \label{equ:diff_diff_fixedpoint}
\end{multline}
where $\expval{\hat{u}}(t,x,\rapidity) = x-v(\rapidity)t + 2\pi \vu{T}\expval{\heightfield(t,x,\rapidity)}$. We believe that it should be possible to derive \eqref{equ:diff_diff_diff_new} from an equation similar to \eqref{equ:diff_diff_fixedpoint}. However, we do not think that \eqref{equ:diff_diff_fixedpoint} as written is correct. This is due to a fundamental problem: while $\expval{\hat{\heightfield}^0}(\hat{x},\rapidity)$ is smooth, the individual configuration $\expval{\hat{\heightfield}^0}(\hat{x},\rapidity) + \delta\hat{\heightfield}^0(\hat{x},\rapidity)$ is rough. Hence, one cannot take its derivatives in $\hat{x}$ as appears in \eqref{equ:diff_diff_fixedpoint}. Simply using \eqref{equ:diff_diff_fixedpoint} however does not give the \eqref{equ:diff_diff_diff_new}, including for hard rods, where we know that \eqref{equ:diff_diff_diff_new} is correct.

\begin{openproblem}
	Show \eqref{equ:diff_diff_diff_new} by expanding \eqref{equ:fixedpoint_fixedpoint} up to order $1/L$. 
\end{openproblem}

In~\cite{PhysRevLett.134.187101} an alternative approach was presented, which is not restricted to integrable models, but applies to all linearly degenerate PT symmetric models. The crucial observation in~\cite{PhysRevLett.134.187101} is that in a PT symmetric system the $1/L$ correction to the expectation value of currents vanishes in a local equilibrium state \eqref{equ:pre_hydro_LES}. However, naively applying an expansion as in \eqref{equ:diff_diff_LD_expansion} gives a $1/(L\Delta x)$ correction as in \eqref{equ:diff_diff_corr_cg}. Hence, in order to apply BMFT on the diffusive scale, this has to be subtracted. Applied onto the current, this cancels the $\delta$ part in the correlations, meaning the only contribution stems from the long range correlations. However, we are lacking a solid physical justification for the appearance of this additional term in BMFT. One possible explanation might come from the fact that in the derivation of BMFT in \cref{sec:pre_hydro_bmft} we implicitly used the microcanonical ensemble in each fluid cell, while in the derivation of hydro (\cref{sec:pre_hydro_euler}) we used the grand canonical ensemble\footnote{While equivalent on the Euler scale these two ensembles differ precisely by the $1/(L\Delta x)$ term of \eqref{equ:diff_diff_corr_cg} on the diffusive scale}.

\begin{openproblem}
	Justify the absence of the contribution from the $1/(L\Delta x)$ term of \eqref{equ:diff_diff_corr_cg} in the diffusive GHD equation \eqref{equ:diff_diff_diff_new}.
\end{openproblem}  

Note that new equations \eqref{equ:diff_diff_diff_new} and \eqref{equ:diff_diff_diff_corr} are time reversal symmetric. This is in stark contrast to \eqref{equ:pre_GHD_diff_GHD}, which increases entropy (and thus cannot be symmetric under time reversal). The time-reversibility stems from the absence of loss of information: unlike \eqref{equ:pre_GHD_diff_GHD}, which discards long range correlations, all information about the initial state remains in \eqref{equ:diff_diff_diff_new}. 

\begin{result}
	The new set of diffusive GHD equations \eqref{equ:diff_diff_diff_new} and \eqref{equ:diff_diff_diff_corr} is time reversible and thus entropy conserving. In particular, it cannot be the cause of thermalization.
\end{result}

\begin{remark}
	Note that, if \eqref{equ:diff_diff_diff_new} is correct, its explicit solution can be obtained by applying the expansion \eqref{equ:diff_diff_LD_expansion} to \eqref{equ:diff_HR_GHD_obs} in hard rods or to \eqref{equ:fixedpoint_fixedpoint} in a general integrable model.
\end{remark}

\subsection{Does this generalize to non-integrable models?}
\label{sec:diff_diff_nonint}
Since we have found that the Navier-Stokes like diffusive equation in integrable systems is incorrect and has to be extended due to the long range correlations, this raises the question whether or not Navier-Stokes \eqref{equ:pre_GHD_diff_GHD} is valid in any system at all.

First, let us discuss one-dimensional non-integrable models. As mentioned in \cref{sec:pre_hydro_shock} the hydrodynamics of generic models will have entropy increasing shocks. In fact, it is known that correlations in those models are not diffusive, but KPZ super-diffusive as predicted by nonlinear fluctuating hydrodynamics~\cite{Spohn2014}. The only exception might be models with linearly degenerate hydrodynamics (which do not produce shocks). It is not clear whether any non-integrable models exists with this property, but if they do then we believe that the diffusive correction could be sum of \eqref{equ:diff_diff_diff_new} and potentially unaccounted intrinsic diffusion described by \eqref{equ:pre_hydro_NS}. 

\begin{openproblem}
	Find non-integrable models with linearly degenerate hydrodynamics (or show their absence) and derive their diffusive correction.
\end{openproblem}

In dimensions $D\geq 2$ we do not expect the long range correlations to have an effect. This is simply because the initial state fluctuations scale as $\expval{\delta q} = 1/L^D$ (thermal fluctuations decay with the volume, not with the length scale), while diffusion remains of order $1/L$. Thus diffusion should dominate. In non-integrable models we believe that there is intrinsic noise, i.e. that if we could compute ``hydrodynamics without averaging'' on a non-integrable model we would find an error scaling as $1/L$. This intrinsic noise should be generated from the non-observable microscopic initial state noise and emerge over time as effective white noise. In integrable models this does not exist because there is no sufficiently strong mixing: particles that start sufficiently close will remain close throughout the Euler evolution. Thus, the noise is simply transported along the GHD characteristics. In non-integrable models trajectories should be chaotic, hence diverging exponentially fast from each other. Thus, details from the initial state in any region will quickly affect any other region, which should give rise to an unpredictable noise.

Therefore, it still seems reasonable to expect the Navier-Stokes equation to emerge in higher dimensions. In particular, it has been proven in hard spheres~\cite{deng2025hilbertssixthproblemderivation} in the Boltzmann limit (however see \cref{rem:pre_hydro_boltzmann}).

\begin{remark}
	Nevertheless, the failure of Navier-Stokes in integrable models highlights that higher order corrections in hydrodynamics are not as simple as expected and that it is always crucial to carefully investigate the validity of assumptions leading to any simplified equation. Even if Navier-Stokes is indeed correct, we are far from understanding why and how it emerges.
\end{remark}

\section{Entropy increase and thermalization}
\label{sec:diff_entropy}
The fact that the new diffusive equation is entropy conserving implies that diffusive GHD cannot be responsible for thermalization. We would now sketch an alternative explanation for thermalization in integrable models from the ``hydrodynamics without averaging'' perspective.

We have seen that the contraction (and thus also the expansion) transformation introduce an error $\max(\Delta x^2, \sqrt{\Delta x/L})$. The non-interacting part of the time-evolution however has an error scaling as $\max(\Delta x+t^2\Delta\rapidity^2, \sqrt{\Delta x+t^2\Delta\rapidity^2}/\sqrt{L})$. Therefore, the approximation will break down if $t = \order{1/\Delta \rapidity}$. At this point particles originating from the same cell will be macroscopically apart, hence coarse-graining breaks down. Recall that \eqref{equ:diff_HR_GHD_obs} is the exact solution to the GHD equation. Therefore, at this point GHD must break down.

\begin{result}
	Euler GHD breaks down on the time scale $t = \order{1/\Delta \rapidity}$. This should hold independently of the precise coarse-graining scheme.
\end{result}

\begin{remark}
	Many ideas and the analysis presented in this section already appeared 2022 in~\cite{Chakraborti_2022}. Therefore, this section should be seen as a reinterpretation of their result in the context of ``hydrodynamics without averaging''.
\end{remark}

\begin{figure}[!h]
	\centering
	\includegraphics{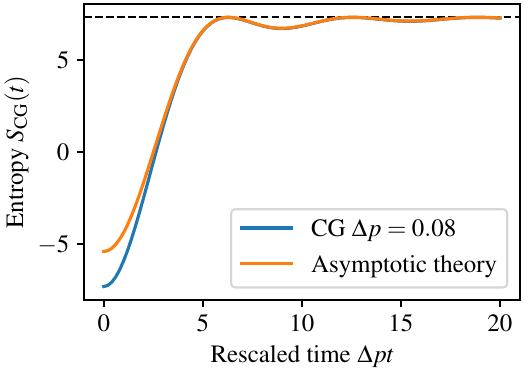}
	\caption[Evolution of coarse-grained entropy]{Evolution of the coarse-grained (classical) entropy starting from $\rho^0(x,\rapidity) = (10-9\sin(x))e^{-\rapidity^2/2}/\sqrt{2\pi}$ with $\Delta \rapidity = 0.08$ and $\Delta x=2\pi/100$ (the system is a periodic box of size $2\pi$) at long times $t = \order{1/\Delta \rapidity} \gg 1$ much beyond Euler scale. We indeed observe thermalization: the entropy approaches its maximum possible value (dashed line), however in an oscillatory fashion well-described by the asymptotic formula \eqref{equ:diff_thermalization_S}. This figure was reproduced from~\cite{hydrowithoutaveraging}.}
	\label{fig:diff_entropy}
\end{figure}

We will now try to understand what happens on this time scale $t = \order{1/\Delta \rapidity}$ in the most simple setting: non-interacting particles. Note that due to the mapping to non-interacting particles, similar arguments should apply to integrable models.  As we are interested in thermalization, we need to study the system in a finite box with periodic boundary conditions\footnote{If we study times of order $t \gg \order{1}$ on the infinite system, then particles of different velocities will become infinitely separated.}. Also, we will ignore the effect of the initial coarse-graining and assume a smooth initial quasi-particle density $\rho^0(x,\rapidity)$. The averaged density at time $t$ in cell $\set{C}_{\alpha,\beta}$ is then
\begin{align}
	\rho_{\alpha,\beta}(t) &= \int_{-1/2}^{1/2}\dd{y}\dd{\rapidityp}\rho^0(x_\alpha-\rapidity_\beta t + y\Delta x - \rapidityp\Delta \rapidity t,\rapidity_\beta+q\Delta \rapidity) = B_t(x_\alpha-\rapidity_\beta t,\rapidity_\beta).\label{equ:diff_thermalization_rho_cg_t}
\end{align}
Let us now understand the asymptotics as $\Delta \rapidity t \to \infty$ of 
\begin{align}
	B_t(x,\rapidity) = \int_{-1/2}^{1/2}\dd{y}\dd{\rapidityp}\rho^0(x + y\Delta x - \rapidityp\Delta \rapidity t,\rapidity+q\Delta \rapidity) \approx  \int_{-1/2}^{1/2}\dd{\rapidityp}\rho^0(x - \rapidityp\Delta \rapidity t,\rapidity).\label{equ:diff_thermalization_rho_cg_t_B}
\end{align}
Here we already send $\Delta x \sim \Delta \rapidity \to 0$ for convenience (this can be done at a later stage, but does not affect the final result). Now expand $\rho^0(x,\rapidity) = \sum_{k=-\infty}^\infty \tilde{\rho}^0_k(\rapidity)e^{ikx}$
\begin{align}
	B_t(x,\rapidity) &\approx \sum_k \int_{-1/2}^{1/2}\dd{\rapidityp} \tilde{\rho}^0_k(\rapidity) e^{ikx -ik\rapidityp\Delta \rapidity t} = \tilde{\rho}^0_0(\rapidity) + \sum_{k\neq 0} \tilde{\rho}^0_k(\rapidity) \frac{\sin(k\Delta \rapidity t/2)}{k\Delta\rapidity t/2} e^{ikx}.\label{equ:diff_thermalization_rho_cg_t_B_asympt}
\end{align}
Already here we observe thermalization: as $\Delta\rapidity t \to \infty$ (as expected) the density converges to the constant value of the GGE $\tilde{\rho}^0_0(\rapidity) = \int_0^{2\pi}\tfrac{\dd{x}}{2\pi}\rho^0(x,\rapidity)$. We can quantify the thermalization with the entropy
\begin{align}
	S\ind{CG}(t) &= -\Delta x\Delta\rapidity\sum_{\alpha,\beta} \gamma(\rho_{\alpha,\beta}(t)) = -\Delta x\Delta\rapidity\sum_{\alpha,\beta} \gamma(B_t(x_\alpha+\rapidity_\beta t, \rapidity_\beta))\\
	&\to -\int\dd{x}\dd{\rapidity} \gamma(B_t(x, \rapidity))= -\int\dd{x}\dd{\rapidity} \gamma\qty(\tilde{\rho}^0_0(\rapidity) +  \sum_{k\neq 0} \tilde{\rho}^0_k(\rapidity) \frac{\sin(k\Delta \rapidity t/2)}{k\Delta\rapidity t/2} e^{ikx})\\
	&= S(\infty) - \tfrac{2\pi}{2} \int\dd{\rapidity} \gamma''\qty(\tilde{\rho}^0_0(\rapidity)) \sum_{k\neq 0} \abs{\tilde{\rho}_k^0(\rapidity)}^2 \qty(\frac{\sin(k\Delta \rapidity t/2)}{k\Delta\rapidity t/2})^2 + \order{1/(\Delta \rapidity t)^3}\label{equ:diff_thermalization_S},
\end{align}
where $S(\infty) = 2\pi\int\dd{\rapidity} \gamma\qty(\tilde{\rho}^0_0(\rapidity))$ is the GGE entropy. The entropy approaches the equilibrium non-monotonously(!) on the time-scale $T\ind{th} \sim 1/\Delta \rapidity$, as expected (see \cref{fig:diff_entropy}).

\begin{result}\label{res:diff_entropy_free}
	In free particles, the coarse-grained entropy $S\ind{CG}(t)$ transits from the initial value $S\ind{CG}(0)$ towards the equilibrium value on the time-scale $T\ind{th} \sim 1/\Delta \rapidity \ll L$. This implies that Euler GHD breaks down at timescale $T\ind{th}$. 
\end{result}

Since all integrable models can be mapped to non-interacting particles, it seems plausible that the same should be true in all models.
\begin{conjecture}\label{conj:diff_entropy}
	\Cref{res:diff_entropy_free} is valid in a general integrable model.
\end{conjecture}

As a corollary of this conjecture we also have
\begin{result}
	Thermalization in integrable models is not due to diffusion (which would imply $T\ind{th} \sim L$), but due to coarse-graining.
\end{result}

\begin{remark}
	This result makes it clear that thermalization to a GGE cannot be meaningfully defined without fixing what the observer can measure about the system (and how precisely). Different values of $\Delta \rapidity$ (and different coarse-graining schemes) lead to different thermalization times. In fact, these results suggest that the way thermalization is perceived is not determined by the system, but by the observer.
\end{remark}

\begin{remark}
	The oscillations and non-monotonicity of \eqref{equ:diff_thermalization_S} clearly emerges from the coarse-graining scheme we used. If another scheme is used, say the one described in \cref{rem:diff_HR_obs_smooth}, the shape of \eqref{equ:diff_thermalization_S} might be different (and potentially grow monotonically).
\end{remark}

\section{Simulation of diffusive GHD in quantum systems}
\label{sec:diff_simulation_fermionic}
We identified the \scbm s in \cref{sec:LL} and then studied its classical properties in \cref{sec:scbm}. In \cref{fig:scbm_GHD} we demonstrated that we can use these particle models to simulate the Euler GHD of the quantum model. However, at that point we concluded that classical and quantum models will differ on the diffusive scale as \eqref{equ:pre_GHD_diff_GHD} depends on the particle statistics. Based on the new insights into diffusion gained in this chapter, we would like to propose a method to simulate the diffusive GHD of a quantum system like Lieb-Liniger using a \scbm. Note that, based on the usual derivation of hydrodynamics, this would be fully absurd: Even if we would start with some however complicated classical state mimicking the quantum state of the quantum model, after a short evolution time, both models would be expected to thermalize locally: the quantum model to a state with local correlations corresponding to fermions $\gamma(n) = n\log n + (1-n)\log(1-n)$, and the classical model to a state with local classical correlations $\gamma(n)=n\log n-n$. Thus, both models should ultimately diffuse differently.

However, in this chapter we found that diffusion is not intrinsic. It emerges from the correlations and the particle statistics dependence of diffusion is only due to different initial correlations. Hence, if we would initialize a \scbm~in a state with fluctuations that mimic those of a quantum system, the evolutions should coincide. Note that we do not need to choose an initial state that has the correct microscopic correlations. As long as the coarse-grained correlations have the correct local statistics it should be sufficient. Even more, we already know that these correlations \eqref{equ:fixedpoint_corr_corr_hatrho_init} are simply diagonal $\expval{\delta\hat{\rho}^0(\hat{x},\rapidity)\delta\hat{\rho}^0(\hat{y},\rapidity)} = \tfrac{1}{2\pi\gamma''(2\pi \hat{\rho}^0(\hat{x},\rapidity))}\delta(\hat{x}-\hat{y})\delta(\rapidity-\rapidityp)$ in contracted coordinates.  

Therefore, let us first describe a way to generate a state with fermionic correlations in contracted coordinates. There are many ways but a simple one is the following. Define cells $\set{C}_{\alpha,\beta}$ (in contracted coordinates) as in \eqref{equ:diff_HR_cells_def} and choose $\Delta x, \Delta \rapidity = 1/\sqrt{L}$ for simplicity. Now, for each of these cells $\set{C}_{\alpha,\beta}$ flip a coin and place a particle in it (say at a uniformly random position), with probability $p_{\alpha,\beta} = \hat{\rho}^0(\hat{x}_\alpha,\rapidity_\beta) L\Delta x\Delta \rapidity = \hat{\rho}^0(\hat{x}_\alpha,\rapidity_\beta)$. Define $n_{\alpha,\beta}=1$, if we place a particle in $\set{C}_{\alpha,\beta}$, otherwise $n_{\alpha,\beta}=0$. Then the moment generating function of $L\expval{\hat{\rho},\hat{\testfunction}} = \sum_{\alpha,\beta} n_{\alpha,\beta}\hat{\testfunction}(\hat{x}_\alpha,\rapidity_\beta)$ is given by
\begin{align}
	\mathbb{E}[e^{L\expval{\hat{\rho},\hat{\testfunction}}}] &= \prod_{\alpha,\beta}\mathbb{E}\qty[e^{n_{\alpha,\beta}\hat{\testfunction}(\hat{x}_\alpha,\rapidity_\beta)}] = \prod_{\alpha,\beta}\qty[(1-\hat{\rho}^0(\hat{x}_\alpha,\rapidity_\beta)) + \hat{\rho}^0(\hat{x}_\alpha,\rapidity_\beta) e^{\hat{\testfunction}(\hat{x}_\alpha,\rapidity_\beta)}]\label{equ:diff_simferm_mgf}
\end{align}
and thus the scaled cumulant generating function becomes
\begin{align}
	\LD{F}[\hat{\testfunction}] &= \tfrac{1}{L} \log \mathbb{E}[e^{L\expval{\hat{\rho},\hat{\testfunction}}}] = \tfrac{1}{L} \sum_{\alpha,\beta} \log \qty[(1-\hat{\rho}^0(\hat{x}_\alpha,\rapidity_\beta)) + \hat{\rho}^0(\hat{x}_\alpha,\rapidity_\beta) e^{\hat{\testfunction}(\hat{x}_\alpha,\rapidity_\beta)}]\\
	&\to \int\dd{\hat{x}}\dd{\rapidity} \log \qty[(1-\hat{\rho}^0(\hat{x},\rapidity)) + \hat{\rho}^0(\hat{x},\rapidity) e^{\hat{\testfunction}(\hat{x},\rapidity)}].\label{equ:diff_simferm_scgf}
\end{align}
Using large deviation theory (see appendix \ref{app:LD}) we can obtain the rate function $\LD{I}[\hat{\rho}] = \max_{\hat{\testfunction}} \expval{\hat{\rho},\hat{\testfunction}} - \LD{F}[\hat{\testfunction}]$ as its Legendre transform. The maximum is attained at
\begin{align}
	\hat{\rho}(\hat{x},\rapidity) &= \fdv{\LD{F}[\hat{\testfunction}]}{\testfunction(\hat{x},\rapidity)} = \frac{\hat{\rho}^0(\hat{x},\rapidity)e^{\hat{\testfunction}(\hat{x},\rapidity)}}{(1-\hat{\rho}^0(\hat{x},\rapidity)) + \hat{\rho}^0(\hat{x},\rapidity) e^{\hat{\testfunction}(\hat{x},\rapidity)}},\label{equ:diff_simferm_legendre}
\end{align}
which gives $e^{\hat{\testfunction}(\hat{x},\rapidity)} = (1-\hat{\rho}^0(\hat{x},\rapidity))/(1-\hat{\rho}(\hat{x},\rapidity))\cdot \hat{\rho}(\hat{x},\rapidity)/\hat{\rho}^0(\hat{x},\rapidity)$ and thus
\begin{align}
	\LD{I}[\hat{\rho}] &= \int\dd{\hat{x}}\dd{\rapidity}\hat{\rho}(\hat{x},\rapidity)\log \hat{\rho}(\hat{x},\rapidity) + (1-\hat{\rho}(\hat{x},\rapidity))\log(1-\hat{\rho}(\hat{x},\rapidity)) \\
	&- \int\dd{\hat{x}}\dd{\rapidity}\hat{\rho}(\hat{x},\rapidity) \log\tfrac{\hat{\rho}^0(\hat{x},\rapidity)}{1-\hat{\rho}^0(\hat{x},\rapidity)} + \log(1-\hat{\rho}^0(\hat{x},\rapidity)).\label{equ:diff_simferm_rate_function}
\end{align}
This is the free energy \eqref{equ:pre_int_GHD_TBA_free_energy} of a non-interacting fermionic particle $\gamma(n) = n\log n + (1-n)\log(1-n)$ with local $\beta(\hat{x},\rapidity) = \log\tfrac{\hat{\rho}^0(\hat{x},\rapidity)}{1-\hat{\rho}^0(\hat{x},\rapidity)}$.

We can extend this to interacting \scbm s as follows: repeating the derivation in \cref{sec:scbm_thermo} but replacing the classical measure $\sum_N \tfrac{1}{N!}\int\dd[N]{\hat{x}}\dd[N]{\rapidity}$ by the measure above with $\hat{\rho}^0(\hat{x},\rapidity) = \tfrac{1}{2}$, we obtain instead of \eqref{equ:scbm_thermo_F}
\begin{align}
	\mathcal{F}[\rho] &= \int\dd{x}\dd{\rapidity} 1\upd{dr} \qty[\tfrac{n(x,\rapidity)}{2\pi} \log\tfrac{n(x,\rapidity)}{2\pi} + (1-\tfrac{n(x,\rapidity)}{2\pi})\log(1-\tfrac{n(x,\rapidity)}{2\pi}) + \tfrac{n(x,\rapidity)}{2\pi} \beta(x,\rapidity)].\label{equ:diff_simferm_free_energy}
\end{align}
This is (up to a factor of $2\pi$, which can be taken away by a rescaling), the large deviation rate function of a Lieb-Liniger local equilibrium state \eqref{equ:pre_hydro_LES}. Numerically sampled configurations from this measure will thus show fermionic correlation functions and therefore the observed diffusive correction under this ensemble should coincide with the diffusive correction of a model with fermionic quasi-particles like Lieb-Liniger.

\begin{conjecture}\label{conj:diff_simulation_fermionic}
	Numerically simulating GHD using \scbm s with the above initial state should have the same diffusive correction as a quantum model with fermionic statistics (initialized in the same state).
\end{conjecture}

Checking this conjecture numerically would be very hard since we are barely able to obtain precise microscopic simulations to check Euler GHD. Since obtaining sufficient statistics was already demanding in hard rods, observing the very small diffusive correction $1/L$ is currently out of reach.

\begin{remark}
	Towards the end of writing the thesis a preprint~\cite{urilyon2025simulatinggeneralisedfluidsinteracting} was published. They also propose \cref{conj:diff_simulation_fermionic} and use it to simulate quantum systems.
\end{remark}

\section{Conclusion}
In this chapter we applied the new concept ``hydrodynamics without averaging'' to hard rods and established that (on each configuration) Euler GHD is accurate even on the diffusive scale. This means that there is no intrinsic diffusive $1/L$ correction to GHD. In the spirit of ``diffusion from convection'', the $1/L$ correction thus purely emerges from transporting fluctuations of the initial state, which in 1D also happen to be of order $1/L$. Interestingly (and in agreement with previous work), shortly after releasing the system from a local equilibrium state, this $1/L$ correction indeed has the form of a Navier-Stokes like equation. Only at later times it will differ from it. Furthermore, the evolution is time reversible, hence the new equation cannot increase entropy and cannot be the origin of thermalization. We discussed an alternative explanation for thermalization which is based on the eventual breakdown of coarse-graining required to define hydrodynamics. 

Another interesting observation is that GHD also works even if the state is microscopically far from local equilibrium, as demonstrated by using initial states constructed from the Ginibre ensemble.

%!TEX root = thesis.tex

\chapter{Conclusion and outlook}
We hope that the work presented in this thesis convinced the reader that the hydrodynamic approximation is an insightful and powerful approach to study integrable models and physical systems in general. The first comment we would like to make is that integrable models can be much more complicated than those we restricted ourselves to. Some changes, like having multiple particle species as in the XXZ chain, can be accounted for easily~\cite{BenGHD,PhysRevLett.117.207201}. Others, like the absence of a good reference state in the XYZ chain, are less trivial to deal with~\cite{10.21468/SciPostPhys.14.6.158}. As already mentioned in the introduction, the point of this thesis is not to give an exhaustive overview of all possibilities, but instead to lay out the intuition and the tools in simple settings. We hope that they will be a starting point for tackling more complicated questions and for identifying many further interesting effects.

A couple of open problems were already identified in the text. Most importantly finishing the new derivation discussed in \cref{sec:LL} by developing tools to tackle the sum over permutations. Once a proper derivation is established, it would be an ideal starting point to understand the diffusive correction in quantum models and also to identify the strength of quantum effects. In GGE states, coherent quantum effects are typically suppressed, but in pure states (for which GHD might also apply) they will present a correction term of some order. In \cref{sec:scbm} it would be interesting to make sense of \scbm s with negative phase-shifts and to try to come up with a similar construction for quantum models. In \cref{sec:fixedpoint} one could attempt to generalize the proof of absence of shocks to more general models, or contrary, to identify models where GHD might develop novel types of singularities. And in \cref{sec:diff} it would be interesting to generalize the idea of ``Hydrodynamics without averaging'' to quantum models and to show that Euler GHD (on each individual configuration) is accurate beyond the diffusive scale more generally. Furthermore, classifying the noise within each cell, which we found was the dominant error of Euler GHD, and understanding its effect might shed light on higher order corrections to GHD.

In general, the work described here often used exact relations which are not valid in the presence of an external potential. Generalizing the many results derived in this thesis, from the quantum derivation of GHD to the diffusive correction, to GHD with an external potential would thus be a much harder task. On the other hand, the explicit integrability breaking of an external potential might allow for more complicated phenomena. For instance, turbulence-like behavior, which is impossible without external potential, was observed numerically in hard rods~\cite{PhysRevResearch.6.023083}. Then there is the question of how integrable systems thermalize towards the thermal state of the trap (see e.g.\ numerical simulations in~\cite{PhysRevResearch.6.023083,10.21468/SciPostPhys.6.6.070}). Previously it was believed to be due to diffusion, but with the new understanding of diffusive GHD from \cref{sec:diff}, this seems to be unrealistic. A particularly interesting open problem in this regard is to understand the failure of thermalization of hard rods in a harmonic trap~\cite{PhysRevLett.120.164101,PhysRevE.108.064130}.

As mentioned in the introduction, the solvability of integrable systems makes them ideal as a starting point to understand also non-integrable models. In particular, ideas like BMFT and ``Hydrodynamics without averaging'' will also be applicable to non-integrable systems (including in higher dimensions). Since no microscopic solutions exists in non-integrable models, making progress will be much harder and these new tools could provide crucial viewpoints. The failure of Navier-Stokes-like diffusion in integrable models raises the natural question whether or not Navier-Stokes is valid in non-integrable models (in particular in higher dimensions). We discussed in \cref{sec:diff_diff_nonint} that long range correlations will be subleading in higher dimensions and thus \eqref{equ:diff_diff_diff_new} will not be the diffusive equation. Nevertheless, the assumptions leading to hydrodynamics should be carefully revisited and checked. This will be beneficial even if Navier-Stokes is found to be correct, as deeper microscopic understanding of its emergence might help to understand its limitations. For instance, it might allow us to finally understand the physics behind turbulence~\cite{Frisch_1995}. It may also provide valuable guidance to investigate properties of solutions to the Navier-Stokes equation (connected to one of the Millenium prize problems), or to rigorously show its emergence (connected to one of the Hilbert problems).

%\include{chapter2}
%\include{chapter3}

% ********************************** Back Matter *******************************
% Backmatter should be commented out, if you are using appendices after References
%\backmatter

% ********************************** Bibliography ******************************
\begin{spacing}{0.9}

% To use the conventional natbib style referencing
% Bibliography style previews: http://nodonn.tipido.net/bibstyle.php
% Reference styles: http://sites.stat.psu.edu/~surajit/present/bib.htm

%\bibliographystyle{apalike}
%\bibliographystyle{unsrt} % Use for unsorted references  
%\bibliographystyle{plainnat} % use this to have URLs listed in References
\cleardoublepage
%\bibliography{references} % Path to your References.bib file

% If you would like to use BibLaTeX for your references, pass `custombib' as
% an option in the document class. The location of 'reference.bib' should be
% specified in the preamble.tex file in the custombib section.
% Comment out the lines related to natbib above and uncomment the following line.

%\printbibliography[heading=bibintoc, title={References}]

\begingroup
\sloppy
% If you want to break on URL numbers
%\setcounter{biburlnumpenalty}{9000}
% If you want to break on URL lower case letters
%\setcounter{biburllcpenalty}{9000}
% If you want to break on URL UPPER CASE letters
%\setcounter{biburlucpenalty}{9000}
%\RaggedRight %% needs package ragged2e
%\raggedright
%\bibliographystyle{...}
\normalem
\printbibliography[heading=bibintoc, title={References}]
\endgroup

\end{spacing}

% ********************************** Appendices ********************************

\begin{appendices} % Using appendices environment for more functunality
\crefalias{section}{appendix}
%\include{appendix1}
%!TEX root = thesis.tex
% ******************************* Thesis Appendix B ********************************

\chapter{Useful relations in GHD}
\label{app:dr}
In this appendix, we summarize mathematical relations in the context of GHD.

\section{Properties of the dressing}
We define the dressing and the transposed dressing to a function $f(\rapidity)$ as solutions to the following equations
\begin{align}
	f\upd{dr}(\rapidity) &= f(\rapidity) + \int\tfrac{\dd{\rapidityp}}{2\pi}\varphi(\rapidity-\rapidityp )n(\rapidityp)f\upd{dr}(\rapidityp)\label{equ:app_dr_dr_eq},\\
	f\upd{drT}(\rapidity) &= f(\rapidity) + n(\rapidity) \int\tfrac{\dd{\rapidityp}}{2\pi}\varphi(\rapidity-\rapidityp )f\upd{drT}(\rapidityp).\label{equ:app_dr_drT_eq}
\end{align}
Here $\varphi(\rapidity)$ is the phase shift of the model (we assume $\varphi(-\rapidity) = \varphi(\rapidity)$) and $n(\rapidity)$ is the occupation function describing the state. This can be compactly written using the integral operator $\vu{T}$ with kernel $T(\rapidity,\rapidityp) = \tfrac{1}{2\pi}\varphi(\rapidity-\rapidityp)$:
\begin{align}
	f\upd{dr} &= f + \vu{T}nf\upd{dr}, & f\upd{drT} &= f + n\vu{T}f\upd{drT}.\label{equ:app_dr_eq_T}
\end{align}
Here we associate $n(\rapidity)$ with its multiplication operator $n[f](\rapidity) = n(\rapidity)f(\rapidity)$. In this notation we can write the solution to the dressing formally as
\begin{align}
	f\upd{dr} &= (\vb{1}-\vu{T}n)^{-1} f, & f\upd{drT} &= (\vb{1}-n\vu{T})^{-1} f.\label{equ:app_dr_sol}
\end{align}
Here we can see that $\cdot\upd{dr} = (\vb{1}-\vu{T}n)^{-1}$ and $\cdot\upd{drT} = (\vb{1}-n\vu{T})^{-1}$ are indeed transposes of each other (in the $L^2$ sense). In particular, we have
\begin{align}
	\int\dd{\rapidity}f(\rapidity)g\upd{dr}(\rapidity) &= \int\dd{\rapidity}f\upd{drT}(\rapidity)g(\rapidity).\label{equ:app_dr_dr_drT}
\end{align}

Furthermore, they satisfy the following useful identities:
\begin{align}
	n(\vb{1}-\vu{T}n)^{-1} &= (\vb{1}-n\vu{T})^{-1}n, &  (\vb{1}-\vu{T}n)^{-1}\vu{T} &= \vu{T}(\vb{1}-n\vu{T})^{-1},\label{equ:app_dr_multiplication_identities}
\end{align}
which are equivalent to
\begin{align}
	nf\upd{dr} &= \qty[nf]\upd{drT}, & [\vu{T}f]\upd{dr} &= \vu{T}f\upd{drT}\label{equ:app_dr_multiplication_identities_f}.
\end{align}
Identities like \eqref{equ:app_dr_multiplication_identities} follow from
\begin{align}
	(\vb{1}-n\vu{T})n(\vb{1}-\vu{T}n)^{-1} = n(\vb{1}-\vu{T}n)^{-1} - n\vu{T}n(\vb{1}-\vu{T}n)^{-1} = n (\vb{1}-\vu{T}n)(\vb{1}-\vu{T}n)^{-1} = n.\label{equ:app_dr_multiplication_identities_deriv}
\end{align}

We will often need to take derivatives of dressed quantities. Note that for any matrix we have $0 = \delta (\vb{A}^{-1}\vb{A}) = \delta \vb{A}^{-1}\vb{A} + \vb{A}^{-1}\delta\vb{A}$ and thus $\delta \vb{A}^{-1} = -\vb{A}^{-1}\delta\vb{A}\vb{A}^{-1}$. Applying this to the dressing we have
\begin{align}
	\delta (\vb{1}-\vu{T}n)^{-1} &= (\vb{1}-\vu{T}n)^{-1}\vu{T}\delta n(\vb{1}-\vu{T}n)^{-1}, & \delta (\vb{1}-n\vu{T})^{-1} &= (\vb{1}-n\vu{T})^{-1}\delta n\vu{T}(\vb{1}-n\vu{T})^{-1}.\label{equ:app_dr_delta}
\end{align}
This means that, for instance, if we take the dressing of a position dependent $f(x,\rapidity)$ we have
\begin{align}
	\partial_x f\upd{dr} &= \qty[\vu{T}\partial_xn f\upd{dr} + \partial_x f]\upd{dr}, & \partial_x f\upd{drT} &= \qty[\partial_xn \vu{T} f\upd{drT} + \partial_x f]\upd{drT}.\label{equ:app_dr_delx}
\end{align}

\section{Relations of densities and currents}
Given an $n(\rapidity)$ we define the quasi-particle density and the current as
\begin{align}
	\rho(\rapidity) &= \frac{1}{2\pi}n(\rapidity)1\upd{dr}(\rapidity), & j(\rapidity) &= \frac{1}{2\pi}n(\rapidity)v\upd{dr}(\rapidity).\label{equ:app_dr_rho_j_def}
\end{align}
From this we have
\begin{align}
	2\pi\vu{T}\rho &= \vu{T}n1\upd{dr} = 1\upd{dr} - 1, & 2\pi\vu{T}j &= v\upd{dr} - v.\label{equ:app_dr_T_rho_j}
\end{align}

If we perturb $n \to n+\delta n$ these quantities change as follows
\begin{align}
	2\pi \delta \rho &= \delta n 1\upd{dr} + n(\vb{1}-\vu{T}n)^{-1} \vu{T} \delta n1\upd{dr} = \qty[\delta n 1\upd{dr}]\upd{drT}, & 2\pi \delta j &= \qty[\delta n v\upd{dr}]\upd{drT}.\label{equ:app_dr_rho_j_delta}
\end{align}

In particular, we can write
\begin{align}
	\delta n &= \frac{2\pi}{1\upd{dr}} (\vb{1}-n\vu{T}) \delta \rho.\label{equ:app_dr_deltan_deltarho}
\end{align}

\section{Contracted coordinates}
We define the map to contracted coordinates as
\begin{align}
	\hat{X}(x,\rapidity) &= x + 2\pi \vu{T} \heightfield(x,\rapidity),\label{equ:app_dr_hatX_def}
\end{align}
where $\heightfield(x,\rapidity) = \int_{-\infty}^x\dd{y}\rho(y,\rapidity)$. Its derivative is given by
\begin{align}
	\partial_x \hat{X}(x,\rapidity) = 1 + 2\pi \vu{T} \rho(x,\rapidity) = 1\upd{dr}(x,\rapidity).\label{equ:app_dr_hatX_delx}
\end{align}
The density in the new coordinates satisfies $\hat{\rho}\dd{\hat{x}} = \rho\dd{x}$, i.e.
\begin{align}
	\rho(x,\rapidity) &= \hat{\rho}(\hat{X}(x,\rapidity),\rapidity) 1\upd{dr}(x,\rapidity).\label{equ:app_dr_rho_hatrho}
\end{align}
Comparing with \eqref{equ:app_dr_rho_j_def} we identify
\begin{align}
	n(x,\rapidity) &= 2\pi \hat{\rho}(\hat{X}(x,\rapidity),\rapidity).\label{equ:app_dr_n_hatrho}
\end{align}
From this we find
\begin{align}
	\delta n(x,\rapidity) &= 2\pi \delta \hat{\rho}(\hat{X}(x,\rapidity),\rapidity) + (2\pi)^2 \partial_{\hat{x}}\hat{\rho}(\hat{X}(x,\rapidity),\rapidity) \vu{T}\delta \Phi(x,\rapidity)\label{equ:app_dr_deltan_deltaPhi},\\
	\partial_x n(x,\rapidity) &= 2\pi 1\upd{dr}(x,\rapidity) \partial_{\hat{x}}\hat{\rho}(\hat{X}(x,\rapidity),\rapidity).\label{equ:app_dr_delxn_delxhatrho}
\end{align}

\section{Expansion of current up to second order}\label{app:dr_hessj}
Perturbing \eqref{equ:app_dr_rho_j_def} with two $\delta \rho$, denoted $\delta_1\rho$ and $\delta_2\rho$ we find
\begin{align}
	2\pi \delta_1\delta_2 j &= (\vb{1}-n\vu{T})^{-1}\delta_2 n \vu{T}(\vb{1}-n\vu{T})^{-1}\delta_1 n v\upd{dr} + (\vb{1}-n\vu{T})^{-1}\delta_1 n \delta_2 v\upd{dr} + (\vb{1}-n\vu{T})^{-1}\delta_1\delta_2 n v\upd{dr}\\
	&= (\vb{1}-n\vu{T})^{-1}\qty[\delta_2 n \delta_1 v\upd{dr} + \delta_1 n \delta_2 v\upd{dr}] + (\vb{1}-n\vu{T})^{-1}\delta_1\delta_2 n v\upd{dr}.\label{equ:app_dr_hessj_deriv1}
\end{align}
Now observe
\begin{align}
	\delta_1\delta_2 n &= \delta_2\qty(\frac{2\pi}{1\upd{dr}} (\vb{1}-n\vu{T}) \delta_1 \rho) = -\frac{2\pi}{1\upd{dr}} \delta_2 n\vu{T} \delta_1 \rho  -\frac{2\pi}{{1\upd{dr}}^2} \delta_21\upd{dr} (\vb{1}-n\vu{T}) \delta_1 \rho\\
	&=-\frac{1}{1\upd{dr}} \qty[\delta_2 n\delta_1 1\upd{dr}  +\delta_1 n\delta_2 1\upd{dr}].
\end{align}
Inserting this into \eqref{equ:app_dr_hessj_deriv1} gives
\begin{align}
		2\pi(\vb{1}-n\vb{T})^{-1}\delta_1\delta_2 j &= \delta_1n \delta_2v\upd{dr} +  \delta_2n \delta_1v\upd{dr} - v\upd{eff} \qty[\delta_1n \delta_21\upd{dr} +  \delta_2n \delta_11\upd{dr}].
\end{align}
Therefore, if we can apply the approximation \eqref{equ:diff_diff_LD_expansion}, we find that averages of currents are given by:
\begin{align}
	\expval{j(\rapidity)} &=  \tfrac{1}{2\pi}v\upd{dr}n + \tfrac{1}{2\pi}(\vb{1}-n\vb{T})^{-1}\qty[\expval{\delta n \delta v\upd{dr}}- v\upd{eff}\expval{\delta n\delta 1\upd{dr}}] + \order{1/L^2}.
\end{align}

\chapter{Large deviation theory}
\label{app:LD}
We do not need many results of large deviation theory (see for instance~\cite{TOUCHETTE20091}), but we still would like to discuss the basics of it; it helps to gain intuition into the meaning of expressions like \eqref{equ:pre_int_TBA_pathintegral} and \eqref{equ:scbm_thermo_Z_final}.

Large deviation theory is concerned with the limit $L\to \infty$ of probability distributions of the form\footnote{There is a precise meaning of $\sim$ (often denoted as $\asymp$), but for this brief review it is not important.} $p(x) \sim e^{-L\LD{I}(x)}$. Here $x$ is a random variable that could be anything: $x \in \mathbb{R}$ could be a number, $x\in\mathbb{R}^n$ could be a vector or $x$ could be a function $\rho(\cdot,\cdot)$ (which we need for expressions like \eqref{equ:pre_int_TBA_pathintegral}).

The idea is that, as $L \to \infty$, the probability will be dominated by the minimum $x_0$ of $\LD{I}(x)$. Formally, this is justified by the Laplace approximation (see appendix \ref{app:SP}), i.e.
\begin{align}
	\mathbb{E}[f(x)] &= \frac{\int\dd{x}f(x)e^{-L\LD{I}(x)}}{\int\dd{x} e^{-L\LD{I}(x)}} \to f(x_0).
\end{align}

Hence, any $\order{1}$ observable can easily be computed as $L \to \infty$. The purpose of large deviation theory is usually to compute the expectation value of observables that are extremely large in regions with extremely low probability. The classic example are observables of the type $e^{L\lambda x}$. The Laplace approximation gives for these expectation values
\begin{align}
	\mathbb{E}[e^{L\lambda x}] &\sim \int\dd{x}e^{-L(\LD{I}(x)-\lambda x)} \sim e^{L\LD{F}(\lambda)},
\end{align}
where the scaled cumulant generating function $\LD{F}(\lambda)$ is given by the Legendre transform
\begin{align}
	\LD{F}(\lambda) = \lim_{L\to \infty}\tfrac{1}{L} \log \mathbb{E}[e^{L\lambda x}] = \max_{x} (\lambda x - \LD{I}(x)).
\end{align}
Note that since Legendre transforms are (under some assumptions) invertible (by again a Legendre transform), knowing either $\LD{F}(\lambda)$ or $\LD{I}(x)$ is sufficient to determine the other (this is often used in large deviation theory).

Since the scaled cumulant generating function is $\order{1}$, the cumulant generating function $\log \mathbb{E}[e^{\lambda x}]$ is $\order{L}$, i.e.
\begin{align}
	\log \mathbb{E}[e^{\lambda x}] = \sum_{n=0}^\infty \kappa_n[x] \tfrac{\lambda^n}{n!} \sim L \LD{F}(\lambda/L).
\end{align}
Here $\kappa_n[x]$ are the cumulants of $x$. Using $\LD{F}(0) = 0$ and expanding we find the large deviation scalings:
\begin{align}
	\kappa_1 = \mathbb{E}[x] &\to \LD{F}'(0)\\
	\kappa_2 = \mathrm{Var}[x] &\to \LD{F}''(0)/L,\\
	\kappa_n &\to \LD{F}^{(n)}(0)/L^{n-1}.
\end{align}
Thus, as $L\to \infty$, the random variable $x$ has $\kappa_{n\geq 2} \to 0$, which means that $x$ becomes deterministic. Alternatively, if we take the leading order correction into account, then $\kappa_{n\geq 3} \to 0$, implying that $x$ is a Gaussian variable with variance $\order{1/L}$.

\begin{remark}
	Large deviation scaling often appears in systems that are self-averaging (like for instance thermodynamics or hydrodynamics). Therefore, it is a common tool to understand fluctuations in such self-averaging systems.   
\end{remark}

\begin{remark}
	An example from physics: in thermodynamics, where $p(E) \sim e^{-\beta E}$, one would typically identify $x = E/L$ as the energy density, $\lambda=\beta$ as the inverse temperature, $\LD{I}(x)=-\LD{S}(x)$ with the entropy and $\LD{F}(\lambda) = -f(\lambda)$ as the free energy.
\end{remark}

\begin{remark}
	Note that in order to describe the system including the leading order $1/L$ correction, in addition to $\mathrm{Var}[x] = \LD{F}''(0)/L$ (which is provided by large deviation theory), one also needs to infer the correction to $\mathbb{E}[x] = \LD{F}'(0) + \order{1/L}$. This correction is not provided by large deviation theory. This is  why, for instance, BMFT is not a priori able to capture the diffusive $1/L$ correction to hydrodynamics. 
\end{remark}

\chapter{The stationary phase approximation}
\label{app:SP}
The stationary phase approximation (see for instance ~\cite[Chap 8]{bleistein1986asymptotic})
\begin{align}
	\int\dd[N]{x} f(\vec{x})e^{iLg(\vec{x})} \sim \sqrt{\frac{(2\pi i)^N}{L^N\det \vb{H}_g(\vec{x}_0)}} e^{iLg(\vec{x}_0)} \qty(f(\vec{x}_0) + \order{1/L}),\label{equ:app_SP_basic}
\end{align}
where $\vec{x}_0$ is a stationary point of $g(\vec{x})$ and $\vb{H}_{g,ij} = \partial_{x_i}\partial_{x_j} g(\vec{x}_0)$ is the Hessian, is closely related to the Laplace approximation
\begin{align}
	\int\dd[N]{x} f(\vec{x})e^{-Lg(\vec{x})} \sim \sqrt{\frac{(2\pi)^N}{L^N\det \vb{H}_g(\vec{x}_0)}} e^{-Lg(\vec{x}_0)} \qty(f(\vec{x}_0) + \order{1/L}).\label{equ:app_SP_laplace}
\end{align}
In both approximations it is essential that $g(\vec{x})$ is a real function\footnote{There exists a version for complex $g(\vec{x})$, the saddle-point approximation or method of steepest descent. However, they are only applicable if $f$ and $g$ are holomorphic.}. In \eqref{equ:app_SP_basic} $\vec{x}_0$ can be the location of a local maximum or a minimum of $g(\vec{x})$\footnote{If there are multiple stationary points, \eqref{equ:app_SP_basic} has to be summed over all $\vec{x}_0$}; in \eqref{equ:app_SP_laplace} only the global minimizer $\vec{x}_0$ is important.

We will only discuss the stationary-phase approximation, the Laplace approximation can be treated similarly. Formally, the stationary phase approximation has some conditions (for instance that $g(\vec{x})$ is smooth), which likely do not apply to the derivation in \cref{sec:LL}. The purpose of this appendix is therefore to gain intuition into when this approximation is actually valid.

The idea behind the stationary phase approximation is that a) fast oscillating phases average to zero and hence points of stationary phase should dominate the integral and b) that around a stationary point the phase can be approximated by a Gaussian. This can be seen best under the rescaling $\vec{x} = \vec{x}_0 + \vec{x}/\sqrt{L}$:
\begin{align}
	\int\dd[N]{x}f(\vec{x})e^{iLg(\vec{x})} &= \int\tfrac{\dd[N]{y}}{\sqrt{L}^N}f(\vec{x}_0+\vec{y}/\sqrt{L})e^{iLg(\vec{x}_0+\vec{y}/\sqrt{L})}.
\end{align}
Now taking the limit $L\to\infty$ we find
\begin{align}
	\int\dd[N]{x}f(\vec{x})e^{iLg(\vec{x})} &= \int\tfrac{\dd[N]{y}}{\sqrt{L}^N}f(\vec{x}_0)e^{iLg(\vec{x}_0)+\tfrac{1}{2}i\vec{y}^T\vb{H}_g(\vec{x}_0)\vec{y}} + \order{1/L}.\label{equ:app_SP_deriv_SP_gauss}
\end{align}
This is now a Gaussian integral over $y$, which gives rise to the determinant in \eqref{equ:app_SP_basic}\footnote{Note that the oscillating integral in \eqref{equ:app_SP_deriv_SP_gauss} is only defined in a Fresnel sense, which requires that $f(\vec{x})$ decays for $\abs{\vec{x}}\to \infty$.}. What we learn from \eqref{equ:app_SP_deriv_SP_gauss} is that the stationary phase approximation is effectively an integral over a region $\vec{x}=\vec{x}_0 + \order{1/\sqrt{L\vb{H}_g(\vec{x}_0)}}$. In this region, the stationary phase approximation has two main assumptions
\begin{enumerate}
	\item For $\vec{x}=\vec{x}_0 + \order{1/\sqrt{L\vb{H}_g(\vec{x}_0)}}$ the function $f(\vec{x})$ is constant.
	\item For $\vec{x}=\vec{x}_0 + \order{1/\sqrt{L\vb{H}_g(\vec{x}_0)}}$ the function $g(\vec{x})$ is well approximated by its second order Taylor polynomial.
\end{enumerate}

Only if this is true the stationary phase approximation is justified.

Realizing this is quite important because depending on the scaling these approximations have to be applied for different objects. For instance, in \cref{sec:scbm_thermo} we will apply the large deviation theory (which is based on the Laplace approximation). What we find there is that the rate function (which is the equivalent of the fast oscillating phase $g(x)$ here) is of order $N$. This means that one cannot take the saddle point over $x_i$: for each $x_i$ we have $\partial_{x_i}^2g(x) = \order{1}$ thus the integral runs over a range $x_i + \order{1}$. Over this range $g(x)$ is clearly not be well approximated by its quadratic Taylor polynomial. That is the reason why we perform the Laplace approximation instead over a density, i.e. an expression like $\rho(x,\rapidity)=\tfrac{1}{L}\sum_i \delta(x-x_i)\delta(\rapidity-\rapidity_i)$. In the neighborhood of $\rho(x,\rapidity)+\order{1/\sqrt{L}}$ the rate function is well approximated by its quadratic Taylor polynomial, hence the approximation is justified.

In the stationary phase approximation of \eqref{equ:LL_SP_psi_macroscopic} the phase is of order $NL \sim L^2$. This means that if we would try to do an integral over a density again, we would need to perform a path integral over densities in a neighborhood $\rho(x,\rapidity)+\order{1/L}$. But a $1/L$ change to a particle density is a very small change. It can be obtained, for instance, by adding or removing a single particle or by moving a particle by an $\order{1}$ distance. Hence, if the phase is of order $L^2$, we need to take into account the microscopic structure of the model. We believe therefore that the integral over $\rho(x,\rapidity)$ can not be performed as a path integral, but instead has to be an integral over microscopic configurations.

\begin{remark}
	Note that this will not change the fact that the stationary point dominates the integral, but it will change the prefactor in \eqref{equ:app_SP_basic} from a determinant to something else. This prefactor is often not important and hence not considered. In \cref{sec:LL}, however, the prefactor is important because the saddle point does not fully fix all degrees of freedom in the system. After the saddle point approximation is taken, we are still left with another large scale integral, for which some parts of the prefactor might still be important.
\end{remark}

\begin{remark}
	By expanding \eqref{equ:app_SP_deriv_SP_gauss} to higher order in $1/\sqrt{L}$, one can systematically compute higher order corrections to \eqref{equ:app_SP_basic}.
\end{remark}

\section{Alternative strategy for the stationary phase approximation}
\label{app:SP_alter}
In sections \ref{sec:LL_deriv_1} and \ref{sec:LL_deriv_2} we use two slightly different mathematical strategies to perform the large scale analysis. One of them is the stationary phase approximation and the other is similar to the example we discuss now. Consider the integral
\begin{align}
	\int\dd{x}\dd{\rapidity} f(x)g(\rapidity) e^{iLxp}.
\end{align}
One way to compute it, is to do the stationary phase approximation, which gives
\begin{align}
	\int\dd{x}\dd{\rapidity} f(x)g(\rapidity) e^{iLx\rapidity} \to \frac{2\pi}{L}f(0)g(0).
\end{align}
Alternatively, it can be computed as follows. Denote the Fourier transform by $\tilde{f}(\rapidityp) = \int\dd{x}f(x)e^{i\rapidityp x}$ and observe
\begin{align}
	\int\dd{x}\dd{\rapidity} f(x)g(\rapidity) e^{iLx\rapidity} &= \int\dd{\rapidity} \tilde{f}(L\rapidity)g(\rapidity) =  \int\frac{\dd{\rapidity}}{L} \tilde{f}(\rapidity)g(\rapidity/L) = \int\frac{\dd{\rapidity}}{L} \tilde{f}(\rapidity)g(0) + \order{1/L^2}\\
	&= \frac{2\pi}{L} f(0)g(0) + \order{1/L^2}.\label{equ:app_SP_alter_formula}
\end{align}
Both ways lead to the same result. However, the approximations are fairly different. In the stationary phase approximation \eqref{equ:app_SP_basic}, we assume that $f$ and $g$ are almost constant over $x, p \sim 1/\sqrt{L}$, while in \eqref{equ:app_SP_alter_formula} we only assume that $g$ is almost constant over $p \sim 1/L$.

How are both results connected? To understand this, let us compute an explicit example $f(x) = \theta(\abs{x} < \Delta x)$ and $g(\rapidity) = \theta(\abs{\rapidity} < \Delta \rapidity)$. We find
\begin{align}
	\int\dd{x}\dd{\rapidity} f(x)g(\rapidity) e^{iLx\rapidity} &= 2\int_{-\Delta\rapidity}^{\Delta\rapidity}\dd{\rapidity} \frac{\sin(L \Delta x\rapidity)}{L\Delta x} = 2\int_{-L\Delta x\Delta\rapidity}^{L\Delta x\Delta\rapidity}\frac{\dd{z}}{L} \frac{\sin(z)}{z}.
\end{align}
This only gives the result of the above approximations if $L\Delta x\Delta\rapidity \gg 1$, implying $\Delta x\Delta\rapidity \gg 1/L$. Hence, we do not necessarily need that $\Delta x,\Delta \rapidity \gg 1/\sqrt{L}$, for instance $\Delta x \gg 1$ and $\Delta \rapidity \sim 1/L$ is also admissible.

This rescaling invariance is a special feature of the phase $g(x,\rapidity) = x\rapidity$. Note that if we rescale $x\to \alpha x$ and $\rapidity \to \rapidity/\alpha$, the phase is invariant.

\chapter{The push-forward of measures}
\label{app:push}
The push-forward of a measure (called image measure in~\cite{bogachev2007measure}) is an important concept in measure theory, specifically probability theory. In probability theory the push-forward $(f_*p)(y)$ of a probability distribution $p(x)$ by $y=f(x)$ gives the probability distribution of the random variable $y$.

More generally its definition is as follows: consider a measure $\rho(x)$ on a set $x\in \set{M}$. One way to think about a measure is a map mapping observables (or test functions) $\testfunction(x)$ onto numbers, in the sense of
\begin{align}
	\testfunction \mapsto \expval{\rho,\testfunction} := \int_{\set{M}}\dd{x}\rho(x)\testfunction(x).\label{equ:app_push_def_measure}
\end{align} 

Imagine we have a map $y=f(x)$ mapping $x \in\set{M}$ to another set $y\in\set{N}$. The push-forward $f_*\rho$ defines a measure on $\set{N}$ in a natural way as follows
\begin{align}
	\testfunction \mapsto \expval{f_*\rho,\testfunction} = \int_{\set{N}}\dd{y}(f_*\rho)(y)\testfunction(y) :=  \int_{\set{M}}\dd{x}\rho(x)\testfunction(f(x)) = \expval{\rho,\testfunction(f(\cdot))}.\label{equ:app_push_def_push}
\end{align}

The idea is that $f$ moves point a $x\in \set{M}$ to $y=f(x)$ before measuring it with $\testfunction(y)$. If $\rho(x)$ is a continuous function and $f$ is invertible, then $(f_*\rho) \dd{y} = \rho\dd{x}$, i.e.
\begin{align}
	(f_*\rho)(y) = \frac{\rho(f^{-1}(y))}{\abs{f'(f^{-1}(y))}}.\label{equ:app_push_local_formula}
\end{align}
This can be derived from \eqref{equ:app_push_def_push} by doing a change of variables $y=f(x)$ and using the fact that \eqref{equ:app_push_def_push} has to hold any $\testfunction(y)$. Eq. \eqref{equ:app_push_local_formula} might be more explicit, however (in the opinion of the author) it is also considerably harder to handle practically than the implicit version \eqref{equ:app_push_def_push}. Hence, we will use \eqref{equ:app_push_def_push} throughout the thesis. 

\begin{remark}\label{rem:app_push_multivalued}
	If $f$ is not invertible, then \eqref{equ:app_push_local_formula} becomes a sum over all $x\in f^{-1}(y)$, i.e.
	\begin{align}
		(f_*\rho)(y) = \sum_{x\in f^{-1}(y)}\frac{\rho(x)}{\abs{f'(x)}}.
	\end{align}
	Note that this way the conservation of total measure  $\int\dd{y}(f_*\rho)(y) = \int\dd{x}\rho(x)$ is ensured.
\end{remark}

\chapter{Details on convexity of the Bethe phase}
\label{app:convexity_of_bethe_phase}
In \cref{sec:LL_SP} we do a stationary phase approximation (in $\rapidity$) on
\begin{align}
	S_t(\vec{x},\vec{\rapidity}) &= \sum_i \rapidity_i (x_i-\hat{x}_i^0) +\tfrac{1}{4L}\sum_{i\neq j}\sgn(x_i-x_j) \phi(\rapidity_i-\rapidity_j)- \sum_i\rapidity_i^2 t,
\end{align}
for which it is important to show that $S_t$ only has a single stationary point. This follows from the fact that $S_t$ is concave in $\rapidity$, as we will show by investigating $\vb{H}_{ij} = \partial_{\rapidity_i}\partial_{\rapidity_j}S_t(\vec{x},\vec{\rapidity})$:
\begin{align}
	\vb{H}_{ij} &= -2t \delta_{ij} + \tfrac{1}{2L} \delta_{ij}\sum_{k\neq i}\sgn(x_i-x_k)\varphi'(\rapidity_i-\rapidity_k) - \tfrac{1}{2L}\sgn(x_i-x_j)\varphi'(\rapidity_i-\rapidity_j).
\end{align}
Hence we can write $\vb{H}_{ij} = -2\delta_{ij} + \sum_k \vb{A}_{ik} - \vb{A}_{ij}$, where $\vb{A}_{ij} = \tfrac{1}{2L}\sgn(x_i-x_j)\varphi'(\rapidity_i-\rapidity_j)$. Note that any single matrix entry of $\abs{\vb{A}_{ij}} \leq \tfrac{1}{2L}\sup_\rapidity\abs{\varphi'(\rapidity)}$ and thus $\vec{v}^T\vb{A}\vec{v} \leq \tfrac{N}{2L}\sup_\rapidity\abs{\varphi'(\rapidity)} \vec{v}^T\vec{v}$. Using this we find
\begin{align}
	\vec{v}^T\vb{H}\vec{v} \leq \qty[-2t + \tfrac{N}{L}\sup_\rapidity\abs{\varphi'(\rapidity)}] \vec{v}^T\vec{v}, 
\end{align}
implying that $\vb{H}$ is negative definite for $t > t\ind{c} = \tfrac{N}{2L}\sup_\rapidity\abs{\varphi'(\rapidity)}$, hence $S_t$ is concave.

\chapter{Details on the scaling of the error of space contraction}
\label{app:diff_err_scaling}
In this appendix we want to analyze the scaling of expression \eqref{equ:diff_HR_contract_diff}. We already know the scaling of $\avg{y}_{\alpha,\beta}$ and $\avg{\rapidityp}_{\alpha,\beta}$, so let us concentrate on the other term which abbreviate as $A$
\begin{align}
	A := \tfrac{d}{2L^2} \sum_\alpha A_\alpha = \tfrac{d}{2L^2} \sum_{\alpha} \sum_\beta \partial_{\hat{x}}\hat{\testfunction}(\hat{X}_\alpha,\rapidity_\beta) \qty[\sum_{i\in \set{C}_{\alpha,\beta}}\sum_{j\in \set{A}_\alpha}\delta_{i\neq j}\sgn(y_i-y_j)].
\end{align}
First, note that due to the antisymmetry of $\sgn(y_i-y_j)$, we can write this as
\begin{align}
	A_\alpha &= \tfrac{d}{2}\sum_{\beta\neq \beta'}  (\partial_{\hat{x}}\hat{\testfunction}(\hat{X}_\alpha,\rapidity_\beta) - \partial_{\hat{x}}\hat{\testfunction}(\hat{X}_\alpha,\rapidity_\beta')) \sum_{i\in \set{C}_{\alpha,\beta}}\sum_{j\in \set{C}_{\alpha,\beta'}}\sgn(y_i-y_j).
\end{align}
It is obvious from the antisymmetry that $\mathbb{E}[A_\alpha] = 0$. To compute its variance, note that \sloppy$\mathbb{E}[\sgn(y_i-y_j)\sgn(y_k-y_l)]$ is only non-zero if $i=k$ or $j=l$, in which case it is
\begin{align}
	\mathbb{E}[\sgn(y_1-y_2)\sgn(y_1-y_3)] &= 1 + 4\int_{-1/2}^{1/2}\dd{y_1}\dd{y_2}\dd{y_3}\theta(y_1-y_2)\theta(y_1-y_3)\\
	&= 1+4\int_{-1/2}^{1/2}\dd{y_1} (y_1+\tfrac{1}{2})^2 = \tfrac{7}{3} = \order{1}
\end{align}
or $\mathbb{E}[\sgn(y_1-y_2)\sgn(y_1-y_2)] = 1$ in case both $i=k,j=l$. From this we find that terms in
\begin{align}
	\mathrm{Var}[A_\alpha] &= \mathbb{E}[A_\alpha^2] =  \tfrac{d^2}{4}\sum_{\beta\neq \beta'} \sum_{\beta_2\neq \beta'_2} (\ldots),
\end{align}
where all $\beta$'s are different will vanish. If $\beta=\beta_2$, we find
\begin{multline}
	\mathrm{Var}[A_\alpha] \sim \sum_{\beta,\beta',\beta'_2} \delta_{\beta\neq \beta'}\delta_{\beta\neq \beta'_2}  (\partial_{\hat{x}}\hat{\testfunction}(\hat{X}_\alpha,\rapidity_\beta) - \partial_{\hat{x}}\hat{\testfunction}(\hat{X}_\alpha,\rapidity_\beta'))\\
	\times(\partial_{\hat{x}}\hat{\testfunction}(\hat{X}_\alpha,\rapidity_\beta) - \partial_{\hat{x}}\hat{\testfunction}(\hat{X}_\alpha,\rapidity_{\beta'_2})) n_{\alpha,\beta}n_{\alpha,\beta'}n_{\alpha,\beta'_2} + \ldots
\end{multline}
and since all of the other terms (for instance $\beta'=\beta'_2$) give similar contributions we conclude
\begin{align}
	\mathrm{Var}[A_\alpha] &\sim L^3\Delta x^3.
\end{align}
Here we used $n_{\alpha,\beta'} \sim L\Delta x\Delta \rapidity$ and each sum over $\beta$ has on the order of $1/\Delta \rapidity$ terms. Since all $A_\alpha$ are independent, we thus finally find
\begin{align}
	\mathrm{Var}[A] \sim \frac{1}{L^4}\sum_\alpha \mathrm{Var}[A_\alpha] \sim \frac{1}{L} \sum_\alpha \Delta x^3 \sim \frac{\Delta x^2}{L}. 
\end{align}
Again, we used that  the sum over $\alpha$ has on the order of $1/\Delta x$ terms. This result is of the same order as the one we had found for $\avg{y}_{\alpha,\beta}$ and $\avg{\rapidityp}_{\alpha,\beta}$.

\end{appendices}

% *************************************** Index ********************************
\printthesisindex % If index is present

\end{document}